\documentclass[twocolumn,showkeys,showpacs,preprintnumbers,prd,superscriptaddress,nofootinbib,aps,10pt]{revtex4-1}
\usepackage{graphicx,epsf,bm,amsmath,amsfonts,amssymb,epstopdf,natbib,verbatim,multirow,appendix,array,diagbox,xcolor,tikz}
\usepackage{hyperref}
\hypersetup{colorlinks=true,urlcolor=blue,citecolor=blue,linkcolor=blue,menucolor=blue,anchorcolor=blue,filecolor=blue}

\newcolumntype{?}{!{\vrule width 3pt}}

\date{\today}

\definecolor{darkgreen}{RGB}{0,100,0}
\definecolor{lime}{HTML}{A6CE39}
\DeclareRobustCommand{\orcidicon}{
\begin{tikzpicture}
\draw[lime, fill=lime] (0,0) 
circle [radius=0.16] 
node[white] {{\fontfamily{qag}\selectfont \tiny ID}};
\draw[white, fill=white] (-0.0625,0.095) 
circle [radius=0.007];
\end{tikzpicture}
\hspace{-3mm}
}

\foreach \x in {A, ..., Z}{\expandafter\xdef\csname orcid\x\endcsname{\noexpand\href{https://orcid.org/\csname orcidauthor\x\endcsname}{\noexpand\orcidicon}}}

\begin{document}

\title{Running into tension: primordial black holes from ultra-slow-roll inflation,\\spectral running, and the Hubble tension}

\author{Miguel A. Sabogal\hspace{-1mm}\orcidA{}}
\email{miguel.sabogalgarcia@unitn.it}
\affiliation{Department of Physics, University of Trento, Via Sommarive 14, 38123 Povo (TN), Italy}
\affiliation{Trento Institute for Fundamental Physics and Applications (TIFPA)-INFN, Via Sommarive 14, 38123 Povo (TN), Italy}

\author{Antonio J. Iovino\hspace{-1mm}\orcidB{}}
\email{a.iovino@nyu.edu}
\affiliation{Center for Astrophysics and Space Science, New York University Abu Dhabi, PO Box 129188, Abu Dhabi, United Arab Emirates \looseness=-1}

\author{Sunny Vagnozzi\hspace{-1mm}\orcidC{}}
\email{sunny.vagnozzi@unitn.it}
\affiliation{Department of Physics, University of Trento, Via Sommarive 14, 38123 Povo (TN), Italy}
\affiliation{Trento Institute for Fundamental Physics and Applications (TIFPA)-INFN, Via Sommarive 14, 38123 Povo (TN), Italy}

\begin{abstract}
\noindent Single-field ultra-slow-roll (USR) inflation is among the most studied mechanisms for primordial black hole (PBH) formation. When a non-stationary inflection point close to CMB scales is present, these models predict a negative spectral running ($\alpha_s<0$), whose magnitude increases with the PBH mass. This is in tension with recent hints for positive running from Atacama Cosmology Telescope (ACT) Cosmic Microwave Background (CMB) data. However, inflationary parameters inferred from CMB data are sensitive to the assumed pre-recombination expansion history, which is precisely where new physics motivated by the Hubble tension should operate. Focusing on axion-like early dark energy (EDE) as a benchmark, we investigate the effect of such pre-recombination new physics on $\alpha_s$, and hence on the viability of USR PBH models, in light of state-of-the-art CMB data from Planck, ACT, and the South Pole Telescope, together with Baryon Acoustic Oscillation data from DESI DR2. Our analysis therefore provides an updated set of constraints on $\alpha_s$ and the running of the running $\beta_s$. For most dataset combinations, moving from $\Lambda$CDM to EDE increases the inferred $\alpha_s$: once the acoustic angular scale $\theta_s$ is fixed, EDE increases the diffusion-to-acoustic angular scale ratio $\theta_d/\theta_s$, and the shift in $\alpha_s$ compensates this extra damping by increasing small-scale power. In this sense, tension calls for tension: taking the Hubble tension seriously as an indication for new physics strengthens the challenges faced by USR PBH models. More broadly, our analysis stresses that inflationary model selection using CMB-inferred inflationary parameters such as $n_s$ and $\alpha_s$ may be premature, especially until the Hubble tension, and more generally the pre-recombination expansion history, is understood.
\end{abstract}

\maketitle

\section{Introduction}
\label{sec:introduction}

Primordial black holes (PBHs)~\cite{Zeldovich:1967lct,Hawking:1974rv,Chapline:1975ojl,Carr:1975qj} have emerged as a remarkably well-motivated target, both from a theoretical and observational standpoint (see e.g.\ Refs.~\cite{Khlopov:2008qy,Carr:2023tpt,Choudhury:2024aji,Khlopov:2024nqp,Shankaranarayanan:2026hnn} for reviews). On the theoretical side, PBHs offer a compelling dark matter candidate, particularly in the asteroid-mass window $M_{\rm PBH}\sim 10^{-17}$--$10^{-12}\,M_{\odot}$, where they can constitute the entirety of the observed dark matter abundance~\cite{Belotsky:2014kca,Carr:2016drx,Green:2020jor,Carr:2026hot}. On the observational side, several of the compact binary mergers detected by the LIGO-Virgo-KAGRA (LVK) collaboration~\cite{LIGOScientific:2018mvr,LIGOScientific:2019kan,LIGOScientific:2020ibl,KAGRA:2021vkt,LIGOScientific:2025slb} feature sub-solar or low-mass components~\cite{LIGOScientific:2021job,LVK:2022ydq,Prunier:2023uoo,Morras:2023jvb,Yuan:2024yyo} whose astrophysical origin remains uncertain, and for which PBHs are a natural candidate~\cite{Bird:2016dcv,Kavanagh:2018ggo,Wong:2020yig,Hutsi:2020sol,DeLuca:2021wjr,Franciolini:2022tfm,Miller:2024fpo,Andres-Carcasona:2024wqk,DeLuca:2025fln,Andres-Carcasona:2026avd}. In addition, a plausible interpretation for the nHz gravitational wave (GW) signal recently detected by pulsar timing array (PTA) experiments~\cite{NANOGrav:2023gor,EPTA:2023fyk,Reardon:2023gzh,Xu:2023wog} is that of a scalar-induced GW (SIGW) background sourced by the same enhanced curvature perturbations which are responsible for PBH formation~\cite{Vaskonen:2020lbd,Chen:2019xse,DeLuca:2020agl,Bhaumik:2020dor,Inomata:2020xad,Kohri:2020qqd,Domenech:2020ers,Vagnozzi:2020gtf,Namba:2020kij,Sugiyama:2020roc,Zhou:2020kkf,Lin:2021vwc,Rezazadeh:2021clf,Kawasaki:2021ycf,Ahmed:2021ucx,Yi:2022ymw,Yi:2022anu,Dandoy:2023jot,Zhao:2023xnh,Ferrante:2023bgz,Cai:2023uhc,Vagnozzi:2023lwo,Franciolini:2023pbf,Wang:2024vfv,Iovino:2024tyg,Yogesh:2025hll,Gouttenoire:2025jxe,Mohammadi:2025avz}. In most of the standard formation scenarios, PBHs form from the gravitational collapse of rare overdense regions seeded by large amplitude peaks in the (dimensionless) primordial curvature power spectrum, which must be enhanced by several orders of magnitude above the nearly scale-invariant Cosmic Microwave Background (CMB) value, $\mathcal{P}_\zeta(k_{\rm CMB})\sim 10^{-9}$, up to $\mathcal{P}_\zeta(k_{\rm pk})\sim 10^{-2}$ at the relevant small scales. Among the simplest and most studied single-field inflationary mechanisms capable of generating such an enhancement is a transient phase of ultra-slow-roll (USR) inflation, during which the inflaton traverses a flat feature in its potential, typically an inflection point, causing a dramatic amplification 
of curvature perturbations~\cite{Ivanov:1994pa,Kinney:1997ne,Garcia-Bellido:2017mdw,Germani:2017bcs,Kannike:2017bxn,Di:2017ndc,Ballesteros:2017fsr,Cicoli:2018asa,Ballesteros:2020qam,Ragavendra:2020sop,Karam:2022nym,Ballesteros:2024zdp,Allegrini:2024ooy,Briaud:2025hra,Totolou:2025dsb}.~\footnote{The level of fine-tuning required to achieve the correct PBH abundance in these models has been recently discussed in Refs.~\cite{Cole:2023wyx,Iovino:2025tcv,Profumo:2026qpn}.}

A key feature of USR models for PBH formation is that the dynamics which generate the enhanced peak in the curvature power spectrum cannot be decoupled from those which control the CMB-scale part of the inflationary trajectory. Such a coupling has important consequences. USR inflation is generically divided into an initial slow-roll (SR) phase which lasts for $N_1$ $e$-folds, during which the CMB modes exit the horizon and the large-scale power spectrum is imprinted, followed by the USR phase which ultimately seeds PBH formation. The position of the resulting enhanced peak is set by $N_1$: the shorter the initial SR phase, the closer the USR feature lies to the CMB window, and the heavier the PBHs produced. The fact that the PBH mass controls the location of the USR feature relative to the CMB scales leads in turn to a mass-dependent imprint on the inflationary observables constrained by CMB experiments: the scalar spectral index $n_s$, the tensor-to-scalar ratio $r$, and the spectral running $\alpha_s$. The closer the feature, the stronger the impact~\cite{Kannike:2017bxn,Ballesteros:2020qam,Allegrini:2025jha}. It is therefore CMB-scale observables, and not just the small-scale power spectrum amplitude, that carry direct information about these models. As a result, precision CMB measurements can play a key role when testing the viability of USR models for PBH formation.

A central prediction of these scenarios is that of negative running ($\alpha_s<0$)\,\cite{Allegrini:2025jha}. This follows from the necessity of an initial SR phase of reduced duration, followed by a successful PBH-producing USR phase: together, these force the inflaton to cross a non-stationary inflection point close to the CMB scales and push the primordial power spectrum away from scale-invariance~\cite{Allegrini:2025jha}. This expectation, however, is in tension with recent primordial power spectrum constraints from the Atacama Cosmology Telescope (\textit{ACT}) CMB data. The latest \textit{ACT} DR6 measurements\,\cite{AtacamaCosmologyTelescope:2025nti}, once combined with large-to-intermediate scale CMB data from the \textit{Planck} satellite, shift $n_s$ towards larger values, closer to the scale-invariant value of $n_s=1$, which has itself led to a renewed interest in inflationary model-building activity (see e.g.\ Refs.~\cite{Kallosh:2025rni,Aoki:2025wld,Gialamas:2025kef,Dioguardi:2025vci,Kim:2025dyi,Gao:2025onc,Drees:2025ngb,Zharov:2025zjg,Yin:2025rrs,Gialamas:2025ofz,Haque:2025uis,Yogesh:2025wak,Addazi:2025qra,Peng:2025bws,Pallis:2025nrv,Odintsov:2025wai,McDonald:2025tfp,Choudhury:2025vso,Ahmed:2025rrg,Pal:2025ewf,Mohammadi:2025gbu,Ahmed:2025sfm,Ketov:2025cqg,Odintsov:2025bmp,Das:2025bws,Zhu:2025twm,Yuennan:2025kde,Odintsov:2025jky,Oikonomou:2025htz,Oikonomou:2025xms,Modak:2025bjv,Odintsov:2025eiv,Aoki:2025ywt,Pallis:2025vxo,Qiu:2025iqm,Ahmed:2025eip,DOnofrio:2025bol,Fu:2025ciy,Singh:2025uyr,Bostan:2025jkt,Serish:2025ian,Yuennan:2025mlg,Choudhury:2025hnu,Addazi:2025agg,Oikonomou:2026gae,Nojiri:2026hij,Oikonomou:2026vrw,Odintsov:2026doe,Santos:2026ojg,Yuennan:2026fcn,Oikonomou:2026mvp,Dimopoulos:2026iwq,Khan:2026nsz,Yogesh:2026esn,Odintsov:2026cxz,Khan:2026pba,Khan:2026doo,Odintsov:2026dss,Oikonomou:2026qkj,Gonuguntla:2026rkw,Yang:2026rzn,Qiu:2026npi,Khan:2026bmf,Ijaz:2026ear,Khan:2026php,Mohammadi:2026iod,Nojiri:2026ish,Kallosh:2026tke,DelGrosso:2026zbg,Cembranos:2026cat,DeAngelis:2026mwx,Iacconi:2026twl,Chudaykin:2026amr,Oikonomou:2026jdg,Seidabadi:2026gnc,Odintsov:2026veu,Kallosh:2026kfx,Kulchoakrungsun:2026pyz}). The same combination, with the further addition of CMB lensing and Baryon Acoustic Oscillations (BAO) data, also leads to $\alpha_s=0.0062 \pm 0.0052$~\cite{AtacamaCosmologyTelescope:2025nti}. This mild preference for positive running points precisely in the opposite direction to that required in USR models for PBH formation: as pointed out in Ref.~\cite{Allegrini:2025jha}, the resulting tension for these models is progressively more pronounced for heavier PBHs (given that the relevant feature lies closer to the CMB scales), especially for those in the solar mass range whose mergers are relevant for ground-based GW experiments.

The previous conclusions rest upon an essential caveat. Constraints on inflationary parameters are often treated as direct, model-independent measurements of the primordial power spectrum, and therefore ultimately on the parameters of the inflationary Lagrangian. In truth, they are not. As is the case with any cosmological parameter, the inferred inflationary parameters depend on the assumed cosmological model (typically the $\Lambda$CDM model), and are particularly sensitive to the assumed pre-recombination expansion history, or more generally the post-inflationary history~\cite{Gerbino:2016sgw,DiValentino:2018zjj,Ye:2021nej,Takahashi:2021bti,Forconi:2021que,Giare:2022rvg,Jiang:2022uyg,Ye:2022efx,RoyChoudhury:2022rva,Jiang:2022qlj,Lin:2022gbl,Hazra:2022rdl,Giare:2023kiv,Giare:2023wzl,Jiang:2023bsz,Peng:2023bik,Forconi:2023hsj,Fu:2023tfo,Giare:2024akf,Giare:2024sdl,Wang:2024tjd,daCosta:2024grm,Forconi:2025zzu,Peng:2025tqt,Balkenhol:2025wms,Yuan:2026xcg,Garny:2026gcs,Oikonomou:2026ris,Giare:2026oti}. These considerations open a potential loophole to the ``USR-ACT tension'', in case the preference for positive running could be removed or at least weakened by introducing new pre-recombination physics. We therefore believe that the tension should not be over-interpreted, at least not until the stability of constraints on $\alpha_s$ under the assumed cosmological model, and in particular under physically motivated modifications of the (pre-recombination) expansion history, is better understood.

The above is not merely an academic question, and is unavoidable if one seriously takes the Hubble tension (see e.g.\ Refs.~\cite{Verde:2019ivm,DiValentino:2020zio,DiValentino:2021izs,Perivolaropoulos:2021jda,Shah:2021onj,Abdalla:2022yfr,DiValentino:2022fjm,Hu:2023jqc,Vagnozzi:2023nrq,Verde:2023lmm,CosmoVerseNetwork:2025alb,Cai:2026swf,Schoneberg:2026eys,Schoneberg:2026vaf,Wang:2026jxd} for reviews) as an indication for new physics~\cite{Mortsell:2018mfj,Vagnozzi:2018jhn,Yang:2018euj,Guo:2018ans,Kreisch:2019yzn,Vagnozzi:2019ezj,Visinelli:2019qqu,DiValentino:2019ffd,DiValentino:2019jae,Hart:2019dxi,Krishnan:2020obg,Alestas:2020mvb,Jedamzik:2020krr,Sekiguchi:2020teg,RoyChoudhury:2020dmd,Brinckmann:2020bcn,Gao:2021xnk,Marra:2021fvf,SolaPeracaula:2021gxi,Dainotti:2021pqg,Krishnan:2021dyb,Hart:2021kad,Cyr-Racine:2021oal,Anchordoqui:2021gji,Akarsu:2021fol,Ren:2022aeo,Nojiri:2022ski,Schoneberg:2022grr,Moshafi:2022mva,Rezazadeh:2022lsf,Banerjee:2022ynv,deSa:2022hsh,Akarsu:2022typ,Lee:2022gzh,Khodadi:2023ezj,Bernui:2023byc,Ben-Dayan:2023rgt,Gomez-Valent:2023hov,Ruchika:2023ugh,Adil:2023exv,Frion:2023xwq,Gomez-Valent:2023uof,Akarsu:2024qiq,Giare:2024ytc,Lynch:2024gmp,Giare:2024smz,Lynch:2024hzh,Nozari:2024wir,Li:2024qso,Escamilla:2024xmz,RoyChoudhury:2024wri,Mirpoorian:2024fka,Gomez-Valent:2024ejh,Li:2025owk,Yang:2025vnm,Lee:2025yah,Wang:2025dzn,Zhang:2025dwu,Chen:2025nfy,Kumar:2025obb,Hogas:2025mii,Li:2026xaz,Bouhmadi-Lopez:2026dte,Dai:2026pvx,Wang:2026rkx,Mukhopadhyay:2026fyk,Pantos:2026rpe,Solanki:2026tpi,Wang:2026kor,Jusufi:2026rfd,Dhyani:2026trw,Li:2026hwq,Akarsu:2026lva,Jia:2026vdt,Li:2026asg}, as opposed to unaccounted systematics~\cite{Efstathiou:2020wxn,Wojtak:2022bct,Giani:2023aor,Perivolaropoulos:2024yxv,Hogas:2026urs}. In this case, modifications to the pre-recombination expansion history are precisely those required to solve the tension, for two reasons~\cite{Bernal:2016gxb,Addison:2017fdm,Lemos:2018smw,Aylor:2018drw,Schoneberg:2019wmt,Knox:2019rjx,Arendse:2019hev,Efstathiou:2021ocp,Cai:2021weh,Keeley:2022ojz}: firstly, consistency with BAO measurements requires a reduction of the sound horizon, which necessarily requires new physics operating prior to recombination; at the same time, post-recombination attempts to solve the tension are severely constrained by BAO and unanchored Type Ia Supernovae (SNeIa) measurements. Pre-recombination solutions to the Hubble tension, such as early dark energy (EDE), are indeed known to strongly affect the inference of inflationary parameters such as $n_s$, which typically increases. The reason is that these models affect the CMB damping tail in such a way as to increase the amount of damping, which in turn can be compensated by an enhanced small-scale power spectrum. It is then natural to ask whether pre-recombination new physics called to solve the Hubble tension can affect $\alpha_s$ in a similar way. With these premises in mind, in this work we address two closely related questions.
\begin{enumerate}
\item \textit{\textbf{From the PBH point of view}}: can Hubble tension-motivated pre-recombination new physics shift current constraints on $\alpha_s$ towards the negative values preferred by USR models for PBH formation, thereby ``saving'' these models?
\item \textit{\textbf{From the cosmological point of view}}: how robust are current constraints on $\alpha_s$, with respect to both the assumed pre-recombination expansion history and adopted CMB dataset?
\end{enumerate}
Answering these two interconnected questions will allow us to determine the extent to which the challenges faced by USR models for PBH formation depend on the assumption of $\Lambda$CDM in the pre-recombination history, as well as the adoption of the \textit{ACT} CMB measurements.

To address these questions, we use axion-like EDE as a benchmark model of pre-recombination new physics. The reason is that it provides a well-studied model of new physics which is relatively successful at alleviating the Hubble tension, while exhibiting the shift towards larger values of $n_s$ due to its impact on the damping tail: it is therefore a useful benchmark for testing the extent to which the ``USR-ACT tension'' is affected by Hubble tension-motivated modifications of the pre-recombination expansion history. To assess the dependence on the adopted CMB dataset, we consider a large set of dataset combinations, involving CMB measurements from Planck, \textit{ACT}, and the South Pole Telescope (\textit{SPT}), alone or in combination with each other, and/or in combination with BAO measurements from the Dark Energy Spectroscopic Instrument (\textit{DESI}) DR2 release. Overall, our analysis includes 32 runs, obtained by combining different cosmological models ($\Lambda$CDM and EDE, allowing for non-zero spectral running $\alpha_s$ and potentially for the running of the running $\beta_s$) and dataset combinations: since several of these dataset combinations had not been previously used to constrain $\alpha_s$, even within $\Lambda$CDM, a useful byproduct of our work is an updated set of limits on $\alpha_s$ in light of state-of-the-art cosmological data (see also Ref.~\cite{K:2025juc}). In short, we find that for most models/dataset combinations, moving from $\Lambda$CDM to EDE increases the inferred $\alpha_s$ (as one could a priori expect, given the impact on $n_s$): in fact, for our most constraining combination involving all 3 CMB datasets, $\alpha_s$ is always positive and larger with respect to the value inferred assuming $\Lambda$CDM. While our analysis focuses on EDE, we argue that this is expected to apply more broadly to any pre-recombination new physics model aimed at solving the Hubble tension, such as models involving extra relativistic components. In this sense, tension calls for tension: because of the model-dependence of the inferred inflationary parameters, taking the Hubble tension seriously strengthens, rather than weakens, the tension faced by USR models for PBH formation.

The rest of this work is then organized as follows. In Sec.~\ref{sec:theory} we briefly review various theoretical aspects which are central to our study: production of PBHs from USR inflation (Sec.~\ref{subsec:pbhfromusr}), the reason why the Hubble tension generically affects the inference of cosmological parameters (Sec.~\ref{subsec:inflationarynewphysics}), and the axion-like EDE model (Sec.~\ref{subsec:ede}). In Sec.~\ref{sec:datasets} we discuss the datasets and methodology adopted to obtain our results. These are presented in Sec.~\ref{sec:results}, and critically discussed with an eye to their broader implications in Sec.~\ref{sec:discussion}. Finally, in Sec.~\ref{sec:conclusions} we close by drawing concluding remarks.

\section{Theory}
\label{sec:theory}

We now review some background material which is useful to understand the premises of our work. We begin by reviewing the formation of PBHs from USR inflation, and why this generally requires $\alpha_s<0$. We then explain why inflationary parameters inferred from cosmological observations are generally model-dependent, and should therefore not be treated as direct, model-independent measurements of the primordial power spectrum. Finally, we review the axion-like EDE model we adopt as benchmark pre-recombination modification aimed at solving the Hubble tension.

\subsection{Primordial black holes from ultra-slow-roll inflation}
\label{subsec:pbhfromusr}

The simplest single-field realization of inflation that can produce a sizeable abundance of PBHs relies on a transient USR phase, during which the inflaton crosses a quasi-stationary feature of the potential, typically an inflection point or a shallow local extremum, that strongly amplifies the curvature power spectrum at small scales~\cite{Ivanov:1994pa,Kinney:1997ne}. The required hierarchy between the curvature power spectrum at CMB scales, $\mathcal{P}_\zeta(k_{\rm CMB})\sim 10^{-9}$, and at PBH scales, $\mathcal{P}_\zeta(k_{\rm pk})\sim 10^{-2}$, must be generated within a finite number of $e$-folds, since heavy non-evaporating PBHs require the USR feature to occur relatively close in $e$-fold time to the CMB window. Schematically, denoting by $N_1$ the number of $e$-folds elapsed between horizon exit of CMB scales and the onset of the USR phase, the typical horizon mass at the spectral peak scales as follows~\cite{Allegrini:2025jha}:
\begin{equation}
M_{\text{PBH}} \simeq \mathcal{O}(1)\,M_{\odot} \times e^{-2(N_1-18)}.
\end{equation}
Hence solar-mass PBHs (relevant for LVK~\cite{LIGOScientific:2018mvr,LIGOScientific:2020ibl,KAGRA:2021vkt,LIGOScientific:2025slb,LIGOScientific:2019kan,Kavanagh:2018ggo,Wong:2020yig,Hutsi:2020sol,DeLuca:2021wjr,Franciolini:2022tfm,Andres-Carcasona:2024wqk}) are produced when the USR feature is close to the scales relevant for CMB spectral distortion constraints from FIRAS~\cite{Chluba:2012we,Chluba:2013dna} (small $N_1$), while asteroid-mass PBHs (relevant for the dark matter window) require larger $N_1\sim[30-40]$. PTA signals associated with USR scenarios, through the SIGW background sourced by the same enhanced curvature spectrum, correspond to a peak scale around $k_{\rm pk}\sim 10^{7}\,{\rm Mpc}^{-1}$, again pointing to relatively small $N_1$ and to PBHs of solar or sub-solar mass~\cite{NANOGrav:2023gor,EPTA:2023fyk,Reardon:2023gzh,Xu:2023wog,Franciolini:2023pbf}.

The reduced duration of the initial SR phase, combined with the necessity of a sizeable spectral peak, push USR models systematically away from scale invariance~\cite{Kannike:2017bxn,Ballesteros:2020qam}. As shown in detail in Ref.~\cite{Allegrini:2025jha}, the requirement of a successful PBH-producing USR phase forces the inflaton to cross a non-stationary inflection point near CMB scales, where the inflationary observables admit a controlled expansion in two parameters: the offset $\delta_\star$ from the inflection point and the parameter $\beta$ measuring the deviation from a stationary one.~\footnote{The fact that \textit{ACT} data is in tension with USR models of PBH formation was first pointed out by Frolovsky and Ketov~\cite{Frolovsky:2025iao}, who showed, working exclusively within a specific modified Starobinsky-like realization, that the new \textit{ACT} preference for positive $\alpha_s$ is in tension with the values typically required to obtain a sizeable PBH peak. Their analysis, unlike that of Ref.~\cite{Allegrini:2025jha}, was confined to a single benchmark model.} The resulting structure is most transparent when expressed at the level of the running of the spectral index, which to leading order takes the schematic form:
\begin{equation}
\alpha_s \simeq -\frac{12\beta}{V_0^2}-\frac{36\delta_{\star}^2}{V_0^2}\,,
\label{eq:alpha_s_usr}
\end{equation}
where $V_0$ denotes the height of the potential at the inflection point. The crucial point is that, in order for the inflaton to actually cross the inflection point and reach the small-field region where the USR feature is located, $\beta$ must be strictly positive: a negative $\beta$ would generate a local maximum that traps the field. Together with the manifestly negative $\delta_\star^{2}$ contribution, this leads to the robust prediction that $\alpha_s<0$ in this entire class of polynomial USR models. The same negative-running tendency is shared by other commonly studied USR realizations, including the well-known Higgs- and fibre-inflation-inspired potentials~\cite{Cicoli:2018asa,Garcia-Bellido:2017mdw,Germani:2017bcs}.

\begin{figure*}[t!]
\centering
\includegraphics[width=0.486\linewidth]{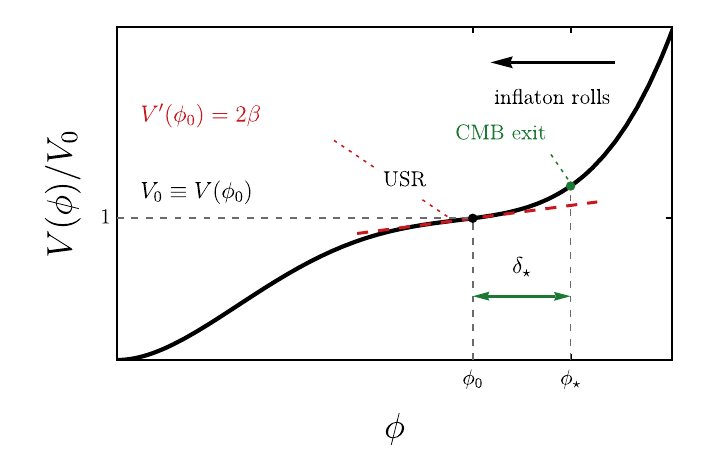} \quad \includegraphics[width=0.486\linewidth]{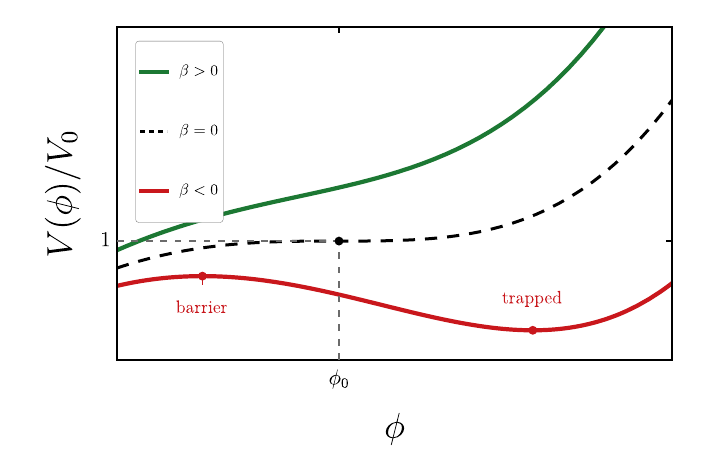}
\caption{Illustration of the quantities entering the right-hand side of Eq.~(\ref{eq:alpha_s_usr}), for the representative polynomial potential $V(\phi)=m^2\phi^2/2-\mu\phi^3/3+\lambda\phi^4/4$ with $\lambda=\mu^2(1+\beta)/(4m^2)$, shown in units $m=\mu=1$. \textit{Left panel:} the inflaton rolls from large to small field values (black arrow) and traverses the near-inflection point at $\phi_0$, whose height defines $V_0$. The CMB modes exit the horizon earlier, at $\phi_\star$, and the offset $\delta_\star \equiv \phi_\star - \phi_0$ (green double arrow) measures how far the inflaton still is from the feature at that moment: heavier PBHs correspond to smaller $N_1$, hence smaller $\delta_\star$ and a larger imprint on CMB-scale observables. The red dashed line is the tangent to the potential at $\phi_0$: for this parametrization one has exactly $V'(\phi_0)=2\beta$ and $V''(\phi_0)=3\beta$, so that $\beta$ measures precisely the deviation of the feature from a stationary inflection point. \textit{Right panel:} zoom on the feature, showing why $\beta>0$ is required. For $\beta>0$ (green) the potential is monotonic and the inflaton crosses the feature, undergoing the USR phase which seeds PBH formation; $\beta=0$ (black dashed) corresponds to an exactly stationary inflection point; for $\beta<0$ (red) a local minimum/maximum pair is generated and the inflaton is trapped before reaching the small-field region, so that no viable PBH-producing background is obtained. The requirement $\beta>0$, combined with the manifestly negative $\delta_\star^2$ term, is what makes $\alpha_s<0$ a robust prediction of this class of models.}
\label{fig:potential}
\end{figure*}

The meaning of the three quantities entering Eq.~(\ref{eq:alpha_s_usr}) is illustrated in Fig.~\ref{fig:potential}, for the representative polynomial potential $V(\phi)=m^2\phi^2/2-\mu\phi^3/3+\lambda\phi^4/4$, with the quartic coupling parametrized as $\lambda=\mu^2(1+\beta)/(4m^2)$. This parametrization is convenient because $\beta=0$ corresponds to an exactly stationary inflection point at $\phi_0=\mu/(2\lambda)$, where $V'(\phi_0)=V''(\phi_0)=0$ and $V_0\equiv V(\phi_0)$, whereas for $\beta\neq 0$ one finds exactly $V'(\phi_0)=2\beta$ and $V''(\phi_0)=3\beta$ (in units $m=\mu=1$). The parameter $\beta$ therefore has a transparent geometrical meaning: it is a direct measure of the residual slope of the potential at the feature, i.e.\ of its departure from stationarity. Its sign is not a matter of choice. For $\beta>0$ the potential remains monotonic and the inflaton traverses the feature, experiencing the transient USR phase which enhances the curvature perturbations; for $\beta<0$, on the other hand, a local minimum and a local maximum are generated on either side of $\phi_0$, and the inflaton becomes trapped before it can reach the small-field region, so that no PBH-producing background exists in the first place. This is the sense in which $\beta>0$ is required, rather than assumed. The remaining quantity, $\delta_\star$, is the displacement of the inflaton from the feature at the time the CMB modes leave the horizon, and enters Eq.~(\ref{eq:alpha_s_usr}) squared, hence with a definite negative sign irrespective of the side from which the feature is approached. The combination of these two facts is what makes the negative sign of $\alpha_s$ robust within the given class of models.

This sign of $\alpha_s$ is in direct tension with recent CMB data from \textit{ACT} which, when combined with \textit{Planck} data, indicates a positive running, $\alpha_s = 0.0062\pm 0.0052$~\cite{AtacamaCosmologyTelescope:2025nti,AtacamaCosmologyTelescope:2025blo}, alongside a higher value of the spectral index, $n_s = 0.9743 \pm 0.0034$. The tension is not uniform across the PBH mass range: the magnitude of $\vert \alpha_s \vert$ predicted by USR models grows with the produced mass of PBHs, and hence becomes more pronounced for solar-ish mass PBHs. As our work will show, this conclusion is sensitive to the assumed CMB dataset.

\subsection{Inflationary parameters and the Hubble tension}
\label{subsec:inflationarynewphysics}

We now review why the type of new physics required to solve the Hubble tension affects the determination of inflationary parameters, typically increasing $n_s$. Our discussion also clarifies why constraints on inflationary parameters cannot be treated as direct, model-independent measurements of the primordial power spectrum.

New physics aiming to address the Hubble tension must do so while keeping the acoustic angular scale $\theta_s$ fixed, as this is measured by the CMB to exquisite precision, with very little model-dependence. This quantity is given by the following:
\begin{equation}
\theta_s \equiv \frac{r_s^{\star}}{D_A^{\star}}\,,
\label{eq:thetas}
\end{equation}
where $r_s^{\star}$ is the comoving sound horizon at recombination, and $D_A^{\star}$ is the comoving angular diameter distance to the surface of last-scattering. In principle one could raise $H_0$ while keeping $\theta_s$ fixed in (at least) two ways:
\begin{enumerate}
\item lowering $r_s^{\star}$, with the corresponding increase in $H_0$ decreasing $D_A^{\star}$ by the same amount: since the integral determining $r_s^{\star}$ runs over the pre-recombination epoch, this requires pre-recombination (early-time) new physics;
\item keeping $r_s^{\star}$ fixed, while altering the shape of the post-recombination expansion rate, i.e.\ $E(z) \equiv H(z)/H_0$, in such a way so as to reach a higher $H_0$ while keeping $D_A^{\star}$ fixed: this is the route followed by post-recombination (late-time) new physics attempts to solve the Hubble tension.
\end{enumerate}
At the geometrical level, CMB measurements cannot distinguish between the above two options towards raising the inferred value of $H_0$.

BAO measurements play a key role because they break this geometrical degeneracy in favor of the first option, i.e.\ the pre-recombination class of solutions reducing $r_s^{\star}$, as extensively discussed in the literature (see for instance Refs.~\cite{Bernal:2016gxb,Addison:2017fdm,Lemos:2018smw,Aylor:2018drw,Schoneberg:2019wmt,Knox:2019rjx,Arendse:2019hev,Efstathiou:2021ocp,Cai:2021weh,Keeley:2022ojz}). BAO essentially measure the angular size of the comoving sound horizon at baryon drag, $r_{\text{drag}}$ (itself closely related to $r_s^{\star} \simeq 1.018 r_{\text{drag}}$), along and transverse to the line-of-sight, as well as a volume-averaged combination of the two. It is easy to show (see e.g.\ Eqs.~(3-5) of Ref.~\cite{Jiang:2024xnu}) that BAO measurements are directly sensitive to the combination $r_{\text{drag}}H_0$. Loosely speaking, current BAO measurements are consistent with a value $r_{\text{drag}}H_0 \approx 10000\,{\text{km}}/{\text{s}}$, which in $\Lambda$CDM corresponds to a ``high'' sound horizon $r_{\text{drag}} \sim 147\,{\text{Mpc}}$ and a ``low'' $H_0 \approx 67\,{\text{km}}/{\text{s}}/{\text{Mpc}}$. Keeping the product $r_{\text{drag}}H_0$ constant requires that an increase in $H_0$ be accompanied by a proportional decrease in $r_{\text{drag}}$ (see e.g.\ Fig.~1 of Ref.~\cite{Knox:2019rjx}), i.e.\ $\delta r_{\text{drag}}/r_{\text{drag}} \simeq -\delta H_0/H_0$. At present, solving the Hubble tension requires a decrease in the sound horizon at the $\simeq 7\%$ level, $\delta r_{\text{drag}}/r_{\text{drag}} \simeq -0.07$. As stressed earlier, this requires new physics operating prior to recombination, as the integral determining $r_{\text{drag}}$ runs across redshifts $z>z_{\text{drag}} \approx 1060$.~\footnote{Aside from the amplitude of BAO measurements and their sensitivity to the product $r_{\text{drag}}H_0$, another important obstruction to late-time solutions to the Hubble tension comes from unanchored SNeIa, which place tight constraints on the shape of the expansion history $E(z)$~\cite{Benevento:2020fev,Teixeira:2025czm,Zhou:2025kws,Pedrotti:2025ccw,Zhou:2025dxo,Bansal:2026axl,Zhou:2026iar}.}

If $r_{\text{drag}}$ decreases by the same relative amount by which $H_0$ increases ($\delta r_{\text{drag}}/r_{\text{drag}} \simeq -\delta H_0/H_0$), while the shape of the post-recombination expansion rate remains unchanged, the CMB constraint on $\theta_s$ is automatically satisfied. In fact, the relation between the two sound horizons implies $\delta r_s^{\star}/r_s^{\star} \simeq \delta r_{\text{drag}}/r_{\text{drag}}$, whereas $\delta D_A^{\star}/D_A^{\star}=-\delta H_0/H_0$, implying that $\delta r_s^{\star}/r_s^{\star} \simeq \delta D_A^{\star}/D_A^{\star}$. Since the numerator and denominator of Eq.~(\ref{eq:thetas}) decrease by the same relative amount, $\theta_s$ remains fixed, as required by CMB observations. However, the CMB damping tail also tightly constrains the angular size of the photon diffusion scale at recombination $\theta_d$, defined as follows:
\begin{equation}
\theta_d \equiv \frac{r_d^{\star}}{D_A^{\star}}\,,
\label{eq:thetad}
\end{equation}
where $r_d^{\star}$ is the comoving photon diffusion scale at recombination and, crucially, the denominator is the same as in Eq.~(\ref{eq:thetas}). Whereas $r_s^{\star}$ is sensitive to the expansion history in the entire decade of scale factor prior to recombination, the sensitivity of $r_d^{\star}$ to the expansion history is much more localized around recombination (compare the blue and orange kernels in Fig.~2 of Ref.~\cite{Knox:2019rjx}). As a result, models which increase the pre-recombination expansion rate in order to reduce $r_s^{\star}$ are only able to reduce $r_d^{\star}$ by a smaller relative amount: $\vert \delta r_d^{\star}/r_d^{\star} \vert<\vert \delta r_s^{\star}/r_s^{\star} \vert$.~\footnote{As discussed in Refs.~\cite{Knox:2019rjx,Toda:2024ncp}, for many models which attempt to solve the Hubble tension by increasing the pre-recombination rate close to matter-radiation equality, the fractional decrease in $r_d^{\star}$ is about half that of $r_s^{\star}$: $\delta r_d^{\star}/r_d^{\star} \simeq 1/2 \delta r_s^{\star}/r_s^{\star}$. The factor of $1/2$ is related to the random walk nature of diffusion damping, because of which $r_d^{\star}$ depends on the integral of $\sqrt{1/H(z)}$, as opposed to $1/H(z)$ in the case of $r_s^{\star}$.} However, it is the same quantity which appears at the denominators of Eqs.~(\ref{eq:thetas},\ref{eq:thetad}). Therefore, a model which can keep $\theta_s$ fixed will in general not be able to keep $\theta_d$ fixed, as $\vert \delta r_d^{\star}/r_d^{\star} \vert<\vert \delta r_s^{\star}/r_s^{\star} \vert \simeq \vert \delta D_A^{\star}/D_A^{\star} \vert$.~\footnote{Stated otherwise, the different sensitivity of $r_s^{\star}$ and $r_d^{\star}$ to the pre-recombination expansion history make it very difficult to keep the diffusion-to-acoustic angular scale ratio $\theta_d/\theta_s$ fixed. As observed in Ref.~\cite{Pedrotti:2026dwj}, this can potentially be achieved by considering modifications to the expansion history which are much more extended compared to the ones considered so far, mostly concentrated around matter-radiation equality.} Since in Eq.~(\ref{eq:thetad}) the numerator decreases more slowly than the denominator, keeping $\theta_s$ fixed will result in $\theta_d$ increasing.

An increase in $\theta_d$ implies that diffusion damping becomes relevant at smaller multipoles, since $\ell \propto 1/\theta$. The damping tail in the CMB temperature power spectrum will therefore be more strongly suppressed, at a given multipole, relative to the reference best-fit $\Lambda$CDM model. To maintain consistency with the measured CMB power spectrum, this extra damping needs to be compensated, and the most economical way to do so is to increase $n_s$, thereby providing additional power to smaller scales. This is the reason why most early-time modifications aimed at solving the Hubble tension do so at the expense of an increase in $n_s$ (alongside shifts in other cosmological parameters, including but not limited to $\omega_c$ and $\omega_b$~\cite{Vagnozzi:2021gjh,Vagnozzi:2021tjv,Poulin:2024ken,Pedrotti:2024kpn,Giovanetti:2026aku}). This behaviour has been extensively discussed, for instance, in the case of EDE~\cite{Jiang:2022uyg,Giare:2024akf}, where the inferred value of $n_s$ can move significantly closer to the scale-invariant value of $n_s=1$, motivating extensive theoretical follow-up.

Summing up, early-time modifications to the expansion rate typically shift the inferred inflationary parameters. These affect the primordial power spectrum so as to compensate the extra damping associated to the increase in $\theta_d$, itself tied to the different responses of $r_d^{\star}$ and $r_s^{\star}$ to modifications to the pre-recombination expansion rate. This discussion clarifies a very important point: \textit{inflationary parameters inferred from cosmological observations depend on the assumed underlying model, and \textbf{cannot} be treated as direct, model-independent measurements of the primordial power spectrum}. Given that attempts to resolve the Hubble tension are generally accompanied by an increase in $n_s$, it is natural to ask whether similar considerations hold for $\alpha_s$ as well. Indeed, increasing $\alpha_s$ increases the amount of small-scale power (at fixed $n_s$ and above the pivot scale), which counteracts the extra damping following from a larger value of $\theta_d$. As we shall see, this expectation is confirmed within our EDE benchmark: $\alpha_s$ does indeed increase, although the increase is much milder compared to that in $n_s$.

\subsection{Axion-like early dark energy}
\label{subsec:ede}

We now turn to EDE as a benchmark model of early-time new physics aimed at addressing the Hubble tension (see e.g.\ Refs.~\cite{Kamionkowski:2022pkx,Poulin:2023lkg,McDonough:2023qcu} for reviews). In general, the term EDE refers to a component which was dynamically relevant at $z \gg 1$, and whose equation of state was $w \simeq -1$ at some point during its evolution. It is often (although not always) the case that EDE takes the form of a scalar field initially frozen in its potential by Hubble friction. In our case, we specialize to the so-called axion-like EDE model, first introduced in the context of the Hubble tension in Refs.~\cite{Karwal:2016vyq,Poulin:2018dzj,Poulin:2018cxd}. Our motivation is that this is a concrete and widely studied pre-recombination new physics model which exhibits the features relevant for our analysis, i.e.\ a decrease in the sound horizon accompanied by an increase in $n_s$ to compensate the impact on the damping tail. For this reason, we adopt it as a useful benchmark for testing the impact of early-time new physics on the inferred value of $\alpha_s$. Nevertheless, we stress that several other EDE models beyond axion-like EDE are being developed and studied, such as (cold and hot) new EDE, AdS EDE, acoustic EDE, and rock 'n' roll EDE among others: we refer the reader to Refs.~\cite{Agrawal:2019lmo,Lin:2019qug,Smith:2019ihp,Niedermann:2019olb,Berghaus:2019cls,Sakstein:2019fmf,Nojiri:2019fft,Ye:2020btb,Zumalacarregui:2020cjh,Ballesteros:2020sik,Braglia:2020iik,Ballardini:2020iws,Gogoi:2020qif,Braglia:2020bym,Gonzalez:2020fdy,Ye:2020oix,Niedermann:2020qbw,Murgia:2020ryi,Smith:2020rxx,CarrilloGonzalez:2020oac,Braglia:2020auw,Adi:2020qqf,Oikonomou:2020qah,Oikonomou:2020oex,Karwal:2021vpk,Jiang:2021bab,Gomez-Valent:2021cbe,Ye:2021iwa,Poulin:2021bjr,Niedermann:2021ijp,Niedermann:2021vgd,Herold:2021ksg,Wang:2022jpo,Smith:2022hwi,Oikonomou:2022yle,Reeves:2022aoi,Gomez-Valent:2022bku,Escudero:2022rbq,Cruz:2022oqk,Wang:2022bmk,Herold:2022iib,Cruz:2023cxy,Goldstein:2023gnw,Eskilt:2023nxm,Cruz:2023lmn,Odintsov:2023cli,Ye:2023zel,Sharma:2023kzr,Efstathiou:2023fbn,Pedreira:2023qqt,Khalife:2023qbu,Wang:2024jug,Garny:2024ums,Chatrchyan:2024xjj,Jiang:2024tll,Jiang:2024nha,Simon:2024jmu,Forconi:2025cwp,Jiang:2025ylr,Stahl:2025czl,Jiang:2025hco,Smith:2025grk,Poulin:2025nfb,Yashiki:2025loj,Garny:2025kqj,Toda:2025kcq,Wang:2025djw,Yin:2026gss,Gonzalez-Fuentes:2026rgu,Jhaveri:2026bla,Bella:2026zuk,Carloni:2026yut,Zhang:2026fzj,Giare:2026tyk,Du:2026qtq} for examples in this direction, as well as various studies on EDE.

The axion-like EDE model consists of an ultralight scalar field $\phi$ (with mass typically $m \lesssim 10^{-27}\,{\text{eV}})$, governed by the following potential~\cite{Poulin:2018dzj,Poulin:2018cxd}:
\begin{equation}
V(\phi) = m^2f^2 \left [ 1- \cos \left ( \frac{\phi}{f} \right ) \right ] ^n\,,
\label{eq:potential}
\end{equation}
where $m$ and $f$ are the field's mass and decay constant respectively, and the potential index $n$ takes integer values.~\footnote{In the context of cosmological studies one often refers to the field as an ``axion'', and therefore to the model as ``axion-like'' EDE, despite the field being actually a scalar rather than a pseudoscalar. The reason is that, for $n=1$, the above potential naturally appears in axion-like/pseudo-Nambu-Goldstone boson settings. Nevertheless, we stress that in the context of our study, the potential should be interpreted as a toy potential, and not as one tied to a fundamental theory.} Initially, EDE is displaced from its minimum and frozen by Hubble friction, resembling a cosmological constant. When the Hubble rate is of order the field's mass, the field becomes dynamical and starts rolling down its potential. Eventually, it oscillates around the minimum of its potential and dilutes, redshifting as an effective fluid with equation of state (after averaging over oscillations) $\langle w \rangle = (n-1)/(n+1)$.

In cosmological analyses, it is common practice to trade the fundamental parameters $m$ and $f$ for two phenomenological parameters: the critical redshift $z_c$ at which EDE becomes dynamical, and $f_{\text{ede}}$, the fractional energy density contributed by EDE at redshift $z_c$. Aside from these parameters, EDE is characterized by the initial misalignment angle $\theta_i=\phi_i/f$, where $\phi_i$ is the initial field value, and the potential index $n$. Without loss of generality, $\theta_i$ can be taken in the range $0 \leq \theta_i \leq \pi$. Since cosmological data are not particularly sensitive to $n$~\cite{Simon:2023hlp}, in what follows we will set $n=3$: with this choice, once the field starts oscillating around the minimum its effective equation of state is $\langle w \rangle = 1/2$, allowing EDE to decay more quickly than the radiation component. As commonly done in such studies, the behaviour of EDE perturbations is followed at the scalar field level (rather than treating it as an effective fluid), with linear fluctuations in $\phi$ evolved consistently with the metric and matter perturbations in the perturbed Einstein-Boltzmann-Klein-Gordon system. We refer the reader to Refs.~\cite{Poulin:2018cxd,Poulin:2023lkg} for further details on the equations governing the perturbations of the axion-like EDE model.

On the model side, constructing well-motivated theoretical realizations of EDE (e.g.\ within string theory~\cite{McDonough:2022pku,Cicoli:2023qri}) has proven challenging since, taken at face value, the potential in Eq.~(\ref{eq:potential}) requires a delicate cancellation between various higher-order instanton corrections. Moreover, the required energy injection has to occur within a relatively narrow redshift window, raising questions about the naturalness of the mechanism. We stress that, while important, these challenges are not unique to EDE, but shared to varying degree with other cosmological models based on scalar fields, such as inflation, dark energy, and fuzzy dark matter. Most importantly, they do not affect the role of axion-like EDE in our work: that of providing a concrete benchmark for assessing how a phenomenological cosmological model which has so far proven relatively successful at alleviating the Hubble tension affects the inferred value of $\alpha_s$.

\section{Datasets and methodology}
\label{sec:datasets}

To determine constraints on $\alpha_s$ and assess the extent to which the ``USR-ACT tension'' depends on the assumed pre-recombination expansion history and adopted CMB dataset, we make use of the following measurements:
\begin{itemize}
\item \textbf{Planck} -- CMB temperature, polarization, and lensing measurements from the \textit{Planck} NPIPE data release (PR4)~\cite{Planck:2020olo}. Specifically, we make use of the \texttt{CamSpec} high-$\ell$ TTTEEE likelihood~\cite{Rosenberg:2022sdy}, the \texttt{Commander} low-$\ell$ TT likelihood~\cite{Planck:2019nip}, the \texttt{Sroll2} low-$\ell$ EE likelihood~\cite{Delouis:2019bub}, and the PR4 lensing likelihood~\cite{Carron:2022eyg}. We refer to the resulting combination as \textit{Planck}.
\item \textbf{ACT} -- CMB temperature, polarization, and lensing measurements from \textit{ACT} DR6. Specifically, we adopt the \textit{ACT} \texttt{lite} (compressed) DR6 TTTEEE likelihood and the \textit{ACT} DR6 lensing likelihood~\cite{AtacamaCosmologyTelescope:2025blo}. Following the \textit{ACT} convention, we include the \textit{Planck} \texttt{Sroll2} low-$\ell$ EE likelihood to provide the large-scale polarization information required to constrain the optical depth to reionization $\tau$. We refer to this combination as \textit{ACT}.
\item \textbf{SPT} -- CMB temperature, polarization, and lensing measurements from the \textit{SPT}-3G D1 observations. Specifically, we include the \textit{SPT}-3G D1 TTTEEE likelihood~\cite{SPT-3G:2025bzu} and the \texttt{MUSE} lensing likelihood~\cite{SPT-3G:2024atg}. Following the same logic as with the \textit{ACT} dataset combination, we add the \textit{Planck} \texttt{Sroll2} low-$\ell$ EE likelihood (note that our approach differs from the \textit{SPT} convention of simply including a Gaussian prior on $\tau$, as it retains the full E-mode polarization information rather than compressing it into a value of $\tau$). We refer to the above combination as \textit{SPT}.
\item \textbf{DESI} -- BAO measurements from the \textit{DESI} DR2 release. This includes measurements of $D_V/r_{\text{drag}}$, $D_M/r_{\text{drag}}$, and $D_H/r_{\text{drag}}$ across 7 redshift bins in the range $0.295<z<2.330$~\cite{DESI:2025zgx}, where $D_V$, $D_M$, and $D_H$ are the volume-averaged distance, transverse comoving distance, and Hubble distance respectively. We refer to the combinations of these measurements as \textit{DESI}.
\end{itemize}
Following recent works, we also combine data from all three CMB experiments to fully take advantage of their complementary scale coverage, and derive the tightest CMB-based constraints on the relevant cosmological parameters. When doing so, following the prescriptions of the \textit{ACT} and \textit{SPT} collaborations, we combine the \textit{Planck} likelihoods with multipole cuts of $\ell<1000$ in TT and $\ell<600$ in EE, the \textit{ACT} DR6 TTTEEE likelihood, the \textit{SPT}-3G D1 TTTEEE likelihood, the \textit{Planck} \texttt{Sroll2} likelihood, the joint \textit{Planck}-\textit{ACT} lensing likelihood, and the \texttt{MUSE} lensing likelihood. The specific choice of scale cuts is designed to minimize the correlations between the \textit{Planck} and \textit{ACT} datasets, themselves assumed to be uncorrelated with the \textit{SPT} dataset, an assumption which is well justified by the relatively small overlap in sky regions. We refer to this combination of CMB datasets as \textit{\textbf{CMB-SPA}}.

In our analysis, we consider eight combinations of the above datasets, i.e.\ the three CMB datasets and their combination, with and without BAO data. Explicitly, we consider these datasets/dataset combinations:
\begin{itemize}
\item \textit{Planck}
\item \textit{ACT}
\item \textit{SPT}
\item \textit{CMB-SPA}
\item \textit{Planck}+\textit{DESI}
\item \textit{ACT}+\textit{DESI}
\item \textit{SPT}+\textit{DESI}
\item \textit{CMB-SPA}+\textit{DESI}
\end{itemize}
This choice of combinations allows us to systematically study the robustness of the inferred values of $\alpha_s$. While the most constraining combinations are those involving \textit{CMB-SPA}, considering also the three CMB experiments individually allows us to determine the extent to which shifts in $\alpha_s$ are driven by a specific CMB dataset. In addition, although BAO data is not directly sensitive to inflationary parameters, the addition of the \textit{DESI} dataset is extremely important. In fact, constraints on $n_s$ are driven in part by a degeneracy between $n_s$ and $\omega_m$ which manifests in CMB data, as well as the mild tension between CMB and BAO data~\cite{McDonough:2025lzo}. Comparing results from the CMB-only and CMB+\textit{DESI} dataset combinations therefore allows us to assess whether the inferred value of $\alpha_s$ is affected by the CMB-BAO tension. We note that the choice of dataset combinations resembles that used in Ref.~\cite{McDonough:2025lzo} (which focused solely on $n_s$), although it differs in the use of \textit{Planck} PR4 data instead of PR3 one, as well as the use of the \texttt{Sroll2} low-$\ell$ EE likelihood when using \textit{SPT} data. En passant, we also note that most of these dataset combinations had not been previously used to constrain $\alpha_s$. For instance, the constraints on $\alpha_s$ explicitly reported by the \textit{ACT} collaboration always make use of \textit{Planck} data as well~\cite{AtacamaCosmologyTelescope:2025nti}. Our analysis thus provides, as a useful byproduct, a set of updated constraints on $\alpha_s$ (as well as $n_s$) in light of various combinations of recent cosmological datasets.

In our analysis we consider four different cosmological models. Our baseline model is the 7-parameter $\Lambda$CDM+$\alpha_s$ model, whose free parameters are the standard 6 $\Lambda$CDM cosmological parameters (the acoustic angular scale $\theta_s$, the physical baryon and cold dark matter densities $\omega_b$ and $\omega_c$, the amplitude and spectral index of the primordial power spectrum of curvature perturbations $A_s$ and $n_s$, and the optical depth to reionization $\tau$), as well as the running of the spectral index $\alpha_s$.~\footnote{We fix the sum of the neutrino masses to $\sum m_{\nu}=0.06\,{\text{eV}}$, given the current upper limits on $\sum m_{\nu}$~\cite{Vagnozzi:2017ovm,Tanseri:2022zfe,Jiang:2024viw,RoyChoudhury:2025dhe,RoyChoudhury:2025iis}.} We also consider the 8-parameter $\Lambda$CDM+$\alpha_s$+$\beta_s$ where the running of the running of the spectral index $\beta_s$ is an additional free parameter. In what follows, when no ambiguity arises, we will refer to both the $\Lambda$CDM+$\alpha_s$ and $\Lambda$CDM+$\alpha_s$+$\beta_s$ models as simply $\Lambda$CDM, as a shorthand to avoid an unnecessarily heavy notation.

The other two cosmological models we consider are extensions of the 9-parameter axion-like EDE model. Beyond the 6 $\Lambda$CDM ones, its 3 additional free parameters are the axion misalignment angle $\theta_i$, as well as the two phenomenological parameters $z_c$ and $f_{\text{EDE}}$. Specifically, we consider the 10-parameter EDE+$\alpha_s$ model, and the 11-parameter EDE+$\alpha_s$+$\beta_s$ model. We fix the potential index to $n=3$, noting that cosmological data is largely insensitive to the value of $n$. As previously, when no ambiguity arises, we will refer to both these models as simply EDE.

To obtain theoretical predictions for the relevant cosmological observables, we use the \texttt{CLASS} Boltzmann solver for the two $\Lambda$CDM-based models, and the \texttt{CLASS\_EDE} Boltzmann solver~\cite{Hill:2020osr} for the two EDE-based models, assuming adiabatic initial conditions in both cases.~\footnote{The \texttt{CLASS} and \texttt{CLASS\_EDE} Boltzmann solvers are both publicly available at \href{https://github.com/lesgourg/class\_public}{https://github.com/lesgourg/class\_public} and \href{https://github.com/mwt5345/class\_ede}{github.com/mwt5345/class\_ede} respectively.} \texttt{CLASS\_EDE} is an extension of the Boltzmann solver \texttt{CLASS}~\cite{Blas:2011rf} which explicitly solves the perturbed Klein-Gordon equation for the axion-like EDE field.

We sample the posterior distributions for the parameters of the four cosmological models using Monte Carlo Markov Chain (MCMC) methods, using the cosmological MCMC sampler \texttt{Cobaya}~\cite{Torrado:2020dgo}.~\footnote{The \texttt{Cobaya} MCMC sampler is publicly available at \href{https://github.com/CobayaSampler/cobaya}{github.com/CobayaSampler/cobaya}.} We set wide, flat priors on all cosmological parameters (except for $z_c$, for which we instead sample $\log_{10}z_c$, consistently with earlier analyses), and verify that our posteriors are not affected by the choice of lower and upper prior boundaries. We assess the convergence of our MCMC chains using the Gelman-Rubin $R-1$ parameter~\cite{Gelman:1992zz}, requiring $R-1<0.03$ for our chains to be considered converged. Our chains are then analyzed using the \texttt{GetDist} package~\cite{Lewis:2019xzd}.~\footnote{The \texttt{GetDist} package for analyzing MCMC chains is publicly available at \href{https://github.com/cmbant/getdist}{github.com/cmbant/getdist}.}

\section{Results}
\label{sec:results}

\begin{figure*}[!htb]
\centering
\includegraphics[width=0.495\linewidth]{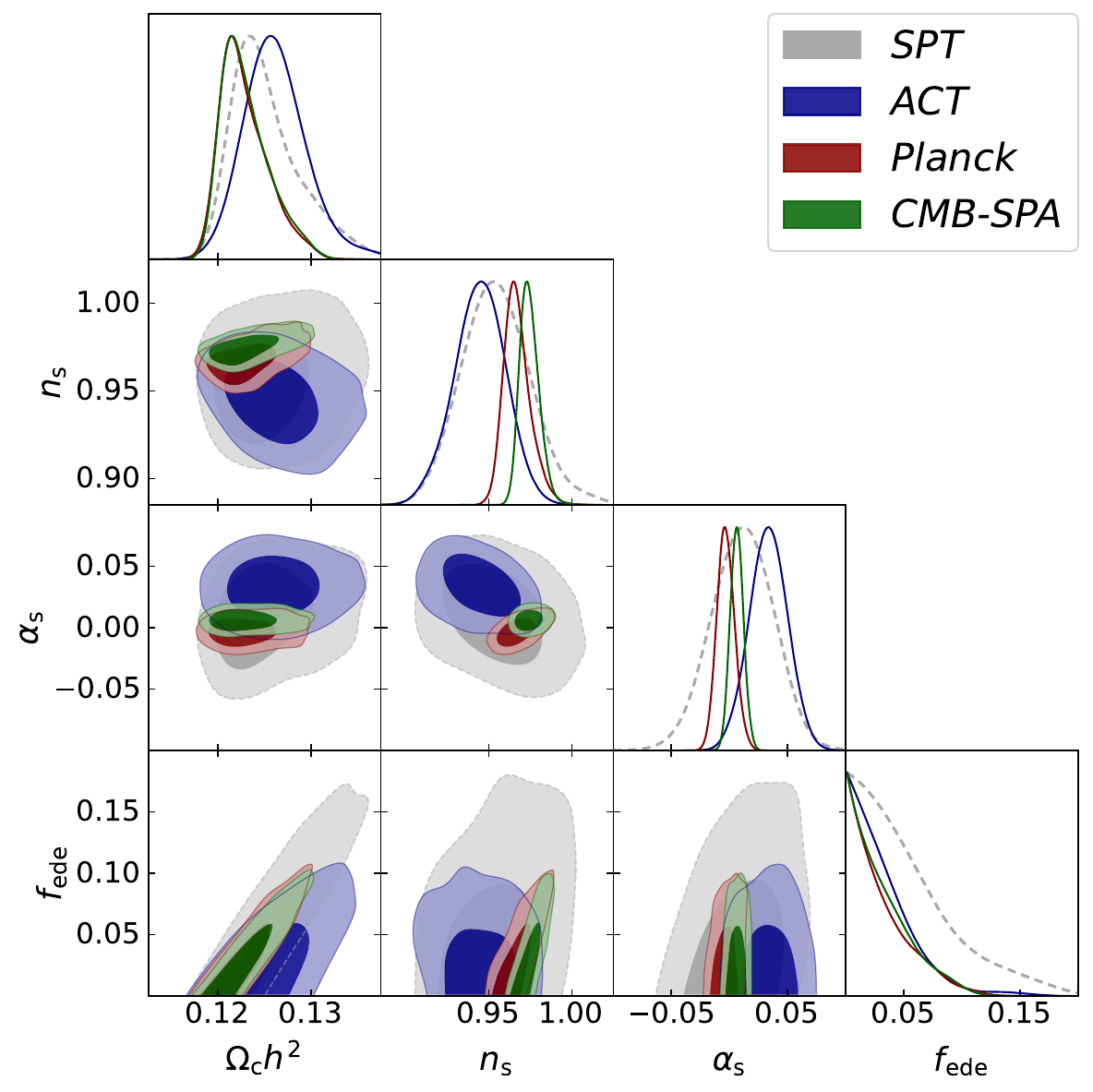}
\includegraphics[width=0.495\linewidth]{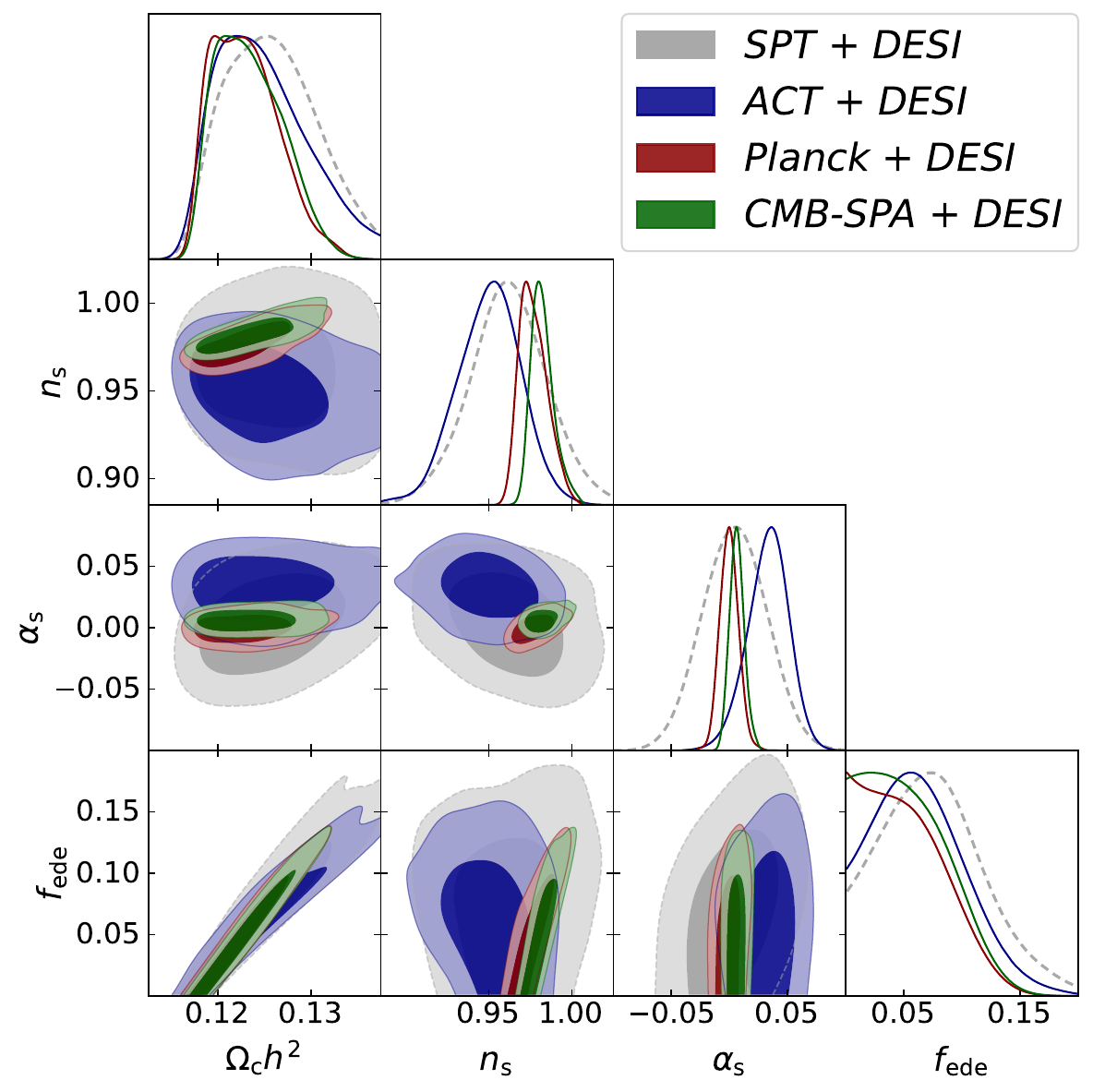}
\caption{Triangular plot showing 2D joint and 1D marginalized posterior probability distributions for the physical cold dark matter density $\omega_c$, the scalar spectral index $n_s$ and its running $\alpha_s$, and the EDE fraction $f_{\text{ede}}$, obtained within the EDE+$\alpha_s$ model in light of the four CMB-only dataset combinations (left panel), and the same combinations supplemented by \textit{DESI} BAO data (right panel). The 2D contours show $68\%$ and $95\%$ credible regions. In the $\alpha_s$-$n_s$ panel, a clear negative correlation between the two parameters is visible, reflecting the fact that a positive running can partially compensate the \textit{ACT} preference for a bluer spectrum. The addition of \textit{DESI} data does not qualitatively alter our conclusions.}
\label{fig:edealphas}
\end{figure*}

\begin{figure*}[!htb]
\centering
\includegraphics[width=0.495\linewidth]{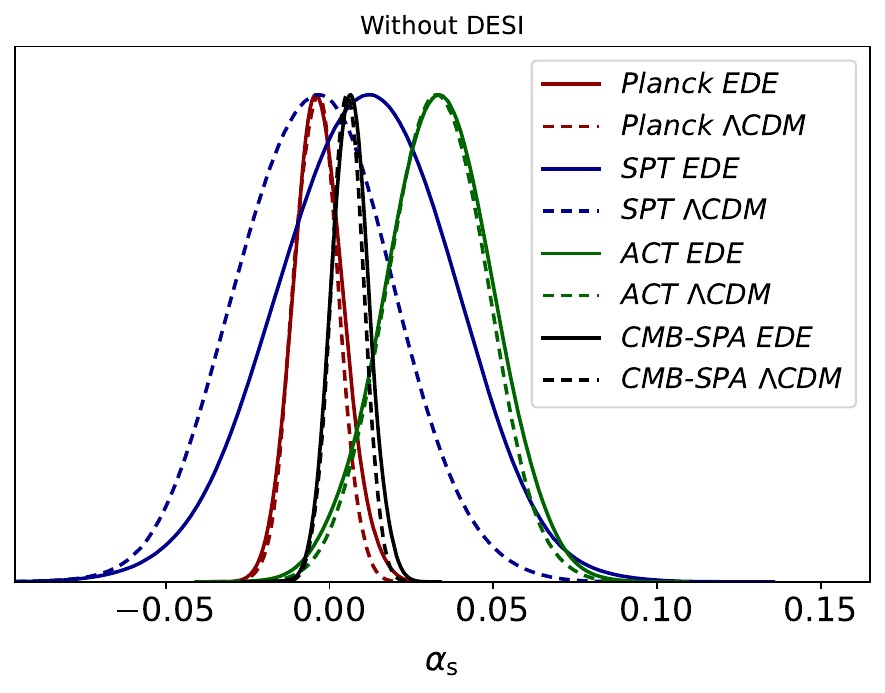}
\includegraphics[width=0.495\linewidth]{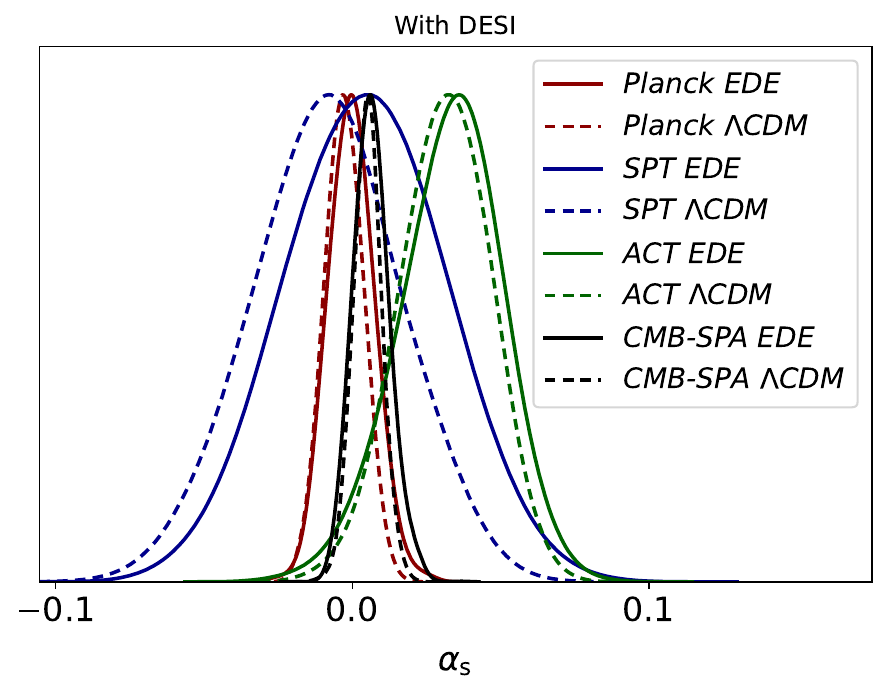}
\caption{1D marginalized posterior distributions for the running of the spectral index $\alpha_s$, obtained within the $\Lambda$CDM+$\alpha_s$ (dashed curves) and EDE+$\alpha_s$ (solid curves) models, in light of all eight dataset combinations considered in this work: the four CMB-only dataset combinations (left panel), and the same dataset combinations supplemented by \textit{DESI} BAO data (right panel). For most dataset combinations, the central value of $\alpha_s$ increases when moving from $\Lambda$CDM to EDE, consistent with the theoretical expectation discussed in  Sec.~\ref{subsec:inflationarynewphysics}. }
\label{fig:alphas}
\end{figure*}

\begin{figure*}[!htb]
\centering
\includegraphics[width=0.495\linewidth]{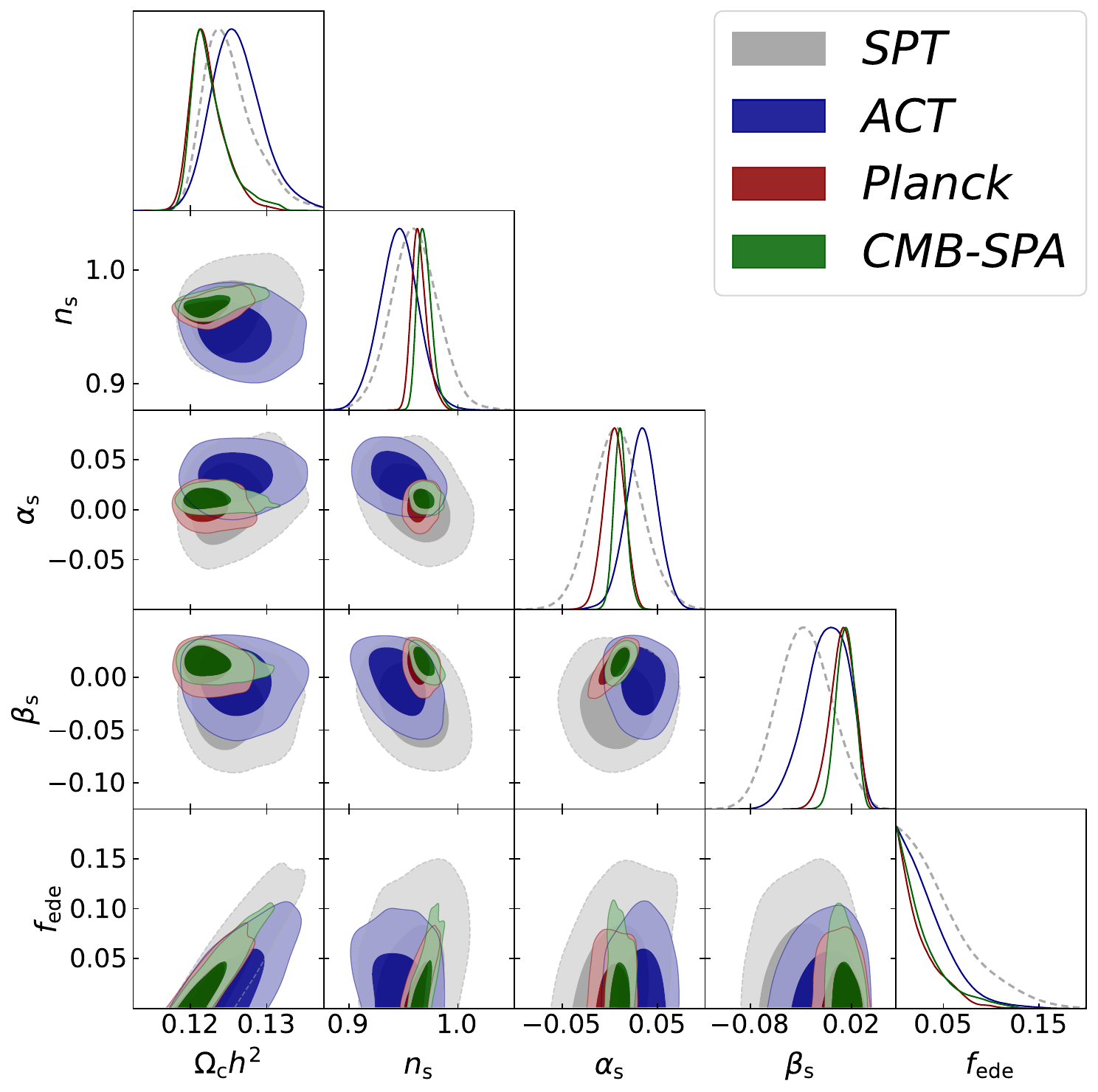}
\includegraphics[width=0.495\linewidth]{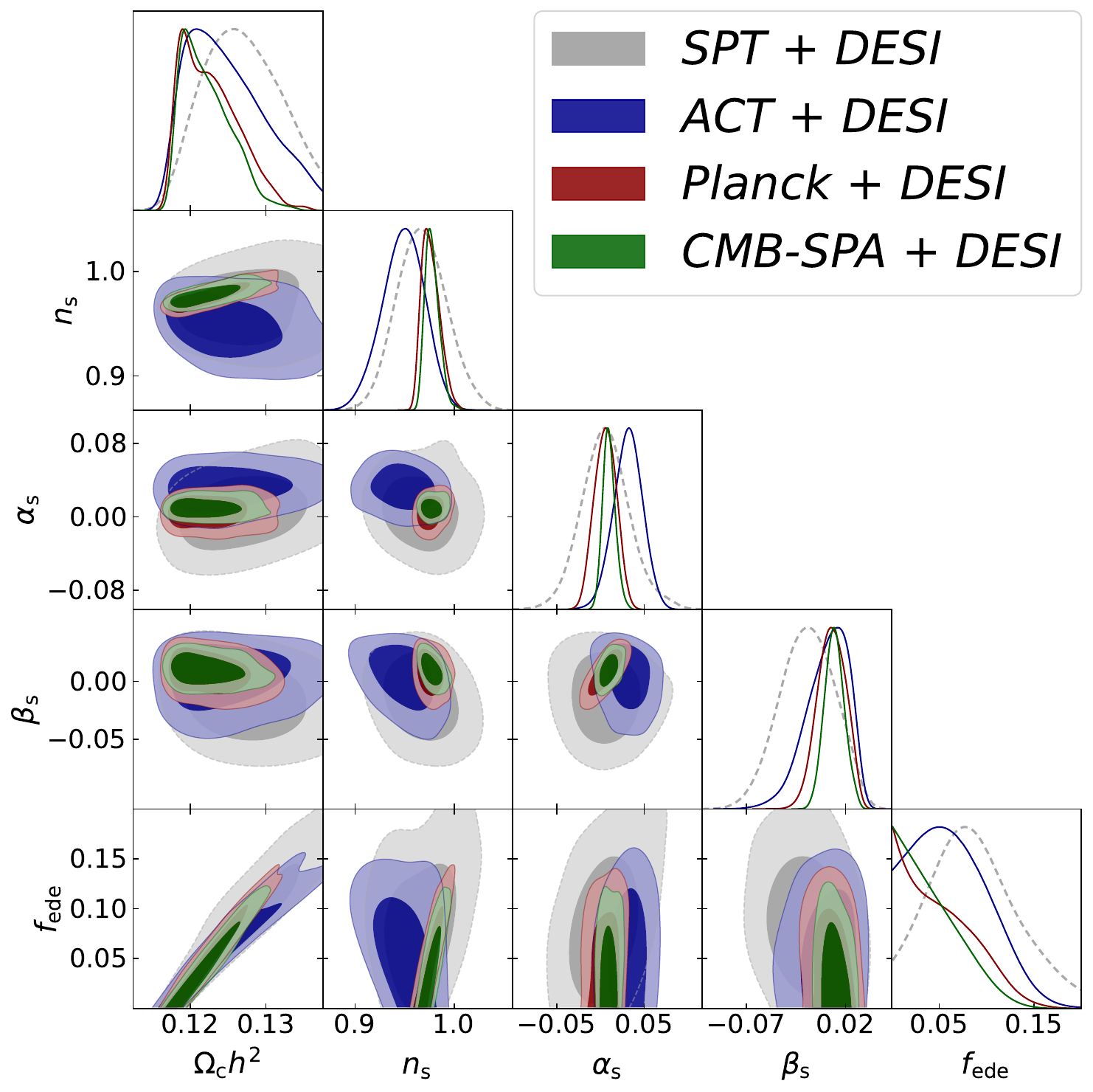}
\caption{Same as in Fig.~\ref{fig:edealphas}, but also including the running of the running $\beta_s$ as as free parameter. Doing so slightly broadens all posteriors (including those of $\alpha_s$) and redistributes the increase in small-scale power between $\alpha_s$ and $\beta_s$, but does not qualitatively alter our conclusions.}
\label{fig:edealphasbetas}
\end{figure*}

Our main results are summarized in four tables and three plots. In particular, constraints on cosmological parameters within the $\Lambda$CDM+$\alpha_s$ and EDE+$\alpha_s$ models are presented in Tab.~\ref{tab:alphasnodesi} for the four CMB-only dataset combinations, and in Tab.~\ref{tab:alphasdesi} when \textit{DESI} data is included. The counterparts of these tables for the cases where $\beta_s$ is varied as well are Tab.~\ref{tab:alphasbetasnodesi} and Tab.~\ref{tab:alphasbetasdesi} respectively. Corner plots showing constraints on selected EDE parameters within the EDE+$\alpha_s$ model, for all eight dataset combinations, are provided in Fig.~\ref{fig:edealphas}. Comparisons between the 1D $\alpha_s$ posteriors within the $\Lambda$CDM+$\alpha_s$ and EDE+$\alpha_s$ models are shown in Fig.~\ref{fig:alphas}. Finally, corner plots showing constraints on selected EDE parameters within the EDE+$\alpha_s$+$\beta_s$ model can be found in Fig.~\ref{fig:edealphasbetas}.

We begin by discussing the results obtained when, in addition to the standard $\Lambda$CDM and EDE parameters, only $\alpha_s$ is allowed to vary. For the four CMB-only dataset combinations, we find the results reported in Tab.~\ref{tab:alphasnodesi}. Within $\Lambda$CDM, and considering the three individual CMB experiments, we see that only \textit{ACT} reports the well-known $\simeq 2.2\sigma$ indication for positive running, $\alpha_s=0.033 \pm 0.015$. On the other hand, neither \textit{Planck} nor \textit{SPT} show a preference for non-zero running, with both central values being negative, and both inferences being consistent with $\alpha_s=0$ well within the $1\sigma$ level: $\alpha_s=-0.0042 \pm 0.0068$ and $\alpha_s=-0.004 \pm 0.023$ respectively (for recent discussions on this mild disagreement regarding the sign of $\alpha_s$ between different CMB probes, see for instance Ref.~\cite{Odintsov:2026dss}). Our most robust constraint on the running is obtained combining all three CMB experiments within the \textit{CMB-SPA} combination, from which we infer $\alpha_s=0.0054 \pm 0.0050$, i.e.\ a factor of $3$ improvement in sensitivity compared to \textit{ACT} alone. While the central value is positive, not only is it almost an order of magnitude smaller than the central value inferred from \textit{ACT} alone but, most importantly, this inference is now consistent with $\alpha_s=0$ at the $1.1\sigma$ level. To the best of our knowledge, this is among the first constraints on the running of the spectral index obtained combining data from all three CMB experiments.

\begin{table*}[!htb]
\footnotesize
\resizebox{\textwidth}{!}{
\begin{tabular}{|c?c|c||c|c||c|c||c|c|}
\hline
\multirow{2}{*}{\textbf{Parameter}} & \multicolumn{2}{c||}{\textbf{\textit{Planck}}} & \multicolumn{2}{c||}{\textbf{\textit{ACT}}} & \multicolumn{2}{c||}{\textbf{\textit{SPT}}} & \multicolumn{2}{c|}{\textbf{\textit{CMB-SPA}}} \\
\cline{2-9}
& $\Lambda$CDM & EDE & $\Lambda$CDM & EDE & $\Lambda$CDM & EDE & $\Lambda$CDM & EDE \\
\hline\hline
$n_s$ & $0.9616\pm 0.0045$ & $0.9670^{+0.0061}_{-0.0090}$ & $0.946\pm 0.012$ & $0.945^{+0.016}_{-0.014}$ & $0.955\pm 0.018$ & $0.955^{+0.019}_{-0.022}$ & $0.9690\pm 0.0033$ & $0.9746^{+0.0047}_{-0.0068}$ \\
$\alpha_s$ & $-0.0042\pm 0.0068$ & $-0.0031\pm 0.0076$ & $0.033\pm 0.015$ & $0.033\pm 0.017$ & $-0.004\pm 0.023$ & $0.011\pm 0.027$ & $0.0054\pm 0.0050$ & $0.0063\pm 0.0057$ \\
$\omega_c$ & $0.1199\pm 0.0011$ & $0.1228^{+0.0017}_{-0.0031}$ & $0.1236\pm 0.0024$ & $0.1262^{+0.0028}_{-0.0037}$ & $0.1209\pm 0.0016$ & $0.1252^{+0.0022}_{-0.0047}$ & $0.11997\pm 0.00098$ & $0.1230^{+0.0016}_{-0.0032}$ \\
$f_{\text{ede}}$ & -- & $< 0.082$ & -- & $< 0.087$ & -- & $< 0.144$ & -- & $< 0.082$ \\
\hline
$H_0\,[{\text{km}/{\text{s}}/{\text{Mpc}}}]$ & $67.19\pm 0.48$ & $68.14^{+0.66}_{-1.2}$ & $66.16\pm 0.87$ & $67.34^{+0.91}_{-1.7}$ & $66.89\pm 0.61$ & $68.44^{+0.68}_{-1.9}$ & $67.29\pm 0.40$ & $68.31^{+0.60}_{-1.1}$ \\
$\Omega_m$ & $0.316\pm 0.007$ & $0.314\pm 0.007$ & $0.335\pm 0.014$ & $0.330\pm 0.016$ & $0.321\pm 0.009$ & $0.317^{+0.013}_{-0.0097}$ & $0.316\pm 0.006$ & $0.313\pm 0.006$ \\
\hline
\end{tabular}}
\caption{$68\%$ credible intervals on the scalar spectral index and its running $n_s$ and $\alpha_s$, the physical cold dark matter density $\omega_c$, and the EDE fraction $f_{\text{ede}}$, obtained within the $\Lambda$CDM+$\alpha_s$ and EDE+$\alpha_s$ models, in light of the four CMB-only dataset combinations. When only upper limits are reported (as is the case for $f_{\text{ede}}$), these should be understood as $95\%$ upper limits.}
\label{tab:alphasnodesi}
\end{table*}

\begin{table*}[!htb]
\footnotesize
\resizebox{\textwidth}{!}{
\begin{tabular}{|c?c|c||c|c||c|c||c|c|}
\hline
\multirow{3}{*}{\textbf{Parameter}} & \multicolumn{8}{c|}{\textbf{\textit{+ DESI}}} \\
\cline{2-9}
& \multicolumn{2}{c||}{\textbf{\textit{Planck}}} & \multicolumn{2}{c||}{\textbf{\textit{ACT}}} & \multicolumn{2}{c||}{\textbf{\textit{SPT}}} & \multicolumn{2}{c|}{\textbf{\textit{CMB-SPA}}} \\
\cline{2-9}
& $\Lambda$CDM & EDE & $\Lambda$CDM & EDE & $\Lambda$CDM & EDE & $\Lambda$CDM & EDE \\
\hline\hline
$n_s$ & $0.9677\pm 0.0038$ & $0.9772^{+0.0067}_{-0.010}$ & $0.960\pm 0.011$ & $0.951\pm 0.018$ & $0.959\pm 0.017$ & $0.962\pm 0.024$ & $0.9739\pm 0.0029$ & $0.9825^{+0.0049}_{-0.0082}$ \\
$\alpha_s$ & $-0.0026\pm 0.0066$ & $-0.0002^{+0.0075}_{-0.0087}$ & $0.031\pm 0.016$ & $0.033^{+0.019}_{-0.017}$ & $-0.009\pm 0.025$ & $0.004\pm 0.027$ & $0.0050\pm 0.0049$ & $0.0064^{+0.0058}_{-0.0067}$ \\
$\omega_c$ & $0.11760\pm 0.00063$ & $0.1229^{+0.0025}_{-0.0047}$ & $0.11731\pm 0.00080$ & $0.1252^{+0.0034}_{-0.0069}$ & $0.11748\pm 0.00080$ & $0.1261^{+0.0040}_{-0.0063}$ & $0.11779\pm 0.00061$ & $0.1234^{+0.0026}_{-0.0046}$ \\
$f_{\text{ede}}$ & -- & $< 0.113$ & -- & $0.066^{+0.024}_{-0.057}$ & -- & $0.078^{+0.033}_{-0.055}$ & -- & $< 0.115$ \\
\hline
$H_0\,[{\text{km}/{\text{s}}/{\text{Mpc}}}]$ & $68.22\pm 0.28$ & $69.71^{+0.80}_{-1.3}$ & $68.50\pm 0.31$ & $70.7^{+1.0}_{-1.9}$ & $68.22\pm 0.32$ & $70.7^{+1.2}_{-1.8}$ & $68.19\pm 0.25$ & $69.82^{+0.86}_{-1.3}$ \\
$\Omega_m$ & $0.302\pm 0.004$ & $0.301\pm 0.004$ & $0.299\pm 0.004$ & $0.297\pm 0.004$ & $0.301\pm 0.004$ & $0.298\pm 0.004$ & $0.303\pm 0.003$ & $0.301\pm 0.004$ \\
\hline
\end{tabular}}
\caption{As in Tab.~\ref{tab:alphasnodesi}, but adding \textit{DESI} data to each of the four CMB-only dataset combinations.}
\label{tab:alphasdesi}
\end{table*}

A comment on the \textit{ACT} results within $\Lambda$CDM is in order before moving on. From Tab.~\ref{tab:alphasnodesi} one sees that, within the $\Lambda$CDM+$\alpha_s$ model, $n_s=0.946 \pm 0.012$ is inferred. This value may appear somewhat surprising, especially in light of the well-known \textit{ACT} preference for a bluer spectrum. Indeed, combining \textit{ACT} data with \textit{Planck} and BAO data leads to $n_s=0.974 \pm 0.003$~\cite{AtacamaCosmologyTelescope:2025blo}. We stress however that this usually quoted ``high $n_s$'' value is obtained \textit{within the baseline $\Lambda$CDM model, where $\alpha_s=0$}. Once varied, $\alpha_s$ is able to partially absorb the preference for larger values of $n_s$, as manifested by a clear negative correlation between $n_s$ and $\alpha_s$ (see e.g.\ Fig.~\ref{fig:edealphas} within EDE). This highlights how the relevant quantity is not simply the value of $n_s$ at \textit{ACT}'s pivot scale, but rather the \textit{convex} shape of the power spectrum preferred by \textit{ACT}, which gets increasingly blue towards the scales which are well probed by \textit{ACT} (or, with a slight abuse of language, the overall scale-dependent effective tilt). Another important point to note is that \textit{ACT} alone lacks measurements of the first acoustic peak, and therefore the lever arm required to robustly constrain $n_s$ and $\alpha_s$, unless it is combined with experiments (such as \textit{Planck}) which do measure large angular scales, as explicitly discussed in Sec.~4.1 of the \textit{ACT} DR6 extended models paper~\cite{AtacamaCosmologyTelescope:2025nti}. The \textit{ACT}-only primordial power spectra shown in Fig.~5 of the same paper further support this interpretation of \textit{ACT} data alone preferring a redder and more convex spectrum. As a sanity check, we have explicitly verified that we can reproduce the results of the \textit{ACT} public chains for exactly the same dataset combinations considered there, and that our treatment of low-multipole polarization (\texttt{Sroll2} likelihood versus Gaussian prior on $\tau$) has a negligible impact on these conclusions. We conclude that the inference of a substantially lower value of $n_s$ within the $\Lambda$CDM+$\alpha_s$ model is a consequence of \textit{ACT}'s scale coverage and intrinsic preference for a somewhat convex spectrum, alongside the $\alpha_s$-$n_s$ degeneracy, rather than an issue with our chains. This discussion highlights the importance of combining all three CMB datasets within the \textit{CMB-SPA} combination, to ensure wide scale coverage and the lever arm required to robustly measure parameters governing the shape of the primordial power spectrum: this is the reason why we regard the results obtained from the \textit{CMB-SPA} and \textit{CMB-SPA}+\textit{DESI} dataset combinations as being our most trustworthy ones.

When moving to EDE (see Fig.~\ref{fig:edealphas} for a corner plot showing our constraints on selected cosmological parameters) we observe that for the \textit{Planck}, \textit{SPT}, and \textit{CMB-SPA} combinations, $\alpha_s$ always increases (whereas for the \textit{ACT} dataset, the central value of $\alpha_s$ remains stable at the positive and relatively large value of $0.033$), although the increase is always very mild. Under the assumption of Gaussianity, which is well motivated (see Fig.~\ref{fig:alphas} for 1D $\alpha_s$ posteriors in $\Lambda$CDM and EDE for all dataset combinations considered), the increase is by $0.1\sigma$, $0.5\sigma$, and $0.1\sigma$ respectively. For reference, while remaining stable within \textit{ACT} and \textit{SPT}, for the \textit{Planck} and \textit{CMB-SPA} dataset combinations, $n_s$ increases by $\approx 0.7\sigma$ and $\approx 1.0\sigma$ respectively. The increases in $\alpha_s$ are also accompanied by a $10$--$20\%$ broadening of the uncertainties, due to the marginalization over the 3 additional EDE parameters ($f_{\text{ede}}$, $z_c$, and $\theta_i$). Our most robust constraint on the running within EDE is obtained within the \textit{CMB-SPA} dataset combination, from which we infer $\alpha_s=0.0063 \pm 0.0057$.

Once \textit{DESI} data is included, we find the results reported in Tab.~\ref{tab:alphasdesi}. Up to a shift in $n_s$ towards larger values, which is consistent with the findings of Ref.~\cite{McDonough:2025lzo} and is inevitably intertwined with \textit{DESI}'s hints for dynamical dark energy, the results we find are qualitatively consistent with those from the \textit{DESI}-free dataset combinations. In particular, for all four dataset combinations we observe a mild increase in $\alpha_s$ (at most at the $0.4\sigma$ level for the \textit{SPT}+\textit{DESI} dataset combination) accompanied by a $10$--$20\%$ broadening of the associated uncertainties. From the \textit{CMB-SPA}+\textit{DESI} dataset combination, we infer $\alpha_s=0.0050 \pm 0.0049$ and $\alpha_s=0.0064^{+0.0058}_{-0.0067}$ within $\Lambda$CDM and EDE respectively. These two figures, which to the best of our knowledge represent the first constraints on $\alpha_s$ from a combination of all available state-of-the-art cosmological data, are essentially consistent with no running within $1\sigma$. We note (see both Tab.~\ref{tab:alphasnodesi} and Tab.~\ref{tab:alphasdesi}) that for all eight dataset combinations considered, no detection of EDE at $\gtrsim 1.5\sigma$ is observed, with our tightest $2\sigma$ upper limit being $f_{\text{ede}}<0.082$ from the \textit{CMB-SPA} dataset combination. For the \textit{CMB-SPA}+\textit{DESI} dataset combination, the value of $H_0=69.82^{+0.86}_{-1.30}\,{\text{km}}/{\text{s}}/{\text{Mpc}}$ we find is in $\simeq 3.1\sigma$ Gaussian tension with the latest local distance network (H0DN) determination of $H_0=(73.50 \pm 0.81)\,{\text{km}}/{\text{s}}/{\text{Mpc}}$~\cite{H0DN:2025lyy}.

\begin{table*}[!htb]
\footnotesize
\resizebox{\textwidth}{!}{
\begin{tabular}{|c?c|c||c|c||c|c||c|c|}
\hline
\multirow{2}{*}{\textbf{Parameter}} & \multicolumn{2}{c||}{\textbf{\textit{Planck}}} & \multicolumn{2}{c||}{\textbf{\textit{ACT}}} & \multicolumn{2}{c||}{\textbf{\textit{SPT}}} & \multicolumn{2}{c|}{\textbf{\textit{CMB-SPA}}} \\
\cline{2-9}
& $\Lambda$CDM & EDE & $\Lambda$CDM & EDE & $\Lambda$CDM & EDE & $\Lambda$CDM & EDE \\
\hline\hline
$n_s$ & $0.9598\pm 0.0048$ & $0.9649^{+0.0056}_{-0.0085}$ & $0.948\pm 0.014$ & $0.948\pm 0.018$ & $0.955\pm 0.017$ & $0.960\pm 0.022$ & $0.9648\pm 0.0042$ & $0.9694^{+0.0054}_{-0.0071}$ \\
$\alpha_s$ & $0.0044^{+0.011}_{-0.0092}$ & $0.004\pm 0.011$ & $0.033\pm 0.015$ & $0.034\pm 0.017$ & $0.003^{+0.024}_{-0.021}$ & $0.008\pm 0.026$ & $0.00996\pm 0.0055$ & $0.0109^{+0.0063}_{-0.0072}$ \\
$\beta_s$ & $0.013^{+0.013}_{-0.011}$ & $0.011^{+0.014}_{-0.010}$ & $-0.006\pm 0.022$ & $-0.005^{+0.026}_{-0.018}$ & $-0.026^{+0.024}_{-0.027}$ & $-0.026\pm 0.026$ & $0.0145\pm 0.0084$ & $0.0138\pm 0.0087$ \\
$\omega_c$ & $0.1200\pm 0.0011$ & $0.1223^{+0.0014}_{-0.0027}$ & $0.1235\pm 0.0024$ & $0.1261^{+0.0027}_{-0.0038}$ & $0.1212\pm 0.0016$ & $0.1253^{+0.0021}_{-0.0042}$ & $0.12019\pm 0.00098$ & $0.1227^{+0.0013}_{-0.0029}$ \\
$f_{\text{ede}}$ & -- & $< 0.068$ & -- & $< 0.092$ & -- & $< 0.123$  & -- & $< 0.082$ \\
\hline
$H_0\,[{\text{km}/{\text{s}}/{\text{Mpc}}}]$ & $67.11^{+0.47}_{-0.52}$ & $67.93^{+0.58}_{-1.1}$ & $66.20\pm 0.87$ & $67.47^{+0.98}_{-1.8}$ & $66.73\pm 0.61$ & $68.15^{+0.73}_{-1.8}$ & $67.19\pm 0.41$ & $68.06^{+0.44}_{-1.1}$ \\
$\Omega_m$ & $0.317\pm 0.007$ & $0.315\pm 0.007$ & $0.335\pm 0.014$ & $0.328\pm 0.016$ & $0.324\pm 0.009$ & $0.319^{+0.013}_{-0.011}$ & $0.317^{+0.005}_{-0.006}$ & $0.315^{+0.007}_{-0.006}$ \\
\hline
\end{tabular}}
\caption{As in Tab.~\ref{tab:alphasnodesi}, but also including the running of the running $\beta_s$ as a free parameter.}
\label{tab:alphasbetasnodesi}
\end{table*}

\begin{table*}[!htb]
\footnotesize
\resizebox{\textwidth}{!}{
\begin{tabular}{|c?c|c||c|c||c|c||c|c|}
\hline
\multirow{3}{*}{\textbf{Parameter}} & \multicolumn{8}{c|}{\textbf{\textit{DESI+}}} \\
\cline{2-9}
& \multicolumn{2}{c||}{\textbf{\textit{Planck}}} & \multicolumn{2}{c||}{\textbf{\textit{ACT}}} & \multicolumn{2}{c||}{\textbf{\textit{SPT}}} & \multicolumn{2}{c|}{\textbf{\textit{CMB-SPA}}} \\
\cline{2-9}
& $\Lambda$CDM & EDE & $\Lambda$CDM & EDE & $\Lambda$CDM & EDE & $\Lambda$CDM & EDE \\
\hline\hline
$n_s$ & $0.9667\pm 0.0038$ & $0.9756^{+0.0072}_{-0.010}$ & $0.960\pm 0.013$ & $0.950\pm 0.020$ & $0.961\pm 0.018$ & $0.967\pm 0.025$ & $0.9704\pm 0.0038$ & $0.9774^{+0.0058}_{-0.0080}$ \\
$\alpha_s$ & $0.0039\pm 0.0096$ & $0.004\pm 0.013$ & $0.031\pm 0.016$ & $0.031^{+0.018}_{-0.015}$ & $-0.011\pm 0.025$ & $0.006^{+0.025}_{-0.028}$  & $0.0088\pm 0.0057$ & $0.0097^{+0.0063}_{-0.0077}$ \\
$\beta_s$ & $0.011\pm 0.012$ & $0.007\pm 0.013$ & $0.002^{+0.025}_{-0.018}$ & $0.004^{+0.023}_{-0.014}$ & $-0.008\pm 0.025$ & $-0.014\pm 0.025$ 
& $0.0119\pm 0.0086$ & $0.0098\pm 0.0089$ \\
$\omega_c$ & $0.11761\pm 0.00063$ & $0.1226^{+0.0024}_{-0.0049}$ & $0.11729\pm 0.00074$ & $0.1247^{+0.0033}_{-0.0065}$ & $0.11747\pm 0.00082$ & $0.1274^{+0.0046}_{-0.0065}$ & $0.11788\pm 0.00062$ & $0.1220^{+0.0020}_{-0.0040}$ \\
$f_{\text{ede}}$ & -- & $< 0.121$ & -- & $0.062^{+0.022}_{-0.055}$ & -- & $0.089^{+0.036}_{-0.058}$ & -- & $< 0.101$ \\
\hline
$H_0\,[{\text{km}/{\text{s}}/{\text{Mpc}}}]$ & $68.20\pm 0.29$ & $69.63^{+0.81}_{-1.4}$ & $68.52\pm 0.30$ & $70.5^{+1.0}_{-1.8}$ & $68.21\pm 0.31$ & $71.0^{+1.3}_{-1.8}$ & $68.15^{+0.25}_{-0.28}$ & $69.38^{+0.63}_{-1.1}$ \\
$\Omega_m$ & $0.302\pm 0.004$ & $0.301\pm 0.004$ & $0.299\pm 0.004$ & $0.297\pm 0.004$ & $0.302\pm 0.004$ & $0.298\pm 0.004$ & $0.304\pm 0.004$ & $0.302\pm 0.004$ \\
\hline
\end{tabular}}
\caption{As in Tab.~\ref{tab:alphasdesi}, but also including the running of the running $\beta_s$ as a free parameter.}
\label{tab:alphasbetasdesi}
\end{table*}

Allowing $\beta_s$ to vary does not qualitatively change our conclusions: see Tab.~\ref{tab:alphasbetasnodesi} and Tab.~\ref{tab:alphasbetasdesi} for the obtained parameter constraints, and the corner plots of Fig.~\ref{fig:edealphasbetas}. We note that once $\beta_s$ is freed, the inferred values of $\alpha_s$ become positive for most dataset combinations, even within $\Lambda$CDM. The trend of $\alpha_s$ increasing when moving from $\Lambda$CDM to EDE overall persists, but is even milder than previously. This is unsurprising, since part of the role previously played by $\alpha_s$ in providing more small-scale power to counteract the extra damping due to the larger $\theta_d$, is now ``redistributed'' between $\alpha_s$ and $\beta_s$. We also note that the $n_s$ trends are altered to a more significant extent, with $n_s$ in some cases decreasing when moving from $\Lambda$CDM to EDE, although not in the case where all three CMB datasets are combined. For what concerns $\beta_s$ itself, we observe no obvious trend when moving from $\Lambda$CDM to EDE, with the shifts depending on the specific dataset adopted. From the \textit{CMB-SPA} dataset combination, we infer $\beta_s=0.0145 \pm 0.0084$ and $\beta_s=0.0138 \pm 0.0087$ within $\Lambda$CDM and EDE respectively, in both cases consistent with $\beta_s=0$ within $\approx 1.5\sigma$. Once \textit{DESI} data is added, the constraints we obtain within $\Lambda$CDM and EDE are $\beta_s=0.0119 \pm 0.0086$ and $\beta_s=0.0098 \pm 0.0089$ respectively, again consistent with no running of the running within $\approx 1.5\sigma$. We therefore conclude that our overall findings are only marginally affected by the choice of varying $\beta_s$, and remain robust against this parameter extension.

Summing up, the overall increase observed in $\alpha_s$ is consistent with the theoretical expectation we laid out previously in Sec.~\ref{subsec:inflationarynewphysics}. In particular, the larger value of the running increases the amount of small-scale power, which compensates the extra damping due to the larger value of $\theta_d$. For our most robust dataset combinations involving all three CMB datasets, despite the positive central values, $\alpha_s$ remains consistent with no running, albeit slightly disfavoring negative values. Compared to the inference of $n_s$, the impact of the CMB-BAO tension, and correspondingly \textit{DESI}'s hints for dynamical dark energy (see Ref.~\cite{McDonough:2025lzo}), are much milder for what concerns $\alpha_s$, as one could have expected. If we were to summarize our results in one simple sentence, this would be: \\
\textit{there is certainly no evidence for non-zero running from state-of-the-art cosmological data; there may be extremely mild hints, if at all, for a positive running}.

\section{Discussion}
\label{sec:discussion}

We now discuss the implications of our results for the viability of USR models of PBH formation, in light of the constraints on $\alpha_s$ obtained across all dataset combinations and cosmological models considered in this work. The central question is whether the tension between the negative running generically predicted by USR inflation and the constraints from CMB data can be alleviated by invoking pre-recombination new physics motivated by the Hubble tension.

Although the precise relation between $\alpha_s$ and the mass of the PBHs produced in a given USR model depends on the specific inflationary potential, Ref.~\cite{Allegrini:2025jha} has shown that for a broad class of polynomial single-field models this relation can be cast in the following remarkably simple power-law form:
\begin{equation}
\alpha_s \simeq a_1 \left(\frac{M_{\rm PBH}}{M_{\odot}}\right)^{a_2}\,,
\label{eq:alphas_mass}
\end{equation}
where the coefficients $(a_1, a_2)$ depend on the specific model considered. Two particularly well-studied cases are minimally coupled (MC) and non-minimally coupled (NMC) polynomial inflation~\cite{Ballesteros:2017fsr,Kannike:2017bxn,Ballesteros:2020qam,Frosina:2023nxu,Allegrini:2025jha,Allegrini:2024ooy}, for which one finds $(a_1,\,a_2) \sim (-0.035,\,-0.046)$ and $(a_1,\,a_2) \sim (-0.018,\,-0.035)$ respectively. In both cases the coefficients $a_1 < 0$ and $a_2 < 0$, reflecting two important physical facts: first, USR models predict a negative running at all masses  ($\alpha_s < 0$ for all $M_{\rm PBH}$); second, the magnitude $\vert \alpha_s \vert$ increases as $M_{\rm PBH}$ increases~\cite{Allegrini:2025jha}. This is a direct consequence of the fact that heavier PBHs require the USR feature to sit closer to the CMB window (smaller $N_1$), therefore leaving a more significant imprint on the large-scale power spectrum.

It is important to stress that the relation shown in Eq.~(\ref{eq:alphas_mass}) carries a degree of model-dependence that goes beyond the choice of $(a_1, a_2)$. In particular, the coefficients reported above were derived under specific assumptions on the amplitude of the curvature power spectrum peak, i.e.\ the requirement that $\mathcal{P}_\zeta(k_{\rm pk}) \sim 10^{-2}$ in order to produce a sizeable PBH abundance. Since the amplitude of the peak is not a free parameter but is rather constrained by the requirement of a given $f_{\rm PBH}$, variations in the assumed PBH dark matter fraction will translate into mild shifts of the coefficients $(a_1, a_2)$. Nevertheless, the qualitative picture, i.e.\ negative running increasing in magnitude towards heavier PBH masses, is robust and model-independent within the class of single-field USR models based on inflection-point potentials.

\begin{figure*}[!htb]
\centering
\includegraphics[width=0.99\linewidth]{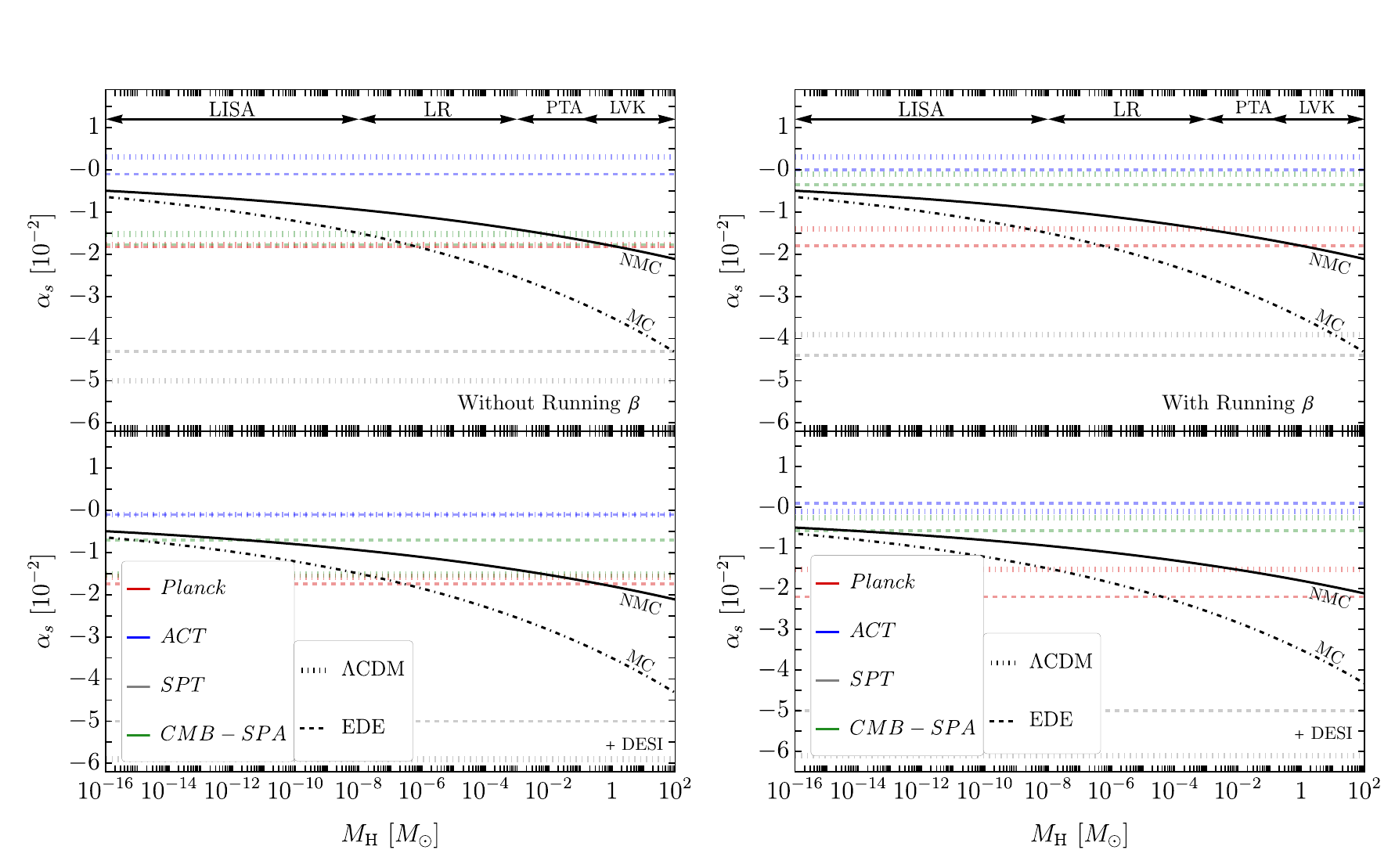}
\caption{Theoretical predictions for the running of the spectral index $\alpha_s$ as a function of the PBH horizon mass $M_H$, for the minimally coupled (MC, black dash-dotted curves) and non-minimally coupled (NMC, black solid curves) polynomial USR inflation models. The horizontal bands show the $2\sigma$ lower limits on $\alpha_s$ inferred from each dataset combination, both within $\Lambda$CDM (dotted horizontal lines) and EDE (dashed horizontal lines), with the legend color coding indicating the CMB dataset adopted. The lower (upper) sub-panels correspond to the cases which include (do not include) \textit{DESI} BAO data, whereas the right (left) sub-panels correspond to the cases which include (do not include) the running of the running $\beta_s$ as a free parameter. The arrows at the top are indicative of the PBH mass ranges probed by different experimental targets: LISA~\cite{Garcia-Bellido:2016dkw,Garcia-Bellido:2017aan,Bartolo:2018evs,Bartolo:2018rku,Iovino:2025cdy,Hong:2026rcl} (asteroid masses, $M_H\sim 10^{-18}-10^{-8}\,M_{\odot}$), laser ranging (LR) experiments~\cite{Blas:2026xws,Foster:2025nzf} ($M_H\sim 10^{-8}-10^{-3}\,M_{\odot}$), pulsar timing arrays (PTA)~\cite{Iovino:2024tyg} ($M_H\sim 10^{-3}-10^{2}\,M_{\odot}$), and LIGO-Virgo-KAGRA ~\cite{LIGOScientific:2018mvr,LIGOScientific:2020ibl,KAGRA:2021vkt,LIGOScientific:2025slb} (LVK, solar masses, $M_H\sim 1-10^{2}\,M_{\odot}$). USR PBH models are viable, in light of a given dataset combination and within a given cosmological model, only within the $M_H$ region where the theoretical curves lie \textit{above} the corresponding $2\sigma$ lower limit on $\alpha_s$. We therefore recommend reading the plot from right to left: starting from the largest masses, a model becomes viable once the theoretical curve crosses the horizontal line of interest, with the viable $M_H$ region lying to the left of this crossing.}
\label{fig:pbhs}
\end{figure*}

In Fig.~\ref{fig:pbhs}, we translate our constraints on $\alpha_s$ into implications for PBH formation, by overplotting the theoretical predictions of Eq.~(\ref{eq:alphas_mass}) for the NMC (black solid curves) and MC (black dash-dotted curves) models as a function of $M_{\text{H}}/M_{\odot}$ (where $M_{\text{H}}$ is the PBH horizon mass) against the $2\sigma$ lower limits on $\alpha_s$ inferred from the different dataset combinations, shown as horizontal bands. The arrows at the top of the figure indicate the approximate PBH mass ranges to which different classes of experiments are sensitive: asteroid-mass PBHs ($M_{\rm PBH} \sim 10^{-18}$--$10^{-8}\,M_{\odot}$) are the target of space-based interferometers such as LISA~\cite{Garcia-Bellido:2016dkw,Garcia-Bellido:2017aan,Bartolo:2018evs,Bartolo:2018rku,Iovino:2025cdy,Hong:2026rcl}; laser ranging (LR) experiments~\cite{Blas:2026xws,Foster:2025nzf} probe the mass range $M_{\rm PBH} \sim 10^{-8}$--$10^{-3}\,M_{\odot}$, while microlensing surveys~\cite{MACHO:2000qbb,Jedamzik:1998hc,Niikura:2019kqi,Niikura:2017zjd,Hawkins:2020zie,Gorton:2022fyb,Hawkins:2022vqo,Hawkins:2025mlo} are sensitive across the broader window $M_{\rm PBH} \sim 10^{-12}$--$1\,M_{\odot}$, thus overlapping with both the LISA, LR and PTA ranges; sub-solar- to solar-mass PBHs ($M_{\rm PBH} \sim 10^{-3}$--$10^{2}\,M_{\odot}$) source a SIGW signal in the nHz band observable by PTAs such as NANOGrav~\cite{Iovino:2024tyg}; finally, solar-mass PBHs ($M_{\rm PBH} \sim 1$--$10^{2}\,M_{\odot}$) are the primary target of ground-based GW interferometers, such as those of LVK~\cite{LIGOScientific:2018mvr,LIGOScientific:2020ibl,KAGRA:2021vkt,LIGOScientific:2025slb}.

A particularly instructive way to read Fig.~\ref{fig:pbhs} is to trace the fate of USR models as a function of PBH mass, moving from right to left along the horizontal axis, and to ask, for each dataset combination and cosmological model, which portion of the PBH mass spectrum remains within the $2\sigma$ lower limits allowed region of $\alpha_s$. USR PBH models are viable, in light of a given dataset combination and within a given cosmological model, only within the $M_H$ region where the theoretical curves lie \textit{above} the corresponding $2\sigma$ lower limit on $\alpha_s$. Starting from the largest masses to the right, a model therefore becomes viable once the theoretical curve crosses the horizontal line of interest, with the viable $M_H$ region lying to the left of this crossing.

We begin with the \textit{Planck} dataset, which provides the most conservative constraints. Within $\Lambda$CDM, the $2\sigma$ lower limit on $\alpha_s$ is consistent with the NMC prediction across nearly the entire mass range shown, excluding only PBHs heavier than the sun, and with the MC prediction up to planetary masses. The inclusion of EDE shifts $\alpha_s$ downward by a small amount, consistent with the mild $0.1\sigma$ shift reported in Sec.~\ref{sec:results}, leaving the allowed mass windows largely unchanged.  The \textit{Planck} dataset therefore allows asteroid-mass USR PBH models as viable dark matter candidates, and leaves a meaningful allowed window in the planetary mass range, irrespective of whether $\Lambda$CDM or EDE is assumed. Moreover, when the running of the running $\beta_s$ is also freed, the situation does not change appreciably for both models.

The \textit{SPT} dataset presents a qualitatively different situation. The uncertainties on $\alpha_s$ from \textit{SPT} alone are sufficiently large, with the $2\sigma$ limits spanning a range of order $\sim 0.05$, such that both the MC and NMC theoretical curves lie within the allowed region across essentially the entire mass range of interest, from asteroid masses up to solar masses. This holds both within $\Lambda$CDM and EDE, and even when $\beta_s$ is freed. The \textit{SPT} dataset is therefore currently not constraining enough to discriminate between viable and excluded USR PBH models across the mass range of interest, and no meaningful exclusion can be derived from it in isolation.

Moving to the \textit{ACT} dataset, the constraints become significantly more stringent. The strong preference for positive running, $\alpha_s = 0.033 \pm 0.015$~\cite{AtacamaCosmologyTelescope:2025nti}, places the $2\sigma$ lower limit well above zero, in severe tension with the negative values predicted by both the MC and NMC models across all masses. Notably, the inclusion of EDE leaves the central value of $\alpha_s$ essentially unchanged within \textit{ACT} (as discussed in Sec.~\ref{sec:results}), so the tension persists regardless of the assumed pre-recombination expansion history. Within \textit{ACT}, USR models for PBH formation are disfavored at more than $2\sigma$ across the entire mass range shown, from asteroid to solar masses, with no meaningful allowed window remaining.

Considering the \textit{CMB-SPA} dataset, within $\Lambda$CDM, the $2\sigma$ lower limit on $\alpha_s$ already excludes both the MC and NMC models in the solar-mass and sub-solar-mass ranges, while leaving the asteroid-mass window marginally open. When EDE is assumed, however, $\alpha_s$ shifts upward with the inclusion of \textit{DESI}, unlike the downward shift observed for most other dataset combinations, by approximately $0.1\sigma$, enough to close even the asteroid-mass window for both models.  Allowing $\beta_s$ to vary within the \textit{CMB-SPA} dataset combination further modifies the picture: the redistribution of small-scale power between $\alpha_s$ and $\beta_s$ shifts the inferred constraints in a way that completely closes the asteroid-mass window within both $\Lambda$CDM and EDE.

Taken together, these results paint a consistent but nuanced picture. The allowed mass window for USR PBH models depends sensitively on the adopted CMB dataset and, to a lesser extent, on the assumed pre-recombination expansion history. The \textit{Planck} dataset is the most permissive, allowing asteroid and sub-solar masses robustly, with the main caveat being the MC model when $\beta_s$ is freed. The \textit{SPT} dataset is too uncertain to be constraining. The \textit{ACT} dataset is the most restrictive excluding the entire range of PBH masses regardless of the pre-recombination expansion history. In the case of \textit{CMB-SPA} the precise prediction depends on the assumed pre-recombination expansion history, with the overall picture that sub-solar mass PBHs, or heavier ones, are excluded in both models.

Before closing, three important caveats of our analysis and a broader methodological point are worthy of further discussions. The first caveat is that, while we have considered a relatively wide range of dataset combinations, our conclusions are still contingent on the set of (CMB and BAO) datasets adopted. Constraints on the running have also been obtained from other datasets, e.g.\ the Lyman-$\alpha$ forest, with varying conclusions~\cite{Fairbairn:2025fko}. While perhaps stating the obvious, we stress that our findings could of course change should one choose to adopt different datasets. At the same time, the breadth of CMB combinations considered in the present work makes our analysis a useful baseline against which constraints on $\alpha_s$ from complementary probes can be compared.

The second caveat is that we have focused on a specific model of Hubble tension-motivated pre-recombination new physics, i.e.\ axion-like EDE, which we have shown generically leads to an increase in $\alpha_s$. The skeptical reader may object that this conclusion is specifically valid within axion-like EDE, and strictly speaking they would not be wrong. However, the argument according to which pre-recombination new physics motivated to address the Hubble tension leads to an increase in $\theta_d$, and with it the amount of damping, is quite generic (see e.g.\ the discussion in Sec.~V of the seminal Ref.~\cite{Knox:2019rjx}). This is true at least for the classes of new physics models considered so far, and can be avoided only by invoking a quite specific shape for the modified expansion history (see Fig.~1 of Ref.~\cite{Pedrotti:2026dwj}), which however does not match any model constructed so far. For this reason, we actually have good reason to expect that our results should apply more broadly beyond axion-like EDE, although checking this model-by-model would be too computationally expensive, and is beyond the scope of our paper.

Nonetheless, as further support to this qualitative argument, it is worth considering the results of Ref.~\cite{Garny:2026gcs}, which studied two related dark radiation extensions (with $\Delta N_{\text{eff}}>0$) injecting extra energy in the pre-recombination era in order to address the Hubble tension. These include a class of models where the dark radiation fluid is initially coupled to dark matter, and undergoes dark radiation-matter decoupling close to matter-radiation equality, as well as the self-interacting dark radiation subset of this class. While the scope of the work was rather different (and arguably complementary) to ours, i.e.\ to show that allowing for $\alpha_s$ and $\beta_s$ to vary substantially loosens the otherwise very tight constraints on $\Delta N_{\text{eff}}$ from \textit{ACT}, their results indeed support the trend we observed in axion-like EDE of $\alpha_s$ increasing in the presence of pre-recombination new physics. In particular, Ref.~\cite{Garny:2026gcs} finds a positive correlation between $\alpha_s$ and $\Delta N_{\text{eff}}$ (see their Fig.~1), which unsurprisingly maps onto a positive correlation between $\alpha_s$ and $H_0$ (see their Fig.~3). This is in line with the physical picture we discussed previously: extra pre-recombination energy injection increases the amount of damping, which can be partially compensated by positive running. These results provide independent, although not conclusive, support for the idea that the upward shift in $\alpha_s$ may not be specific to axion-like EDE, but rather a broader feature of pre-recombination solutions which increase the photon diffusion angular scale: a definitive statement would require a model-by-model analysis, which is beyond the scope of this work.

The third caveat, and arguably the most important one for what concerns the PBH side of our analysis, concerns the class of inflationary models to which our conclusions apply. As anticipated in Sec.~\ref{sec:introduction}, the prediction $\alpha_s<0$ is not a statement about PBH formation in general. It rests on two assumptions: that a single field is dynamically relevant during inflation, and that the enhancement of the curvature power spectrum originates from a (near-)inflection point of the potential which the inflaton must traverse, rather than from a feature engineered directly at small scales. It is precisely these assumptions that tie the small-scale peak to the CMB-scale trajectory: the same Taylor coefficients of the potential around the feature govern both, and cannot be adjusted independently. We stress that this is not a peculiarity of polynomial potentials, which we adopted merely as a concrete and calculable representative: the expansion leading to Eq.~(\ref{eq:alpha_s_usr}) only requires that the feature be a non-stationary inflection point, and applies equally to the Higgs- and fibre-inflation-inspired realizations~\cite{Cicoli:2018asa,Garcia-Bellido:2017mdw,Germani:2017bcs}, as well as to the modified Starobinsky construction of Ref.~\cite{Frolovsky:2025iao}. Conversely, our conclusions can be evaded, and we make no claim about, models which break the link between small and large scales. Two obvious examples are multi-field scenarios, most notably the curvaton~\cite{Linde:1996gt,Enqvist:2001zp,Lyth:2001nq,Moroi:2001ct,Allegrini:2026jqt}, where the curvature perturbation on CMB scales and the small-scale peak originate from different fields, and single-field constructions in which a localized bump or dip is added by hand to an otherwise featureless potential. In the latter case the feature can in principle be placed arbitrarily far from the CMB window without affecting large-scale observables, although the required shape is typically postulated at the level of the effective potential rather than derived, and its ultraviolet origin is usually left unspecified. We therefore regard the class of models we consider as the physically best-motivated and most widely studied one, but not as exhaustive, and we caution the reader against extrapolating our conclusions beyond it.

This brings us to a broader methodological point. Constraints on inflationary parameters such as $n_s$ and $\alpha_s$ should \textit{not} be interpreted as direct measurements of the primordial power spectrum independently of the assumed cosmological model and primordial power spectrum parametrization, and adopted dataset. Nevertheless, it is useful to distinguish between three effects which are sometimes mixed up in the literature:~\footnote{In the literature, there has been some confusion particularly for what concerns the first two points. This is perhaps understandable, since the strong \textit{ACT} DR4 preference for larger values of $n_s$ and the possible preference for EDE in the same dataset appeared almost simultaneously. These two effects are not necessarily different aspects of the same physical trend, and are in principle logically distinct. The distinction has become clearer with \textit{ACT} DR6 data, where the preference for non-zero $f_{\text{ede}}$ has disappeared, while the preference for larger values of $n_s$ somewhat survived, albeit less strongly than previously.}
\begin{enumerate}
\item the dataset dependence of constraints on inflationary parameters, e.g.\ the \textit{ACT} preference for larger $n_s$ (more evident in \textit{ACT} DR4, but still evident in \textit{ACT} DR6);
\item the model-dependence, or more precisely early-time physics dependence, of constraints on inflationary parameters;
\item the more general dependence of inflationary constraints on the assumed primordial power spectrum, for instance the choice of whether $\alpha_s$ and $\beta_s$ are allowed to vary.
\end{enumerate}
As our work explicitly shows, constraints on inflationary parameters are particularly sensitive to the pre-recombination expansion history, which is itself a major open issue in cosmology at the moment, in light of the Hubble tension. Similar considerations hold for the CMB-BAO tension, whose impact on $n_s$ constraints has been extensively discussed in Ref.~\cite{McDonough:2025lzo}: there it has been shown that current constraints on $n_s$ obtained from combinations including \textit{DESI} data are inevitably entangled with \textit{DESI}'s preference for dynamical DE, and more generally the tension between CMB and BAO data, which manifests also, though not only, in differences among the preferred values of $\Omega_m$~\cite{Colgain:2024ksa,Weiner:2026sfm,Shlivko:2026jxa}. Although we have stressed these points several times throughout our work, they bear repeating, as they are particularly relevant for the use of CMB constraints in inflationary model building. To be as blunt as possible, we believe that \textit{inflationary model selection based on $n_s$, $\alpha_s$, and $\beta_s$ is premature until the cause of the Hubble tension (as well as the CMB-BAO tension) is understood}, a point which has also been stressed in other works in the cosmology literature~\cite{Gerbino:2016sgw,DiValentino:2018zjj,Ye:2021nej,Takahashi:2021bti,Forconi:2021que,Giare:2022rvg,Ye:2022efx,RoyChoudhury:2022rva,Jiang:2022uyg,Jiang:2022qlj,Lin:2022gbl,Hazra:2022rdl,Giare:2023kiv,Giare:2023wzl,Jiang:2023bsz,Peng:2023bik,Bostan:2023ped,Forconi:2023hsj,Fu:2023tfo,Giare:2024akf,Giare:2024sdl,Wang:2024tjd,daCosta:2024grm,Forconi:2025zzu,Peng:2025tqt,Balkenhol:2025wms,Yuan:2026xcg,Garny:2026gcs,Oikonomou:2026ris,Giare:2026oti}. Such model selection is, at best, conditional on the cosmological model assumed in obtaining these constraints, and this is especially important in the inflationary model-building literature. While $n_s$, $\alpha_s$, and $\beta_s$ are without doubt extremely powerful probes of inflation, their interpretation as setting constraints on the inflationary Lagrangian requires specifying the assumed post-inflationary cosmological model.

\section{Conclusions}
\label{sec:conclusions}

Single-field ultra-slow roll inflation is one of the most studied PBH formation mechanisms. This class of models has been argued to predict a negative running of the spectral index ($\alpha_s<0$), in potential tension with recent \textit{ACT} data showing a $\simeq 2.2\sigma$ preference for $\alpha_s>0$~\cite{AtacamaCosmologyTelescope:2025nti}. However, inflationary parameters (such as $n_s$ and $\alpha_s$) inferred from cosmological observations are especially sensitive to the assumed pre-recombination expansion history, which is exactly where modifications are required to address the Hubble tension, if this is indeed to be taken as an indication of new physics~\cite{Bernal:2016gxb,Addison:2017fdm,Lemos:2018smw,Aylor:2018drw,Schoneberg:2019wmt,Knox:2019rjx,Arendse:2019hev,Efstathiou:2021ocp,Cai:2021weh,Keeley:2022ojz}. With these premises in mind, the goal of our work has been a simple one: to study the impact of the Hubble tension, and in particular of pre-recombination new physics, on the tension between single-field USR models of PBH formation and current CMB constraints on $\alpha_s$. On the model side, we have focused on axion-like EDE, as a concrete and widely studied benchmark model for Hubble tension-motivated pre-recombination new physics. On the data side, we have considered CMB data from \textit{Planck}, \textit{ACT}, \textit{SPT}, and the joint \textit{CMB-SPA} combination, with and without \textit{DESI} DR2 BAO data. Overall, our analysis includes 32 MCMC runs (eight dataset combinations, within $\Lambda$CDM and axion-like EDE, with half of the runs treating the running of the running $\beta_s$ as a free parameter), so a byproduct of our work is a set of up-to-date constraints on $\alpha_s$ from recent cosmological data.

Our main findings can be summarized as follows. Within $\Lambda$CDM, only \textit{ACT} shows the well-known $\simeq 2.2\sigma$ preference for positive running, while \textit{Planck} and \textit{SPT} are individually consistent with $\alpha_s=0$ well within $1\sigma$. Combining all three CMB experiments, we find $\alpha_s = 0.0054 \pm 0.0050$, consistent with no running at the $1.1\sigma$ level. Adding \textit{DESI} data or freeing up $\beta_s$ does not qualitatively alter the picture, and from our most complete dataset combination (\textit{CMB-SPA}+\textit{DESI}) we infer $\alpha_s=0.0050 \pm 0.0049$. We stress that these are among the first constraints on $\alpha_s$ obtained by combining data from all three state-of-the-art CMB experiments. When moving from $\Lambda$CDM to EDE, the overall trend we observe is one where $\alpha_s$ increases mildly, thereby increasing the tension with USR models for PBH formation. This is consistent with our theoretical expectations: raising the pre-recombination expansion rate increases the angular size of the photon diffusion scale at recombination $\theta_d$, and the additional damping in the CMB can be compensated by enhancing the small-scale primordial power spectrum, explaining the increase in $\alpha_s$, alongside the well-known and widely discussed increase in $n_s$.

When we translate our $\alpha_s$ constraints into implications for PBH formation, the picture we find is a consistent but nuanced one. \textit{SPT} is the most permissive dataset, allowing the entire PBH mass range regardless of whether $\Lambda$CDM or EDE is assumed, while \textit{Planck} allows only asteroid- to sub-solar-mass PBHs from USR, and \textit{ACT} excludes the entire USR PBH mass spectrum at $>2\sigma$ regardless of the pre-recombination expansion history. Assuming instead the complete \textit{CMB-SPA} combination within $\Lambda$CDM excludes sub-solar-mass and heavier masses, while leaving the asteroid-mass window marginally open: crucially, even this last remaining window is closed when moving to EDE, given the upward shift in $\alpha_s$. Consistently with our theoretical expectation, we find that in no dataset combination does EDE widen the allowed mass window with respect to $\Lambda$CDM. The same qualitative conclusions persist overall when including \textit{DESI} BAO data, or when $\beta_s$ is freed.

Our results illustrate a broader principle: inflationary parameters inferred from CMB data are model-dependent, and are especially sensitive to the assumed pre-recombination expansion history. Given the different responses of the sound horizon and diffusion scale to changes in the pre-recombination expansion history, any successful pre-recombination solution to the Hubble tension which reduces the sound horizon while keeping the acoustic angular scale $\theta_s$ fixed will almost inevitably increase the angular size of the photon diffusion scale at recombination $\theta_d$: this can be compensated by enhanced small-scale power, generically shifting $n_s$ and $\alpha_s$ upwards. This argument applies, with varying degree, to most pre-recombination new physics models aiming to address the Hubble tension, including those based on extra relativistic species. Taking the Hubble tension seriously as evidence for new physics therefore strengthens, rather than weakens, the challenges faced by USR models of PBH formation~\cite{Allegrini:2025jha}. The solar- and sub-solar-mass PBH candidates directly relevant to current and upcoming GW phenomenology, including LVK mergers, the SIGW interpretation of PTA signals, and future Einstein Telescope observations, are thus disfavored by \textit{ACT}: this conclusion only becomes more robust, not less, when the pre-recombination model uncertainty due to the Hubble tension is taken into account. Finally, an important message which our analysis stresses once more is that inflationary parameters such as $n_s$ and $\alpha_s$ should \textit{not} be interpreted as direct measurements of the primordial power spectrum independently of the assumed cosmological model and adopted dataset, and inflationary model selection based on these parameters is premature until the cause of the Hubble tension and the CMB-BAO tension are well understood.

We can identify several interesting directions for future studies. First, the dataset-dependence of our results highlights the importance of resolving the mild discrepancies currently observed between CMB experiments, and future CMB data~\cite{SimonsObservatory:2018koc,SimonsObservatory:2019qwx} will be especially crucial in this respect. Second, while axion-like EDE is a useful and widely studied benchmark for our study, it is worth exploring other pre-recombination solutions to the Hubble tension, even within the wide EDE class: we expect our qualitative conclusions to be robust against this choice, but quantitative differences are to be expected. Third, the mass-running relation of Eq.~(\ref{eq:alphas_mass}) carries a degree of model-dependence tied to the assumed PBH abundance, and the specific USR model under consideration. However, the take-home message is clear: \textit{future analyses of USR models for PBH formation must account for the joint model-dependence of both $n_s$ and $\alpha_s$ when assessing their viability in light of CMB data.}

\begin{acknowledgments}
\noindent M.A.S.\ acknowledges the use of the Sci-Mind computing infrastructure developed by the Centro Brasileiro de Pesquisas F\'{i}sicas (CBPF, Brazilian Center for Research in Physics) LAB-IA team for some of the numerical calculations performed in this work, and thanks Paulo Jos\'{e} Russano and Marcelo Portes de Albuquerque for extensive support on infrastructure-related matters. S.V.\ thanks William Giar\`{e} for several useful discussions. M.A.S.\ and S.V.\ acknowledge support from the Istituto Nazionale di Fisica Nucleare (INFN) through the Commissione Scientifica Nazionale 4 (CSN4) Iniziativa Specifica ``Quantum Fields in Gravity, Cosmology and Black Holes'' (FLAG). A.J.I.\ is funded by Tamkeen under the research grant to NYUAD ADHPG-AD457. This publication is based upon work from the COST Action CA21136 ``Addressing observational tensions in cosmology with systematics and fundamental physics'' (CosmoVerse), supported by COST (European Cooperation in Science and Technology).
\end{acknowledgments}
\appendix

\bibliographystyle{apsrev4-1}
\bibliography{runningspectralindex}

\begin{thebibliography}{450}%
\makeatletter
\providecommand \@ifxundefined [1]{%
 \@ifx{#1\undefined}
}%
\providecommand \@ifnum [1]{%
 \ifnum #1\expandafter \@firstoftwo
 \else \expandafter \@secondoftwo
 \fi
}%
\providecommand \@ifx [1]{%
 \ifx #1\expandafter \@firstoftwo
 \else \expandafter \@secondoftwo
 \fi
}%
\providecommand \natexlab [1]{#1}%
\providecommand \enquote  [1]{``#1''}%
\providecommand \bibnamefont  [1]{#1}%
\providecommand \bibfnamefont [1]{#1}%
\providecommand \citenamefont [1]{#1}%
\providecommand \href@noop [0]{\@secondoftwo}%
\providecommand \href [0]{\begingroup \@sanitize@url \@href}%
\providecommand \@href[1]{\@@startlink{#1}\@@href}%
\providecommand \@@href[1]{\endgroup#1\@@endlink}%
\providecommand \@sanitize@url [0]{\catcode `\\12\catcode `\$12\catcode
  `\&12\catcode `\#12\catcode `\^12\catcode `\_12\catcode `\%12\relax}%
\providecommand \@@startlink[1]{}%
\providecommand \@@endlink[0]{}%
\providecommand \url  [0]{\begingroup\@sanitize@url \@url }%
\providecommand \@url [1]{\endgroup\@href {#1}{\urlprefix }}%
\providecommand \urlprefix  [0]{URL }%
\providecommand \Eprint [0]{\href }%
\providecommand \doibase [0]{http://dx.doi.org/}%
\providecommand \selectlanguage [0]{\@gobble}%
\providecommand \bibinfo  [0]{\@secondoftwo}%
\providecommand \bibfield  [0]{\@secondoftwo}%
\providecommand \translation [1]{[#1]}%
\providecommand \BibitemOpen [0]{}%
\providecommand \bibitemStop [0]{}%
\providecommand \bibitemNoStop [0]{.\EOS\space}%
\providecommand \EOS [0]{\spacefactor3000\relax}%
\providecommand \BibitemShut  [1]{\csname bibitem#1\endcsname}%
\let\auto@bib@innerbib\@empty
\bibitem [{\citenamefont {Zel'dovich}\ and\ \citenamefont
  {Novikov}(1967)}]{Zeldovich:1967lct}%
  \BibitemOpen
  \bibfield  {author} {\bibinfo {author} {\bibfnamefont {Y.~B.}\ \bibnamefont
  {Zel'dovich}}\ and\ \bibinfo {author} {\bibfnamefont {I.~D.}\ \bibnamefont
  {Novikov}},\ }\href@noop {} {\bibfield  {journal} {\bibinfo  {journal} {Sov.
  Astron.}\ }\textbf {\bibinfo {volume} {10}},\ \bibinfo {pages} {602}
  (\bibinfo {year} {1967})}\BibitemShut {NoStop}%
\bibitem [{\citenamefont {Hawking}(1974)}]{Hawking:1974rv}%
  \BibitemOpen
  \bibfield  {author} {\bibinfo {author} {\bibfnamefont {S.~W.}\ \bibnamefont
  {Hawking}},\ }\href {\doibase 10.1038/248030a0} {\bibfield  {journal}
  {\bibinfo  {journal} {Nature}\ }\textbf {\bibinfo {volume} {248}},\ \bibinfo
  {pages} {30} (\bibinfo {year} {1974})}\BibitemShut {NoStop}%
\bibitem [{\citenamefont {Chapline}(1975)}]{Chapline:1975ojl}%
  \BibitemOpen
  \bibfield  {author} {\bibinfo {author} {\bibfnamefont {G.~F.}\ \bibnamefont
  {Chapline}},\ }\href {\doibase 10.1038/253251a0} {\bibfield  {journal}
  {\bibinfo  {journal} {Nature}\ }\textbf {\bibinfo {volume} {253}},\ \bibinfo
  {pages} {251} (\bibinfo {year} {1975})}\BibitemShut {NoStop}%
\bibitem [{\citenamefont {Carr}(1975)}]{Carr:1975qj}%
  \BibitemOpen
  \bibfield  {author} {\bibinfo {author} {\bibfnamefont {B.~J.}\ \bibnamefont
  {Carr}},\ }\href {\doibase 10.1086/153853} {\bibfield  {journal} {\bibinfo
  {journal} {Astrophys. J.}\ }\textbf {\bibinfo {volume} {201}},\ \bibinfo
  {pages} {1} (\bibinfo {year} {1975})}\BibitemShut {NoStop}%
\bibitem [{\citenamefont {Khlopov}(2010)}]{Khlopov:2008qy}%
  \BibitemOpen
  \bibfield  {author} {\bibinfo {author} {\bibfnamefont {M.~Y.}\ \bibnamefont
  {Khlopov}},\ }\href {\doibase 10.1088/1674-4527/10/6/001} {\bibfield
  {journal} {\bibinfo  {journal} {Res. Astron. Astrophys.}\ }\textbf {\bibinfo
  {volume} {10}},\ \bibinfo {pages} {495} (\bibinfo {year} {2010})},\ \Eprint
  {http://arxiv.org/abs/0801.0116} {arXiv:0801.0116 [astro-ph]} \BibitemShut
  {NoStop}%
\bibitem [{\citenamefont {Carr}\ \emph {et~al.}(2024)\citenamefont {Carr},
  \citenamefont {Clesse}, \citenamefont {Garcia-Bellido}, \citenamefont
  {Hawkins},\ and\ \citenamefont {Kuhnel}}]{Carr:2023tpt}%
  \BibitemOpen
  \bibfield  {author} {\bibinfo {author} {\bibfnamefont {B.}~\bibnamefont
  {Carr}}, \bibinfo {author} {\bibfnamefont {S.}~\bibnamefont {Clesse}},
  \bibinfo {author} {\bibfnamefont {J.}~\bibnamefont {Garcia-Bellido}},
  \bibinfo {author} {\bibfnamefont {M.}~\bibnamefont {Hawkins}}, \ and\
  \bibinfo {author} {\bibfnamefont {F.}~\bibnamefont {Kuhnel}},\ }\href
  {\doibase 10.1016/j.physrep.2023.11.005} {\bibfield  {journal} {\bibinfo
  {journal} {Phys. Rept.}\ }\textbf {\bibinfo {volume} {1054}},\ \bibinfo
  {pages} {1} (\bibinfo {year} {2024})},\ \Eprint
  {http://arxiv.org/abs/2306.03903} {arXiv:2306.03903 [astro-ph.CO]}
  \BibitemShut {NoStop}%
\bibitem [{\citenamefont {Choudhury}\ and\ \citenamefont
  {Sami}(2025)}]{Choudhury:2024aji}%
  \BibitemOpen
  \bibfield  {author} {\bibinfo {author} {\bibfnamefont {S.}~\bibnamefont
  {Choudhury}}\ and\ \bibinfo {author} {\bibfnamefont {M.}~\bibnamefont
  {Sami}},\ }\href {\doibase 10.1016/j.physrep.2024.10.007} {\bibfield
  {journal} {\bibinfo  {journal} {Phys. Rept.}\ }\textbf {\bibinfo {volume}
  {1103}},\ \bibinfo {pages} {1} (\bibinfo {year} {2025})},\ \Eprint
  {http://arxiv.org/abs/2407.17006} {arXiv:2407.17006 [gr-qc]} \BibitemShut
  {NoStop}%
\bibitem [{\citenamefont {Khlopov}(2024)}]{Khlopov:2024nqp}%
  \BibitemOpen
  \bibfield  {author} {\bibinfo {author} {\bibfnamefont {M.}~\bibnamefont
  {Khlopov}},\ }\href {\doibase 10.3390/sym16111487} {\bibfield  {journal}
  {\bibinfo  {journal} {Symmetry}\ }\textbf {\bibinfo {volume} {16}},\ \bibinfo
  {pages} {1487} (\bibinfo {year} {2024})}\BibitemShut {NoStop}%
\bibitem [{\citenamefont {Shankaranarayanan}\ \emph {et~al.}(2026)\citenamefont
  {Shankaranarayanan}, \citenamefont {Bhattacharya},\ and\ \citenamefont
  {Vidyarthi}}]{Shankaranarayanan:2026hnn}%
  \BibitemOpen
  \bibfield  {author} {\bibinfo {author} {\bibfnamefont {S.}~\bibnamefont
  {Shankaranarayanan}}, \bibinfo {author} {\bibfnamefont {S.}~\bibnamefont
  {Bhattacharya}}, \ and\ \bibinfo {author} {\bibfnamefont {A.}~\bibnamefont
  {Vidyarthi}},\ }\href@noop {} {\  (\bibinfo {year} {2026})},\ \Eprint
  {http://arxiv.org/abs/2606.23846} {arXiv:2606.23846 [gr-qc]} \BibitemShut
  {NoStop}%
\bibitem [{\citenamefont {Belotsky}\ \emph {et~al.}(2014)\citenamefont
  {Belotsky}, \citenamefont {Dmitriev}, \citenamefont {Esipova}, \citenamefont
  {Gani}, \citenamefont {Grobov}, \citenamefont {Khlopov}, \citenamefont
  {Kirillov}, \citenamefont {Rubin},\ and\ \citenamefont
  {Svadkovsky}}]{Belotsky:2014kca}%
  \BibitemOpen
  \bibfield  {author} {\bibinfo {author} {\bibfnamefont {K.~M.}\ \bibnamefont
  {Belotsky}}, \bibinfo {author} {\bibfnamefont {A.~D.}\ \bibnamefont
  {Dmitriev}}, \bibinfo {author} {\bibfnamefont {E.~A.}\ \bibnamefont
  {Esipova}}, \bibinfo {author} {\bibfnamefont {V.~A.}\ \bibnamefont {Gani}},
  \bibinfo {author} {\bibfnamefont {A.~V.}\ \bibnamefont {Grobov}}, \bibinfo
  {author} {\bibfnamefont {M.~Y.}\ \bibnamefont {Khlopov}}, \bibinfo {author}
  {\bibfnamefont {A.~A.}\ \bibnamefont {Kirillov}}, \bibinfo {author}
  {\bibfnamefont {S.~G.}\ \bibnamefont {Rubin}}, \ and\ \bibinfo {author}
  {\bibfnamefont {I.~V.}\ \bibnamefont {Svadkovsky}},\ }\href {\doibase
  10.1142/S0217732314400057} {\bibfield  {journal} {\bibinfo  {journal} {Mod.
  Phys. Lett. A}\ }\textbf {\bibinfo {volume} {29}},\ \bibinfo {pages}
  {1440005} (\bibinfo {year} {2014})},\ \Eprint
  {http://arxiv.org/abs/1410.0203} {arXiv:1410.0203 [astro-ph.CO]} \BibitemShut
  {NoStop}%
\bibitem [{\citenamefont {Carr}\ \emph {et~al.}(2016)\citenamefont {Carr},
  \citenamefont {Kuhnel},\ and\ \citenamefont {Sandstad}}]{Carr:2016drx}%
  \BibitemOpen
  \bibfield  {author} {\bibinfo {author} {\bibfnamefont {B.}~\bibnamefont
  {Carr}}, \bibinfo {author} {\bibfnamefont {F.}~\bibnamefont {Kuhnel}}, \ and\
  \bibinfo {author} {\bibfnamefont {M.}~\bibnamefont {Sandstad}},\ }\href
  {\doibase 10.1103/PhysRevD.94.083504} {\bibfield  {journal} {\bibinfo
  {journal} {Phys. Rev. D}\ }\textbf {\bibinfo {volume} {94}},\ \bibinfo
  {pages} {083504} (\bibinfo {year} {2016})},\ \Eprint
  {http://arxiv.org/abs/1607.06077} {arXiv:1607.06077 [astro-ph.CO]}
  \BibitemShut {NoStop}%
\bibitem [{\citenamefont {Green}\ and\ \citenamefont
  {Kavanagh}(2021)}]{Green:2020jor}%
  \BibitemOpen
  \bibfield  {author} {\bibinfo {author} {\bibfnamefont {A.~M.}\ \bibnamefont
  {Green}}\ and\ \bibinfo {author} {\bibfnamefont {B.~J.}\ \bibnamefont
  {Kavanagh}},\ }\href {\doibase 10.1088/1361-6471/abc534} {\bibfield
  {journal} {\bibinfo  {journal} {J. Phys. G}\ }\textbf {\bibinfo {volume}
  {48}},\ \bibinfo {pages} {043001} (\bibinfo {year} {2021})},\ \Eprint
  {http://arxiv.org/abs/2007.10722} {arXiv:2007.10722 [astro-ph.CO]}
  \BibitemShut {NoStop}%
\bibitem [{\citenamefont {Carr}\ \emph {et~al.}(2026)\citenamefont {Carr},
  \citenamefont {Iovino}, \citenamefont {Perna}, \citenamefont {Vaskonen},\
  and\ \citenamefont {Veerm{\"a}e}}]{Carr:2026hot}%
  \BibitemOpen
  \bibfield  {author} {\bibinfo {author} {\bibfnamefont {B.}~\bibnamefont
  {Carr}}, \bibinfo {author} {\bibfnamefont {A.~J.}\ \bibnamefont {Iovino}},
  \bibinfo {author} {\bibfnamefont {G.}~\bibnamefont {Perna}}, \bibinfo
  {author} {\bibfnamefont {V.}~\bibnamefont {Vaskonen}}, \ and\ \bibinfo
  {author} {\bibfnamefont {H.}~\bibnamefont {Veerm{\"a}e}},\ }\href {\doibase
  10.1007/s40766-026-00080-z} {\bibfield  {journal} {\bibinfo  {journal} {Riv.
  Nuovo Cim.}\ }\textbf {\bibinfo {volume} {49}},\ \bibinfo {pages} {225}
  (\bibinfo {year} {2026})},\ \Eprint {http://arxiv.org/abs/2601.06024}
  {arXiv:2601.06024 [astro-ph.CO]} \BibitemShut {NoStop}%
\bibitem [{\citenamefont {Abbott}\ \emph
  {et~al.}(2019{\natexlab{a}})\citenamefont {Abbott} \emph
  {et~al.}}]{LIGOScientific:2018mvr}%
  \BibitemOpen
  \bibfield  {author} {\bibinfo {author} {\bibfnamefont {B.~P.}\ \bibnamefont
  {Abbott}} \emph {et~al.} (\bibinfo {collaboration} {LIGO Scientific,
  Virgo}),\ }\href {\doibase 10.1103/PhysRevX.9.031040} {\bibfield  {journal}
  {\bibinfo  {journal} {Phys. Rev. X}\ }\textbf {\bibinfo {volume} {9}},\
  \bibinfo {pages} {031040} (\bibinfo {year} {2019}{\natexlab{a}})},\ \Eprint
  {http://arxiv.org/abs/1811.12907} {arXiv:1811.12907 [astro-ph.HE]}
  \BibitemShut {NoStop}%
\bibitem [{\citenamefont {Abbott}\ \emph
  {et~al.}(2019{\natexlab{b}})\citenamefont {Abbott} \emph
  {et~al.}}]{LIGOScientific:2019kan}%
  \BibitemOpen
  \bibfield  {author} {\bibinfo {author} {\bibfnamefont {B.~P.}\ \bibnamefont
  {Abbott}} \emph {et~al.} (\bibinfo {collaboration} {LIGO Scientific,
  Virgo}),\ }\href {\doibase 10.1103/PhysRevLett.123.161102} {\bibfield
  {journal} {\bibinfo  {journal} {Phys. Rev. Lett.}\ }\textbf {\bibinfo
  {volume} {123}},\ \bibinfo {pages} {161102} (\bibinfo {year}
  {2019}{\natexlab{b}})},\ \Eprint {http://arxiv.org/abs/1904.08976}
  {arXiv:1904.08976 [astro-ph.CO]} \BibitemShut {NoStop}%
\bibitem [{\citenamefont {Abbott}\ \emph {et~al.}(2021)\citenamefont {Abbott}
  \emph {et~al.}}]{LIGOScientific:2020ibl}%
  \BibitemOpen
  \bibfield  {author} {\bibinfo {author} {\bibfnamefont {R.}~\bibnamefont
  {Abbott}} \emph {et~al.} (\bibinfo {collaboration} {LIGO Scientific,
  Virgo}),\ }\href {\doibase 10.1103/PhysRevX.11.021053} {\bibfield  {journal}
  {\bibinfo  {journal} {Phys. Rev. X}\ }\textbf {\bibinfo {volume} {11}},\
  \bibinfo {pages} {021053} (\bibinfo {year} {2021})},\ \Eprint
  {http://arxiv.org/abs/2010.14527} {arXiv:2010.14527 [gr-qc]} \BibitemShut
  {NoStop}%
\bibitem [{\citenamefont {Abbott}\ \emph
  {et~al.}(2023{\natexlab{a}})\citenamefont {Abbott} \emph
  {et~al.}}]{KAGRA:2021vkt}%
  \BibitemOpen
  \bibfield  {author} {\bibinfo {author} {\bibfnamefont {R.}~\bibnamefont
  {Abbott}} \emph {et~al.} (\bibinfo {collaboration} {KAGRA, VIRGO, LIGO
  Scientific}),\ }\href {\doibase 10.1103/PhysRevX.13.041039} {\bibfield
  {journal} {\bibinfo  {journal} {Phys. Rev. X}\ }\textbf {\bibinfo {volume}
  {13}},\ \bibinfo {pages} {041039} (\bibinfo {year} {2023}{\natexlab{a}})},\
  \Eprint {http://arxiv.org/abs/2111.03606} {arXiv:2111.03606 [gr-qc]}
  \BibitemShut {NoStop}%
\bibitem [{\citenamefont {Abac}\ \emph {et~al.}(2025)\citenamefont {Abac} \emph
  {et~al.}}]{LIGOScientific:2025slb}%
  \BibitemOpen
  \bibfield  {author} {\bibinfo {author} {\bibfnamefont {A.~G.}\ \bibnamefont
  {Abac}} \emph {et~al.} (\bibinfo {collaboration} {LIGO Scientific, VIRGO,
  KAGRA}),\ }\href@noop {} {\  (\bibinfo {year} {2025})},\ \Eprint
  {http://arxiv.org/abs/2508.18082} {arXiv:2508.18082 [gr-qc]} \BibitemShut
  {NoStop}%
\bibitem [{\citenamefont {Abbott}\ \emph {et~al.}(2022)\citenamefont {Abbott}
  \emph {et~al.}}]{LIGOScientific:2021job}%
  \BibitemOpen
  \bibfield  {author} {\bibinfo {author} {\bibfnamefont {R.}~\bibnamefont
  {Abbott}} \emph {et~al.} (\bibinfo {collaboration} {LIGO Scientific, VIRGO,
  KAGRA}),\ }\href {\doibase 10.1103/PhysRevLett.129.061104} {\bibfield
  {journal} {\bibinfo  {journal} {Phys. Rev. Lett.}\ }\textbf {\bibinfo
  {volume} {129}},\ \bibinfo {pages} {061104} (\bibinfo {year} {2022})},\
  \Eprint {http://arxiv.org/abs/2109.12197} {arXiv:2109.12197 [astro-ph.CO]}
  \BibitemShut {NoStop}%
\bibitem [{\citenamefont {Abbott}\ \emph
  {et~al.}(2023{\natexlab{b}})\citenamefont {Abbott} \emph
  {et~al.}}]{LVK:2022ydq}%
  \BibitemOpen
  \bibfield  {author} {\bibinfo {author} {\bibfnamefont {R.}~\bibnamefont
  {Abbott}} \emph {et~al.} (\bibinfo {collaboration} {LVK}),\ }\href {\doibase
  10.1093/mnras/stad588} {\bibfield  {journal} {\bibinfo  {journal} {Mon. Not.
  Roy. Astron. Soc.}\ }\textbf {\bibinfo {volume} {524}},\ \bibinfo {pages}
  {5984} (\bibinfo {year} {2023}{\natexlab{b}})},\ \bibinfo {note} {[Erratum:
  Mon.Not.Roy.Astron.Soc. 526, 6234 (2023)]},\ \Eprint
  {http://arxiv.org/abs/2212.01477} {arXiv:2212.01477 [astro-ph.HE]}
  \BibitemShut {NoStop}%
\bibitem [{\citenamefont {Prunier}\ \emph {et~al.}(2024)\citenamefont
  {Prunier}, \citenamefont {Morr{\'a}s}, \citenamefont {Siles}, \citenamefont
  {Clesse}, \citenamefont {Garc{\'\i}a-Bellido},\ and\ \citenamefont
  {Ruiz~Morales}}]{Prunier:2023uoo}%
  \BibitemOpen
  \bibfield  {author} {\bibinfo {author} {\bibfnamefont {M.}~\bibnamefont
  {Prunier}}, \bibinfo {author} {\bibfnamefont {G.}~\bibnamefont {Morr{\'a}s}},
  \bibinfo {author} {\bibfnamefont {J.~F.~N.}\ \bibnamefont {Siles}}, \bibinfo
  {author} {\bibfnamefont {S.}~\bibnamefont {Clesse}}, \bibinfo {author}
  {\bibfnamefont {J.}~\bibnamefont {Garc{\'\i}a-Bellido}}, \ and\ \bibinfo
  {author} {\bibfnamefont {E.}~\bibnamefont {Ruiz~Morales}},\ }\href {\doibase
  10.1016/j.dark.2024.101582} {\bibfield  {journal} {\bibinfo  {journal} {Phys.
  Dark Univ.}\ }\textbf {\bibinfo {volume} {46}},\ \bibinfo {pages} {101582}
  (\bibinfo {year} {2024})},\ \Eprint {http://arxiv.org/abs/2311.16085}
  {arXiv:2311.16085 [gr-qc]} \BibitemShut {NoStop}%
\bibitem [{\citenamefont {Morras}\ \emph {et~al.}(2023)\citenamefont {Morras}
  \emph {et~al.}}]{Morras:2023jvb}%
  \BibitemOpen
  \bibfield  {author} {\bibinfo {author} {\bibfnamefont {G.}~\bibnamefont
  {Morras}} \emph {et~al.},\ }\href {\doibase 10.1016/j.dark.2023.101285}
  {\bibfield  {journal} {\bibinfo  {journal} {Phys. Dark Univ.}\ }\textbf
  {\bibinfo {volume} {42}},\ \bibinfo {pages} {101285} (\bibinfo {year}
  {2023})},\ \Eprint {http://arxiv.org/abs/2301.11619} {arXiv:2301.11619
  [gr-qc]} \BibitemShut {NoStop}%
\bibitem [{\citenamefont {Yuan}\ and\ \citenamefont
  {Huang}(2024)}]{Yuan:2024yyo}%
  \BibitemOpen
  \bibfield  {author} {\bibinfo {author} {\bibfnamefont {C.}~\bibnamefont
  {Yuan}}\ and\ \bibinfo {author} {\bibfnamefont {Q.-G.}\ \bibnamefont
  {Huang}},\ }\href {\doibase 10.1088/1475-7516/2024/09/051} {\bibfield
  {journal} {\bibinfo  {journal} {JCAP}\ }\textbf {\bibinfo {volume} {09}},\
  \bibinfo {pages} {051} (\bibinfo {year} {2024})},\ \Eprint
  {http://arxiv.org/abs/2404.03328} {arXiv:2404.03328 [astro-ph.CO]}
  \BibitemShut {NoStop}%
\bibitem [{\citenamefont {Bird}\ \emph {et~al.}(2016)\citenamefont {Bird},
  \citenamefont {Cholis}, \citenamefont {Mu{\~n}oz}, \citenamefont
  {Ali-Ha{\"\i}moud}, \citenamefont {Kamionkowski}, \citenamefont {Kovetz},
  \citenamefont {Raccanelli},\ and\ \citenamefont {Riess}}]{Bird:2016dcv}%
  \BibitemOpen
  \bibfield  {author} {\bibinfo {author} {\bibfnamefont {S.}~\bibnamefont
  {Bird}}, \bibinfo {author} {\bibfnamefont {I.}~\bibnamefont {Cholis}},
  \bibinfo {author} {\bibfnamefont {J.~B.}\ \bibnamefont {Mu{\~n}oz}}, \bibinfo
  {author} {\bibfnamefont {Y.}~\bibnamefont {Ali-Ha{\"\i}moud}}, \bibinfo
  {author} {\bibfnamefont {M.}~\bibnamefont {Kamionkowski}}, \bibinfo {author}
  {\bibfnamefont {E.~D.}\ \bibnamefont {Kovetz}}, \bibinfo {author}
  {\bibfnamefont {A.}~\bibnamefont {Raccanelli}}, \ and\ \bibinfo {author}
  {\bibfnamefont {A.~G.}\ \bibnamefont {Riess}},\ }\href {\doibase
  10.1103/PhysRevLett.116.201301} {\bibfield  {journal} {\bibinfo  {journal}
  {Phys. Rev. Lett.}\ }\textbf {\bibinfo {volume} {116}},\ \bibinfo {pages}
  {201301} (\bibinfo {year} {2016})},\ \Eprint
  {http://arxiv.org/abs/1603.00464} {arXiv:1603.00464 [astro-ph.CO]}
  \BibitemShut {NoStop}%
\bibitem [{\citenamefont {Kavanagh}\ \emph {et~al.}(2018)\citenamefont
  {Kavanagh}, \citenamefont {Gaggero},\ and\ \citenamefont
  {Bertone}}]{Kavanagh:2018ggo}%
  \BibitemOpen
  \bibfield  {author} {\bibinfo {author} {\bibfnamefont {B.~J.}\ \bibnamefont
  {Kavanagh}}, \bibinfo {author} {\bibfnamefont {D.}~\bibnamefont {Gaggero}}, \
  and\ \bibinfo {author} {\bibfnamefont {G.}~\bibnamefont {Bertone}},\ }\href
  {\doibase 10.1103/PhysRevD.98.023536} {\bibfield  {journal} {\bibinfo
  {journal} {Phys. Rev. D}\ }\textbf {\bibinfo {volume} {98}},\ \bibinfo
  {pages} {023536} (\bibinfo {year} {2018})},\ \Eprint
  {http://arxiv.org/abs/1805.09034} {arXiv:1805.09034 [astro-ph.CO]}
  \BibitemShut {NoStop}%
\bibitem [{\citenamefont {Wong}\ \emph {et~al.}(2021)\citenamefont {Wong},
  \citenamefont {Franciolini}, \citenamefont {De~Luca}, \citenamefont
  {Baibhav}, \citenamefont {Berti}, \citenamefont {Pani},\ and\ \citenamefont
  {Riotto}}]{Wong:2020yig}%
  \BibitemOpen
  \bibfield  {author} {\bibinfo {author} {\bibfnamefont {K.~W.~K.}\
  \bibnamefont {Wong}}, \bibinfo {author} {\bibfnamefont {G.}~\bibnamefont
  {Franciolini}}, \bibinfo {author} {\bibfnamefont {V.}~\bibnamefont
  {De~Luca}}, \bibinfo {author} {\bibfnamefont {V.}~\bibnamefont {Baibhav}},
  \bibinfo {author} {\bibfnamefont {E.}~\bibnamefont {Berti}}, \bibinfo
  {author} {\bibfnamefont {P.}~\bibnamefont {Pani}}, \ and\ \bibinfo {author}
  {\bibfnamefont {A.}~\bibnamefont {Riotto}},\ }\href {\doibase
  10.1103/PhysRevD.103.023026} {\bibfield  {journal} {\bibinfo  {journal}
  {Phys. Rev. D}\ }\textbf {\bibinfo {volume} {103}},\ \bibinfo {pages}
  {023026} (\bibinfo {year} {2021})},\ \Eprint
  {http://arxiv.org/abs/2011.01865} {arXiv:2011.01865 [gr-qc]} \BibitemShut
  {NoStop}%
\bibitem [{\citenamefont {H\"utsi}\ \emph {et~al.}(2021)\citenamefont
  {H\"utsi}, \citenamefont {Raidal}, \citenamefont {Vaskonen},\ and\
  \citenamefont {Veerm\"ae}}]{Hutsi:2020sol}%
  \BibitemOpen
  \bibfield  {author} {\bibinfo {author} {\bibfnamefont {G.}~\bibnamefont
  {H\"utsi}}, \bibinfo {author} {\bibfnamefont {M.}~\bibnamefont {Raidal}},
  \bibinfo {author} {\bibfnamefont {V.}~\bibnamefont {Vaskonen}}, \ and\
  \bibinfo {author} {\bibfnamefont {H.}~\bibnamefont {Veerm\"ae}},\ }\href
  {\doibase 10.1088/1475-7516/2021/03/068} {\bibfield  {journal} {\bibinfo
  {journal} {JCAP}\ }\textbf {\bibinfo {volume} {03}},\ \bibinfo {pages} {068}
  (\bibinfo {year} {2021})},\ \Eprint {http://arxiv.org/abs/2012.02786}
  {arXiv:2012.02786 [astro-ph.CO]} \BibitemShut {NoStop}%
\bibitem [{\citenamefont {De~Luca}\ \emph
  {et~al.}(2021{\natexlab{a}})\citenamefont {De~Luca}, \citenamefont
  {Franciolini}, \citenamefont {Pani},\ and\ \citenamefont
  {Riotto}}]{DeLuca:2021wjr}%
  \BibitemOpen
  \bibfield  {author} {\bibinfo {author} {\bibfnamefont {V.}~\bibnamefont
  {De~Luca}}, \bibinfo {author} {\bibfnamefont {G.}~\bibnamefont
  {Franciolini}}, \bibinfo {author} {\bibfnamefont {P.}~\bibnamefont {Pani}}, \
  and\ \bibinfo {author} {\bibfnamefont {A.}~\bibnamefont {Riotto}},\ }\href
  {\doibase 10.1088/1475-7516/2021/05/003} {\bibfield  {journal} {\bibinfo
  {journal} {JCAP}\ }\textbf {\bibinfo {volume} {05}},\ \bibinfo {pages} {003}
  (\bibinfo {year} {2021}{\natexlab{a}})},\ \Eprint
  {http://arxiv.org/abs/2102.03809} {arXiv:2102.03809 [astro-ph.CO]}
  \BibitemShut {NoStop}%
\bibitem [{\citenamefont {Franciolini}\ \emph {et~al.}(2022)\citenamefont
  {Franciolini}, \citenamefont {Musco}, \citenamefont {Pani},\ and\
  \citenamefont {Urbano}}]{Franciolini:2022tfm}%
  \BibitemOpen
  \bibfield  {author} {\bibinfo {author} {\bibfnamefont {G.}~\bibnamefont
  {Franciolini}}, \bibinfo {author} {\bibfnamefont {I.}~\bibnamefont {Musco}},
  \bibinfo {author} {\bibfnamefont {P.}~\bibnamefont {Pani}}, \ and\ \bibinfo
  {author} {\bibfnamefont {A.}~\bibnamefont {Urbano}},\ }\href {\doibase
  10.1103/PhysRevD.106.123526} {\bibfield  {journal} {\bibinfo  {journal}
  {Phys. Rev. D}\ }\textbf {\bibinfo {volume} {106}},\ \bibinfo {pages}
  {123526} (\bibinfo {year} {2022})},\ \Eprint
  {http://arxiv.org/abs/2209.05959} {arXiv:2209.05959 [astro-ph.CO]}
  \BibitemShut {NoStop}%
\bibitem [{\citenamefont {Miller}\ \emph {et~al.}(2024)\citenamefont {Miller},
  \citenamefont {Aggarwal}, \citenamefont {Clesse}, \citenamefont {De~Lillo},
  \citenamefont {Sachdev}, \citenamefont {Astone}, \citenamefont {Palomba},
  \citenamefont {Piccinni},\ and\ \citenamefont {Pierini}}]{Miller:2024fpo}%
  \BibitemOpen
  \bibfield  {author} {\bibinfo {author} {\bibfnamefont {A.~L.}\ \bibnamefont
  {Miller}}, \bibinfo {author} {\bibfnamefont {N.}~\bibnamefont {Aggarwal}},
  \bibinfo {author} {\bibfnamefont {S.}~\bibnamefont {Clesse}}, \bibinfo
  {author} {\bibfnamefont {F.}~\bibnamefont {De~Lillo}}, \bibinfo {author}
  {\bibfnamefont {S.}~\bibnamefont {Sachdev}}, \bibinfo {author} {\bibfnamefont
  {P.}~\bibnamefont {Astone}}, \bibinfo {author} {\bibfnamefont
  {C.}~\bibnamefont {Palomba}}, \bibinfo {author} {\bibfnamefont {O.~J.}\
  \bibnamefont {Piccinni}}, \ and\ \bibinfo {author} {\bibfnamefont
  {L.}~\bibnamefont {Pierini}},\ }\href {\doibase
  10.1103/PhysRevLett.133.111401} {\bibfield  {journal} {\bibinfo  {journal}
  {Phys. Rev. Lett.}\ }\textbf {\bibinfo {volume} {133}},\ \bibinfo {pages}
  {111401} (\bibinfo {year} {2024})},\ \Eprint
  {http://arxiv.org/abs/2402.19468} {arXiv:2402.19468 [gr-qc]} \BibitemShut
  {NoStop}%
\bibitem [{\citenamefont {Andr\'es-Carcasona}\ \emph
  {et~al.}(2024)\citenamefont {Andr\'es-Carcasona}, \citenamefont {Iovino},
  \citenamefont {Vaskonen}, \citenamefont {Veerm\"ae}, \citenamefont
  {Mart\'\i{}nez}, \citenamefont {Pujol\`as},\ and\ \citenamefont
  {Mir}}]{Andres-Carcasona:2024wqk}%
  \BibitemOpen
  \bibfield  {author} {\bibinfo {author} {\bibfnamefont {M.}~\bibnamefont
  {Andr\'es-Carcasona}}, \bibinfo {author} {\bibfnamefont {A.~J.}\ \bibnamefont
  {Iovino}}, \bibinfo {author} {\bibfnamefont {V.}~\bibnamefont {Vaskonen}},
  \bibinfo {author} {\bibfnamefont {H.}~\bibnamefont {Veerm\"ae}}, \bibinfo
  {author} {\bibfnamefont {M.}~\bibnamefont {Mart\'\i{}nez}}, \bibinfo {author}
  {\bibfnamefont {O.}~\bibnamefont {Pujol\`as}}, \ and\ \bibinfo {author}
  {\bibfnamefont {L.~M.}\ \bibnamefont {Mir}},\ }\href {\doibase
  10.1103/PhysRevD.110.023040} {\bibfield  {journal} {\bibinfo  {journal}
  {Phys. Rev. D}\ }\textbf {\bibinfo {volume} {110}},\ \bibinfo {pages}
  {023040} (\bibinfo {year} {2024})},\ \Eprint
  {http://arxiv.org/abs/2405.05732} {arXiv:2405.05732 [astro-ph.CO]}
  \BibitemShut {NoStop}%
\bibitem [{\citenamefont {De~Luca}\ \emph {et~al.}(2025)\citenamefont
  {De~Luca}, \citenamefont {Franciolini},\ and\ \citenamefont
  {Riotto}}]{DeLuca:2025fln}%
  \BibitemOpen
  \bibfield  {author} {\bibinfo {author} {\bibfnamefont {V.}~\bibnamefont
  {De~Luca}}, \bibinfo {author} {\bibfnamefont {G.}~\bibnamefont
  {Franciolini}}, \ and\ \bibinfo {author} {\bibfnamefont {A.}~\bibnamefont
  {Riotto}},\ }\href@noop {} {\  (\bibinfo {year} {2025})},\ \Eprint
  {http://arxiv.org/abs/2508.09965} {arXiv:2508.09965 [astro-ph.CO]}
  \BibitemShut {NoStop}%
\bibitem [{\citenamefont {Andr{\'e}s-Carcasona}\ \emph
  {et~al.}(2026)\citenamefont {Andr{\'e}s-Carcasona}, \citenamefont {Iovino},
  \citenamefont {Vallejo-Pag{\`e}s}, \citenamefont {Vaskonen}, \citenamefont
  {Veerm{\"a}e}, \citenamefont {Mart{\'\i}nez},\ and\ \citenamefont
  {Mir}}]{Andres-Carcasona:2026avd}%
  \BibitemOpen
  \bibfield  {author} {\bibinfo {author} {\bibfnamefont {M.}~\bibnamefont
  {Andr{\'e}s-Carcasona}}, \bibinfo {author} {\bibfnamefont {A.~J.}\
  \bibnamefont {Iovino}}, \bibinfo {author} {\bibfnamefont {E.}~\bibnamefont
  {Vallejo-Pag{\`e}s}}, \bibinfo {author} {\bibfnamefont {V.}~\bibnamefont
  {Vaskonen}}, \bibinfo {author} {\bibfnamefont {H.}~\bibnamefont
  {Veerm{\"a}e}}, \bibinfo {author} {\bibfnamefont {M.}~\bibnamefont
  {Mart{\'\i}nez}}, \ and\ \bibinfo {author} {\bibfnamefont {L.~M.}\
  \bibnamefont {Mir}},\ }\href@noop {} {\  (\bibinfo {year} {2026})},\ \Eprint
  {http://arxiv.org/abs/2605.15749} {arXiv:2605.15749 [astro-ph.CO]}
  \BibitemShut {NoStop}%
\bibitem [{\citenamefont {Agazie}\ \emph {et~al.}(2023)\citenamefont {Agazie}
  \emph {et~al.}}]{NANOGrav:2023gor}%
  \BibitemOpen
  \bibfield  {author} {\bibinfo {author} {\bibfnamefont {G.}~\bibnamefont
  {Agazie}} \emph {et~al.} (\bibinfo {collaboration} {NANOGrav}),\ }\href
  {\doibase 10.3847/2041-8213/acdac6} {\bibfield  {journal} {\bibinfo
  {journal} {Astrophys. J. Lett.}\ }\textbf {\bibinfo {volume} {951}},\
  \bibinfo {pages} {L8} (\bibinfo {year} {2023})},\ \Eprint
  {http://arxiv.org/abs/2306.16213} {arXiv:2306.16213 [astro-ph.HE]}
  \BibitemShut {NoStop}%
\bibitem [{\citenamefont {Antoniadis}\ \emph {et~al.}(2023)\citenamefont
  {Antoniadis} \emph {et~al.}}]{EPTA:2023fyk}%
  \BibitemOpen
  \bibfield  {author} {\bibinfo {author} {\bibfnamefont {J.}~\bibnamefont
  {Antoniadis}} \emph {et~al.} (\bibinfo {collaboration} {EPTA, InPTA:}),\
  }\href {\doibase 10.1051/0004-6361/202346844} {\bibfield  {journal} {\bibinfo
   {journal} {Astron. Astrophys.}\ }\textbf {\bibinfo {volume} {678}},\
  \bibinfo {pages} {A50} (\bibinfo {year} {2023})},\ \Eprint
  {http://arxiv.org/abs/2306.16214} {arXiv:2306.16214 [astro-ph.HE]}
  \BibitemShut {NoStop}%
\bibitem [{\citenamefont {Reardon}\ \emph {et~al.}(2023)\citenamefont {Reardon}
  \emph {et~al.}}]{Reardon:2023gzh}%
  \BibitemOpen
  \bibfield  {author} {\bibinfo {author} {\bibfnamefont {D.~J.}\ \bibnamefont
  {Reardon}} \emph {et~al.},\ }\href {\doibase 10.3847/2041-8213/acdd02}
  {\bibfield  {journal} {\bibinfo  {journal} {Astrophys. J. Lett.}\ }\textbf
  {\bibinfo {volume} {951}},\ \bibinfo {pages} {L6} (\bibinfo {year} {2023})},\
  \Eprint {http://arxiv.org/abs/2306.16215} {arXiv:2306.16215 [astro-ph.HE]}
  \BibitemShut {NoStop}%
\bibitem [{\citenamefont {Xu}\ \emph {et~al.}(2023)\citenamefont {Xu} \emph
  {et~al.}}]{Xu:2023wog}%
  \BibitemOpen
  \bibfield  {author} {\bibinfo {author} {\bibfnamefont {H.}~\bibnamefont {Xu}}
  \emph {et~al.},\ }\href {\doibase 10.1088/1674-4527/acdfa5} {\bibfield
  {journal} {\bibinfo  {journal} {Res. Astron. Astrophys.}\ }\textbf {\bibinfo
  {volume} {23}},\ \bibinfo {pages} {075024} (\bibinfo {year} {2023})},\
  \Eprint {http://arxiv.org/abs/2306.16216} {arXiv:2306.16216 [astro-ph.HE]}
  \BibitemShut {NoStop}%
\bibitem [{\citenamefont {Vaskonen}\ and\ \citenamefont
  {Veerm{\"a}e}(2021)}]{Vaskonen:2020lbd}%
  \BibitemOpen
  \bibfield  {author} {\bibinfo {author} {\bibfnamefont {V.}~\bibnamefont
  {Vaskonen}}\ and\ \bibinfo {author} {\bibfnamefont {H.}~\bibnamefont
  {Veerm{\"a}e}},\ }\href {\doibase 10.1103/PhysRevLett.126.051303} {\bibfield
  {journal} {\bibinfo  {journal} {Phys. Rev. Lett.}\ }\textbf {\bibinfo
  {volume} {126}},\ \bibinfo {pages} {051303} (\bibinfo {year} {2021})},\
  \Eprint {http://arxiv.org/abs/2009.07832} {arXiv:2009.07832 [astro-ph.CO]}
  \BibitemShut {NoStop}%
\bibitem [{\citenamefont {Chen}\ \emph {et~al.}(2020)\citenamefont {Chen},
  \citenamefont {Yuan},\ and\ \citenamefont {Huang}}]{Chen:2019xse}%
  \BibitemOpen
  \bibfield  {author} {\bibinfo {author} {\bibfnamefont {Z.-C.}\ \bibnamefont
  {Chen}}, \bibinfo {author} {\bibfnamefont {C.}~\bibnamefont {Yuan}}, \ and\
  \bibinfo {author} {\bibfnamefont {Q.-G.}\ \bibnamefont {Huang}},\ }\href
  {\doibase 10.1103/PhysRevLett.124.251101} {\bibfield  {journal} {\bibinfo
  {journal} {Phys. Rev. Lett.}\ }\textbf {\bibinfo {volume} {124}},\ \bibinfo
  {pages} {251101} (\bibinfo {year} {2020})},\ \Eprint
  {http://arxiv.org/abs/1910.12239} {arXiv:1910.12239 [astro-ph.CO]}
  \BibitemShut {NoStop}%
\bibitem [{\citenamefont {De~Luca}\ \emph
  {et~al.}(2021{\natexlab{b}})\citenamefont {De~Luca}, \citenamefont
  {Franciolini},\ and\ \citenamefont {Riotto}}]{DeLuca:2020agl}%
  \BibitemOpen
  \bibfield  {author} {\bibinfo {author} {\bibfnamefont {V.}~\bibnamefont
  {De~Luca}}, \bibinfo {author} {\bibfnamefont {G.}~\bibnamefont
  {Franciolini}}, \ and\ \bibinfo {author} {\bibfnamefont {A.}~\bibnamefont
  {Riotto}},\ }\href {\doibase 10.1103/PhysRevLett.126.041303} {\bibfield
  {journal} {\bibinfo  {journal} {Phys. Rev. Lett.}\ }\textbf {\bibinfo
  {volume} {126}},\ \bibinfo {pages} {041303} (\bibinfo {year}
  {2021}{\natexlab{b}})},\ \Eprint {http://arxiv.org/abs/2009.08268}
  {arXiv:2009.08268 [astro-ph.CO]} \BibitemShut {NoStop}%
\bibitem [{\citenamefont {Bhaumik}\ and\ \citenamefont
  {Jain}(2021)}]{Bhaumik:2020dor}%
  \BibitemOpen
  \bibfield  {author} {\bibinfo {author} {\bibfnamefont {N.}~\bibnamefont
  {Bhaumik}}\ and\ \bibinfo {author} {\bibfnamefont {R.~K.}\ \bibnamefont
  {Jain}},\ }\href {\doibase 10.1103/PhysRevD.104.023531} {\bibfield  {journal}
  {\bibinfo  {journal} {Phys. Rev. D}\ }\textbf {\bibinfo {volume} {104}},\
  \bibinfo {pages} {023531} (\bibinfo {year} {2021})},\ \Eprint
  {http://arxiv.org/abs/2009.10424} {arXiv:2009.10424 [astro-ph.CO]}
  \BibitemShut {NoStop}%
\bibitem [{\citenamefont {Inomata}\ \emph {et~al.}(2021)\citenamefont
  {Inomata}, \citenamefont {Kawasaki}, \citenamefont {Mukaida},\ and\
  \citenamefont {Yanagida}}]{Inomata:2020xad}%
  \BibitemOpen
  \bibfield  {author} {\bibinfo {author} {\bibfnamefont {K.}~\bibnamefont
  {Inomata}}, \bibinfo {author} {\bibfnamefont {M.}~\bibnamefont {Kawasaki}},
  \bibinfo {author} {\bibfnamefont {K.}~\bibnamefont {Mukaida}}, \ and\
  \bibinfo {author} {\bibfnamefont {T.~T.}\ \bibnamefont {Yanagida}},\ }\href
  {\doibase 10.1103/PhysRevLett.126.131301} {\bibfield  {journal} {\bibinfo
  {journal} {Phys. Rev. Lett.}\ }\textbf {\bibinfo {volume} {126}},\ \bibinfo
  {pages} {131301} (\bibinfo {year} {2021})},\ \Eprint
  {http://arxiv.org/abs/2011.01270} {arXiv:2011.01270 [astro-ph.CO]}
  \BibitemShut {NoStop}%
\bibitem [{\citenamefont {Kohri}\ and\ \citenamefont
  {Terada}(2021)}]{Kohri:2020qqd}%
  \BibitemOpen
  \bibfield  {author} {\bibinfo {author} {\bibfnamefont {K.}~\bibnamefont
  {Kohri}}\ and\ \bibinfo {author} {\bibfnamefont {T.}~\bibnamefont {Terada}},\
  }\href {\doibase 10.1016/j.physletb.2020.136040} {\bibfield  {journal}
  {\bibinfo  {journal} {Phys. Lett. B}\ }\textbf {\bibinfo {volume} {813}},\
  \bibinfo {pages} {136040} (\bibinfo {year} {2021})},\ \Eprint
  {http://arxiv.org/abs/2009.11853} {arXiv:2009.11853 [astro-ph.CO]}
  \BibitemShut {NoStop}%
\bibitem [{\citenamefont {Dom\`enech}\ and\ \citenamefont
  {Pi}(2022)}]{Domenech:2020ers}%
  \BibitemOpen
  \bibfield  {author} {\bibinfo {author} {\bibfnamefont {G.}~\bibnamefont
  {Dom\`enech}}\ and\ \bibinfo {author} {\bibfnamefont {S.}~\bibnamefont
  {Pi}},\ }\href {\doibase 10.1007/s11433-021-1839-6} {\bibfield  {journal}
  {\bibinfo  {journal} {Sci. China Phys. Mech. Astron.}\ }\textbf {\bibinfo
  {volume} {65}},\ \bibinfo {pages} {230411} (\bibinfo {year} {2022})},\
  \Eprint {http://arxiv.org/abs/2010.03976} {arXiv:2010.03976 [astro-ph.CO]}
  \BibitemShut {NoStop}%
\bibitem [{\citenamefont {Vagnozzi}(2021{\natexlab{a}})}]{Vagnozzi:2020gtf}%
  \BibitemOpen
  \bibfield  {author} {\bibinfo {author} {\bibfnamefont {S.}~\bibnamefont
  {Vagnozzi}},\ }\href {\doibase 10.1093/mnrasl/slaa203} {\bibfield  {journal}
  {\bibinfo  {journal} {Mon. Not. Roy. Astron. Soc.}\ }\textbf {\bibinfo
  {volume} {502}},\ \bibinfo {pages} {L11} (\bibinfo {year}
  {2021}{\natexlab{a}})},\ \Eprint {http://arxiv.org/abs/2009.13432}
  {arXiv:2009.13432 [astro-ph.CO]} \BibitemShut {NoStop}%
\bibitem [{\citenamefont {Namba}\ and\ \citenamefont
  {Suzuki}(2020)}]{Namba:2020kij}%
  \BibitemOpen
  \bibfield  {author} {\bibinfo {author} {\bibfnamefont {R.}~\bibnamefont
  {Namba}}\ and\ \bibinfo {author} {\bibfnamefont {M.}~\bibnamefont {Suzuki}},\
  }\href {\doibase 10.1103/PhysRevD.102.123527} {\bibfield  {journal} {\bibinfo
   {journal} {Phys. Rev. D}\ }\textbf {\bibinfo {volume} {102}},\ \bibinfo
  {pages} {123527} (\bibinfo {year} {2020})},\ \Eprint
  {http://arxiv.org/abs/2009.13909} {arXiv:2009.13909 [astro-ph.CO]}
  \BibitemShut {NoStop}%
\bibitem [{\citenamefont {Sugiyama}\ \emph {et~al.}(2021)\citenamefont
  {Sugiyama}, \citenamefont {Takhistov}, \citenamefont {Vitagliano},
  \citenamefont {Kusenko}, \citenamefont {Sasaki},\ and\ \citenamefont
  {Takada}}]{Sugiyama:2020roc}%
  \BibitemOpen
  \bibfield  {author} {\bibinfo {author} {\bibfnamefont {S.}~\bibnamefont
  {Sugiyama}}, \bibinfo {author} {\bibfnamefont {V.}~\bibnamefont {Takhistov}},
  \bibinfo {author} {\bibfnamefont {E.}~\bibnamefont {Vitagliano}}, \bibinfo
  {author} {\bibfnamefont {A.}~\bibnamefont {Kusenko}}, \bibinfo {author}
  {\bibfnamefont {M.}~\bibnamefont {Sasaki}}, \ and\ \bibinfo {author}
  {\bibfnamefont {M.}~\bibnamefont {Takada}},\ }\href {\doibase
  10.1016/j.physletb.2021.136097} {\bibfield  {journal} {\bibinfo  {journal}
  {Phys. Lett. B}\ }\textbf {\bibinfo {volume} {814}},\ \bibinfo {pages}
  {136097} (\bibinfo {year} {2021})},\ \Eprint
  {http://arxiv.org/abs/2010.02189} {arXiv:2010.02189 [astro-ph.CO]}
  \BibitemShut {NoStop}%
\bibitem [{\citenamefont {Zhou}\ \emph {et~al.}(2020)\citenamefont {Zhou},
  \citenamefont {Jiang}, \citenamefont {Cai}, \citenamefont {Sasaki},\ and\
  \citenamefont {Pi}}]{Zhou:2020kkf}%
  \BibitemOpen
  \bibfield  {author} {\bibinfo {author} {\bibfnamefont {Z.}~\bibnamefont
  {Zhou}}, \bibinfo {author} {\bibfnamefont {J.}~\bibnamefont {Jiang}},
  \bibinfo {author} {\bibfnamefont {Y.-F.}\ \bibnamefont {Cai}}, \bibinfo
  {author} {\bibfnamefont {M.}~\bibnamefont {Sasaki}}, \ and\ \bibinfo {author}
  {\bibfnamefont {S.}~\bibnamefont {Pi}},\ }\href {\doibase
  10.1103/PhysRevD.102.103527} {\bibfield  {journal} {\bibinfo  {journal}
  {Phys. Rev. D}\ }\textbf {\bibinfo {volume} {102}},\ \bibinfo {pages}
  {103527} (\bibinfo {year} {2020})},\ \Eprint
  {http://arxiv.org/abs/2010.03537} {arXiv:2010.03537 [astro-ph.CO]}
  \BibitemShut {NoStop}%
\bibitem [{\citenamefont {Lin}\ \emph {et~al.}(2023)\citenamefont {Lin},
  \citenamefont {Gao}, \citenamefont {Gong}, \citenamefont {Lu}, \citenamefont
  {Wang},\ and\ \citenamefont {Zhang}}]{Lin:2021vwc}%
  \BibitemOpen
  \bibfield  {author} {\bibinfo {author} {\bibfnamefont {J.}~\bibnamefont
  {Lin}}, \bibinfo {author} {\bibfnamefont {S.}~\bibnamefont {Gao}}, \bibinfo
  {author} {\bibfnamefont {Y.}~\bibnamefont {Gong}}, \bibinfo {author}
  {\bibfnamefont {Y.}~\bibnamefont {Lu}}, \bibinfo {author} {\bibfnamefont
  {Z.}~\bibnamefont {Wang}}, \ and\ \bibinfo {author} {\bibfnamefont
  {F.}~\bibnamefont {Zhang}},\ }\href {\doibase 10.1103/PhysRevD.107.043517}
  {\bibfield  {journal} {\bibinfo  {journal} {Phys. Rev. D}\ }\textbf {\bibinfo
  {volume} {107}},\ \bibinfo {pages} {043517} (\bibinfo {year} {2023})},\
  \Eprint {http://arxiv.org/abs/2111.01362} {arXiv:2111.01362 [gr-qc]}
  \BibitemShut {NoStop}%
\bibitem [{\citenamefont {Rezazadeh}\ \emph {et~al.}(2022)\citenamefont
  {Rezazadeh}, \citenamefont {Teimoori}, \citenamefont {Karimi},\ and\
  \citenamefont {Karami}}]{Rezazadeh:2021clf}%
  \BibitemOpen
  \bibfield  {author} {\bibinfo {author} {\bibfnamefont {K.}~\bibnamefont
  {Rezazadeh}}, \bibinfo {author} {\bibfnamefont {Z.}~\bibnamefont {Teimoori}},
  \bibinfo {author} {\bibfnamefont {S.}~\bibnamefont {Karimi}}, \ and\ \bibinfo
  {author} {\bibfnamefont {K.}~\bibnamefont {Karami}},\ }\href {\doibase
  10.1140/epjc/s10052-022-10735-w} {\bibfield  {journal} {\bibinfo  {journal}
  {Eur. Phys. J. C}\ }\textbf {\bibinfo {volume} {82}},\ \bibinfo {pages} {758}
  (\bibinfo {year} {2022})},\ \Eprint {http://arxiv.org/abs/2110.01482}
  {arXiv:2110.01482 [gr-qc]} \BibitemShut {NoStop}%
\bibitem [{\citenamefont {Kawasaki}\ and\ \citenamefont
  {Nakatsuka}(2021)}]{Kawasaki:2021ycf}%
  \BibitemOpen
  \bibfield  {author} {\bibinfo {author} {\bibfnamefont {M.}~\bibnamefont
  {Kawasaki}}\ and\ \bibinfo {author} {\bibfnamefont {H.}~\bibnamefont
  {Nakatsuka}},\ }\href {\doibase 10.1088/1475-7516/2021/05/023} {\bibfield
  {journal} {\bibinfo  {journal} {JCAP}\ }\textbf {\bibinfo {volume} {05}},\
  \bibinfo {pages} {023} (\bibinfo {year} {2021})},\ \Eprint
  {http://arxiv.org/abs/2101.11244} {arXiv:2101.11244 [astro-ph.CO]}
  \BibitemShut {NoStop}%
\bibitem [{\citenamefont {Ahmed}\ \emph {et~al.}(2022)\citenamefont {Ahmed},
  \citenamefont {Junaid},\ and\ \citenamefont {Zubair}}]{Ahmed:2021ucx}%
  \BibitemOpen
  \bibfield  {author} {\bibinfo {author} {\bibfnamefont {W.}~\bibnamefont
  {Ahmed}}, \bibinfo {author} {\bibfnamefont {M.}~\bibnamefont {Junaid}}, \
  and\ \bibinfo {author} {\bibfnamefont {U.}~\bibnamefont {Zubair}},\ }\href
  {\doibase 10.1016/j.nuclphysb.2022.115968} {\bibfield  {journal} {\bibinfo
  {journal} {Nucl. Phys. B}\ }\textbf {\bibinfo {volume} {984}},\ \bibinfo
  {pages} {115968} (\bibinfo {year} {2022})},\ \Eprint
  {http://arxiv.org/abs/2109.14838} {arXiv:2109.14838 [astro-ph.CO]}
  \BibitemShut {NoStop}%
\bibitem [{\citenamefont {Yi}\ and\ \citenamefont {Fei}(2023)}]{Yi:2022ymw}%
  \BibitemOpen
  \bibfield  {author} {\bibinfo {author} {\bibfnamefont {Z.}~\bibnamefont
  {Yi}}\ and\ \bibinfo {author} {\bibfnamefont {Q.}~\bibnamefont {Fei}},\
  }\href {\doibase 10.1140/epjc/s10052-023-11233-3} {\bibfield  {journal}
  {\bibinfo  {journal} {Eur. Phys. J. C}\ }\textbf {\bibinfo {volume} {83}},\
  \bibinfo {pages} {82} (\bibinfo {year} {2023})},\ \Eprint
  {http://arxiv.org/abs/2210.03641} {arXiv:2210.03641 [astro-ph.CO]}
  \BibitemShut {NoStop}%
\bibitem [{\citenamefont {Yi}(2023)}]{Yi:2022anu}%
  \BibitemOpen
  \bibfield  {author} {\bibinfo {author} {\bibfnamefont {Z.}~\bibnamefont
  {Yi}},\ }\href {\doibase 10.1088/1475-7516/2023/03/048} {\bibfield  {journal}
  {\bibinfo  {journal} {JCAP}\ }\textbf {\bibinfo {volume} {03}},\ \bibinfo
  {pages} {048} (\bibinfo {year} {2023})},\ \Eprint
  {http://arxiv.org/abs/2206.01039} {arXiv:2206.01039 [gr-qc]} \BibitemShut
  {NoStop}%
\bibitem [{\citenamefont {Dandoy}\ \emph {et~al.}(2023)\citenamefont {Dandoy},
  \citenamefont {Domcke},\ and\ \citenamefont {Rompineve}}]{Dandoy:2023jot}%
  \BibitemOpen
  \bibfield  {author} {\bibinfo {author} {\bibfnamefont {V.}~\bibnamefont
  {Dandoy}}, \bibinfo {author} {\bibfnamefont {V.}~\bibnamefont {Domcke}}, \
  and\ \bibinfo {author} {\bibfnamefont {F.}~\bibnamefont {Rompineve}},\ }\href
  {\doibase 10.21468/SciPostPhysCore.6.3.060} {\bibfield  {journal} {\bibinfo
  {journal} {SciPost Phys. Core}\ }\textbf {\bibinfo {volume} {6}},\ \bibinfo
  {pages} {060} (\bibinfo {year} {2023})},\ \Eprint
  {http://arxiv.org/abs/2302.07901} {arXiv:2302.07901 [astro-ph.CO]}
  \BibitemShut {NoStop}%
\bibitem [{\citenamefont {Zhao}\ \emph {et~al.}(2023)\citenamefont {Zhao},
  \citenamefont {Liu},\ and\ \citenamefont {Li}}]{Zhao:2023xnh}%
  \BibitemOpen
  \bibfield  {author} {\bibinfo {author} {\bibfnamefont {J.-X.}\ \bibnamefont
  {Zhao}}, \bibinfo {author} {\bibfnamefont {X.-H.}\ \bibnamefont {Liu}}, \
  and\ \bibinfo {author} {\bibfnamefont {N.}~\bibnamefont {Li}},\ }\href
  {\doibase 10.1103/PhysRevD.107.043515} {\bibfield  {journal} {\bibinfo
  {journal} {Phys. Rev. D}\ }\textbf {\bibinfo {volume} {107}},\ \bibinfo
  {pages} {043515} (\bibinfo {year} {2023})},\ \Eprint
  {http://arxiv.org/abs/2302.06886} {arXiv:2302.06886 [astro-ph.CO]}
  \BibitemShut {NoStop}%
\bibitem [{\citenamefont {Ferrante}\ \emph {et~al.}(2023)\citenamefont
  {Ferrante}, \citenamefont {Franciolini}, \citenamefont {Iovino},\ and\
  \citenamefont {Urbano}}]{Ferrante:2023bgz}%
  \BibitemOpen
  \bibfield  {author} {\bibinfo {author} {\bibfnamefont {G.}~\bibnamefont
  {Ferrante}}, \bibinfo {author} {\bibfnamefont {G.}~\bibnamefont
  {Franciolini}}, \bibinfo {author} {\bibfnamefont {A.}~\bibnamefont {Iovino},
  \bibfnamefont {Junior.}}, \ and\ \bibinfo {author} {\bibfnamefont
  {A.}~\bibnamefont {Urbano}},\ }\href {\doibase 10.1088/1475-7516/2023/06/057}
  {\bibfield  {journal} {\bibinfo  {journal} {JCAP}\ }\textbf {\bibinfo
  {volume} {06}},\ \bibinfo {pages} {057} (\bibinfo {year} {2023})},\ \Eprint
  {http://arxiv.org/abs/2305.13382} {arXiv:2305.13382 [astro-ph.CO]}
  \BibitemShut {NoStop}%
\bibitem [{\citenamefont {Cai}\ \emph {et~al.}(2024)\citenamefont {Cai},
  \citenamefont {Zhu},\ and\ \citenamefont {Piao}}]{Cai:2023uhc}%
  \BibitemOpen
  \bibfield  {author} {\bibinfo {author} {\bibfnamefont {Y.}~\bibnamefont
  {Cai}}, \bibinfo {author} {\bibfnamefont {M.}~\bibnamefont {Zhu}}, \ and\
  \bibinfo {author} {\bibfnamefont {Y.-S.}\ \bibnamefont {Piao}},\ }\href
  {\doibase 10.1103/PhysRevLett.133.021001} {\bibfield  {journal} {\bibinfo
  {journal} {Phys. Rev. Lett.}\ }\textbf {\bibinfo {volume} {133}},\ \bibinfo
  {pages} {021001} (\bibinfo {year} {2024})},\ \Eprint
  {http://arxiv.org/abs/2305.10933} {arXiv:2305.10933 [gr-qc]} \BibitemShut
  {NoStop}%
\bibitem [{\citenamefont {Vagnozzi}(2023{\natexlab{a}})}]{Vagnozzi:2023lwo}%
  \BibitemOpen
  \bibfield  {author} {\bibinfo {author} {\bibfnamefont {S.}~\bibnamefont
  {Vagnozzi}},\ }\href {\doibase 10.1016/j.jheap.2023.07.001} {\bibfield
  {journal} {\bibinfo  {journal} {JHEAp}\ }\textbf {\bibinfo {volume} {39}},\
  \bibinfo {pages} {81} (\bibinfo {year} {2023}{\natexlab{a}})},\ \Eprint
  {http://arxiv.org/abs/2306.16912} {arXiv:2306.16912 [astro-ph.CO]}
  \BibitemShut {NoStop}%
\bibitem [{\citenamefont {Franciolini}\ \emph {et~al.}(2023)\citenamefont
  {Franciolini}, \citenamefont {Iovino}, \citenamefont {Vaskonen},\ and\
  \citenamefont {Veermae}}]{Franciolini:2023pbf}%
  \BibitemOpen
  \bibfield  {author} {\bibinfo {author} {\bibfnamefont {G.}~\bibnamefont
  {Franciolini}}, \bibinfo {author} {\bibfnamefont {A.}~\bibnamefont {Iovino},
  \bibfnamefont {Junior.}}, \bibinfo {author} {\bibfnamefont {V.}~\bibnamefont
  {Vaskonen}}, \ and\ \bibinfo {author} {\bibfnamefont {H.}~\bibnamefont
  {Veermae}},\ }\href {\doibase 10.1103/PhysRevLett.131.201401} {\bibfield
  {journal} {\bibinfo  {journal} {Phys. Rev. Lett.}\ }\textbf {\bibinfo
  {volume} {131}},\ \bibinfo {pages} {201401} (\bibinfo {year} {2023})},\
  \Eprint {http://arxiv.org/abs/2306.17149} {arXiv:2306.17149 [astro-ph.CO]}
  \BibitemShut {NoStop}%
\bibitem [{\citenamefont {Wang}\ \emph {et~al.}(2024)\citenamefont {Wang},
  \citenamefont {Zhang},\ and\ \citenamefont {Sasaki}}]{Wang:2024vfv}%
  \BibitemOpen
  \bibfield  {author} {\bibinfo {author} {\bibfnamefont {X.}~\bibnamefont
  {Wang}}, \bibinfo {author} {\bibfnamefont {Y.-l.}\ \bibnamefont {Zhang}}, \
  and\ \bibinfo {author} {\bibfnamefont {M.}~\bibnamefont {Sasaki}},\ }\href
  {\doibase 10.1088/1475-7516/2024/07/076} {\bibfield  {journal} {\bibinfo
  {journal} {JCAP}\ }\textbf {\bibinfo {volume} {07}},\ \bibinfo {pages} {076}
  (\bibinfo {year} {2024})},\ \Eprint {http://arxiv.org/abs/2404.02492}
  {arXiv:2404.02492 [astro-ph.CO]} \BibitemShut {NoStop}%
\bibitem [{\citenamefont {Iovino}\ \emph {et~al.}(2024)\citenamefont {Iovino},
  \citenamefont {Perna}, \citenamefont {Riotto},\ and\ \citenamefont
  {Veerm{\"a}e}}]{Iovino:2024tyg}%
  \BibitemOpen
  \bibfield  {author} {\bibinfo {author} {\bibfnamefont {A.~J.}\ \bibnamefont
  {Iovino}}, \bibinfo {author} {\bibfnamefont {G.}~\bibnamefont {Perna}},
  \bibinfo {author} {\bibfnamefont {A.}~\bibnamefont {Riotto}}, \ and\ \bibinfo
  {author} {\bibfnamefont {H.}~\bibnamefont {Veerm{\"a}e}},\ }\href {\doibase
  10.1088/1475-7516/2024/10/050} {\bibfield  {journal} {\bibinfo  {journal}
  {JCAP}\ }\textbf {\bibinfo {volume} {10}},\ \bibinfo {pages} {050} (\bibinfo
  {year} {2024})},\ \Eprint {http://arxiv.org/abs/2406.20089} {arXiv:2406.20089
  [astro-ph.CO]} \BibitemShut {NoStop}%
\bibitem [{\citenamefont {Yogesh}\ and\ \citenamefont
  {Mohammadi}(2025)}]{Yogesh:2025hll}%
  \BibitemOpen
  \bibfield  {author} {\bibinfo {author} {\bibnamefont {Yogesh}}\ and\ \bibinfo
  {author} {\bibfnamefont {A.}~\bibnamefont {Mohammadi}},\ }\href {\doibase
  10.3847/1538-4357/adcee5} {\bibfield  {journal} {\bibinfo  {journal}
  {Astrophys. J.}\ }\textbf {\bibinfo {volume} {986}},\ \bibinfo {pages} {35}
  (\bibinfo {year} {2025})},\ \Eprint {http://arxiv.org/abs/2501.01867}
  {arXiv:2501.01867 [gr-qc]} \BibitemShut {NoStop}%
\bibitem [{\citenamefont {Gouttenoire}\ \emph {et~al.}(2026)\citenamefont
  {Gouttenoire}, \citenamefont {Trifinopoulos},\ and\ \citenamefont
  {Vanvlasselaer}}]{Gouttenoire:2025jxe}%
  \BibitemOpen
  \bibfield  {author} {\bibinfo {author} {\bibfnamefont {Y.}~\bibnamefont
  {Gouttenoire}}, \bibinfo {author} {\bibfnamefont {S.}~\bibnamefont
  {Trifinopoulos}}, \ and\ \bibinfo {author} {\bibfnamefont {M.}~\bibnamefont
  {Vanvlasselaer}},\ }\href {\doibase 10.1088/1475-7516/2026/02/072} {\bibfield
   {journal} {\bibinfo  {journal} {JCAP}\ }\textbf {\bibinfo {volume} {02}},\
  \bibinfo {pages} {072} (\bibinfo {year} {2026})},\ \Eprint
  {http://arxiv.org/abs/2508.19328} {arXiv:2508.19328 [astro-ph.CO]}
  \BibitemShut {NoStop}%
\bibitem [{\citenamefont {Mohammadi}\ \emph
  {et~al.}(2026{\natexlab{a}})\citenamefont {Mohammadi}, \citenamefont
  {Yogesh}, \citenamefont {Wu},\ and\ \citenamefont {Zhu}}]{Mohammadi:2025avz}%
  \BibitemOpen
  \bibfield  {author} {\bibinfo {author} {\bibfnamefont {A.}~\bibnamefont
  {Mohammadi}}, \bibinfo {author} {\bibnamefont {Yogesh}}, \bibinfo {author}
  {\bibfnamefont {Q.}~\bibnamefont {Wu}}, \ and\ \bibinfo {author}
  {\bibfnamefont {T.}~\bibnamefont {Zhu}},\ }\href {\doibase
  10.3847/1538-4357/ae47ed} {\bibfield  {journal} {\bibinfo  {journal}
  {Astrophys. J.}\ }\textbf {\bibinfo {volume} {1000}},\ \bibinfo {pages} {101}
  (\bibinfo {year} {2026}{\natexlab{a}})},\ \Eprint
  {http://arxiv.org/abs/2512.05435} {arXiv:2512.05435 [astro-ph.CO]}
  \BibitemShut {NoStop}%
\bibitem [{\citenamefont {Ivanov}\ \emph {et~al.}(1994)\citenamefont {Ivanov},
  \citenamefont {Naselsky},\ and\ \citenamefont {Novikov}}]{Ivanov:1994pa}%
  \BibitemOpen
  \bibfield  {author} {\bibinfo {author} {\bibfnamefont {P.}~\bibnamefont
  {Ivanov}}, \bibinfo {author} {\bibfnamefont {P.}~\bibnamefont {Naselsky}}, \
  and\ \bibinfo {author} {\bibfnamefont {I.}~\bibnamefont {Novikov}},\ }\href
  {\doibase 10.1103/PhysRevD.50.7173} {\bibfield  {journal} {\bibinfo
  {journal} {Phys. Rev. D}\ }\textbf {\bibinfo {volume} {50}},\ \bibinfo
  {pages} {7173} (\bibinfo {year} {1994})}\BibitemShut {NoStop}%
\bibitem [{\citenamefont {Kinney}(1997)}]{Kinney:1997ne}%
  \BibitemOpen
  \bibfield  {author} {\bibinfo {author} {\bibfnamefont {W.~H.}\ \bibnamefont
  {Kinney}},\ }\href {\doibase 10.1103/PhysRevD.56.2002} {\bibfield  {journal}
  {\bibinfo  {journal} {Phys. Rev. D}\ }\textbf {\bibinfo {volume} {56}},\
  \bibinfo {pages} {2002} (\bibinfo {year} {1997})},\ \Eprint
  {http://arxiv.org/abs/hep-ph/9702427} {arXiv:hep-ph/9702427} \BibitemShut
  {NoStop}%
\bibitem [{\citenamefont {Garcia-Bellido}\ and\ \citenamefont
  {Ruiz~Morales}(2017)}]{Garcia-Bellido:2017mdw}%
  \BibitemOpen
  \bibfield  {author} {\bibinfo {author} {\bibfnamefont {J.}~\bibnamefont
  {Garcia-Bellido}}\ and\ \bibinfo {author} {\bibfnamefont {E.}~\bibnamefont
  {Ruiz~Morales}},\ }\href {\doibase 10.1016/j.dark.2017.09.007} {\bibfield
  {journal} {\bibinfo  {journal} {Phys. Dark Univ.}\ }\textbf {\bibinfo
  {volume} {18}},\ \bibinfo {pages} {47} (\bibinfo {year} {2017})},\ \Eprint
  {http://arxiv.org/abs/1702.03901} {arXiv:1702.03901 [astro-ph.CO]}
  \BibitemShut {NoStop}%
\bibitem [{\citenamefont {Germani}\ and\ \citenamefont
  {Prokopec}(2017)}]{Germani:2017bcs}%
  \BibitemOpen
  \bibfield  {author} {\bibinfo {author} {\bibfnamefont {C.}~\bibnamefont
  {Germani}}\ and\ \bibinfo {author} {\bibfnamefont {T.}~\bibnamefont
  {Prokopec}},\ }\href {\doibase 10.1016/j.dark.2017.09.001} {\bibfield
  {journal} {\bibinfo  {journal} {Phys. Dark Univ.}\ }\textbf {\bibinfo
  {volume} {18}},\ \bibinfo {pages} {6} (\bibinfo {year} {2017})},\ \Eprint
  {http://arxiv.org/abs/1706.04226} {arXiv:1706.04226 [astro-ph.CO]}
  \BibitemShut {NoStop}%
\bibitem [{\citenamefont {Kannike}\ \emph {et~al.}(2017)\citenamefont
  {Kannike}, \citenamefont {Marzola}, \citenamefont {Raidal},\ and\
  \citenamefont {Veerm{\"a}e}}]{Kannike:2017bxn}%
  \BibitemOpen
  \bibfield  {author} {\bibinfo {author} {\bibfnamefont {K.}~\bibnamefont
  {Kannike}}, \bibinfo {author} {\bibfnamefont {L.}~\bibnamefont {Marzola}},
  \bibinfo {author} {\bibfnamefont {M.}~\bibnamefont {Raidal}}, \ and\ \bibinfo
  {author} {\bibfnamefont {H.}~\bibnamefont {Veerm{\"a}e}},\ }\href {\doibase
  10.1088/1475-7516/2017/09/020} {\bibfield  {journal} {\bibinfo  {journal}
  {JCAP}\ }\textbf {\bibinfo {volume} {09}},\ \bibinfo {pages} {020} (\bibinfo
  {year} {2017})},\ \Eprint {http://arxiv.org/abs/1705.06225} {arXiv:1705.06225
  [astro-ph.CO]} \BibitemShut {NoStop}%
\bibitem [{\citenamefont {Di}\ and\ \citenamefont {Gong}(2018)}]{Di:2017ndc}%
  \BibitemOpen
  \bibfield  {author} {\bibinfo {author} {\bibfnamefont {H.}~\bibnamefont
  {Di}}\ and\ \bibinfo {author} {\bibfnamefont {Y.}~\bibnamefont {Gong}},\
  }\href {\doibase 10.1088/1475-7516/2018/07/007} {\bibfield  {journal}
  {\bibinfo  {journal} {JCAP}\ }\textbf {\bibinfo {volume} {07}},\ \bibinfo
  {pages} {007} (\bibinfo {year} {2018})},\ \Eprint
  {http://arxiv.org/abs/1707.09578} {arXiv:1707.09578 [astro-ph.CO]}
  \BibitemShut {NoStop}%
\bibitem [{\citenamefont {Ballesteros}\ and\ \citenamefont
  {Taoso}(2018)}]{Ballesteros:2017fsr}%
  \BibitemOpen
  \bibfield  {author} {\bibinfo {author} {\bibfnamefont {G.}~\bibnamefont
  {Ballesteros}}\ and\ \bibinfo {author} {\bibfnamefont {M.}~\bibnamefont
  {Taoso}},\ }\href {\doibase 10.1103/PhysRevD.97.023501} {\bibfield  {journal}
  {\bibinfo  {journal} {Phys. Rev. D}\ }\textbf {\bibinfo {volume} {97}},\
  \bibinfo {pages} {023501} (\bibinfo {year} {2018})},\ \Eprint
  {http://arxiv.org/abs/1709.05565} {arXiv:1709.05565 [hep-ph]} \BibitemShut
  {NoStop}%
\bibitem [{\citenamefont {Cicoli}\ \emph {et~al.}(2018)\citenamefont {Cicoli},
  \citenamefont {Diaz},\ and\ \citenamefont {Pedro}}]{Cicoli:2018asa}%
  \BibitemOpen
  \bibfield  {author} {\bibinfo {author} {\bibfnamefont {M.}~\bibnamefont
  {Cicoli}}, \bibinfo {author} {\bibfnamefont {V.~A.}\ \bibnamefont {Diaz}}, \
  and\ \bibinfo {author} {\bibfnamefont {F.~G.}\ \bibnamefont {Pedro}},\ }\href
  {\doibase 10.1088/1475-7516/2018/06/034} {\bibfield  {journal} {\bibinfo
  {journal} {JCAP}\ }\textbf {\bibinfo {volume} {06}},\ \bibinfo {pages} {034}
  (\bibinfo {year} {2018})},\ \Eprint {http://arxiv.org/abs/1803.02837}
  {arXiv:1803.02837 [hep-th]} \BibitemShut {NoStop}%
\bibitem [{\citenamefont {Ballesteros}\ \emph
  {et~al.}(2020{\natexlab{a}})\citenamefont {Ballesteros}, \citenamefont {Rey},
  \citenamefont {Taoso},\ and\ \citenamefont {Urbano}}]{Ballesteros:2020qam}%
  \BibitemOpen
  \bibfield  {author} {\bibinfo {author} {\bibfnamefont {G.}~\bibnamefont
  {Ballesteros}}, \bibinfo {author} {\bibfnamefont {J.}~\bibnamefont {Rey}},
  \bibinfo {author} {\bibfnamefont {M.}~\bibnamefont {Taoso}}, \ and\ \bibinfo
  {author} {\bibfnamefont {A.}~\bibnamefont {Urbano}},\ }\href {\doibase
  10.1088/1475-7516/2020/07/025} {\bibfield  {journal} {\bibinfo  {journal}
  {JCAP}\ }\textbf {\bibinfo {volume} {07}},\ \bibinfo {pages} {025} (\bibinfo
  {year} {2020}{\natexlab{a}})},\ \Eprint {http://arxiv.org/abs/2001.08220}
  {arXiv:2001.08220 [astro-ph.CO]} \BibitemShut {NoStop}%
\bibitem [{\citenamefont {Ragavendra}\ \emph {et~al.}(2021)\citenamefont
  {Ragavendra}, \citenamefont {Saha}, \citenamefont {Sriramkumar},\ and\
  \citenamefont {Silk}}]{Ragavendra:2020sop}%
  \BibitemOpen
  \bibfield  {author} {\bibinfo {author} {\bibfnamefont {H.~V.}\ \bibnamefont
  {Ragavendra}}, \bibinfo {author} {\bibfnamefont {P.}~\bibnamefont {Saha}},
  \bibinfo {author} {\bibfnamefont {L.}~\bibnamefont {Sriramkumar}}, \ and\
  \bibinfo {author} {\bibfnamefont {J.}~\bibnamefont {Silk}},\ }\href {\doibase
  10.1103/PhysRevD.103.083510} {\bibfield  {journal} {\bibinfo  {journal}
  {Phys. Rev. D}\ }\textbf {\bibinfo {volume} {103}},\ \bibinfo {pages}
  {083510} (\bibinfo {year} {2021})},\ \Eprint
  {http://arxiv.org/abs/2008.12202} {arXiv:2008.12202 [astro-ph.CO]}
  \BibitemShut {NoStop}%
\bibitem [{\citenamefont {Karam}\ \emph {et~al.}(2023)\citenamefont {Karam},
  \citenamefont {Koivunen}, \citenamefont {Tomberg}, \citenamefont {Vaskonen},\
  and\ \citenamefont {Veerm{\"a}e}}]{Karam:2022nym}%
  \BibitemOpen
  \bibfield  {author} {\bibinfo {author} {\bibfnamefont {A.}~\bibnamefont
  {Karam}}, \bibinfo {author} {\bibfnamefont {N.}~\bibnamefont {Koivunen}},
  \bibinfo {author} {\bibfnamefont {E.}~\bibnamefont {Tomberg}}, \bibinfo
  {author} {\bibfnamefont {V.}~\bibnamefont {Vaskonen}}, \ and\ \bibinfo
  {author} {\bibfnamefont {H.}~\bibnamefont {Veerm{\"a}e}},\ }\href {\doibase
  10.1088/1475-7516/2023/03/013} {\bibfield  {journal} {\bibinfo  {journal}
  {JCAP}\ }\textbf {\bibinfo {volume} {03}},\ \bibinfo {pages} {013} (\bibinfo
  {year} {2023})},\ \Eprint {http://arxiv.org/abs/2205.13540} {arXiv:2205.13540
  [astro-ph.CO]} \BibitemShut {NoStop}%
\bibitem [{\citenamefont {Ballesteros}\ and\ \citenamefont
  {Gamb{\'\i}n~Egea}(2024)}]{Ballesteros:2024zdp}%
  \BibitemOpen
  \bibfield  {author} {\bibinfo {author} {\bibfnamefont {G.}~\bibnamefont
  {Ballesteros}}\ and\ \bibinfo {author} {\bibfnamefont {J.}~\bibnamefont
  {Gamb{\'\i}n~Egea}},\ }\href {\doibase 10.1088/1475-7516/2024/07/052}
  {\bibfield  {journal} {\bibinfo  {journal} {JCAP}\ }\textbf {\bibinfo
  {volume} {07}},\ \bibinfo {pages} {052} (\bibinfo {year} {2024})},\ \Eprint
  {http://arxiv.org/abs/2404.07196} {arXiv:2404.07196 [astro-ph.CO]}
  \BibitemShut {NoStop}%
\bibitem [{\citenamefont {Allegrini}\ \emph {et~al.}(2025)\citenamefont
  {Allegrini}, \citenamefont {Del~Grosso}, \citenamefont {Iovino},\ and\
  \citenamefont {Urbano}}]{Allegrini:2024ooy}%
  \BibitemOpen
  \bibfield  {author} {\bibinfo {author} {\bibfnamefont {S.}~\bibnamefont
  {Allegrini}}, \bibinfo {author} {\bibfnamefont {L.}~\bibnamefont
  {Del~Grosso}}, \bibinfo {author} {\bibfnamefont {A.~J.}\ \bibnamefont
  {Iovino}}, \ and\ \bibinfo {author} {\bibfnamefont {A.}~\bibnamefont
  {Urbano}},\ }\href {\doibase 10.1103/9rr3-ltbt} {\bibfield  {journal}
  {\bibinfo  {journal} {Phys. Rev. D}\ }\textbf {\bibinfo {volume} {111}},\
  \bibinfo {pages} {123557} (\bibinfo {year} {2025})},\ \Eprint
  {http://arxiv.org/abs/2412.14049} {arXiv:2412.14049 [astro-ph.CO]}
  \BibitemShut {NoStop}%
\bibitem [{\citenamefont {Briaud}\ \emph {et~al.}(2025)\citenamefont {Briaud},
  \citenamefont {Karam}, \citenamefont {Koivunen}, \citenamefont {Tomberg},
  \citenamefont {Veerm{\"a}e},\ and\ \citenamefont {Vennin}}]{Briaud:2025hra}%
  \BibitemOpen
  \bibfield  {author} {\bibinfo {author} {\bibfnamefont {V.}~\bibnamefont
  {Briaud}}, \bibinfo {author} {\bibfnamefont {A.}~\bibnamefont {Karam}},
  \bibinfo {author} {\bibfnamefont {N.}~\bibnamefont {Koivunen}}, \bibinfo
  {author} {\bibfnamefont {E.}~\bibnamefont {Tomberg}}, \bibinfo {author}
  {\bibfnamefont {H.}~\bibnamefont {Veerm{\"a}e}}, \ and\ \bibinfo {author}
  {\bibfnamefont {V.}~\bibnamefont {Vennin}},\ }\href {\doibase
  10.1088/1475-7516/2025/05/097} {\bibfield  {journal} {\bibinfo  {journal}
  {JCAP}\ }\textbf {\bibinfo {volume} {05}},\ \bibinfo {pages} {097} (\bibinfo
  {year} {2025})},\ \Eprint {http://arxiv.org/abs/2501.14681} {arXiv:2501.14681
  [astro-ph.CO]} \BibitemShut {NoStop}%
\bibitem [{\citenamefont {Totolou}\ \emph {et~al.}(2025)\citenamefont
  {Totolou}, \citenamefont {Papanikolaou},\ and\ \citenamefont
  {Saridakis}}]{Totolou:2025dsb}%
  \BibitemOpen
  \bibfield  {author} {\bibinfo {author} {\bibfnamefont {D.}~\bibnamefont
  {Totolou}}, \bibinfo {author} {\bibfnamefont {T.}~\bibnamefont
  {Papanikolaou}}, \ and\ \bibinfo {author} {\bibfnamefont {E.~N.}\
  \bibnamefont {Saridakis}},\ }\href@noop {} {\  (\bibinfo {year} {2025})},\
  \Eprint {http://arxiv.org/abs/2512.25044} {arXiv:2512.25044 [gr-qc]}
  \BibitemShut {NoStop}%
\bibitem [{\citenamefont {Cole}\ \emph {et~al.}(2023)\citenamefont {Cole},
  \citenamefont {Gow}, \citenamefont {Byrnes},\ and\ \citenamefont
  {Patil}}]{Cole:2023wyx}%
  \BibitemOpen
  \bibfield  {author} {\bibinfo {author} {\bibfnamefont {P.~S.}\ \bibnamefont
  {Cole}}, \bibinfo {author} {\bibfnamefont {A.~D.}\ \bibnamefont {Gow}},
  \bibinfo {author} {\bibfnamefont {C.~T.}\ \bibnamefont {Byrnes}}, \ and\
  \bibinfo {author} {\bibfnamefont {S.~P.}\ \bibnamefont {Patil}},\ }\href
  {\doibase 10.1088/1475-7516/2023/08/031} {\bibfield  {journal} {\bibinfo
  {journal} {JCAP}\ }\textbf {\bibinfo {volume} {08}},\ \bibinfo {pages} {031}
  (\bibinfo {year} {2023})},\ \Eprint {http://arxiv.org/abs/2304.01997}
  {arXiv:2304.01997 [astro-ph.CO]} \BibitemShut {NoStop}%
\bibitem [{\citenamefont {Iovino}\ and\ \citenamefont
  {Riotto}(2025)}]{Iovino:2025tcv}%
  \BibitemOpen
  \bibfield  {author} {\bibinfo {author} {\bibfnamefont {A.~J.}\ \bibnamefont
  {Iovino}}\ and\ \bibinfo {author} {\bibfnamefont {A.}~\bibnamefont
  {Riotto}},\ }\href@noop {} {\  (\bibinfo {year} {2025})},\ \Eprint
  {http://arxiv.org/abs/2512.19668} {arXiv:2512.19668 [astro-ph.CO]}
  \BibitemShut {NoStop}%
\bibitem [{\citenamefont {Profumo}(2026)}]{Profumo:2026qpn}%
  \BibitemOpen
  \bibfield  {author} {\bibinfo {author} {\bibfnamefont {S.}~\bibnamefont
  {Profumo}},\ }\href@noop {} {\  (\bibinfo {year} {2026})},\ \Eprint
  {http://arxiv.org/abs/2606.12775} {arXiv:2606.12775 [hep-ph]} \BibitemShut
  {NoStop}%
\bibitem [{\citenamefont {Allegrini}\ \emph
  {et~al.}(2026{\natexlab{a}})\citenamefont {Allegrini}, \citenamefont
  {Iovino},\ and\ \citenamefont {Veerm{\"a}e}}]{Allegrini:2025jha}%
  \BibitemOpen
  \bibfield  {author} {\bibinfo {author} {\bibfnamefont {S.}~\bibnamefont
  {Allegrini}}, \bibinfo {author} {\bibfnamefont {A.~J.}\ \bibnamefont
  {Iovino}}, \ and\ \bibinfo {author} {\bibfnamefont {H.}~\bibnamefont
  {Veerm{\"a}e}},\ }\href {\doibase 10.1103/shz2-xzcy} {\bibfield  {journal}
  {\bibinfo  {journal} {Phys. Rev. D}\ }\textbf {\bibinfo {volume} {113}},\
  \bibinfo {pages} {043530} (\bibinfo {year} {2026}{\natexlab{a}})},\ \Eprint
  {http://arxiv.org/abs/2510.18791} {arXiv:2510.18791 [astro-ph.CO]}
  \BibitemShut {NoStop}%
\bibitem [{\citenamefont {Calabrese}\ \emph {et~al.}(2025)\citenamefont
  {Calabrese} \emph {et~al.}}]{AtacamaCosmologyTelescope:2025nti}%
  \BibitemOpen
  \bibfield  {author} {\bibinfo {author} {\bibfnamefont {E.}~\bibnamefont
  {Calabrese}} \emph {et~al.} (\bibinfo {collaboration} {Atacama Cosmology
  Telescope}),\ }\href {\doibase 10.1088/1475-7516/2025/11/063} {\bibfield
  {journal} {\bibinfo  {journal} {JCAP}\ }\textbf {\bibinfo {volume} {11}},\
  \bibinfo {pages} {063} (\bibinfo {year} {2025})},\ \Eprint
  {http://arxiv.org/abs/2503.14454} {arXiv:2503.14454 [astro-ph.CO]}
  \BibitemShut {NoStop}%
\bibitem [{\citenamefont {Kallosh}\ \emph {et~al.}(2025)\citenamefont
  {Kallosh}, \citenamefont {Linde},\ and\ \citenamefont
  {Roest}}]{Kallosh:2025rni}%
  \BibitemOpen
  \bibfield  {author} {\bibinfo {author} {\bibfnamefont {R.}~\bibnamefont
  {Kallosh}}, \bibinfo {author} {\bibfnamefont {A.}~\bibnamefont {Linde}}, \
  and\ \bibinfo {author} {\bibfnamefont {D.}~\bibnamefont {Roest}},\ }\href
  {\doibase 10.1103/d6gn-78hn} {\bibfield  {journal} {\bibinfo  {journal}
  {Phys. Rev. Lett.}\ }\textbf {\bibinfo {volume} {135}},\ \bibinfo {pages}
  {161001} (\bibinfo {year} {2025})},\ \Eprint
  {http://arxiv.org/abs/2503.21030} {arXiv:2503.21030 [hep-th]} \BibitemShut
  {NoStop}%
\bibitem [{\citenamefont {Aoki}\ \emph
  {et~al.}(2025{\natexlab{a}})\citenamefont {Aoki}, \citenamefont {Otsuka},\
  and\ \citenamefont {Yanagita}}]{Aoki:2025wld}%
  \BibitemOpen
  \bibfield  {author} {\bibinfo {author} {\bibfnamefont {S.}~\bibnamefont
  {Aoki}}, \bibinfo {author} {\bibfnamefont {H.}~\bibnamefont {Otsuka}}, \ and\
  \bibinfo {author} {\bibfnamefont {R.}~\bibnamefont {Yanagita}},\ }\href
  {\doibase 10.1103/v4z9-676d} {\bibfield  {journal} {\bibinfo  {journal}
  {Phys. Rev. D}\ }\textbf {\bibinfo {volume} {112}},\ \bibinfo {pages}
  {043505} (\bibinfo {year} {2025}{\natexlab{a}})},\ \Eprint
  {http://arxiv.org/abs/2504.01622} {arXiv:2504.01622 [hep-ph]} \BibitemShut
  {NoStop}%
\bibitem [{\citenamefont {Gialamas}\ \emph
  {et~al.}(2025{\natexlab{a}})\citenamefont {Gialamas}, \citenamefont {Karam},
  \citenamefont {Racioppi},\ and\ \citenamefont {Raidal}}]{Gialamas:2025kef}%
  \BibitemOpen
  \bibfield  {author} {\bibinfo {author} {\bibfnamefont {I.~D.}\ \bibnamefont
  {Gialamas}}, \bibinfo {author} {\bibfnamefont {A.}~\bibnamefont {Karam}},
  \bibinfo {author} {\bibfnamefont {A.}~\bibnamefont {Racioppi}}, \ and\
  \bibinfo {author} {\bibfnamefont {M.}~\bibnamefont {Raidal}},\ }\href
  {\doibase 10.1103/6fpc-67s1} {\bibfield  {journal} {\bibinfo  {journal}
  {Phys. Rev. D}\ }\textbf {\bibinfo {volume} {112}},\ \bibinfo {pages}
  {103544} (\bibinfo {year} {2025}{\natexlab{a}})},\ \Eprint
  {http://arxiv.org/abs/2504.06002} {arXiv:2504.06002 [astro-ph.CO]}
  \BibitemShut {NoStop}%
\bibitem [{\citenamefont {Dioguardi}\ \emph {et~al.}(2025)\citenamefont
  {Dioguardi}, \citenamefont {Iovino},\ and\ \citenamefont
  {Racioppi}}]{Dioguardi:2025vci}%
  \BibitemOpen
  \bibfield  {author} {\bibinfo {author} {\bibfnamefont {C.}~\bibnamefont
  {Dioguardi}}, \bibinfo {author} {\bibfnamefont {A.~J.}\ \bibnamefont
  {Iovino}}, \ and\ \bibinfo {author} {\bibfnamefont {A.}~\bibnamefont
  {Racioppi}},\ }\href {\doibase 10.1016/j.physletb.2025.139664} {\bibfield
  {journal} {\bibinfo  {journal} {Phys. Lett. B}\ }\textbf {\bibinfo {volume}
  {868}},\ \bibinfo {pages} {139664} (\bibinfo {year} {2025})},\ \Eprint
  {http://arxiv.org/abs/2504.02809} {arXiv:2504.02809 [gr-qc]} \BibitemShut
  {NoStop}%
\bibitem [{\citenamefont {Kim}\ \emph {et~al.}(2025)\citenamefont {Kim},
  \citenamefont {Wang}, \citenamefont {Zhang},\ and\ \citenamefont
  {Ren}}]{Kim:2025dyi}%
  \BibitemOpen
  \bibfield  {author} {\bibinfo {author} {\bibfnamefont {J.}~\bibnamefont
  {Kim}}, \bibinfo {author} {\bibfnamefont {X.}~\bibnamefont {Wang}}, \bibinfo
  {author} {\bibfnamefont {Y.-l.}\ \bibnamefont {Zhang}}, \ and\ \bibinfo
  {author} {\bibfnamefont {Z.}~\bibnamefont {Ren}},\ }\href {\doibase
  10.1088/1475-7516/2025/09/011} {\bibfield  {journal} {\bibinfo  {journal}
  {JCAP}\ }\textbf {\bibinfo {volume} {09}},\ \bibinfo {pages} {011} (\bibinfo
  {year} {2025})},\ \Eprint {http://arxiv.org/abs/2504.12035} {arXiv:2504.12035
  [astro-ph.CO]} \BibitemShut {NoStop}%
\bibitem [{\citenamefont {Gao}\ \emph {et~al.}(2025)\citenamefont {Gao},
  \citenamefont {Gong}, \citenamefont {Yi},\ and\ \citenamefont
  {Zhang}}]{Gao:2025onc}%
  \BibitemOpen
  \bibfield  {author} {\bibinfo {author} {\bibfnamefont {Q.}~\bibnamefont
  {Gao}}, \bibinfo {author} {\bibfnamefont {Y.}~\bibnamefont {Gong}}, \bibinfo
  {author} {\bibfnamefont {Z.}~\bibnamefont {Yi}}, \ and\ \bibinfo {author}
  {\bibfnamefont {F.}~\bibnamefont {Zhang}},\ }\href {\doibase
  10.1016/j.dark.2025.102106} {\bibfield  {journal} {\bibinfo  {journal} {Phys.
  Dark Univ.}\ }\textbf {\bibinfo {volume} {50}},\ \bibinfo {pages} {102106}
  (\bibinfo {year} {2025})},\ \Eprint {http://arxiv.org/abs/2504.15218}
  {arXiv:2504.15218 [astro-ph.CO]} \BibitemShut {NoStop}%
\bibitem [{\citenamefont {Drees}\ and\ \citenamefont
  {Xu}(2025)}]{Drees:2025ngb}%
  \BibitemOpen
  \bibfield  {author} {\bibinfo {author} {\bibfnamefont {M.}~\bibnamefont
  {Drees}}\ and\ \bibinfo {author} {\bibfnamefont {Y.}~\bibnamefont {Xu}},\
  }\href {\doibase 10.1016/j.physletb.2025.139612} {\bibfield  {journal}
  {\bibinfo  {journal} {Phys. Lett. B}\ }\textbf {\bibinfo {volume} {867}},\
  \bibinfo {pages} {139612} (\bibinfo {year} {2025})},\ \Eprint
  {http://arxiv.org/abs/2504.20757} {arXiv:2504.20757 [astro-ph.CO]}
  \BibitemShut {NoStop}%
\bibitem [{\citenamefont {Zharov}\ \emph {et~al.}(2025)\citenamefont {Zharov},
  \citenamefont {Sobol},\ and\ \citenamefont {Vilchinskii}}]{Zharov:2025zjg}%
  \BibitemOpen
  \bibfield  {author} {\bibinfo {author} {\bibfnamefont {D.~S.}\ \bibnamefont
  {Zharov}}, \bibinfo {author} {\bibfnamefont {O.~O.}\ \bibnamefont {Sobol}}, \
  and\ \bibinfo {author} {\bibfnamefont {S.~I.}\ \bibnamefont {Vilchinskii}},\
  }\href {\doibase 10.1103/km3q-rm34} {\bibfield  {journal} {\bibinfo
  {journal} {Phys. Rev. D}\ }\textbf {\bibinfo {volume} {112}},\ \bibinfo
  {pages} {023544} (\bibinfo {year} {2025})},\ \Eprint
  {http://arxiv.org/abs/2505.01129} {arXiv:2505.01129 [astro-ph.CO]}
  \BibitemShut {NoStop}%
\bibitem [{\citenamefont {Yin}(2025)}]{Yin:2025rrs}%
  \BibitemOpen
  \bibfield  {author} {\bibinfo {author} {\bibfnamefont {W.}~\bibnamefont
  {Yin}},\ }\href {\doibase 10.1088/1475-7516/2025/09/062} {\bibfield
  {journal} {\bibinfo  {journal} {JCAP}\ }\textbf {\bibinfo {volume} {09}},\
  \bibinfo {pages} {062} (\bibinfo {year} {2025})},\ \Eprint
  {http://arxiv.org/abs/2505.03004} {arXiv:2505.03004 [hep-ph]} \BibitemShut
  {NoStop}%
\bibitem [{\citenamefont {Gialamas}\ \emph
  {et~al.}(2025{\natexlab{b}})\citenamefont {Gialamas}, \citenamefont
  {Katsoulas},\ and\ \citenamefont {Tamvakis}}]{Gialamas:2025ofz}%
  \BibitemOpen
  \bibfield  {author} {\bibinfo {author} {\bibfnamefont {I.~D.}\ \bibnamefont
  {Gialamas}}, \bibinfo {author} {\bibfnamefont {T.}~\bibnamefont {Katsoulas}},
  \ and\ \bibinfo {author} {\bibfnamefont {K.}~\bibnamefont {Tamvakis}},\
  }\href {\doibase 10.1088/1475-7516/2025/09/060} {\bibfield  {journal}
  {\bibinfo  {journal} {JCAP}\ }\textbf {\bibinfo {volume} {09}},\ \bibinfo
  {pages} {060} (\bibinfo {year} {2025}{\natexlab{b}})},\ \Eprint
  {http://arxiv.org/abs/2505.03608} {arXiv:2505.03608 [gr-qc]} \BibitemShut
  {NoStop}%
\bibitem [{\citenamefont {Haque}\ \emph {et~al.}(2025)\citenamefont {Haque},
  \citenamefont {Pal},\ and\ \citenamefont {Paul}}]{Haque:2025uis}%
  \BibitemOpen
  \bibfield  {author} {\bibinfo {author} {\bibfnamefont {M.~R.}\ \bibnamefont
  {Haque}}, \bibinfo {author} {\bibfnamefont {S.}~\bibnamefont {Pal}}, \ and\
  \bibinfo {author} {\bibfnamefont {D.}~\bibnamefont {Paul}},\ }\href {\doibase
  10.1016/j.physletb.2025.139852} {\bibfield  {journal} {\bibinfo  {journal}
  {Phys. Lett. B}\ }\textbf {\bibinfo {volume} {869}},\ \bibinfo {pages}
  {139852} (\bibinfo {year} {2025})},\ \Eprint
  {http://arxiv.org/abs/2505.04615} {arXiv:2505.04615 [astro-ph.CO]}
  \BibitemShut {NoStop}%
\bibitem [{\citenamefont {Yogesh}\ \emph {et~al.}(2025)\citenamefont {Yogesh},
  \citenamefont {Mohammadi}, \citenamefont {Wu},\ and\ \citenamefont
  {Zhu}}]{Yogesh:2025wak}%
  \BibitemOpen
  \bibfield  {author} {\bibinfo {author} {\bibnamefont {Yogesh}}, \bibinfo
  {author} {\bibfnamefont {A.}~\bibnamefont {Mohammadi}}, \bibinfo {author}
  {\bibfnamefont {Q.}~\bibnamefont {Wu}}, \ and\ \bibinfo {author}
  {\bibfnamefont {T.}~\bibnamefont {Zhu}},\ }\href {\doibase
  10.1088/1475-7516/2025/10/010} {\bibfield  {journal} {\bibinfo  {journal}
  {JCAP}\ }\textbf {\bibinfo {volume} {10}},\ \bibinfo {pages} {010} (\bibinfo
  {year} {2025})},\ \Eprint {http://arxiv.org/abs/2505.05363} {arXiv:2505.05363
  [astro-ph.CO]} \BibitemShut {NoStop}%
\bibitem [{\citenamefont {Addazi}\ \emph
  {et~al.}(2025{\natexlab{a}})\citenamefont {Addazi}, \citenamefont
  {Aldabergenov},\ and\ \citenamefont {Ketov}}]{Addazi:2025qra}%
  \BibitemOpen
  \bibfield  {author} {\bibinfo {author} {\bibfnamefont {A.}~\bibnamefont
  {Addazi}}, \bibinfo {author} {\bibfnamefont {Y.}~\bibnamefont
  {Aldabergenov}}, \ and\ \bibinfo {author} {\bibfnamefont {S.~V.}\
  \bibnamefont {Ketov}},\ }\href {\doibase 10.1016/j.physletb.2025.139883}
  {\bibfield  {journal} {\bibinfo  {journal} {Phys. Lett. B}\ }\textbf
  {\bibinfo {volume} {869}},\ \bibinfo {pages} {139883} (\bibinfo {year}
  {2025}{\natexlab{a}})},\ \Eprint {http://arxiv.org/abs/2505.10305}
  {arXiv:2505.10305 [gr-qc]} \BibitemShut {NoStop}%
\bibitem [{\citenamefont {Peng}\ \emph {et~al.}(2026)\citenamefont {Peng},
  \citenamefont {Chen},\ and\ \citenamefont {Liu}}]{Peng:2025bws}%
  \BibitemOpen
  \bibfield  {author} {\bibinfo {author} {\bibfnamefont {Z.-Z.}\ \bibnamefont
  {Peng}}, \bibinfo {author} {\bibfnamefont {Z.-C.}\ \bibnamefont {Chen}}, \
  and\ \bibinfo {author} {\bibfnamefont {L.}~\bibnamefont {Liu}},\ }\href
  {\doibase 10.1103/hzcf-q2rk} {\bibfield  {journal} {\bibinfo  {journal}
  {Phys. Rev. D}\ }\textbf {\bibinfo {volume} {113}},\ \bibinfo {pages}
  {063527} (\bibinfo {year} {2026})},\ \Eprint
  {http://arxiv.org/abs/2505.12816} {arXiv:2505.12816 [astro-ph.CO]}
  \BibitemShut {NoStop}%
\bibitem [{\citenamefont {Pallis}(2025)}]{Pallis:2025nrv}%
  \BibitemOpen
  \bibfield  {author} {\bibinfo {author} {\bibfnamefont {C.}~\bibnamefont
  {Pallis}},\ }\href {\doibase 10.1016/j.physletb.2025.139739} {\bibfield
  {journal} {\bibinfo  {journal} {Phys. Lett. B}\ }\textbf {\bibinfo {volume}
  {868}},\ \bibinfo {pages} {139739} (\bibinfo {year} {2025})},\ \Eprint
  {http://arxiv.org/abs/2505.23243} {arXiv:2505.23243 [hep-ph]} \BibitemShut
  {NoStop}%
\bibitem [{\citenamefont {Odintsov}\ and\ \citenamefont
  {Oikonomou}(2025{\natexlab{a}})}]{Odintsov:2025wai}%
  \BibitemOpen
  \bibfield  {author} {\bibinfo {author} {\bibfnamefont {S.~D.}\ \bibnamefont
  {Odintsov}}\ and\ \bibinfo {author} {\bibfnamefont {V.~K.}\ \bibnamefont
  {Oikonomou}},\ }\href {\doibase 10.1016/j.physletb.2025.139779} {\bibfield
  {journal} {\bibinfo  {journal} {Phys. Lett. B}\ }\textbf {\bibinfo {volume}
  {868}},\ \bibinfo {pages} {139779} (\bibinfo {year} {2025}{\natexlab{a}})},\
  \Eprint {http://arxiv.org/abs/2506.08193} {arXiv:2506.08193 [gr-qc]}
  \BibitemShut {NoStop}%
\bibitem [{\citenamefont {McDonald}(2025)}]{McDonald:2025tfp}%
  \BibitemOpen
  \bibfield  {author} {\bibinfo {author} {\bibfnamefont {J.}~\bibnamefont
  {McDonald}},\ }\href {\doibase 10.1103/k3r6-klxs} {\bibfield  {journal}
  {\bibinfo  {journal} {Phys. Rev. D}\ }\textbf {\bibinfo {volume} {112}},\
  \bibinfo {pages} {123525} (\bibinfo {year} {2025})},\ \Eprint
  {http://arxiv.org/abs/2506.12916} {arXiv:2506.12916 [hep-ph]} \BibitemShut
  {NoStop}%
\bibitem [{\citenamefont {Choudhury}\ \emph {et~al.}(2026)\citenamefont
  {Choudhury}, \citenamefont {Bauyrzhan}, \citenamefont {Singh},\ and\
  \citenamefont {Yerzhanov}}]{Choudhury:2025vso}%
  \BibitemOpen
  \bibfield  {author} {\bibinfo {author} {\bibfnamefont {S.}~\bibnamefont
  {Choudhury}}, \bibinfo {author} {\bibfnamefont {G.}~\bibnamefont
  {Bauyrzhan}}, \bibinfo {author} {\bibfnamefont {S.~K.}\ \bibnamefont
  {Singh}}, \ and\ \bibinfo {author} {\bibfnamefont {K.}~\bibnamefont
  {Yerzhanov}},\ }\href {\doibase 10.1016/j.jheap.2026.100656} {\bibfield
  {journal} {\bibinfo  {journal} {JHEAp}\ }\textbf {\bibinfo {volume} {54}},\
  \bibinfo {pages} {100656} (\bibinfo {year} {2026})},\ \Eprint
  {http://arxiv.org/abs/2506.15407} {arXiv:2506.15407 [astro-ph.CO]}
  \BibitemShut {NoStop}%
\bibitem [{\citenamefont {Ahmed}\ and\ \citenamefont
  {Rehman}(2025)}]{Ahmed:2025rrg}%
  \BibitemOpen
  \bibfield  {author} {\bibinfo {author} {\bibfnamefont {W.}~\bibnamefont
  {Ahmed}}\ and\ \bibinfo {author} {\bibfnamefont {M.~U.}\ \bibnamefont
  {Rehman}},\ }\href {\doibase 10.1103/7qhl-bskx} {\bibfield  {journal}
  {\bibinfo  {journal} {Phys. Rev. D}\ }\textbf {\bibinfo {volume} {112}},\
  \bibinfo {pages} {063519} (\bibinfo {year} {2025})},\ \Eprint
  {http://arxiv.org/abs/2506.18077} {arXiv:2506.18077 [astro-ph.CO]}
  \BibitemShut {NoStop}%
\bibitem [{\citenamefont {Pal}(2025)}]{Pal:2025ewf}%
  \BibitemOpen
  \bibfield  {author} {\bibinfo {author} {\bibfnamefont {B.~K.}\ \bibnamefont
  {Pal}},\ }\href {\doibase 10.1140/epjc/s10052-025-15087-9} {\bibfield
  {journal} {\bibinfo  {journal} {Eur. Phys. J. C}\ }\textbf {\bibinfo {volume}
  {85}},\ \bibinfo {pages} {1379} (\bibinfo {year} {2025})},\ \Eprint
  {http://arxiv.org/abs/2506.20744} {arXiv:2506.20744 [astro-ph.CO]}
  \BibitemShut {NoStop}%
\bibitem [{\citenamefont {Mohammadi}\ \emph
  {et~al.}(2026{\natexlab{b}})\citenamefont {Mohammadi}, \citenamefont
  {Yogesh},\ and\ \citenamefont {Wang}}]{Mohammadi:2025gbu}%
  \BibitemOpen
  \bibfield  {author} {\bibinfo {author} {\bibfnamefont {A.}~\bibnamefont
  {Mohammadi}}, \bibinfo {author} {\bibnamefont {Yogesh}}, \ and\ \bibinfo
  {author} {\bibfnamefont {A.}~\bibnamefont {Wang}},\ }\href {\doibase
  10.1016/j.physletb.2025.140054} {\bibfield  {journal} {\bibinfo  {journal}
  {Phys. Lett. B}\ }\textbf {\bibinfo {volume} {872}},\ \bibinfo {pages}
  {140054} (\bibinfo {year} {2026}{\natexlab{b}})},\ \Eprint
  {http://arxiv.org/abs/2507.06544} {arXiv:2507.06544 [astro-ph.CO]}
  \BibitemShut {NoStop}%
\bibitem [{\citenamefont {Ahmed}\ \emph
  {et~al.}(2026{\natexlab{a}})\citenamefont {Ahmed}, \citenamefont {Allehabi},\
  and\ \citenamefont {Rehman}}]{Ahmed:2025sfm}%
  \BibitemOpen
  \bibfield  {author} {\bibinfo {author} {\bibfnamefont {W.}~\bibnamefont
  {Ahmed}}, \bibinfo {author} {\bibfnamefont {S.~O.}\ \bibnamefont {Allehabi}},
  \ and\ \bibinfo {author} {\bibfnamefont {M.~U.}\ \bibnamefont {Rehman}},\
  }\href {\doibase 10.1103/jxg5-khj2} {\bibfield  {journal} {\bibinfo
  {journal} {Phys. Rev. D}\ }\textbf {\bibinfo {volume} {113}},\ \bibinfo
  {pages} {043532} (\bibinfo {year} {2026}{\natexlab{a}})},\ \Eprint
  {http://arxiv.org/abs/2508.01998} {arXiv:2508.01998 [hep-ph]} \BibitemShut
  {NoStop}%
\bibitem [{\citenamefont {Ketov}\ \emph {et~al.}(2025)\citenamefont {Ketov},
  \citenamefont {Pozdeeva},\ and\ \citenamefont {Vernov}}]{Ketov:2025cqg}%
  \BibitemOpen
  \bibfield  {author} {\bibinfo {author} {\bibfnamefont {S.~V.}\ \bibnamefont
  {Ketov}}, \bibinfo {author} {\bibfnamefont {E.~O.}\ \bibnamefont {Pozdeeva}},
  \ and\ \bibinfo {author} {\bibfnamefont {S.~Y.}\ \bibnamefont {Vernov}},\
  }\href {\doibase 10.1088/1475-7516/2025/12/040} {\bibfield  {journal}
  {\bibinfo  {journal} {JCAP}\ }\textbf {\bibinfo {volume} {12}},\ \bibinfo
  {pages} {040} (\bibinfo {year} {2025})},\ \Eprint
  {http://arxiv.org/abs/2508.08927} {arXiv:2508.08927 [gr-qc]} \BibitemShut
  {NoStop}%
\bibitem [{\citenamefont {Odintsov}\ and\ \citenamefont
  {Paul}(2025)}]{Odintsov:2025bmp}%
  \BibitemOpen
  \bibfield  {author} {\bibinfo {author} {\bibfnamefont {S.~D.}\ \bibnamefont
  {Odintsov}}\ and\ \bibinfo {author} {\bibfnamefont {T.}~\bibnamefont
  {Paul}},\ }\href {\doibase 10.1016/j.physletb.2025.139930} {\bibfield
  {journal} {\bibinfo  {journal} {Phys. Lett. B}\ }\textbf {\bibinfo {volume}
  {870}},\ \bibinfo {pages} {139930} (\bibinfo {year} {2025})},\ \Eprint
  {http://arxiv.org/abs/2508.11377} {arXiv:2508.11377 [gr-qc]} \BibitemShut
  {NoStop}%
\bibitem [{\citenamefont {Das}(2025)}]{Das:2025bws}%
  \BibitemOpen
  \bibfield  {author} {\bibinfo {author} {\bibfnamefont {S.}~\bibnamefont
  {Das}},\ }\href {\doibase 10.1103/rytr-blks} {\bibfield  {journal} {\bibinfo
  {journal} {Phys. Rev. D}\ }\textbf {\bibinfo {volume} {112}},\ \bibinfo
  {pages} {023543} (\bibinfo {year} {2025})},\ \Eprint
  {http://arxiv.org/abs/2508.14602} {arXiv:2508.14602 [astro-ph.CO]}
  \BibitemShut {NoStop}%
\bibitem [{\citenamefont {Zhu}\ \emph {et~al.}(2025)\citenamefont {Zhu},
  \citenamefont {Gao}, \citenamefont {Gong},\ and\ \citenamefont
  {Yi}}]{Zhu:2025twm}%
  \BibitemOpen
  \bibfield  {author} {\bibinfo {author} {\bibfnamefont {Y.}~\bibnamefont
  {Zhu}}, \bibinfo {author} {\bibfnamefont {Q.}~\bibnamefont {Gao}}, \bibinfo
  {author} {\bibfnamefont {Y.}~\bibnamefont {Gong}}, \ and\ \bibinfo {author}
  {\bibfnamefont {Z.}~\bibnamefont {Yi}},\ }\href {\doibase
  10.1140/epjc/s10052-025-14969-2} {\bibfield  {journal} {\bibinfo  {journal}
  {Eur. Phys. J. C}\ }\textbf {\bibinfo {volume} {85}},\ \bibinfo {pages}
  {1227} (\bibinfo {year} {2025})},\ \Eprint {http://arxiv.org/abs/2508.09707}
  {arXiv:2508.09707 [astro-ph.CO]} \BibitemShut {NoStop}%
\bibitem [{\citenamefont {Yuennan}\ \emph {et~al.}(2025)\citenamefont
  {Yuennan}, \citenamefont {Koad}, \citenamefont {Atamurotov},\ and\
  \citenamefont {Channuie}}]{Yuennan:2025kde}%
  \BibitemOpen
  \bibfield  {author} {\bibinfo {author} {\bibfnamefont {J.}~\bibnamefont
  {Yuennan}}, \bibinfo {author} {\bibfnamefont {P.}~\bibnamefont {Koad}},
  \bibinfo {author} {\bibfnamefont {F.}~\bibnamefont {Atamurotov}}, \ and\
  \bibinfo {author} {\bibfnamefont {P.}~\bibnamefont {Channuie}},\ }\href
  {\doibase 10.1140/epjc/s10052-025-15060-6} {\bibfield  {journal} {\bibinfo
  {journal} {Eur. Phys. J. C}\ }\textbf {\bibinfo {volume} {85}},\ \bibinfo
  {pages} {1307} (\bibinfo {year} {2025})},\ \Eprint
  {http://arxiv.org/abs/2508.17263} {arXiv:2508.17263 [astro-ph.CO]}
  \BibitemShut {NoStop}%
\bibitem [{\citenamefont {Odintsov}\ and\ \citenamefont
  {Oikonomou}(2025{\natexlab{b}})}]{Odintsov:2025jky}%
  \BibitemOpen
  \bibfield  {author} {\bibinfo {author} {\bibfnamefont {S.~D.}\ \bibnamefont
  {Odintsov}}\ and\ \bibinfo {author} {\bibfnamefont {V.~K.}\ \bibnamefont
  {Oikonomou}},\ }\href {\doibase 10.1016/j.physletb.2025.139909} {\bibfield
  {journal} {\bibinfo  {journal} {Phys. Lett. B}\ }\textbf {\bibinfo {volume}
  {870}},\ \bibinfo {pages} {139909} (\bibinfo {year} {2025}{\natexlab{b}})},\
  \Eprint {http://arxiv.org/abs/2508.17358} {arXiv:2508.17358 [gr-qc]}
  \BibitemShut {NoStop}%
\bibitem [{\citenamefont {Oikonomou}(2025)}]{Oikonomou:2025htz}%
  \BibitemOpen
  \bibfield  {author} {\bibinfo {author} {\bibfnamefont {V.~K.}\ \bibnamefont
  {Oikonomou}},\ }\href {\doibase 10.1016/j.physletb.2025.139972} {\bibfield
  {journal} {\bibinfo  {journal} {Phys. Lett. B}\ }\textbf {\bibinfo {volume}
  {871}},\ \bibinfo {pages} {139972} (\bibinfo {year} {2025})},\ \Eprint
  {http://arxiv.org/abs/2508.17363} {arXiv:2508.17363 [gr-qc]} \BibitemShut
  {NoStop}%
\bibitem [{\citenamefont {Oikonomou}(2026{\natexlab{a}})}]{Oikonomou:2025xms}%
  \BibitemOpen
  \bibfield  {author} {\bibinfo {author} {\bibfnamefont {V.~K.}\ \bibnamefont
  {Oikonomou}},\ }\href {\doibase 10.1016/j.nuclphysb.2026.117437} {\bibfield
  {journal} {\bibinfo  {journal} {Nucl. Phys. B}\ }\textbf {\bibinfo {volume}
  {1026}},\ \bibinfo {pages} {117437} (\bibinfo {year} {2026}{\natexlab{a}})},\
  \Eprint {http://arxiv.org/abs/2508.19196} {arXiv:2508.19196 [gr-qc]}
  \BibitemShut {NoStop}%
\bibitem [{\citenamefont {Modak}(2025)}]{Modak:2025bjv}%
  \BibitemOpen
  \bibfield  {author} {\bibinfo {author} {\bibfnamefont {T.}~\bibnamefont
  {Modak}},\ }\href {\doibase 10.1103/srpt-jd6s} {\bibfield  {journal}
  {\bibinfo  {journal} {Phys. Rev. D}\ }\textbf {\bibinfo {volume} {112}},\
  \bibinfo {pages} {115006} (\bibinfo {year} {2025})},\ \Eprint
  {http://arxiv.org/abs/2509.02979} {arXiv:2509.02979 [hep-ph]} \BibitemShut
  {NoStop}%
\bibitem [{\citenamefont {Odintsov}\ and\ \citenamefont
  {Oikonomou}(2025{\natexlab{c}})}]{Odintsov:2025eiv}%
  \BibitemOpen
  \bibfield  {author} {\bibinfo {author} {\bibfnamefont {S.~D.}\ \bibnamefont
  {Odintsov}}\ and\ \bibinfo {author} {\bibfnamefont {V.~K.}\ \bibnamefont
  {Oikonomou}},\ }\href {\doibase 10.1016/j.physletb.2025.139907} {\bibfield
  {journal} {\bibinfo  {journal} {Phys. Lett. B}\ }\textbf {\bibinfo {volume}
  {870}},\ \bibinfo {pages} {139907} (\bibinfo {year} {2025}{\natexlab{c}})},\
  \Eprint {http://arxiv.org/abs/2509.06251} {arXiv:2509.06251 [gr-qc]}
  \BibitemShut {NoStop}%
\bibitem [{\citenamefont {Aoki}\ \emph
  {et~al.}(2025{\natexlab{b}})\citenamefont {Aoki}, \citenamefont {Otsuka},\
  and\ \citenamefont {Yanagita}}]{Aoki:2025ywt}%
  \BibitemOpen
  \bibfield  {author} {\bibinfo {author} {\bibfnamefont {S.}~\bibnamefont
  {Aoki}}, \bibinfo {author} {\bibfnamefont {H.}~\bibnamefont {Otsuka}}, \ and\
  \bibinfo {author} {\bibfnamefont {R.}~\bibnamefont {Yanagita}},\ }\href
  {\doibase 10.1088/1475-7516/2025/11/088} {\bibfield  {journal} {\bibinfo
  {journal} {JCAP}\ }\textbf {\bibinfo {volume} {11}},\ \bibinfo {pages} {088}
  (\bibinfo {year} {2025}{\natexlab{b}})},\ \Eprint
  {http://arxiv.org/abs/2509.06739} {arXiv:2509.06739 [hep-ph]} \BibitemShut
  {NoStop}%
\bibitem [{\citenamefont {Pallis}(2026)}]{Pallis:2025vxo}%
  \BibitemOpen
  \bibfield  {author} {\bibinfo {author} {\bibfnamefont {C.}~\bibnamefont
  {Pallis}},\ }\href {\doibase 10.1103/h1p2-c333} {\bibfield  {journal}
  {\bibinfo  {journal} {Phys. Rev. D}\ }\textbf {\bibinfo {volume} {113}},\
  \bibinfo {pages} {015033} (\bibinfo {year} {2026})},\ \Eprint
  {http://arxiv.org/abs/2510.02083} {arXiv:2510.02083 [hep-ph]} \BibitemShut
  {NoStop}%
\bibitem [{\citenamefont {Qiu}\ \emph {et~al.}(2026)\citenamefont {Qiu},
  \citenamefont {Pang},\ and\ \citenamefont {Huang}}]{Qiu:2025iqm}%
  \BibitemOpen
  \bibfield  {author} {\bibinfo {author} {\bibfnamefont {Z.}~\bibnamefont
  {Qiu}}, \bibinfo {author} {\bibfnamefont {Y.}~\bibnamefont {Pang}}, \ and\
  \bibinfo {author} {\bibfnamefont {Q.}~\bibnamefont {Huang}},\ }\href
  {\doibase 10.1007/s11433-025-2934-8} {\bibfield  {journal} {\bibinfo
  {journal} {Sci. China Phys. Mech. Astron.}\ }\textbf {\bibinfo {volume}
  {69}},\ \bibinfo {pages} {260413} (\bibinfo {year} {2026})},\ \Eprint
  {http://arxiv.org/abs/2510.18320} {arXiv:2510.18320 [astro-ph.CO]}
  \BibitemShut {NoStop}%
\bibitem [{\citenamefont {Ahmed}\ \emph
  {et~al.}(2026{\natexlab{b}})\citenamefont {Ahmed}, \citenamefont {Pallis},\
  and\ \citenamefont {Rehman}}]{Ahmed:2025eip}%
  \BibitemOpen
  \bibfield  {author} {\bibinfo {author} {\bibfnamefont {W.}~\bibnamefont
  {Ahmed}}, \bibinfo {author} {\bibfnamefont {C.}~\bibnamefont {Pallis}}, \
  and\ \bibinfo {author} {\bibfnamefont {M.~U.}\ \bibnamefont {Rehman}},\
  }\href {\doibase 10.1088/1475-7516/2026/06/048} {\bibfield  {journal}
  {\bibinfo  {journal} {JCAP}\ }\textbf {\bibinfo {volume} {06}},\ \bibinfo
  {pages} {048} (\bibinfo {year} {2026}{\natexlab{b}})},\ \Eprint
  {http://arxiv.org/abs/2510.20478} {arXiv:2510.20478 [hep-ph]} \BibitemShut
  {NoStop}%
\bibitem [{\citenamefont {D'Onofrio}\ \emph {et~al.}(2026)\citenamefont
  {D'Onofrio}, \citenamefont {Odintsov},\ and\ \citenamefont
  {Paul}}]{DOnofrio:2025bol}%
  \BibitemOpen
  \bibfield  {author} {\bibinfo {author} {\bibfnamefont {S.}~\bibnamefont
  {D'Onofrio}}, \bibinfo {author} {\bibfnamefont {S.}~\bibnamefont {Odintsov}},
  \ and\ \bibinfo {author} {\bibfnamefont {T.}~\bibnamefont {Paul}},\ }\href
  {\doibase 10.1103/ktt2-lm19} {\bibfield  {journal} {\bibinfo  {journal}
  {Phys. Rev. D}\ }\textbf {\bibinfo {volume} {113}},\ \bibinfo {pages}
  {043527} (\bibinfo {year} {2026})},\ \Eprint
  {http://arxiv.org/abs/2510.20484} {arXiv:2510.20484 [gr-qc]} \BibitemShut
  {NoStop}%
\bibitem [{\citenamefont {Fu}\ \emph {et~al.}(2026)\citenamefont {Fu},
  \citenamefont {Lu},\ and\ \citenamefont {Wang}}]{Fu:2025ciy}%
  \BibitemOpen
  \bibfield  {author} {\bibinfo {author} {\bibfnamefont {C.}~\bibnamefont
  {Fu}}, \bibinfo {author} {\bibfnamefont {D.}~\bibnamefont {Lu}}, \ and\
  \bibinfo {author} {\bibfnamefont {S.-J.}\ \bibnamefont {Wang}},\ }\href
  {\doibase 10.1103/cm1c-kfpn} {\bibfield  {journal} {\bibinfo  {journal}
  {Phys. Rev. D}\ }\textbf {\bibinfo {volume} {113}},\ \bibinfo {pages}
  {L081304} (\bibinfo {year} {2026})},\ \Eprint
  {http://arxiv.org/abs/2510.24682} {arXiv:2510.24682 [astro-ph.CO]}
  \BibitemShut {NoStop}%
\bibitem [{\citenamefont {Singh}(2025)}]{Singh:2025uyr}%
  \BibitemOpen
  \bibfield  {author} {\bibinfo {author} {\bibfnamefont {S.~K.}\ \bibnamefont
  {Singh}},\ }\href@noop {} {\  (\bibinfo {year} {2025})},\ \Eprint
  {http://arxiv.org/abs/2511.05545} {arXiv:2511.05545 [hep-ph]} \BibitemShut
  {NoStop}%
\bibitem [{\citenamefont {Bostan}\ \emph {et~al.}(2026)\citenamefont {Bostan},
  \citenamefont {Dejrah},\ and\ \citenamefont {Ghoshal}}]{Bostan:2025jkt}%
  \BibitemOpen
  \bibfield  {author} {\bibinfo {author} {\bibfnamefont {N.}~\bibnamefont
  {Bostan}}, \bibinfo {author} {\bibfnamefont {R.~H.}\ \bibnamefont {Dejrah}},
  \ and\ \bibinfo {author} {\bibfnamefont {A.}~\bibnamefont {Ghoshal}},\ }\href
  {\doibase 10.1103/nx7w-xstd} {\bibfield  {journal} {\bibinfo  {journal}
  {Phys. Rev. D}\ }\textbf {\bibinfo {volume} {113}},\ \bibinfo {pages}
  {055042} (\bibinfo {year} {2026})},\ \Eprint
  {http://arxiv.org/abs/2511.05673} {arXiv:2511.05673 [astro-ph.CO]}
  \BibitemShut {NoStop}%
\bibitem [{\citenamefont {Serish}\ \emph {et~al.}(2026)\citenamefont {Serish},
  \citenamefont {Hosseini~Mansoori}, \citenamefont {Felegary}, \citenamefont
  {Akarsu},\ and\ \citenamefont {Sami}}]{Serish:2025ian}%
  \BibitemOpen
  \bibfield  {author} {\bibinfo {author} {\bibfnamefont {T.~F.}\ \bibnamefont
  {Serish}}, \bibinfo {author} {\bibfnamefont {S.~A.}\ \bibnamefont
  {Hosseini~Mansoori}}, \bibinfo {author} {\bibfnamefont {F.}~\bibnamefont
  {Felegary}}, \bibinfo {author} {\bibfnamefont {{\"O}.}~\bibnamefont
  {Akarsu}}, \ and\ \bibinfo {author} {\bibfnamefont {M.}~\bibnamefont
  {Sami}},\ }\href {\doibase 10.1088/1475-7516/2026/04/031} {\bibfield
  {journal} {\bibinfo  {journal} {JCAP}\ }\textbf {\bibinfo {volume} {04}},\
  \bibinfo {pages} {031} (\bibinfo {year} {2026})},\ \Eprint
  {http://arxiv.org/abs/2511.16621} {arXiv:2511.16621 [gr-qc]} \BibitemShut
  {NoStop}%
\bibitem [{\citenamefont {Yuennan}\ \emph
  {et~al.}(2026{\natexlab{a}})\citenamefont {Yuennan}, \citenamefont
  {Eadkhong}, \citenamefont {Atamurotov},\ and\ \citenamefont
  {Channuie}}]{Yuennan:2025mlg}%
  \BibitemOpen
  \bibfield  {author} {\bibinfo {author} {\bibfnamefont {J.}~\bibnamefont
  {Yuennan}}, \bibinfo {author} {\bibfnamefont {T.}~\bibnamefont {Eadkhong}},
  \bibinfo {author} {\bibfnamefont {F.}~\bibnamefont {Atamurotov}}, \ and\
  \bibinfo {author} {\bibfnamefont {P.}~\bibnamefont {Channuie}},\ }\href
  {\doibase 10.1016/j.dark.2026.102282} {\bibfield  {journal} {\bibinfo
  {journal} {Phys. Dark Univ.}\ }\textbf {\bibinfo {volume} {52}},\ \bibinfo
  {pages} {102282} (\bibinfo {year} {2026}{\natexlab{a}})},\ \Eprint
  {http://arxiv.org/abs/2511.17216} {arXiv:2511.17216 [astro-ph.CO]}
  \BibitemShut {NoStop}%
\bibitem [{\citenamefont {Choudhury}\ \emph {et~al.}(2025)\citenamefont
  {Choudhury}, \citenamefont {Singh},\ and\ \citenamefont
  {Sahoo}}]{Choudhury:2025hnu}%
  \BibitemOpen
  \bibfield  {author} {\bibinfo {author} {\bibfnamefont {S.}~\bibnamefont
  {Choudhury}}, \bibinfo {author} {\bibfnamefont {S.~K.}\ \bibnamefont
  {Singh}}, \ and\ \bibinfo {author} {\bibfnamefont {S.~K.}\ \bibnamefont
  {Sahoo}},\ }\href@noop {} {\  (\bibinfo {year} {2025})},\ \Eprint
  {http://arxiv.org/abs/2511.19898} {arXiv:2511.19898 [gr-qc]} \BibitemShut
  {NoStop}%
\bibitem [{\citenamefont {Addazi}\ \emph
  {et~al.}(2025{\natexlab{b}})\citenamefont {Addazi}, \citenamefont
  {Aldabergenov}, \citenamefont {Berkimbayev},\ and\ \citenamefont
  {Cai}}]{Addazi:2025agg}%
  \BibitemOpen
  \bibfield  {author} {\bibinfo {author} {\bibfnamefont {A.}~\bibnamefont
  {Addazi}}, \bibinfo {author} {\bibfnamefont {Y.}~\bibnamefont
  {Aldabergenov}}, \bibinfo {author} {\bibfnamefont {D.}~\bibnamefont
  {Berkimbayev}}, \ and\ \bibinfo {author} {\bibfnamefont {Y.}~\bibnamefont
  {Cai}},\ }\href@noop {} {\  (\bibinfo {year} {2025}{\natexlab{b}})},\ \Eprint
  {http://arxiv.org/abs/2512.21167} {arXiv:2512.21167 [gr-qc]} \BibitemShut
  {NoStop}%
\bibitem [{\citenamefont {Oikonomou}\ \emph {et~al.}(2026)\citenamefont
  {Oikonomou}, \citenamefont {Manouri},\ and\ \citenamefont
  {Konstantellos}}]{Oikonomou:2026gae}%
  \BibitemOpen
  \bibfield  {author} {\bibinfo {author} {\bibfnamefont {V.~K.}\ \bibnamefont
  {Oikonomou}}, \bibinfo {author} {\bibfnamefont {E.~I.}\ \bibnamefont
  {Manouri}}, \ and\ \bibinfo {author} {\bibfnamefont {G.}~\bibnamefont
  {Konstantellos}},\ }\href@noop {} {\  (\bibinfo {year} {2026})},\ \Eprint
  {http://arxiv.org/abs/2601.06946} {arXiv:2601.06946 [gr-qc]} \BibitemShut
  {NoStop}%
\bibitem [{\citenamefont {Nojiri}\ \emph {et~al.}(2026)\citenamefont {Nojiri},
  \citenamefont {Odintsov},\ and\ \citenamefont {Oikonomou}}]{Nojiri:2026hij}%
  \BibitemOpen
  \bibfield  {author} {\bibinfo {author} {\bibfnamefont {S.}~\bibnamefont
  {Nojiri}}, \bibinfo {author} {\bibfnamefont {S.}~\bibnamefont {Odintsov}}, \
  and\ \bibinfo {author} {\bibfnamefont {V.~K.}\ \bibnamefont {Oikonomou}},\
  }\href {\doibase 10.1016/j.physletb.2026.140290} {\bibfield  {journal}
  {\bibinfo  {journal} {Phys. Lett. B}\ }\textbf {\bibinfo {volume} {874}},\
  \bibinfo {pages} {140290} (\bibinfo {year} {2026})},\ \Eprint
  {http://arxiv.org/abs/2601.07879} {arXiv:2601.07879 [gr-qc]} \BibitemShut
  {NoStop}%
\bibitem [{\citenamefont {Oikonomou}(2026{\natexlab{b}})}]{Oikonomou:2026vrw}%
  \BibitemOpen
  \bibfield  {author} {\bibinfo {author} {\bibfnamefont {V.~K.}\ \bibnamefont
  {Oikonomou}},\ }\href {\doibase 10.1016/j.physletb.2026.140430} {\bibfield
  {journal} {\bibinfo  {journal} {Phys. Lett. B}\ }\textbf {\bibinfo {volume}
  {876}},\ \bibinfo {pages} {140430} (\bibinfo {year} {2026}{\natexlab{b}})},\
  \Eprint {http://arxiv.org/abs/2601.20053} {arXiv:2601.20053 [gr-qc]}
  \BibitemShut {NoStop}%
\bibitem [{\citenamefont {Odintsov}\ and\ \citenamefont
  {Oikonomou}(2026{\natexlab{a}})}]{Odintsov:2026doe}%
  \BibitemOpen
  \bibfield  {author} {\bibinfo {author} {\bibfnamefont {S.~D.}\ \bibnamefont
  {Odintsov}}\ and\ \bibinfo {author} {\bibfnamefont {V.~K.}\ \bibnamefont
  {Oikonomou}},\ }\href {\doibase 10.1016/j.nuclphysb.2026.117384} {\bibfield
  {journal} {\bibinfo  {journal} {Nucl. Phys. B}\ }\textbf {\bibinfo {volume}
  {1025}},\ \bibinfo {pages} {117384} (\bibinfo {year} {2026}{\natexlab{a}})},\
  \Eprint {http://arxiv.org/abs/2601.21364} {arXiv:2601.21364 [gr-qc]}
  \BibitemShut {NoStop}%
\bibitem [{\citenamefont {Santos}\ \emph {et~al.}(2026)\citenamefont {Santos},
  \citenamefont {Rodrigues}, \citenamefont {Rodrigues}, \citenamefont
  {Siqueira},\ and\ \citenamefont {Alcaniz}}]{Santos:2026ojg}%
  \BibitemOpen
  \bibfield  {author} {\bibinfo {author} {\bibfnamefont {F.~B. M.~d.}\
  \bibnamefont {Santos}}, \bibinfo {author} {\bibfnamefont {J.~G.}\
  \bibnamefont {Rodrigues}}, \bibinfo {author} {\bibfnamefont {G.}~\bibnamefont
  {Rodrigues}}, \bibinfo {author} {\bibfnamefont {C.}~\bibnamefont {Siqueira}},
  \ and\ \bibinfo {author} {\bibfnamefont {J.~S.}\ \bibnamefont {Alcaniz}},\
  }\href {\doibase 10.1103/tddw-26tl} {\bibfield  {journal} {\bibinfo
  {journal} {Phys. Rev. D}\ }\textbf {\bibinfo {volume} {113}},\ \bibinfo
  {pages} {123545} (\bibinfo {year} {2026})},\ \Eprint
  {http://arxiv.org/abs/2602.06027} {arXiv:2602.06027 [astro-ph.CO]}
  \BibitemShut {NoStop}%
\bibitem [{\citenamefont {Yuennan}\ \emph
  {et~al.}(2026{\natexlab{b}})\citenamefont {Yuennan}, \citenamefont
  {Atamurotov}, \citenamefont {Capozziello},\ and\ \citenamefont
  {Channuie}}]{Yuennan:2026fcn}%
  \BibitemOpen
  \bibfield  {author} {\bibinfo {author} {\bibfnamefont {J.}~\bibnamefont
  {Yuennan}}, \bibinfo {author} {\bibfnamefont {F.}~\bibnamefont {Atamurotov}},
  \bibinfo {author} {\bibfnamefont {S.}~\bibnamefont {Capozziello}}, \ and\
  \bibinfo {author} {\bibfnamefont {P.}~\bibnamefont {Channuie}},\ }\href
  {\doibase 10.1140/epjc/s10052-026-15461-1} {\bibfield  {journal} {\bibinfo
  {journal} {Eur. Phys. J. C}\ }\textbf {\bibinfo {volume} {86}},\ \bibinfo
  {pages} {237} (\bibinfo {year} {2026}{\natexlab{b}})},\ \Eprint
  {http://arxiv.org/abs/2602.17380} {arXiv:2602.17380 [gr-qc]} \BibitemShut
  {NoStop}%
\bibitem [{\citenamefont {Oikonomou}(2026{\natexlab{c}})}]{Oikonomou:2026mvp}%
  \BibitemOpen
  \bibfield  {author} {\bibinfo {author} {\bibfnamefont {V.~K.}\ \bibnamefont
  {Oikonomou}},\ }\href@noop {} {\  (\bibinfo {year} {2026}{\natexlab{c}})},\
  \Eprint {http://arxiv.org/abs/2603.11794} {arXiv:2603.11794 [gr-qc]}
  \BibitemShut {NoStop}%
\bibitem [{\citenamefont {Dimopoulos}\ \emph {et~al.}(2026)\citenamefont
  {Dimopoulos}, \citenamefont {Dioguardi}, \citenamefont {Gialamas},\ and\
  \citenamefont {Racioppi}}]{Dimopoulos:2026iwq}%
  \BibitemOpen
  \bibfield  {author} {\bibinfo {author} {\bibfnamefont {K.}~\bibnamefont
  {Dimopoulos}}, \bibinfo {author} {\bibfnamefont {C.}~\bibnamefont
  {Dioguardi}}, \bibinfo {author} {\bibfnamefont {I.~D.}\ \bibnamefont
  {Gialamas}}, \ and\ \bibinfo {author} {\bibfnamefont {A.}~\bibnamefont
  {Racioppi}},\ }\href@noop {} {\  (\bibinfo {year} {2026})},\ \Eprint
  {http://arxiv.org/abs/2603.15776} {arXiv:2603.15776 [gr-qc]} \BibitemShut
  {NoStop}%
\bibitem [{\citenamefont {Khan}\ \emph
  {et~al.}(2026{\natexlab{a}})\citenamefont {Khan}, \citenamefont {Pirzada},\
  and\ \citenamefont {Mustafa}}]{Khan:2026nsz}%
  \BibitemOpen
  \bibfield  {author} {\bibinfo {author} {\bibfnamefont {I.}~\bibnamefont
  {Khan}}, \bibinfo {author} {\bibnamefont {Pirzada}}, \ and\ \bibinfo {author}
  {\bibfnamefont {G.}~\bibnamefont {Mustafa}},\ }\href@noop {} {\  (\bibinfo
  {year} {2026}{\natexlab{a}})},\ \Eprint {http://arxiv.org/abs/2604.07044}
  {arXiv:2604.07044 [hep-ph]} \BibitemShut {NoStop}%
\bibitem [{\citenamefont {Yogesh}\ \emph {et~al.}(2026)\citenamefont {Yogesh},
  \citenamefont {Bhat}, \citenamefont {Gangopadhyay},\ and\ \citenamefont
  {Sami}}]{Yogesh:2026esn}%
  \BibitemOpen
  \bibfield  {author} {\bibinfo {author} {\bibnamefont {Yogesh}}, \bibinfo
  {author} {\bibfnamefont {I.~A.}\ \bibnamefont {Bhat}}, \bibinfo {author}
  {\bibfnamefont {M.~R.}\ \bibnamefont {Gangopadhyay}}, \ and\ \bibinfo
  {author} {\bibfnamefont {M.}~\bibnamefont {Sami}},\ }\href@noop {} {\
  (\bibinfo {year} {2026})},\ \Eprint {http://arxiv.org/abs/2604.14659}
  {arXiv:2604.14659 [astro-ph.CO]} \BibitemShut {NoStop}%
\bibitem [{\citenamefont {Odintsov}\ \emph {et~al.}(2026)\citenamefont
  {Odintsov}, \citenamefont {Oikonomou}, \citenamefont {Tsyba}, \citenamefont
  {Razina},\ and\ \citenamefont {Rakhatov}}]{Odintsov:2026cxz}%
  \BibitemOpen
  \bibfield  {author} {\bibinfo {author} {\bibfnamefont {S.~D.}\ \bibnamefont
  {Odintsov}}, \bibinfo {author} {\bibfnamefont {V.~K.}\ \bibnamefont
  {Oikonomou}}, \bibinfo {author} {\bibfnamefont {P.}~\bibnamefont {Tsyba}},
  \bibinfo {author} {\bibfnamefont {O.}~\bibnamefont {Razina}}, \ and\ \bibinfo
  {author} {\bibfnamefont {D.}~\bibnamefont {Rakhatov}},\ }\href {\doibase
  10.1016/j.dark.2026.102348} {\bibfield  {journal} {\bibinfo  {journal} {Phys.
  Dark Univ.}\ }\textbf {\bibinfo {volume} {52}},\ \bibinfo {pages} {102348}
  (\bibinfo {year} {2026})},\ \Eprint {http://arxiv.org/abs/2604.18861}
  {arXiv:2604.18861 [gr-qc]} \BibitemShut {NoStop}%
\bibitem [{\citenamefont {Khan}\ \emph
  {et~al.}(2026{\natexlab{b}})\citenamefont {Khan}, \citenamefont {Mustafa},
  \citenamefont {Zafar}, \citenamefont {Atamurotov}, \citenamefont
  {Abdujabbarov},\ and\ \citenamefont {Yuan}}]{Khan:2026pba}%
  \BibitemOpen
  \bibfield  {author} {\bibinfo {author} {\bibfnamefont {I.}~\bibnamefont
  {Khan}}, \bibinfo {author} {\bibfnamefont {G.}~\bibnamefont {Mustafa}},
  \bibinfo {author} {\bibfnamefont {J.}~\bibnamefont {Zafar}}, \bibinfo
  {author} {\bibfnamefont {F.}~\bibnamefont {Atamurotov}}, \bibinfo {author}
  {\bibfnamefont {A.}~\bibnamefont {Abdujabbarov}}, \ and\ \bibinfo {author}
  {\bibfnamefont {C.}~\bibnamefont {Yuan}},\ }\href@noop {} {\  (\bibinfo
  {year} {2026}{\natexlab{b}})},\ \Eprint {http://arxiv.org/abs/2604.20481}
  {arXiv:2604.20481 [hep-ph]} \BibitemShut {NoStop}%
\bibitem [{\citenamefont {Khan}\ \emph
  {et~al.}(2026{\natexlab{c}})\citenamefont {Khan}, \citenamefont {Pirzada},
  \citenamefont {Mustafa}, \citenamefont {Atamurotov},\ and\ \citenamefont
  {Yuan}}]{Khan:2026doo}%
  \BibitemOpen
  \bibfield  {author} {\bibinfo {author} {\bibfnamefont {I.}~\bibnamefont
  {Khan}}, \bibinfo {author} {\bibnamefont {Pirzada}}, \bibinfo {author}
  {\bibfnamefont {G.}~\bibnamefont {Mustafa}}, \bibinfo {author} {\bibfnamefont
  {F.}~\bibnamefont {Atamurotov}}, \ and\ \bibinfo {author} {\bibfnamefont
  {C.}~\bibnamefont {Yuan}},\ }\href@noop {} {\  (\bibinfo {year}
  {2026}{\natexlab{c}})},\ \Eprint {http://arxiv.org/abs/2605.04166}
  {arXiv:2605.04166 [hep-ph]} \BibitemShut {NoStop}%
\bibitem [{\citenamefont {Odintsov}\ and\ \citenamefont
  {Oikonomou}(2026{\natexlab{b}})}]{Odintsov:2026dss}%
  \BibitemOpen
  \bibfield  {author} {\bibinfo {author} {\bibfnamefont {S.~D.}\ \bibnamefont
  {Odintsov}}\ and\ \bibinfo {author} {\bibfnamefont {V.~K.}\ \bibnamefont
  {Oikonomou}},\ }\href@noop {} {\  (\bibinfo {year} {2026}{\natexlab{b}})},\
  \Eprint {http://arxiv.org/abs/2605.17813} {arXiv:2605.17813 [gr-qc]}
  \BibitemShut {NoStop}%
\bibitem [{\citenamefont {Oikonomou}(2026{\natexlab{d}})}]{Oikonomou:2026qkj}%
  \BibitemOpen
  \bibfield  {author} {\bibinfo {author} {\bibfnamefont {V.~K.}\ \bibnamefont
  {Oikonomou}},\ }\href {\doibase 10.1016/j.aop.2026.170547} {\bibfield
  {journal} {\bibinfo  {journal} {Annals Phys.}\ }\textbf {\bibinfo {volume}
  {492}},\ \bibinfo {pages} {170547} (\bibinfo {year} {2026}{\natexlab{d}})},\
  \Eprint {http://arxiv.org/abs/2605.27727} {arXiv:2605.27727 [gr-qc]}
  \BibitemShut {NoStop}%
\bibitem [{\citenamefont {Gonuguntla}\ \emph {et~al.}(2026)\citenamefont
  {Gonuguntla}, \citenamefont {Modak},\ and\ \citenamefont
  {Samanta}}]{Gonuguntla:2026rkw}%
  \BibitemOpen
  \bibfield  {author} {\bibinfo {author} {\bibfnamefont {H.}~\bibnamefont
  {Gonuguntla}}, \bibinfo {author} {\bibfnamefont {T.}~\bibnamefont {Modak}}, \
  and\ \bibinfo {author} {\bibfnamefont {A.}~\bibnamefont {Samanta}},\
  }\href@noop {} {\  (\bibinfo {year} {2026})},\ \Eprint
  {http://arxiv.org/abs/2606.11929} {arXiv:2606.11929 [astro-ph.CO]}
  \BibitemShut {NoStop}%
\bibitem [{\citenamefont {Yang}\ \emph {et~al.}(2026)\citenamefont {Yang},
  \citenamefont {Tao}, \citenamefont {Wang},\ and\ \citenamefont
  {Zhu}}]{Yang:2026rzn}%
  \BibitemOpen
  \bibfield  {author} {\bibinfo {author} {\bibfnamefont {R.}~\bibnamefont
  {Yang}}, \bibinfo {author} {\bibfnamefont {J.}~\bibnamefont {Tao}}, \bibinfo
  {author} {\bibfnamefont {P.}~\bibnamefont {Wang}}, \ and\ \bibinfo {author}
  {\bibfnamefont {M.}~\bibnamefont {Zhu}},\ }\href@noop {} {\  (\bibinfo {year}
  {2026})},\ \Eprint {http://arxiv.org/abs/2606.16711} {arXiv:2606.16711
  [astro-ph.CO]} \BibitemShut {NoStop}%
\bibitem [{\citenamefont {Qiu}\ and\ \citenamefont
  {Huang}(2026)}]{Qiu:2026npi}%
  \BibitemOpen
  \bibfield  {author} {\bibinfo {author} {\bibfnamefont {Z.-C.}\ \bibnamefont
  {Qiu}}\ and\ \bibinfo {author} {\bibfnamefont {Q.-G.}\ \bibnamefont
  {Huang}},\ }\href@noop {} {\  (\bibinfo {year} {2026})},\ \Eprint
  {http://arxiv.org/abs/2606.24474} {arXiv:2606.24474 [astro-ph.CO]}
  \BibitemShut {NoStop}%
\bibitem [{\citenamefont {Khan}\ \emph
  {et~al.}(2026{\natexlab{d}})\citenamefont {Khan}, \citenamefont
  {Capozziello}, \citenamefont {Mustafa}, \citenamefont {Yuan},\ and\
  \citenamefont {Atamurotov}}]{Khan:2026bmf}%
  \BibitemOpen
  \bibfield  {author} {\bibinfo {author} {\bibfnamefont {I.}~\bibnamefont
  {Khan}}, \bibinfo {author} {\bibfnamefont {S.}~\bibnamefont {Capozziello}},
  \bibinfo {author} {\bibfnamefont {G.}~\bibnamefont {Mustafa}}, \bibinfo
  {author} {\bibfnamefont {C.}~\bibnamefont {Yuan}}, \ and\ \bibinfo {author}
  {\bibfnamefont {F.}~\bibnamefont {Atamurotov}},\ }\href@noop {} {\  (\bibinfo
  {year} {2026}{\natexlab{d}})},\ \Eprint {http://arxiv.org/abs/2606.29197}
  {arXiv:2606.29197 [hep-ph]} \BibitemShut {NoStop}%
\bibitem [{\citenamefont {Ijaz}\ \emph {et~al.}(2026)\citenamefont {Ijaz},
  \citenamefont {Pirzada},\ and\ \citenamefont {Rehman}}]{Ijaz:2026ear}%
  \BibitemOpen
  \bibfield  {author} {\bibinfo {author} {\bibfnamefont {N.}~\bibnamefont
  {Ijaz}}, \bibinfo {author} {\bibnamefont {Pirzada}}, \ and\ \bibinfo {author}
  {\bibfnamefont {M.~U.}\ \bibnamefont {Rehman}},\ }\href@noop {} {\  (\bibinfo
  {year} {2026})},\ \Eprint {http://arxiv.org/abs/2607.04504} {arXiv:2607.04504
  [hep-ph]} \BibitemShut {NoStop}%
\bibitem [{\citenamefont {Khan}\ \emph
  {et~al.}(2026{\natexlab{e}})\citenamefont {Khan}, \citenamefont
  {Capozziello}, \citenamefont {Mustafa}, \citenamefont {Ullah},\ and\
  \citenamefont {Atamurotov}}]{Khan:2026php}%
  \BibitemOpen
  \bibfield  {author} {\bibinfo {author} {\bibfnamefont {I.}~\bibnamefont
  {Khan}}, \bibinfo {author} {\bibfnamefont {S.}~\bibnamefont {Capozziello}},
  \bibinfo {author} {\bibfnamefont {G.}~\bibnamefont {Mustafa}}, \bibinfo
  {author} {\bibfnamefont {N.}~\bibnamefont {Ullah}}, \ and\ \bibinfo {author}
  {\bibfnamefont {F.}~\bibnamefont {Atamurotov}},\ }\href@noop {} {\  (\bibinfo
  {year} {2026}{\natexlab{e}})},\ \Eprint {http://arxiv.org/abs/2607.05309}
  {arXiv:2607.05309 [hep-ph]} \BibitemShut {NoStop}%
\bibitem [{\citenamefont {Mohammadi}\ \emph
  {et~al.}(2026{\natexlab{c}})\citenamefont {Mohammadi}, \citenamefont
  {Yogesh}, \citenamefont {Tan},\ and\ \citenamefont
  {Sami}}]{Mohammadi:2026iod}%
  \BibitemOpen
  \bibfield  {author} {\bibinfo {author} {\bibfnamefont {A.}~\bibnamefont
  {Mohammadi}}, \bibinfo {author} {\bibnamefont {Yogesh}}, \bibinfo {author}
  {\bibfnamefont {H.}~\bibnamefont {Tan}}, \ and\ \bibinfo {author}
  {\bibfnamefont {M.}~\bibnamefont {Sami}},\ }\href@noop {} {\  (\bibinfo
  {year} {2026}{\natexlab{c}})},\ \Eprint {http://arxiv.org/abs/2607.06112}
  {arXiv:2607.06112 [gr-qc]} \BibitemShut {NoStop}%
\bibitem [{\citenamefont {Nojiri}\ and\ \citenamefont
  {Odintsov}(2026)}]{Nojiri:2026ish}%
  \BibitemOpen
  \bibfield  {author} {\bibinfo {author} {\bibfnamefont {S.}~\bibnamefont
  {Nojiri}}\ and\ \bibinfo {author} {\bibfnamefont {S.~D.}\ \bibnamefont
  {Odintsov}},\ }\href@noop {} {\  (\bibinfo {year} {2026})},\ \Eprint
  {http://arxiv.org/abs/2607.06582} {arXiv:2607.06582 [gr-qc]} \BibitemShut
  {NoStop}%
\bibitem [{\citenamefont {Kallosh}\ and\ \citenamefont
  {Linde}(2026)}]{Kallosh:2026tke}%
  \BibitemOpen
  \bibfield  {author} {\bibinfo {author} {\bibfnamefont {R.}~\bibnamefont
  {Kallosh}}\ and\ \bibinfo {author} {\bibfnamefont {A.}~\bibnamefont
  {Linde}},\ }\href@noop {} {\  (\bibinfo {year} {2026})},\ \Eprint
  {http://arxiv.org/abs/2607.07684} {arXiv:2607.07684 [hep-th]} \BibitemShut
  {NoStop}%
\bibitem [{\citenamefont {Del~Grosso}\ \emph {et~al.}(2026)\citenamefont
  {Del~Grosso}, \citenamefont {Urbano},\ and\ \citenamefont
  {Ziccolella}}]{DelGrosso:2026zbg}%
  \BibitemOpen
  \bibfield  {author} {\bibinfo {author} {\bibfnamefont {L.}~\bibnamefont
  {Del~Grosso}}, \bibinfo {author} {\bibfnamefont {A.}~\bibnamefont {Urbano}},
  \ and\ \bibinfo {author} {\bibfnamefont {M.}~\bibnamefont {Ziccolella}},\
  }\href@noop {} {\  (\bibinfo {year} {2026})},\ \Eprint
  {http://arxiv.org/abs/2607.09808} {arXiv:2607.09808 [hep-ph]} \BibitemShut
  {NoStop}%
\bibitem [{\citenamefont {Cembranos}\ \emph {et~al.}(2026)\citenamefont
  {Cembranos}, \citenamefont {Luque},\ and\ \citenamefont
  {Rubio}}]{Cembranos:2026cat}%
  \BibitemOpen
  \bibfield  {author} {\bibinfo {author} {\bibfnamefont {J.~A.~R.}\
  \bibnamefont {Cembranos}}, \bibinfo {author} {\bibfnamefont {J.}~\bibnamefont
  {Luque}}, \ and\ \bibinfo {author} {\bibfnamefont {J.}~\bibnamefont
  {Rubio}},\ }\href@noop {} {\  (\bibinfo {year} {2026})},\ \Eprint
  {http://arxiv.org/abs/2607.15354} {arXiv:2607.15354 [astro-ph.CO]}
  \BibitemShut {NoStop}%
\bibitem [{\citenamefont {De~Angelis}\ and\ \citenamefont
  {Rubio}(2026)}]{DeAngelis:2026mwx}%
  \BibitemOpen
  \bibfield  {author} {\bibinfo {author} {\bibfnamefont {M.}~\bibnamefont
  {De~Angelis}}\ and\ \bibinfo {author} {\bibfnamefont {J.}~\bibnamefont
  {Rubio}},\ }\href@noop {} {\  (\bibinfo {year} {2026})},\ \Eprint
  {http://arxiv.org/abs/2607.16405} {arXiv:2607.16405 [astro-ph.CO]}
  \BibitemShut {NoStop}%
\bibitem [{\citenamefont {Iacconi}(2026)}]{Iacconi:2026twl}%
  \BibitemOpen
  \bibfield  {author} {\bibinfo {author} {\bibfnamefont {L.}~\bibnamefont
  {Iacconi}},\ }\href@noop {} {\  (\bibinfo {year} {2026})},\ \Eprint
  {http://arxiv.org/abs/2607.28502} {arXiv:2607.28502 [astro-ph.CO]}
  \BibitemShut {NoStop}%
\bibitem [{\citenamefont {Chudaykin}\ \emph {et~al.}(2026)\citenamefont
  {Chudaykin}, \citenamefont {Ivanov}, \citenamefont {Kallosh}, \citenamefont
  {Linde}, \citenamefont {Philcox},\ and\ \citenamefont
  {Yamada}}]{Chudaykin:2026amr}%
  \BibitemOpen
  \bibfield  {author} {\bibinfo {author} {\bibfnamefont {A.}~\bibnamefont
  {Chudaykin}}, \bibinfo {author} {\bibfnamefont {M.~M.}\ \bibnamefont
  {Ivanov}}, \bibinfo {author} {\bibfnamefont {R.}~\bibnamefont {Kallosh}},
  \bibinfo {author} {\bibfnamefont {A.}~\bibnamefont {Linde}}, \bibinfo
  {author} {\bibfnamefont {O.~H.~E.}\ \bibnamefont {Philcox}}, \ and\ \bibinfo
  {author} {\bibfnamefont {Y.}~\bibnamefont {Yamada}},\ }\href@noop {} {\
  (\bibinfo {year} {2026})},\ \Eprint {http://arxiv.org/abs/2607.28445}
  {arXiv:2607.28445 [astro-ph.CO]} \BibitemShut {NoStop}%
\bibitem [{\citenamefont {Oikonomou}(2026{\natexlab{e}})}]{Oikonomou:2026jdg}%
  \BibitemOpen
  \bibfield  {author} {\bibinfo {author} {\bibfnamefont {V.~K.}\ \bibnamefont
  {Oikonomou}},\ }\href@noop {} {\  (\bibinfo {year} {2026}{\natexlab{e}})},\
  \Eprint {http://arxiv.org/abs/2608.01165} {arXiv:2608.01165 [gr-qc]}
  \BibitemShut {NoStop}%
\bibitem [{\citenamefont {Seidabadi}\ \emph {et~al.}(2026)\citenamefont
  {Seidabadi}, \citenamefont {Saghafi},\ and\ \citenamefont
  {Nozari}}]{Seidabadi:2026gnc}%
  \BibitemOpen
  \bibfield  {author} {\bibinfo {author} {\bibfnamefont {A.}~\bibnamefont
  {Seidabadi}}, \bibinfo {author} {\bibfnamefont {S.}~\bibnamefont {Saghafi}},
  \ and\ \bibinfo {author} {\bibfnamefont {K.}~\bibnamefont {Nozari}},\
  }\href@noop {} {\  (\bibinfo {year} {2026})},\ \Eprint
  {http://arxiv.org/abs/2608.02506} {arXiv:2608.02506 [astro-ph.CO]}
  \BibitemShut {NoStop}%
\bibitem [{\citenamefont {Odintsov}\ and\ \citenamefont
  {Oikonomou}(2026{\natexlab{c}})}]{Odintsov:2026veu}%
  \BibitemOpen
  \bibfield  {author} {\bibinfo {author} {\bibfnamefont {S.~D.}\ \bibnamefont
  {Odintsov}}\ and\ \bibinfo {author} {\bibfnamefont {V.~K.}\ \bibnamefont
  {Oikonomou}},\ }\href@noop {} {\  (\bibinfo {year} {2026}{\natexlab{c}})},\
  \Eprint {http://arxiv.org/abs/2608.09783} {arXiv:2608.09783 [gr-qc]}
  \BibitemShut {NoStop}%
\bibitem [{\citenamefont {Kallosh}\ \emph {et~al.}(2026)\citenamefont
  {Kallosh}, \citenamefont {Linde},\ and\ \citenamefont
  {Yamada}}]{Kallosh:2026kfx}%
  \BibitemOpen
  \bibfield  {author} {\bibinfo {author} {\bibfnamefont {R.}~\bibnamefont
  {Kallosh}}, \bibinfo {author} {\bibfnamefont {A.}~\bibnamefont {Linde}}, \
  and\ \bibinfo {author} {\bibfnamefont {Y.}~\bibnamefont {Yamada}},\
  }\href@noop {} {\  (\bibinfo {year} {2026})},\ \Eprint
  {http://arxiv.org/abs/2608.12819} {arXiv:2608.12819 [astro-ph.CO]}
  \BibitemShut {NoStop}%
\bibitem [{\citenamefont {Kulchoakrungsun}\ \emph {et~al.}(2026)\citenamefont
  {Kulchoakrungsun}, \citenamefont {Channuie},\ and\ \citenamefont
  {Samart}}]{Kulchoakrungsun:2026pyz}%
  \BibitemOpen
  \bibfield  {author} {\bibinfo {author} {\bibfnamefont {E.}~\bibnamefont
  {Kulchoakrungsun}}, \bibinfo {author} {\bibfnamefont {P.}~\bibnamefont
  {Channuie}}, \ and\ \bibinfo {author} {\bibfnamefont {D.}~\bibnamefont
  {Samart}},\ }\href@noop {} {\  (\bibinfo {year} {2026})},\ \Eprint
  {http://arxiv.org/abs/2608.17215} {arXiv:2608.17215 [gr-qc]} \BibitemShut
  {NoStop}%
\bibitem [{\citenamefont {Gerbino}\ \emph {et~al.}(2017)\citenamefont
  {Gerbino}, \citenamefont {Freese}, \citenamefont {Vagnozzi}, \citenamefont
  {Lattanzi}, \citenamefont {Mena}, \citenamefont {Giusarma},\ and\
  \citenamefont {Ho}}]{Gerbino:2016sgw}%
  \BibitemOpen
  \bibfield  {author} {\bibinfo {author} {\bibfnamefont {M.}~\bibnamefont
  {Gerbino}}, \bibinfo {author} {\bibfnamefont {K.}~\bibnamefont {Freese}},
  \bibinfo {author} {\bibfnamefont {S.}~\bibnamefont {Vagnozzi}}, \bibinfo
  {author} {\bibfnamefont {M.}~\bibnamefont {Lattanzi}}, \bibinfo {author}
  {\bibfnamefont {O.}~\bibnamefont {Mena}}, \bibinfo {author} {\bibfnamefont
  {E.}~\bibnamefont {Giusarma}}, \ and\ \bibinfo {author} {\bibfnamefont
  {S.}~\bibnamefont {Ho}},\ }\href {\doibase 10.1103/PhysRevD.95.043512}
  {\bibfield  {journal} {\bibinfo  {journal} {Phys. Rev. D}\ }\textbf {\bibinfo
  {volume} {95}},\ \bibinfo {pages} {043512} (\bibinfo {year} {2017})},\
  \Eprint {http://arxiv.org/abs/1610.08830} {arXiv:1610.08830 [astro-ph.CO]}
  \BibitemShut {NoStop}%
\bibitem [{\citenamefont {Di~Valentino}\ \emph {et~al.}(2018)\citenamefont
  {Di~Valentino}, \citenamefont {Melchiorri}, \citenamefont {Fantaye},\ and\
  \citenamefont {Heavens}}]{DiValentino:2018zjj}%
  \BibitemOpen
  \bibfield  {author} {\bibinfo {author} {\bibfnamefont {E.}~\bibnamefont
  {Di~Valentino}}, \bibinfo {author} {\bibfnamefont {A.}~\bibnamefont
  {Melchiorri}}, \bibinfo {author} {\bibfnamefont {Y.}~\bibnamefont {Fantaye}},
  \ and\ \bibinfo {author} {\bibfnamefont {A.}~\bibnamefont {Heavens}},\ }\href
  {\doibase 10.1103/PhysRevD.98.063508} {\bibfield  {journal} {\bibinfo
  {journal} {Phys. Rev. D}\ }\textbf {\bibinfo {volume} {98}},\ \bibinfo
  {pages} {063508} (\bibinfo {year} {2018})},\ \Eprint
  {http://arxiv.org/abs/1808.09201} {arXiv:1808.09201 [astro-ph.CO]}
  \BibitemShut {NoStop}%
\bibitem [{\citenamefont {Ye}\ \emph {et~al.}(2021)\citenamefont {Ye},
  \citenamefont {Hu},\ and\ \citenamefont {Piao}}]{Ye:2021nej}%
  \BibitemOpen
  \bibfield  {author} {\bibinfo {author} {\bibfnamefont {G.}~\bibnamefont
  {Ye}}, \bibinfo {author} {\bibfnamefont {B.}~\bibnamefont {Hu}}, \ and\
  \bibinfo {author} {\bibfnamefont {Y.-S.}\ \bibnamefont {Piao}},\ }\href
  {\doibase 10.1103/PhysRevD.104.063510} {\bibfield  {journal} {\bibinfo
  {journal} {Phys. Rev. D}\ }\textbf {\bibinfo {volume} {104}},\ \bibinfo
  {pages} {063510} (\bibinfo {year} {2021})},\ \Eprint
  {http://arxiv.org/abs/2103.09729} {arXiv:2103.09729 [astro-ph.CO]}
  \BibitemShut {NoStop}%
\bibitem [{\citenamefont {Takahashi}\ and\ \citenamefont
  {Yin}(2022)}]{Takahashi:2021bti}%
  \BibitemOpen
  \bibfield  {author} {\bibinfo {author} {\bibfnamefont {F.}~\bibnamefont
  {Takahashi}}\ and\ \bibinfo {author} {\bibfnamefont {W.}~\bibnamefont
  {Yin}},\ }\href {\doibase 10.1016/j.physletb.2022.137143} {\bibfield
  {journal} {\bibinfo  {journal} {Phys. Lett. B}\ }\textbf {\bibinfo {volume}
  {830}},\ \bibinfo {pages} {137143} (\bibinfo {year} {2022})},\ \Eprint
  {http://arxiv.org/abs/2112.06710} {arXiv:2112.06710 [astro-ph.CO]}
  \BibitemShut {NoStop}%
\bibitem [{\citenamefont {Forconi}\ \emph {et~al.}(2021)\citenamefont
  {Forconi}, \citenamefont {Giar{\`e}}, \citenamefont {Di~Valentino},\ and\
  \citenamefont {Melchiorri}}]{Forconi:2021que}%
  \BibitemOpen
  \bibfield  {author} {\bibinfo {author} {\bibfnamefont {M.}~\bibnamefont
  {Forconi}}, \bibinfo {author} {\bibfnamefont {W.}~\bibnamefont {Giar{\`e}}},
  \bibinfo {author} {\bibfnamefont {E.}~\bibnamefont {Di~Valentino}}, \ and\
  \bibinfo {author} {\bibfnamefont {A.}~\bibnamefont {Melchiorri}},\ }\href
  {\doibase 10.1103/PhysRevD.104.103528} {\bibfield  {journal} {\bibinfo
  {journal} {Phys. Rev. D}\ }\textbf {\bibinfo {volume} {104}},\ \bibinfo
  {pages} {103528} (\bibinfo {year} {2021})},\ \Eprint
  {http://arxiv.org/abs/2110.01695} {arXiv:2110.01695 [astro-ph.CO]}
  \BibitemShut {NoStop}%
\bibitem [{\citenamefont {Giar{\`e}}\ \emph
  {et~al.}(2023{\natexlab{a}})\citenamefont {Giar{\`e}}, \citenamefont {Renzi},
  \citenamefont {Mena}, \citenamefont {Di~Valentino},\ and\ \citenamefont
  {Melchiorri}}]{Giare:2022rvg}%
  \BibitemOpen
  \bibfield  {author} {\bibinfo {author} {\bibfnamefont {W.}~\bibnamefont
  {Giar{\`e}}}, \bibinfo {author} {\bibfnamefont {F.}~\bibnamefont {Renzi}},
  \bibinfo {author} {\bibfnamefont {O.}~\bibnamefont {Mena}}, \bibinfo {author}
  {\bibfnamefont {E.}~\bibnamefont {Di~Valentino}}, \ and\ \bibinfo {author}
  {\bibfnamefont {A.}~\bibnamefont {Melchiorri}},\ }\href {\doibase
  10.1093/mnras/stad724} {\bibfield  {journal} {\bibinfo  {journal} {Mon. Not.
  Roy. Astron. Soc.}\ }\textbf {\bibinfo {volume} {521}},\ \bibinfo {pages}
  {2911} (\bibinfo {year} {2023}{\natexlab{a}})},\ \Eprint
  {http://arxiv.org/abs/2210.09018} {arXiv:2210.09018 [astro-ph.CO]}
  \BibitemShut {NoStop}%
\bibitem [{\citenamefont {Jiang}\ and\ \citenamefont
  {Piao}(2022)}]{Jiang:2022uyg}%
  \BibitemOpen
  \bibfield  {author} {\bibinfo {author} {\bibfnamefont {J.-Q.}\ \bibnamefont
  {Jiang}}\ and\ \bibinfo {author} {\bibfnamefont {Y.-S.}\ \bibnamefont
  {Piao}},\ }\href {\doibase 10.1103/PhysRevD.105.103514} {\bibfield  {journal}
  {\bibinfo  {journal} {Phys. Rev. D}\ }\textbf {\bibinfo {volume} {105}},\
  \bibinfo {pages} {103514} (\bibinfo {year} {2022})},\ \Eprint
  {http://arxiv.org/abs/2202.13379} {arXiv:2202.13379 [astro-ph.CO]}
  \BibitemShut {NoStop}%
\bibitem [{\citenamefont {Ye}\ \emph {et~al.}(2022)\citenamefont {Ye},
  \citenamefont {Jiang},\ and\ \citenamefont {Piao}}]{Ye:2022efx}%
  \BibitemOpen
  \bibfield  {author} {\bibinfo {author} {\bibfnamefont {G.}~\bibnamefont
  {Ye}}, \bibinfo {author} {\bibfnamefont {J.-Q.}\ \bibnamefont {Jiang}}, \
  and\ \bibinfo {author} {\bibfnamefont {Y.-S.}\ \bibnamefont {Piao}},\ }\href
  {\doibase 10.1103/PhysRevD.106.103528} {\bibfield  {journal} {\bibinfo
  {journal} {Phys. Rev. D}\ }\textbf {\bibinfo {volume} {106}},\ \bibinfo
  {pages} {103528} (\bibinfo {year} {2022})},\ \Eprint
  {http://arxiv.org/abs/2205.02478} {arXiv:2205.02478 [astro-ph.CO]}
  \BibitemShut {NoStop}%
\bibitem [{\citenamefont {Roy~Choudhury}\ \emph {et~al.}(2022)\citenamefont
  {Roy~Choudhury}, \citenamefont {Hannestad},\ and\ \citenamefont
  {Tram}}]{RoyChoudhury:2022rva}%
  \BibitemOpen
  \bibfield  {author} {\bibinfo {author} {\bibfnamefont {S.}~\bibnamefont
  {Roy~Choudhury}}, \bibinfo {author} {\bibfnamefont {S.}~\bibnamefont
  {Hannestad}}, \ and\ \bibinfo {author} {\bibfnamefont {T.}~\bibnamefont
  {Tram}},\ }\href {\doibase 10.1088/1475-7516/2022/10/018} {\bibfield
  {journal} {\bibinfo  {journal} {JCAP}\ }\textbf {\bibinfo {volume} {10}},\
  \bibinfo {pages} {018} (\bibinfo {year} {2022})},\ \Eprint
  {http://arxiv.org/abs/2207.07142} {arXiv:2207.07142 [astro-ph.CO]}
  \BibitemShut {NoStop}%
\bibitem [{\citenamefont {Jiang}\ \emph {et~al.}(2023)\citenamefont {Jiang},
  \citenamefont {Ye},\ and\ \citenamefont {Piao}}]{Jiang:2022qlj}%
  \BibitemOpen
  \bibfield  {author} {\bibinfo {author} {\bibfnamefont {J.-Q.}\ \bibnamefont
  {Jiang}}, \bibinfo {author} {\bibfnamefont {G.}~\bibnamefont {Ye}}, \ and\
  \bibinfo {author} {\bibfnamefont {Y.-S.}\ \bibnamefont {Piao}},\ }\href
  {\doibase 10.1093/mnrasl/slad137} {\bibfield  {journal} {\bibinfo  {journal}
  {Mon. Not. Roy. Astron. Soc.}\ }\textbf {\bibinfo {volume} {527}},\ \bibinfo
  {pages} {L54} (\bibinfo {year} {2023})},\ \Eprint
  {http://arxiv.org/abs/2210.06125} {arXiv:2210.06125 [astro-ph.CO]}
  \BibitemShut {NoStop}%
\bibitem [{\citenamefont {Lin}(2022)}]{Lin:2022gbl}%
  \BibitemOpen
  \bibfield  {author} {\bibinfo {author} {\bibfnamefont {C.-M.}\ \bibnamefont
  {Lin}},\ }\href {\doibase 10.1103/PhysRevD.106.103511} {\bibfield  {journal}
  {\bibinfo  {journal} {Phys. Rev. D}\ }\textbf {\bibinfo {volume} {106}},\
  \bibinfo {pages} {103511} (\bibinfo {year} {2022})},\ \Eprint
  {http://arxiv.org/abs/2204.10475} {arXiv:2204.10475 [hep-th]} \BibitemShut
  {NoStop}%
\bibitem [{\citenamefont {Hazra}\ \emph {et~al.}(2022)\citenamefont {Hazra},
  \citenamefont {Antony},\ and\ \citenamefont {Shafieloo}}]{Hazra:2022rdl}%
  \BibitemOpen
  \bibfield  {author} {\bibinfo {author} {\bibfnamefont {D.~K.}\ \bibnamefont
  {Hazra}}, \bibinfo {author} {\bibfnamefont {A.}~\bibnamefont {Antony}}, \
  and\ \bibinfo {author} {\bibfnamefont {A.}~\bibnamefont {Shafieloo}},\ }\href
  {\doibase 10.1088/1475-7516/2022/08/063} {\bibfield  {journal} {\bibinfo
  {journal} {JCAP}\ }\textbf {\bibinfo {volume} {08}},\ \bibinfo {pages} {063}
  (\bibinfo {year} {2022})},\ \Eprint {http://arxiv.org/abs/2201.12000}
  {arXiv:2201.12000 [astro-ph.CO]} \BibitemShut {NoStop}%
\bibitem [{\citenamefont {Giar{\`e}}\ \emph
  {et~al.}(2023{\natexlab{b}})\citenamefont {Giar{\`e}}, \citenamefont
  {De~Angelis}, \citenamefont {van~de Bruck},\ and\ \citenamefont
  {Di~Valentino}}]{Giare:2023kiv}%
  \BibitemOpen
  \bibfield  {author} {\bibinfo {author} {\bibfnamefont {W.}~\bibnamefont
  {Giar{\`e}}}, \bibinfo {author} {\bibfnamefont {M.}~\bibnamefont
  {De~Angelis}}, \bibinfo {author} {\bibfnamefont {C.}~\bibnamefont {van~de
  Bruck}}, \ and\ \bibinfo {author} {\bibfnamefont {E.}~\bibnamefont
  {Di~Valentino}},\ }\href {\doibase 10.1088/1475-7516/2023/12/014} {\bibfield
  {journal} {\bibinfo  {journal} {JCAP}\ }\textbf {\bibinfo {volume} {12}},\
  \bibinfo {pages} {014} (\bibinfo {year} {2023}{\natexlab{b}})},\ \Eprint
  {http://arxiv.org/abs/2306.12414} {arXiv:2306.12414 [astro-ph.CO]}
  \BibitemShut {NoStop}%
\bibitem [{\citenamefont {Giar{\`e}}\ \emph
  {et~al.}(2023{\natexlab{c}})\citenamefont {Giar{\`e}}, \citenamefont {Pan},
  \citenamefont {Di~Valentino}, \citenamefont {Yang}, \citenamefont {de~Haro},\
  and\ \citenamefont {Melchiorri}}]{Giare:2023wzl}%
  \BibitemOpen
  \bibfield  {author} {\bibinfo {author} {\bibfnamefont {W.}~\bibnamefont
  {Giar{\`e}}}, \bibinfo {author} {\bibfnamefont {S.}~\bibnamefont {Pan}},
  \bibinfo {author} {\bibfnamefont {E.}~\bibnamefont {Di~Valentino}}, \bibinfo
  {author} {\bibfnamefont {W.}~\bibnamefont {Yang}}, \bibinfo {author}
  {\bibfnamefont {J.}~\bibnamefont {de~Haro}}, \ and\ \bibinfo {author}
  {\bibfnamefont {A.}~\bibnamefont {Melchiorri}},\ }\href {\doibase
  10.1088/1475-7516/2023/09/019} {\bibfield  {journal} {\bibinfo  {journal}
  {JCAP}\ }\textbf {\bibinfo {volume} {09}},\ \bibinfo {pages} {019} (\bibinfo
  {year} {2023}{\natexlab{c}})},\ \Eprint {http://arxiv.org/abs/2305.15378}
  {arXiv:2305.15378 [astro-ph.CO]} \BibitemShut {NoStop}%
\bibitem [{\citenamefont {Jiang}\ \emph
  {et~al.}(2024{\natexlab{a}})\citenamefont {Jiang}, \citenamefont {Ye},\ and\
  \citenamefont {Piao}}]{Jiang:2023bsz}%
  \BibitemOpen
  \bibfield  {author} {\bibinfo {author} {\bibfnamefont {J.-Q.}\ \bibnamefont
  {Jiang}}, \bibinfo {author} {\bibfnamefont {G.}~\bibnamefont {Ye}}, \ and\
  \bibinfo {author} {\bibfnamefont {Y.-S.}\ \bibnamefont {Piao}},\ }\href
  {\doibase 10.1016/j.physletb.2024.138588} {\bibfield  {journal} {\bibinfo
  {journal} {Phys. Lett. B}\ }\textbf {\bibinfo {volume} {851}},\ \bibinfo
  {pages} {138588} (\bibinfo {year} {2024}{\natexlab{a}})},\ \Eprint
  {http://arxiv.org/abs/2303.12345} {arXiv:2303.12345 [astro-ph.CO]}
  \BibitemShut {NoStop}%
\bibitem [{\citenamefont {Peng}\ and\ \citenamefont
  {Piao}(2024)}]{Peng:2023bik}%
  \BibitemOpen
  \bibfield  {author} {\bibinfo {author} {\bibfnamefont {Z.-Y.}\ \bibnamefont
  {Peng}}\ and\ \bibinfo {author} {\bibfnamefont {Y.-S.}\ \bibnamefont
  {Piao}},\ }\href {\doibase 10.1103/PhysRevD.109.023519} {\bibfield  {journal}
  {\bibinfo  {journal} {Phys. Rev. D}\ }\textbf {\bibinfo {volume} {109}},\
  \bibinfo {pages} {023519} (\bibinfo {year} {2024})},\ \Eprint
  {http://arxiv.org/abs/2308.01012} {arXiv:2308.01012 [astro-ph.CO]}
  \BibitemShut {NoStop}%
\bibitem [{\citenamefont {Forconi}\ \emph {et~al.}(2024)\citenamefont
  {Forconi}, \citenamefont {Giar{\`e}}, \citenamefont {Mena}, \citenamefont
  {Ruchika}, \citenamefont {Di~Valentino}, \citenamefont {Melchiorri},\ and\
  \citenamefont {Nunes}}]{Forconi:2023hsj}%
  \BibitemOpen
  \bibfield  {author} {\bibinfo {author} {\bibfnamefont {M.}~\bibnamefont
  {Forconi}}, \bibinfo {author} {\bibfnamefont {W.}~\bibnamefont {Giar{\`e}}},
  \bibinfo {author} {\bibfnamefont {O.}~\bibnamefont {Mena}}, \bibinfo {author}
  {\bibnamefont {Ruchika}}, \bibinfo {author} {\bibfnamefont {E.}~\bibnamefont
  {Di~Valentino}}, \bibinfo {author} {\bibfnamefont {A.}~\bibnamefont
  {Melchiorri}}, \ and\ \bibinfo {author} {\bibfnamefont {R.~C.}\ \bibnamefont
  {Nunes}},\ }\href {\doibase 10.1088/1475-7516/2024/05/097} {\bibfield
  {journal} {\bibinfo  {journal} {JCAP}\ }\textbf {\bibinfo {volume} {05}},\
  \bibinfo {pages} {097} (\bibinfo {year} {2024})},\ \Eprint
  {http://arxiv.org/abs/2312.11074} {arXiv:2312.11074 [astro-ph.CO]}
  \BibitemShut {NoStop}%
\bibitem [{\citenamefont {Fu}\ and\ \citenamefont {Wang}(2024)}]{Fu:2023tfo}%
  \BibitemOpen
  \bibfield  {author} {\bibinfo {author} {\bibfnamefont {C.}~\bibnamefont
  {Fu}}\ and\ \bibinfo {author} {\bibfnamefont {S.-J.}\ \bibnamefont {Wang}},\
  }\href {\doibase 10.1103/PhysRevD.109.L041304} {\bibfield  {journal}
  {\bibinfo  {journal} {Phys. Rev. D}\ }\textbf {\bibinfo {volume} {109}},\
  \bibinfo {pages} {L041304} (\bibinfo {year} {2024})},\ \Eprint
  {http://arxiv.org/abs/2310.12932} {arXiv:2310.12932 [astro-ph.CO]}
  \BibitemShut {NoStop}%
\bibitem [{\citenamefont {Giar{\`e}}(2024)}]{Giare:2024akf}%
  \BibitemOpen
  \bibfield  {author} {\bibinfo {author} {\bibfnamefont {W.}~\bibnamefont
  {Giar{\`e}}},\ }\href {\doibase 10.1103/PhysRevD.109.123545} {\bibfield
  {journal} {\bibinfo  {journal} {Phys. Rev. D}\ }\textbf {\bibinfo {volume}
  {109}},\ \bibinfo {pages} {123545} (\bibinfo {year} {2024})},\ \Eprint
  {http://arxiv.org/abs/2404.12779} {arXiv:2404.12779 [astro-ph.CO]}
  \BibitemShut {NoStop}%
\bibitem [{\citenamefont {Giar{\`e}}\ \emph
  {et~al.}(2024{\natexlab{a}})\citenamefont {Giar{\`e}}, \citenamefont
  {Di~Valentino}, \citenamefont {Linder},\ and\ \citenamefont
  {Specogna}}]{Giare:2024sdl}%
  \BibitemOpen
  \bibfield  {author} {\bibinfo {author} {\bibfnamefont {W.}~\bibnamefont
  {Giar{\`e}}}, \bibinfo {author} {\bibfnamefont {E.}~\bibnamefont
  {Di~Valentino}}, \bibinfo {author} {\bibfnamefont {E.~V.}\ \bibnamefont
  {Linder}}, \ and\ \bibinfo {author} {\bibfnamefont {E.}~\bibnamefont
  {Specogna}},\ }\href {\doibase 10.1016/j.dark.2024.101713} {\bibfield
  {journal} {\bibinfo  {journal} {Phys. Dark Univ.}\ }\textbf {\bibinfo
  {volume} {46}},\ \bibinfo {pages} {101713} (\bibinfo {year}
  {2024}{\natexlab{a}})},\ \Eprint {http://arxiv.org/abs/2402.01560}
  {arXiv:2402.01560 [astro-ph.CO]} \BibitemShut {NoStop}%
\bibitem [{\citenamefont {Wang}\ \emph {et~al.}(2025)\citenamefont {Wang},
  \citenamefont {Ye}, \citenamefont {Jiang},\ and\ \citenamefont
  {Piao}}]{Wang:2024tjd}%
  \BibitemOpen
  \bibfield  {author} {\bibinfo {author} {\bibfnamefont {H.}~\bibnamefont
  {Wang}}, \bibinfo {author} {\bibfnamefont {G.}~\bibnamefont {Ye}}, \bibinfo
  {author} {\bibfnamefont {J.-Q.}\ \bibnamefont {Jiang}}, \ and\ \bibinfo
  {author} {\bibfnamefont {Y.-S.}\ \bibnamefont {Piao}},\ }\href {\doibase
  10.1103/w19x-trrq} {\bibfield  {journal} {\bibinfo  {journal} {Phys. Rev. D}\
  }\textbf {\bibinfo {volume} {111}},\ \bibinfo {pages} {123505} (\bibinfo
  {year} {2025})},\ \Eprint {http://arxiv.org/abs/2409.17879} {arXiv:2409.17879
  [astro-ph.CO]} \BibitemShut {NoStop}%
\bibitem [{\citenamefont {da~Costa}(2025)}]{daCosta:2024grm}%
  \BibitemOpen
  \bibfield  {author} {\bibinfo {author} {\bibfnamefont {S.~S.}\ \bibnamefont
  {da~Costa}},\ }\href {\doibase 10.1016/j.dark.2024.101791} {\bibfield
  {journal} {\bibinfo  {journal} {Phys. Dark Univ.}\ }\textbf {\bibinfo
  {volume} {47}},\ \bibinfo {pages} {101791} (\bibinfo {year} {2025})},\
  \Eprint {http://arxiv.org/abs/2412.14290} {arXiv:2412.14290 [astro-ph.CO]}
  \BibitemShut {NoStop}%
\bibitem [{\citenamefont {Forconi}\ and\ \citenamefont
  {DI~Valentino}(2025)}]{Forconi:2025zzu}%
  \BibitemOpen
  \bibfield  {author} {\bibinfo {author} {\bibfnamefont {M.}~\bibnamefont
  {Forconi}}\ and\ \bibinfo {author} {\bibfnamefont {E.}~\bibnamefont
  {DI~Valentino}},\ }\href {\doibase 10.1016/j.dark.2025.101904} {\bibfield
  {journal} {\bibinfo  {journal} {Phys. Dark Univ.}\ }\textbf {\bibinfo
  {volume} {48}},\ \bibinfo {pages} {101904} (\bibinfo {year} {2025})},\
  \Eprint {http://arxiv.org/abs/2503.04705} {arXiv:2503.04705 [astro-ph.CO]}
  \BibitemShut {NoStop}%
\bibitem [{\citenamefont {Peng}\ \emph {et~al.}(2025)\citenamefont {Peng},
  \citenamefont {Jiang}, \citenamefont {Wang},\ and\ \citenamefont
  {Piao}}]{Peng:2025tqt}%
  \BibitemOpen
  \bibfield  {author} {\bibinfo {author} {\bibfnamefont {Z.-Y.}\ \bibnamefont
  {Peng}}, \bibinfo {author} {\bibfnamefont {J.-Q.}\ \bibnamefont {Jiang}},
  \bibinfo {author} {\bibfnamefont {H.}~\bibnamefont {Wang}}, \ and\ \bibinfo
  {author} {\bibfnamefont {Y.-S.}\ \bibnamefont {Piao}},\ }\href {\doibase
  10.1103/ys4m-3qws} {\bibfield  {journal} {\bibinfo  {journal} {Phys. Rev. D}\
  }\textbf {\bibinfo {volume} {112}},\ \bibinfo {pages} {123519} (\bibinfo
  {year} {2025})},\ \Eprint {http://arxiv.org/abs/2509.11902} {arXiv:2509.11902
  [astro-ph.CO]} \BibitemShut {NoStop}%
\bibitem [{\citenamefont {Balkenhol}\ \emph {et~al.}(2025)\citenamefont
  {Balkenhol} \emph {et~al.}}]{Balkenhol:2025wms}%
  \BibitemOpen
  \bibfield  {author} {\bibinfo {author} {\bibfnamefont {L.}~\bibnamefont
  {Balkenhol}} \emph {et~al.},\ }\href@noop {} {\  (\bibinfo {year} {2025})},\
  \Eprint {http://arxiv.org/abs/2512.10613} {arXiv:2512.10613 [astro-ph.CO]}
  \BibitemShut {NoStop}%
\bibitem [{\citenamefont {Yuan}\ \emph {et~al.}(2026)\citenamefont {Yuan},
  \citenamefont {Peng},\ and\ \citenamefont {Piao}}]{Yuan:2026xcg}%
  \BibitemOpen
  \bibfield  {author} {\bibinfo {author} {\bibfnamefont {H.-S.}\ \bibnamefont
  {Yuan}}, \bibinfo {author} {\bibfnamefont {Z.-Y.}\ \bibnamefont {Peng}}, \
  and\ \bibinfo {author} {\bibfnamefont {Y.-S.}\ \bibnamefont {Piao}},\
  }\href@noop {} {\  (\bibinfo {year} {2026})},\ \Eprint
  {http://arxiv.org/abs/2604.02823} {arXiv:2604.02823 [astro-ph.CO]}
  \BibitemShut {NoStop}%
\bibitem [{\citenamefont {Garny}\ \emph {et~al.}(2026)\citenamefont {Garny},
  \citenamefont {Niedermann},\ and\ \citenamefont {Sloth}}]{Garny:2026gcs}%
  \BibitemOpen
  \bibfield  {author} {\bibinfo {author} {\bibfnamefont {M.}~\bibnamefont
  {Garny}}, \bibinfo {author} {\bibfnamefont {F.}~\bibnamefont {Niedermann}}, \
  and\ \bibinfo {author} {\bibfnamefont {M.~S.}\ \bibnamefont {Sloth}},\
  }\href@noop {} {\  (\bibinfo {year} {2026})},\ \Eprint
  {http://arxiv.org/abs/2604.26541} {arXiv:2604.26541 [astro-ph.CO]}
  \BibitemShut {NoStop}%
\bibitem [{\citenamefont {Oikonomou}(2026{\natexlab{f}})}]{Oikonomou:2026ris}%
  \BibitemOpen
  \bibfield  {author} {\bibinfo {author} {\bibfnamefont {V.~K.}\ \bibnamefont
  {Oikonomou}},\ }\href@noop {} {\  (\bibinfo {year} {2026}{\natexlab{f}})},\
  \Eprint {http://arxiv.org/abs/2605.19336} {arXiv:2605.19336 [astro-ph.CO]}
  \BibitemShut {NoStop}%
\bibitem [{\citenamefont {Giar{\`e}}\ \emph {et~al.}(2026)\citenamefont
  {Giar{\`e}}, \citenamefont {Lee},\ and\ \citenamefont
  {Di~Valentino}}]{Giare:2026oti}%
  \BibitemOpen
  \bibfield  {author} {\bibinfo {author} {\bibfnamefont {W.}~\bibnamefont
  {Giar{\`e}}}, \bibinfo {author} {\bibfnamefont {D.~H.}\ \bibnamefont {Lee}},
  \ and\ \bibinfo {author} {\bibfnamefont {E.}~\bibnamefont {Di~Valentino}},\
  }\href {\doibase 10.1016/j.newar.2026.101759} {\bibfield  {journal} {\bibinfo
   {journal} {New Astron. Rev.}\ }\textbf {\bibinfo {volume} {103}},\ \bibinfo
  {pages} {101759} (\bibinfo {year} {2026})},\ \Eprint
  {http://arxiv.org/abs/2607.01226} {arXiv:2607.01226 [astro-ph.CO]}
  \BibitemShut {NoStop}%
\bibitem [{\citenamefont {Verde}\ \emph {et~al.}(2019)\citenamefont {Verde},
  \citenamefont {Treu},\ and\ \citenamefont {Riess}}]{Verde:2019ivm}%
  \BibitemOpen
  \bibfield  {author} {\bibinfo {author} {\bibfnamefont {L.}~\bibnamefont
  {Verde}}, \bibinfo {author} {\bibfnamefont {T.}~\bibnamefont {Treu}}, \ and\
  \bibinfo {author} {\bibfnamefont {A.~G.}\ \bibnamefont {Riess}},\ }\href
  {\doibase 10.1038/s41550-019-0902-0} {\bibfield  {journal} {\bibinfo
  {journal} {Nature Astron.}\ }\textbf {\bibinfo {volume} {3}},\ \bibinfo
  {pages} {891} (\bibinfo {year} {2019})},\ \Eprint
  {http://arxiv.org/abs/1907.10625} {arXiv:1907.10625 [astro-ph.CO]}
  \BibitemShut {NoStop}%
\bibitem [{\citenamefont {Di~Valentino}\ \emph
  {et~al.}(2021{\natexlab{a}})\citenamefont {Di~Valentino} \emph
  {et~al.}}]{DiValentino:2020zio}%
  \BibitemOpen
  \bibfield  {author} {\bibinfo {author} {\bibfnamefont {E.}~\bibnamefont
  {Di~Valentino}} \emph {et~al.},\ }\href {\doibase
  10.1016/j.astropartphys.2021.102605} {\bibfield  {journal} {\bibinfo
  {journal} {Astropart. Phys.}\ }\textbf {\bibinfo {volume} {131}},\ \bibinfo
  {pages} {102605} (\bibinfo {year} {2021}{\natexlab{a}})},\ \Eprint
  {http://arxiv.org/abs/2008.11284} {arXiv:2008.11284 [astro-ph.CO]}
  \BibitemShut {NoStop}%
\bibitem [{\citenamefont {Di~Valentino}\ \emph
  {et~al.}(2021{\natexlab{b}})\citenamefont {Di~Valentino}, \citenamefont
  {Mena}, \citenamefont {Pan}, \citenamefont {Visinelli}, \citenamefont {Yang},
  \citenamefont {Melchiorri}, \citenamefont {Mota}, \citenamefont {Riess},\
  and\ \citenamefont {Silk}}]{DiValentino:2021izs}%
  \BibitemOpen
  \bibfield  {author} {\bibinfo {author} {\bibfnamefont {E.}~\bibnamefont
  {Di~Valentino}}, \bibinfo {author} {\bibfnamefont {O.}~\bibnamefont {Mena}},
  \bibinfo {author} {\bibfnamefont {S.}~\bibnamefont {Pan}}, \bibinfo {author}
  {\bibfnamefont {L.}~\bibnamefont {Visinelli}}, \bibinfo {author}
  {\bibfnamefont {W.}~\bibnamefont {Yang}}, \bibinfo {author} {\bibfnamefont
  {A.}~\bibnamefont {Melchiorri}}, \bibinfo {author} {\bibfnamefont {D.~F.}\
  \bibnamefont {Mota}}, \bibinfo {author} {\bibfnamefont {A.~G.}\ \bibnamefont
  {Riess}}, \ and\ \bibinfo {author} {\bibfnamefont {J.}~\bibnamefont {Silk}},\
  }\href {\doibase 10.1088/1361-6382/ac086d} {\bibfield  {journal} {\bibinfo
  {journal} {Class. Quant. Grav.}\ }\textbf {\bibinfo {volume} {38}},\ \bibinfo
  {pages} {153001} (\bibinfo {year} {2021}{\natexlab{b}})},\ \Eprint
  {http://arxiv.org/abs/2103.01183} {arXiv:2103.01183 [astro-ph.CO]}
  \BibitemShut {NoStop}%
\bibitem [{\citenamefont {Perivolaropoulos}\ and\ \citenamefont
  {Skara}(2022)}]{Perivolaropoulos:2021jda}%
  \BibitemOpen
  \bibfield  {author} {\bibinfo {author} {\bibfnamefont {L.}~\bibnamefont
  {Perivolaropoulos}}\ and\ \bibinfo {author} {\bibfnamefont {F.}~\bibnamefont
  {Skara}},\ }\href {\doibase 10.1016/j.newar.2022.101659} {\bibfield
  {journal} {\bibinfo  {journal} {New Astron. Rev.}\ }\textbf {\bibinfo
  {volume} {95}},\ \bibinfo {pages} {101659} (\bibinfo {year} {2022})},\
  \Eprint {http://arxiv.org/abs/2105.05208} {arXiv:2105.05208 [astro-ph.CO]}
  \BibitemShut {NoStop}%
\bibitem [{\citenamefont {Shah}\ \emph {et~al.}(2021)\citenamefont {Shah},
  \citenamefont {Lemos},\ and\ \citenamefont {Lahav}}]{Shah:2021onj}%
  \BibitemOpen
  \bibfield  {author} {\bibinfo {author} {\bibfnamefont {P.}~\bibnamefont
  {Shah}}, \bibinfo {author} {\bibfnamefont {P.}~\bibnamefont {Lemos}}, \ and\
  \bibinfo {author} {\bibfnamefont {O.}~\bibnamefont {Lahav}},\ }\href
  {\doibase 10.1007/s00159-021-00137-4} {\bibfield  {journal} {\bibinfo
  {journal} {Astron. Astrophys. Rev.}\ }\textbf {\bibinfo {volume} {29}},\
  \bibinfo {pages} {9} (\bibinfo {year} {2021})},\ \Eprint
  {http://arxiv.org/abs/2109.01161} {arXiv:2109.01161 [astro-ph.CO]}
  \BibitemShut {NoStop}%
\bibitem [{\citenamefont {Abdalla}\ \emph {et~al.}(2022)\citenamefont {Abdalla}
  \emph {et~al.}}]{Abdalla:2022yfr}%
  \BibitemOpen
  \bibfield  {author} {\bibinfo {author} {\bibfnamefont {E.}~\bibnamefont
  {Abdalla}} \emph {et~al.},\ }\href {\doibase 10.1016/j.jheap.2022.04.002}
  {\bibfield  {journal} {\bibinfo  {journal} {JHEAp}\ }\textbf {\bibinfo
  {volume} {34}},\ \bibinfo {pages} {49} (\bibinfo {year} {2022})},\ \Eprint
  {http://arxiv.org/abs/2203.06142} {arXiv:2203.06142 [astro-ph.CO]}
  \BibitemShut {NoStop}%
\bibitem [{\citenamefont {Di~Valentino}(2022)}]{DiValentino:2022fjm}%
  \BibitemOpen
  \bibfield  {author} {\bibinfo {author} {\bibfnamefont {E.}~\bibnamefont
  {Di~Valentino}},\ }\href {\doibase 10.3390/universe8080399} {\bibfield
  {journal} {\bibinfo  {journal} {Universe}\ }\textbf {\bibinfo {volume} {8}},\
  \bibinfo {pages} {399} (\bibinfo {year} {2022})}\BibitemShut {NoStop}%
\bibitem [{\citenamefont {Hu}\ and\ \citenamefont {Wang}(2023)}]{Hu:2023jqc}%
  \BibitemOpen
  \bibfield  {author} {\bibinfo {author} {\bibfnamefont {J.-P.}\ \bibnamefont
  {Hu}}\ and\ \bibinfo {author} {\bibfnamefont {F.-Y.}\ \bibnamefont {Wang}},\
  }\href {\doibase 10.3390/universe9020094} {\bibfield  {journal} {\bibinfo
  {journal} {Universe}\ }\textbf {\bibinfo {volume} {9}},\ \bibinfo {pages}
  {94} (\bibinfo {year} {2023})},\ \Eprint {http://arxiv.org/abs/2302.05709}
  {arXiv:2302.05709 [astro-ph.CO]} \BibitemShut {NoStop}%
\bibitem [{\citenamefont {Vagnozzi}(2023{\natexlab{b}})}]{Vagnozzi:2023nrq}%
  \BibitemOpen
  \bibfield  {author} {\bibinfo {author} {\bibfnamefont {S.}~\bibnamefont
  {Vagnozzi}},\ }\href {\doibase 10.3390/universe9090393} {\bibfield  {journal}
  {\bibinfo  {journal} {Universe}\ }\textbf {\bibinfo {volume} {9}},\ \bibinfo
  {pages} {393} (\bibinfo {year} {2023}{\natexlab{b}})},\ \Eprint
  {http://arxiv.org/abs/2308.16628} {arXiv:2308.16628 [astro-ph.CO]}
  \BibitemShut {NoStop}%
\bibitem [{\citenamefont {Verde}\ \emph {et~al.}(2024)\citenamefont {Verde},
  \citenamefont {Sch{\"o}neberg},\ and\ \citenamefont
  {Gil-Mar{\'\i}n}}]{Verde:2023lmm}%
  \BibitemOpen
  \bibfield  {author} {\bibinfo {author} {\bibfnamefont {L.}~\bibnamefont
  {Verde}}, \bibinfo {author} {\bibfnamefont {N.}~\bibnamefont
  {Sch{\"o}neberg}}, \ and\ \bibinfo {author} {\bibfnamefont {H.}~\bibnamefont
  {Gil-Mar{\'\i}n}},\ }\href {\doibase 10.1146/annurev-astro-052622-033813}
  {\bibfield  {journal} {\bibinfo  {journal} {Ann. Rev. Astron. Astrophys.}\
  }\textbf {\bibinfo {volume} {62}},\ \bibinfo {pages} {287} (\bibinfo {year}
  {2024})},\ \Eprint {http://arxiv.org/abs/2311.13305} {arXiv:2311.13305
  [astro-ph.CO]} \BibitemShut {NoStop}%
\bibitem [{\citenamefont {Di~Valentino}\ \emph {et~al.}(2025)\citenamefont
  {Di~Valentino} \emph {et~al.}}]{CosmoVerseNetwork:2025alb}%
  \BibitemOpen
  \bibfield  {author} {\bibinfo {author} {\bibfnamefont {E.}~\bibnamefont
  {Di~Valentino}} \emph {et~al.} (\bibinfo {collaboration} {CosmoVerse
  Network}),\ }\href {\doibase 10.1016/j.dark.2025.101965} {\bibfield
  {journal} {\bibinfo  {journal} {Phys. Dark Univ.}\ }\textbf {\bibinfo
  {volume} {49}},\ \bibinfo {pages} {101965} (\bibinfo {year} {2025})},\
  \Eprint {http://arxiv.org/abs/2504.01669} {arXiv:2504.01669 [astro-ph.CO]}
  \BibitemShut {NoStop}%
\bibitem [{\citenamefont {Cai}\ and\ \citenamefont {Wang}(2026)}]{Cai:2026swf}%
  \BibitemOpen
  \bibfield  {author} {\bibinfo {author} {\bibfnamefont {R.-G.}\ \bibnamefont
  {Cai}}\ and\ \bibinfo {author} {\bibfnamefont {S.-J.}\ \bibnamefont {Wang}},\
  }\href {\doibase 10.1088/1674-4527/ae842f} {\bibfield  {journal} {\bibinfo
  {journal} {Res. Astron. Astrophys.}\ }\textbf {\bibinfo {volume} {26}},\
  \bibinfo {pages} {084011} (\bibinfo {year} {2026})},\ \Eprint
  {http://arxiv.org/abs/2606.20434} {arXiv:2606.20434 [astro-ph.CO]}
  \BibitemShut {NoStop}%
\bibitem [{\citenamefont {Sch{\"o}neberg}\ \emph
  {et~al.}(2026{\natexlab{a}})\citenamefont {Sch{\"o}neberg}, \citenamefont
  {Poulin}, \citenamefont {Ferrari}, \citenamefont {Finelli}, \citenamefont
  {Lesgourgues}, \citenamefont {Morelli}, \citenamefont {Mosbech},
  \citenamefont {Sharma},\ and\ \citenamefont {Simon}}]{Schoneberg:2026eys}%
  \BibitemOpen
  \bibfield  {author} {\bibinfo {author} {\bibfnamefont {N.}~\bibnamefont
  {Sch{\"o}neberg}}, \bibinfo {author} {\bibfnamefont {V.}~\bibnamefont
  {Poulin}}, \bibinfo {author} {\bibfnamefont {A.~G.}\ \bibnamefont {Ferrari}},
  \bibinfo {author} {\bibfnamefont {F.}~\bibnamefont {Finelli}}, \bibinfo
  {author} {\bibfnamefont {J.}~\bibnamefont {Lesgourgues}}, \bibinfo {author}
  {\bibfnamefont {L.}~\bibnamefont {Morelli}}, \bibinfo {author} {\bibfnamefont
  {M.~R.}\ \bibnamefont {Mosbech}}, \bibinfo {author} {\bibfnamefont {R.~K.}\
  \bibnamefont {Sharma}}, \ and\ \bibinfo {author} {\bibfnamefont
  {T.}~\bibnamefont {Simon}},\ }\href@noop {} {\  (\bibinfo {year}
  {2026}{\natexlab{a}})},\ \Eprint {http://arxiv.org/abs/2607.13282}
  {arXiv:2607.13282 [astro-ph.CO]} \BibitemShut {NoStop}%
\bibitem [{\citenamefont {Sch{\"o}neberg}\ \emph
  {et~al.}(2026{\natexlab{b}})\citenamefont {Sch{\"o}neberg}, \citenamefont
  {Poulin}, \citenamefont {Ferrari}, \citenamefont {Finelli}, \citenamefont
  {Lesgourgues}, \citenamefont {Morelli}, \citenamefont {Mosbech},
  \citenamefont {Sharma},\ and\ \citenamefont {Simon}}]{Schoneberg:2026vaf}%
  \BibitemOpen
  \bibfield  {author} {\bibinfo {author} {\bibfnamefont {N.}~\bibnamefont
  {Sch{\"o}neberg}}, \bibinfo {author} {\bibfnamefont {V.}~\bibnamefont
  {Poulin}}, \bibinfo {author} {\bibfnamefont {A.~G.}\ \bibnamefont {Ferrari}},
  \bibinfo {author} {\bibfnamefont {F.}~\bibnamefont {Finelli}}, \bibinfo
  {author} {\bibfnamefont {J.}~\bibnamefont {Lesgourgues}}, \bibinfo {author}
  {\bibfnamefont {L.}~\bibnamefont {Morelli}}, \bibinfo {author} {\bibfnamefont
  {M.~R.}\ \bibnamefont {Mosbech}}, \bibinfo {author} {\bibfnamefont {R.~K.}\
  \bibnamefont {Sharma}}, \ and\ \bibinfo {author} {\bibfnamefont
  {T.}~\bibnamefont {Simon}},\ }\href@noop {} {\  (\bibinfo {year}
  {2026}{\natexlab{b}})},\ \Eprint {http://arxiv.org/abs/2607.13283}
  {arXiv:2607.13283 [astro-ph.CO]} \BibitemShut {NoStop}%
\bibitem [{\citenamefont {Wang}\ and\ \citenamefont
  {Guo}(2026)}]{Wang:2026jxd}%
  \BibitemOpen
  \bibfield  {author} {\bibinfo {author} {\bibfnamefont {J.-Q.}\ \bibnamefont
  {Wang}}\ and\ \bibinfo {author} {\bibfnamefont {Z.-K.}\ \bibnamefont {Guo}},\
  }\href {\doibase 10.1088/1674-4527/ae842d} {\bibfield  {journal} {\bibinfo
  {journal} {Res. Astron. Astrophys.}\ }\textbf {\bibinfo {volume} {26}},\
  \bibinfo {pages} {084009} (\bibinfo {year} {2026})},\ \Eprint
  {http://arxiv.org/abs/2607.21497} {arXiv:2607.21497 [astro-ph.CO]}
  \BibitemShut {NoStop}%
\bibitem [{\citenamefont {M{\"o}rtsell}\ and\ \citenamefont
  {Dhawan}(2018)}]{Mortsell:2018mfj}%
  \BibitemOpen
  \bibfield  {author} {\bibinfo {author} {\bibfnamefont {E.}~\bibnamefont
  {M{\"o}rtsell}}\ and\ \bibinfo {author} {\bibfnamefont {S.}~\bibnamefont
  {Dhawan}},\ }\href {\doibase 10.1088/1475-7516/2018/09/025} {\bibfield
  {journal} {\bibinfo  {journal} {JCAP}\ }\textbf {\bibinfo {volume} {09}},\
  \bibinfo {pages} {025} (\bibinfo {year} {2018})},\ \Eprint
  {http://arxiv.org/abs/1801.07260} {arXiv:1801.07260 [astro-ph.CO]}
  \BibitemShut {NoStop}%
\bibitem [{\citenamefont {Vagnozzi}\ \emph {et~al.}(2018)\citenamefont
  {Vagnozzi}, \citenamefont {Dhawan}, \citenamefont {Gerbino}, \citenamefont
  {Freese}, \citenamefont {Goobar},\ and\ \citenamefont
  {Mena}}]{Vagnozzi:2018jhn}%
  \BibitemOpen
  \bibfield  {author} {\bibinfo {author} {\bibfnamefont {S.}~\bibnamefont
  {Vagnozzi}}, \bibinfo {author} {\bibfnamefont {S.}~\bibnamefont {Dhawan}},
  \bibinfo {author} {\bibfnamefont {M.}~\bibnamefont {Gerbino}}, \bibinfo
  {author} {\bibfnamefont {K.}~\bibnamefont {Freese}}, \bibinfo {author}
  {\bibfnamefont {A.}~\bibnamefont {Goobar}}, \ and\ \bibinfo {author}
  {\bibfnamefont {O.}~\bibnamefont {Mena}},\ }\href {\doibase
  10.1103/PhysRevD.98.083501} {\bibfield  {journal} {\bibinfo  {journal} {Phys.
  Rev. D}\ }\textbf {\bibinfo {volume} {98}},\ \bibinfo {pages} {083501}
  (\bibinfo {year} {2018})},\ \Eprint {http://arxiv.org/abs/1801.08553}
  {arXiv:1801.08553 [astro-ph.CO]} \BibitemShut {NoStop}%
\bibitem [{\citenamefont {Yang}\ \emph {et~al.}(2018)\citenamefont {Yang},
  \citenamefont {Pan}, \citenamefont {Di~Valentino}, \citenamefont {Nunes},
  \citenamefont {Vagnozzi},\ and\ \citenamefont {Mota}}]{Yang:2018euj}%
  \BibitemOpen
  \bibfield  {author} {\bibinfo {author} {\bibfnamefont {W.}~\bibnamefont
  {Yang}}, \bibinfo {author} {\bibfnamefont {S.}~\bibnamefont {Pan}}, \bibinfo
  {author} {\bibfnamefont {E.}~\bibnamefont {Di~Valentino}}, \bibinfo {author}
  {\bibfnamefont {R.~C.}\ \bibnamefont {Nunes}}, \bibinfo {author}
  {\bibfnamefont {S.}~\bibnamefont {Vagnozzi}}, \ and\ \bibinfo {author}
  {\bibfnamefont {D.~F.}\ \bibnamefont {Mota}},\ }\href {\doibase
  10.1088/1475-7516/2018/09/019} {\bibfield  {journal} {\bibinfo  {journal}
  {JCAP}\ }\textbf {\bibinfo {volume} {09}},\ \bibinfo {pages} {019} (\bibinfo
  {year} {2018})},\ \Eprint {http://arxiv.org/abs/1805.08252} {arXiv:1805.08252
  [astro-ph.CO]} \BibitemShut {NoStop}%
\bibitem [{\citenamefont {Guo}\ \emph {et~al.}(2019)\citenamefont {Guo},
  \citenamefont {Zhang},\ and\ \citenamefont {Zhang}}]{Guo:2018ans}%
  \BibitemOpen
  \bibfield  {author} {\bibinfo {author} {\bibfnamefont {R.-Y.}\ \bibnamefont
  {Guo}}, \bibinfo {author} {\bibfnamefont {J.-F.}\ \bibnamefont {Zhang}}, \
  and\ \bibinfo {author} {\bibfnamefont {X.}~\bibnamefont {Zhang}},\ }\href
  {\doibase 10.1088/1475-7516/2019/02/054} {\bibfield  {journal} {\bibinfo
  {journal} {JCAP}\ }\textbf {\bibinfo {volume} {02}},\ \bibinfo {pages} {054}
  (\bibinfo {year} {2019})},\ \Eprint {http://arxiv.org/abs/1809.02340}
  {arXiv:1809.02340 [astro-ph.CO]} \BibitemShut {NoStop}%
\bibitem [{\citenamefont {Kreisch}\ \emph {et~al.}(2020)\citenamefont
  {Kreisch}, \citenamefont {Cyr-Racine},\ and\ \citenamefont
  {Dor{\'e}}}]{Kreisch:2019yzn}%
  \BibitemOpen
  \bibfield  {author} {\bibinfo {author} {\bibfnamefont {C.~D.}\ \bibnamefont
  {Kreisch}}, \bibinfo {author} {\bibfnamefont {F.-Y.}\ \bibnamefont
  {Cyr-Racine}}, \ and\ \bibinfo {author} {\bibfnamefont {O.}~\bibnamefont
  {Dor{\'e}}},\ }\href {\doibase 10.1103/PhysRevD.101.123505} {\bibfield
  {journal} {\bibinfo  {journal} {Phys. Rev. D}\ }\textbf {\bibinfo {volume}
  {101}},\ \bibinfo {pages} {123505} (\bibinfo {year} {2020})},\ \Eprint
  {http://arxiv.org/abs/1902.00534} {arXiv:1902.00534 [astro-ph.CO]}
  \BibitemShut {NoStop}%
\bibitem [{\citenamefont {Vagnozzi}(2020)}]{Vagnozzi:2019ezj}%
  \BibitemOpen
  \bibfield  {author} {\bibinfo {author} {\bibfnamefont {S.}~\bibnamefont
  {Vagnozzi}},\ }\href {\doibase 10.1103/PhysRevD.102.023518} {\bibfield
  {journal} {\bibinfo  {journal} {Phys. Rev. D}\ }\textbf {\bibinfo {volume}
  {102}},\ \bibinfo {pages} {023518} (\bibinfo {year} {2020})},\ \Eprint
  {http://arxiv.org/abs/1907.07569} {arXiv:1907.07569 [astro-ph.CO]}
  \BibitemShut {NoStop}%
\bibitem [{\citenamefont {Visinelli}\ \emph {et~al.}(2019)\citenamefont
  {Visinelli}, \citenamefont {Vagnozzi},\ and\ \citenamefont
  {Danielsson}}]{Visinelli:2019qqu}%
  \BibitemOpen
  \bibfield  {author} {\bibinfo {author} {\bibfnamefont {L.}~\bibnamefont
  {Visinelli}}, \bibinfo {author} {\bibfnamefont {S.}~\bibnamefont {Vagnozzi}},
  \ and\ \bibinfo {author} {\bibfnamefont {U.}~\bibnamefont {Danielsson}},\
  }\href {\doibase 10.3390/sym11081035} {\bibfield  {journal} {\bibinfo
  {journal} {Symmetry}\ }\textbf {\bibinfo {volume} {11}},\ \bibinfo {pages}
  {1035} (\bibinfo {year} {2019})},\ \Eprint {http://arxiv.org/abs/1907.07953}
  {arXiv:1907.07953 [astro-ph.CO]} \BibitemShut {NoStop}%
\bibitem [{\citenamefont {Di~Valentino}\ \emph
  {et~al.}(2020{\natexlab{a}})\citenamefont {Di~Valentino}, \citenamefont
  {Melchiorri}, \citenamefont {Mena},\ and\ \citenamefont
  {Vagnozzi}}]{DiValentino:2019ffd}%
  \BibitemOpen
  \bibfield  {author} {\bibinfo {author} {\bibfnamefont {E.}~\bibnamefont
  {Di~Valentino}}, \bibinfo {author} {\bibfnamefont {A.}~\bibnamefont
  {Melchiorri}}, \bibinfo {author} {\bibfnamefont {O.}~\bibnamefont {Mena}}, \
  and\ \bibinfo {author} {\bibfnamefont {S.}~\bibnamefont {Vagnozzi}},\ }\href
  {\doibase 10.1016/j.dark.2020.100666} {\bibfield  {journal} {\bibinfo
  {journal} {Phys. Dark Univ.}\ }\textbf {\bibinfo {volume} {30}},\ \bibinfo
  {pages} {100666} (\bibinfo {year} {2020}{\natexlab{a}})},\ \Eprint
  {http://arxiv.org/abs/1908.04281} {arXiv:1908.04281 [astro-ph.CO]}
  \BibitemShut {NoStop}%
\bibitem [{\citenamefont {Di~Valentino}\ \emph
  {et~al.}(2020{\natexlab{b}})\citenamefont {Di~Valentino}, \citenamefont
  {Melchiorri}, \citenamefont {Mena},\ and\ \citenamefont
  {Vagnozzi}}]{DiValentino:2019jae}%
  \BibitemOpen
  \bibfield  {author} {\bibinfo {author} {\bibfnamefont {E.}~\bibnamefont
  {Di~Valentino}}, \bibinfo {author} {\bibfnamefont {A.}~\bibnamefont
  {Melchiorri}}, \bibinfo {author} {\bibfnamefont {O.}~\bibnamefont {Mena}}, \
  and\ \bibinfo {author} {\bibfnamefont {S.}~\bibnamefont {Vagnozzi}},\ }\href
  {\doibase 10.1103/PhysRevD.101.063502} {\bibfield  {journal} {\bibinfo
  {journal} {Phys. Rev. D}\ }\textbf {\bibinfo {volume} {101}},\ \bibinfo
  {pages} {063502} (\bibinfo {year} {2020}{\natexlab{b}})},\ \Eprint
  {http://arxiv.org/abs/1910.09853} {arXiv:1910.09853 [astro-ph.CO]}
  \BibitemShut {NoStop}%
\bibitem [{\citenamefont {Hart}\ and\ \citenamefont
  {Chluba}(2020)}]{Hart:2019dxi}%
  \BibitemOpen
  \bibfield  {author} {\bibinfo {author} {\bibfnamefont {L.}~\bibnamefont
  {Hart}}\ and\ \bibinfo {author} {\bibfnamefont {J.}~\bibnamefont {Chluba}},\
  }\href {\doibase 10.1093/mnras/staa412} {\bibfield  {journal} {\bibinfo
  {journal} {Mon. Not. Roy. Astron. Soc.}\ }\textbf {\bibinfo {volume} {493}},\
  \bibinfo {pages} {3255} (\bibinfo {year} {2020})},\ \Eprint
  {http://arxiv.org/abs/1912.03986} {arXiv:1912.03986 [astro-ph.CO]}
  \BibitemShut {NoStop}%
\bibitem [{\citenamefont {Krishnan}\ \emph {et~al.}(2020)\citenamefont
  {Krishnan}, \citenamefont {Colg{\'a}in}, \citenamefont {Ruchika},
  \citenamefont {Sen}, \citenamefont {Sheikh-Jabbari},\ and\ \citenamefont
  {Yang}}]{Krishnan:2020obg}%
  \BibitemOpen
  \bibfield  {author} {\bibinfo {author} {\bibfnamefont {C.}~\bibnamefont
  {Krishnan}}, \bibinfo {author} {\bibfnamefont {E.~{\'O}.}\ \bibnamefont
  {Colg{\'a}in}}, \bibinfo {author} {\bibnamefont {Ruchika}}, \bibinfo {author}
  {\bibfnamefont {A.~A.}\ \bibnamefont {Sen}}, \bibinfo {author} {\bibfnamefont
  {M.~M.}\ \bibnamefont {Sheikh-Jabbari}}, \ and\ \bibinfo {author}
  {\bibfnamefont {T.}~\bibnamefont {Yang}},\ }\href {\doibase
  10.1103/PhysRevD.102.103525} {\bibfield  {journal} {\bibinfo  {journal}
  {Phys. Rev. D}\ }\textbf {\bibinfo {volume} {102}},\ \bibinfo {pages}
  {103525} (\bibinfo {year} {2020})},\ \Eprint
  {http://arxiv.org/abs/2002.06044} {arXiv:2002.06044 [astro-ph.CO]}
  \BibitemShut {NoStop}%
\bibitem [{\citenamefont {Alestas}\ \emph {et~al.}(2020)\citenamefont
  {Alestas}, \citenamefont {Kazantzidis},\ and\ \citenamefont
  {Perivolaropoulos}}]{Alestas:2020mvb}%
  \BibitemOpen
  \bibfield  {author} {\bibinfo {author} {\bibfnamefont {G.}~\bibnamefont
  {Alestas}}, \bibinfo {author} {\bibfnamefont {L.}~\bibnamefont
  {Kazantzidis}}, \ and\ \bibinfo {author} {\bibfnamefont {L.}~\bibnamefont
  {Perivolaropoulos}},\ }\href {\doibase 10.1103/PhysRevD.101.123516}
  {\bibfield  {journal} {\bibinfo  {journal} {Phys. Rev. D}\ }\textbf {\bibinfo
  {volume} {101}},\ \bibinfo {pages} {123516} (\bibinfo {year} {2020})},\
  \Eprint {http://arxiv.org/abs/2004.08363} {arXiv:2004.08363 [astro-ph.CO]}
  \BibitemShut {NoStop}%
\bibitem [{\citenamefont {Jedamzik}\ and\ \citenamefont
  {Pogosian}(2020)}]{Jedamzik:2020krr}%
  \BibitemOpen
  \bibfield  {author} {\bibinfo {author} {\bibfnamefont {K.}~\bibnamefont
  {Jedamzik}}\ and\ \bibinfo {author} {\bibfnamefont {L.}~\bibnamefont
  {Pogosian}},\ }\href {\doibase 10.1103/PhysRevLett.125.181302} {\bibfield
  {journal} {\bibinfo  {journal} {Phys. Rev. Lett.}\ }\textbf {\bibinfo
  {volume} {125}},\ \bibinfo {pages} {181302} (\bibinfo {year} {2020})},\
  \Eprint {http://arxiv.org/abs/2004.09487} {arXiv:2004.09487 [astro-ph.CO]}
  \BibitemShut {NoStop}%
\bibitem [{\citenamefont {Sekiguchi}\ and\ \citenamefont
  {Takahashi}(2021)}]{Sekiguchi:2020teg}%
  \BibitemOpen
  \bibfield  {author} {\bibinfo {author} {\bibfnamefont {T.}~\bibnamefont
  {Sekiguchi}}\ and\ \bibinfo {author} {\bibfnamefont {T.}~\bibnamefont
  {Takahashi}},\ }\href {\doibase 10.1103/PhysRevD.103.083507} {\bibfield
  {journal} {\bibinfo  {journal} {Phys. Rev. D}\ }\textbf {\bibinfo {volume}
  {103}},\ \bibinfo {pages} {083507} (\bibinfo {year} {2021})},\ \Eprint
  {http://arxiv.org/abs/2007.03381} {arXiv:2007.03381 [astro-ph.CO]}
  \BibitemShut {NoStop}%
\bibitem [{\citenamefont {Roy~Choudhury}\ \emph {et~al.}(2021)\citenamefont
  {Roy~Choudhury}, \citenamefont {Hannestad},\ and\ \citenamefont
  {Tram}}]{RoyChoudhury:2020dmd}%
  \BibitemOpen
  \bibfield  {author} {\bibinfo {author} {\bibfnamefont {S.}~\bibnamefont
  {Roy~Choudhury}}, \bibinfo {author} {\bibfnamefont {S.}~\bibnamefont
  {Hannestad}}, \ and\ \bibinfo {author} {\bibfnamefont {T.}~\bibnamefont
  {Tram}},\ }\href {\doibase 10.1088/1475-7516/2021/03/084} {\bibfield
  {journal} {\bibinfo  {journal} {JCAP}\ }\textbf {\bibinfo {volume} {03}},\
  \bibinfo {pages} {084} (\bibinfo {year} {2021})},\ \Eprint
  {http://arxiv.org/abs/2012.07519} {arXiv:2012.07519 [astro-ph.CO]}
  \BibitemShut {NoStop}%
\bibitem [{\citenamefont {Brinckmann}\ \emph {et~al.}(2021)\citenamefont
  {Brinckmann}, \citenamefont {Chang},\ and\ \citenamefont
  {LoVerde}}]{Brinckmann:2020bcn}%
  \BibitemOpen
  \bibfield  {author} {\bibinfo {author} {\bibfnamefont {T.}~\bibnamefont
  {Brinckmann}}, \bibinfo {author} {\bibfnamefont {J.~H.}\ \bibnamefont
  {Chang}}, \ and\ \bibinfo {author} {\bibfnamefont {M.}~\bibnamefont
  {LoVerde}},\ }\href {\doibase 10.1103/PhysRevD.104.063523} {\bibfield
  {journal} {\bibinfo  {journal} {Phys. Rev. D}\ }\textbf {\bibinfo {volume}
  {104}},\ \bibinfo {pages} {063523} (\bibinfo {year} {2021})},\ \Eprint
  {http://arxiv.org/abs/2012.11830} {arXiv:2012.11830 [astro-ph.CO]}
  \BibitemShut {NoStop}%
\bibitem [{\citenamefont {Gao}\ \emph {et~al.}(2021)\citenamefont {Gao},
  \citenamefont {Zhao}, \citenamefont {Xue},\ and\ \citenamefont
  {Zhang}}]{Gao:2021xnk}%
  \BibitemOpen
  \bibfield  {author} {\bibinfo {author} {\bibfnamefont {L.-Y.}\ \bibnamefont
  {Gao}}, \bibinfo {author} {\bibfnamefont {Z.-W.}\ \bibnamefont {Zhao}},
  \bibinfo {author} {\bibfnamefont {S.-S.}\ \bibnamefont {Xue}}, \ and\
  \bibinfo {author} {\bibfnamefont {X.}~\bibnamefont {Zhang}},\ }\href
  {\doibase 10.1088/1475-7516/2021/07/005} {\bibfield  {journal} {\bibinfo
  {journal} {JCAP}\ }\textbf {\bibinfo {volume} {07}},\ \bibinfo {pages} {005}
  (\bibinfo {year} {2021})},\ \Eprint {http://arxiv.org/abs/2101.10714}
  {arXiv:2101.10714 [astro-ph.CO]} \BibitemShut {NoStop}%
\bibitem [{\citenamefont {Marra}\ and\ \citenamefont
  {Perivolaropoulos}(2021)}]{Marra:2021fvf}%
  \BibitemOpen
  \bibfield  {author} {\bibinfo {author} {\bibfnamefont {V.}~\bibnamefont
  {Marra}}\ and\ \bibinfo {author} {\bibfnamefont {L.}~\bibnamefont
  {Perivolaropoulos}},\ }\href {\doibase 10.1103/PhysRevD.104.L021303}
  {\bibfield  {journal} {\bibinfo  {journal} {Phys. Rev. D}\ }\textbf {\bibinfo
  {volume} {104}},\ \bibinfo {pages} {L021303} (\bibinfo {year} {2021})},\
  \Eprint {http://arxiv.org/abs/2102.06012} {arXiv:2102.06012 [astro-ph.CO]}
  \BibitemShut {NoStop}%
\bibitem [{\citenamefont {Sol{\`a}~Peracaula}\ \emph
  {et~al.}(2021)\citenamefont {Sol{\`a}~Peracaula}, \citenamefont
  {G{\'o}mez-Valent}, \citenamefont {de~Cruz~Perez},\ and\ \citenamefont
  {Moreno-Pulido}}]{SolaPeracaula:2021gxi}%
  \BibitemOpen
  \bibfield  {author} {\bibinfo {author} {\bibfnamefont {J.}~\bibnamefont
  {Sol{\`a}~Peracaula}}, \bibinfo {author} {\bibfnamefont {A.}~\bibnamefont
  {G{\'o}mez-Valent}}, \bibinfo {author} {\bibfnamefont {J.}~\bibnamefont
  {de~Cruz~Perez}}, \ and\ \bibinfo {author} {\bibfnamefont {C.}~\bibnamefont
  {Moreno-Pulido}},\ }\href {\doibase 10.1209/0295-5075/134/19001} {\bibfield
  {journal} {\bibinfo  {journal} {EPL}\ }\textbf {\bibinfo {volume} {134}},\
  \bibinfo {pages} {19001} (\bibinfo {year} {2021})},\ \Eprint
  {http://arxiv.org/abs/2102.12758} {arXiv:2102.12758 [astro-ph.CO]}
  \BibitemShut {NoStop}%
\bibitem [{\citenamefont {Dainotti}\ \emph {et~al.}(2021)\citenamefont
  {Dainotti}, \citenamefont {De~Simone}, \citenamefont {Schiavone},
  \citenamefont {Montani}, \citenamefont {Rinaldi},\ and\ \citenamefont
  {Lambiase}}]{Dainotti:2021pqg}%
  \BibitemOpen
  \bibfield  {author} {\bibinfo {author} {\bibfnamefont {M.~G.}\ \bibnamefont
  {Dainotti}}, \bibinfo {author} {\bibfnamefont {B.}~\bibnamefont {De~Simone}},
  \bibinfo {author} {\bibfnamefont {T.}~\bibnamefont {Schiavone}}, \bibinfo
  {author} {\bibfnamefont {G.}~\bibnamefont {Montani}}, \bibinfo {author}
  {\bibfnamefont {E.}~\bibnamefont {Rinaldi}}, \ and\ \bibinfo {author}
  {\bibfnamefont {G.}~\bibnamefont {Lambiase}},\ }\href {\doibase
  10.3847/1538-4357/abeb73} {\bibfield  {journal} {\bibinfo  {journal}
  {Astrophys. J.}\ }\textbf {\bibinfo {volume} {912}},\ \bibinfo {pages} {150}
  (\bibinfo {year} {2021})},\ \Eprint {http://arxiv.org/abs/2103.02117}
  {arXiv:2103.02117 [astro-ph.CO]} \BibitemShut {NoStop}%
\bibitem [{\citenamefont {Krishnan}\ \emph {et~al.}(2021)\citenamefont
  {Krishnan}, \citenamefont {Mohayaee}, \citenamefont {Colg{\'a}in},
  \citenamefont {Sheikh-Jabbari},\ and\ \citenamefont
  {Yin}}]{Krishnan:2021dyb}%
  \BibitemOpen
  \bibfield  {author} {\bibinfo {author} {\bibfnamefont {C.}~\bibnamefont
  {Krishnan}}, \bibinfo {author} {\bibfnamefont {R.}~\bibnamefont {Mohayaee}},
  \bibinfo {author} {\bibfnamefont {E.~{\'O}.}\ \bibnamefont {Colg{\'a}in}},
  \bibinfo {author} {\bibfnamefont {M.~M.}\ \bibnamefont {Sheikh-Jabbari}}, \
  and\ \bibinfo {author} {\bibfnamefont {L.}~\bibnamefont {Yin}},\ }\href
  {\doibase 10.1088/1361-6382/ac1a81} {\bibfield  {journal} {\bibinfo
  {journal} {Class. Quant. Grav.}\ }\textbf {\bibinfo {volume} {38}},\ \bibinfo
  {pages} {184001} (\bibinfo {year} {2021})},\ \Eprint
  {http://arxiv.org/abs/2105.09790} {arXiv:2105.09790 [astro-ph.CO]}
  \BibitemShut {NoStop}%
\bibitem [{\citenamefont {Hart}\ and\ \citenamefont
  {Chluba}(2022)}]{Hart:2021kad}%
  \BibitemOpen
  \bibfield  {author} {\bibinfo {author} {\bibfnamefont {L.}~\bibnamefont
  {Hart}}\ and\ \bibinfo {author} {\bibfnamefont {J.}~\bibnamefont {Chluba}},\
  }\href {\doibase 10.1093/mnras/stab2777} {\bibfield  {journal} {\bibinfo
  {journal} {Mon. Not. Roy. Astron. Soc.}\ }\textbf {\bibinfo {volume} {510}},\
  \bibinfo {pages} {2206} (\bibinfo {year} {2022})},\ \Eprint
  {http://arxiv.org/abs/2107.12465} {arXiv:2107.12465 [astro-ph.CO]}
  \BibitemShut {NoStop}%
\bibitem [{\citenamefont {Cyr-Racine}\ \emph {et~al.}(2022)\citenamefont
  {Cyr-Racine}, \citenamefont {Ge},\ and\ \citenamefont
  {Knox}}]{Cyr-Racine:2021oal}%
  \BibitemOpen
  \bibfield  {author} {\bibinfo {author} {\bibfnamefont {F.-Y.}\ \bibnamefont
  {Cyr-Racine}}, \bibinfo {author} {\bibfnamefont {F.}~\bibnamefont {Ge}}, \
  and\ \bibinfo {author} {\bibfnamefont {L.}~\bibnamefont {Knox}},\ }\href
  {\doibase 10.1103/PhysRevLett.128.201301} {\bibfield  {journal} {\bibinfo
  {journal} {Phys. Rev. Lett.}\ }\textbf {\bibinfo {volume} {128}},\ \bibinfo
  {pages} {201301} (\bibinfo {year} {2022})},\ \Eprint
  {http://arxiv.org/abs/2107.13000} {arXiv:2107.13000 [astro-ph.CO]}
  \BibitemShut {NoStop}%
\bibitem [{\citenamefont {Anchordoqui}\ \emph {et~al.}(2021)\citenamefont
  {Anchordoqui}, \citenamefont {Di~Valentino}, \citenamefont {Pan},\ and\
  \citenamefont {Yang}}]{Anchordoqui:2021gji}%
  \BibitemOpen
  \bibfield  {author} {\bibinfo {author} {\bibfnamefont {L.~A.}\ \bibnamefont
  {Anchordoqui}}, \bibinfo {author} {\bibfnamefont {E.}~\bibnamefont
  {Di~Valentino}}, \bibinfo {author} {\bibfnamefont {S.}~\bibnamefont {Pan}}, \
  and\ \bibinfo {author} {\bibfnamefont {W.}~\bibnamefont {Yang}},\ }\href
  {\doibase 10.1016/j.jheap.2021.08.001} {\bibfield  {journal} {\bibinfo
  {journal} {JHEAp}\ }\textbf {\bibinfo {volume} {32}},\ \bibinfo {pages} {28}
  (\bibinfo {year} {2021})},\ \Eprint {http://arxiv.org/abs/2107.13932}
  {arXiv:2107.13932 [astro-ph.CO]} \BibitemShut {NoStop}%
\bibitem [{\citenamefont {Akarsu}\ \emph {et~al.}(2021)\citenamefont {Akarsu},
  \citenamefont {Kumar}, \citenamefont {{\"O}z{\"u}lker},\ and\ \citenamefont
  {Vazquez}}]{Akarsu:2021fol}%
  \BibitemOpen
  \bibfield  {author} {\bibinfo {author} {\bibfnamefont {{\"O}.}~\bibnamefont
  {Akarsu}}, \bibinfo {author} {\bibfnamefont {S.}~\bibnamefont {Kumar}},
  \bibinfo {author} {\bibfnamefont {E.}~\bibnamefont {{\"O}z{\"u}lker}}, \ and\
  \bibinfo {author} {\bibfnamefont {J.~A.}\ \bibnamefont {Vazquez}},\ }\href
  {\doibase 10.1103/PhysRevD.104.123512} {\bibfield  {journal} {\bibinfo
  {journal} {Phys. Rev. D}\ }\textbf {\bibinfo {volume} {104}},\ \bibinfo
  {pages} {123512} (\bibinfo {year} {2021})},\ \Eprint
  {http://arxiv.org/abs/2108.09239} {arXiv:2108.09239 [astro-ph.CO]}
  \BibitemShut {NoStop}%
\bibitem [{\citenamefont {Ren}\ \emph {et~al.}(2022)\citenamefont {Ren},
  \citenamefont {Yan}, \citenamefont {Zhao}, \citenamefont {Cai},\ and\
  \citenamefont {Saridakis}}]{Ren:2022aeo}%
  \BibitemOpen
  \bibfield  {author} {\bibinfo {author} {\bibfnamefont {X.}~\bibnamefont
  {Ren}}, \bibinfo {author} {\bibfnamefont {S.-F.}\ \bibnamefont {Yan}},
  \bibinfo {author} {\bibfnamefont {Y.}~\bibnamefont {Zhao}}, \bibinfo {author}
  {\bibfnamefont {Y.-F.}\ \bibnamefont {Cai}}, \ and\ \bibinfo {author}
  {\bibfnamefont {E.~N.}\ \bibnamefont {Saridakis}},\ }\href {\doibase
  10.3847/1538-4357/ac6ba5} {\bibfield  {journal} {\bibinfo  {journal}
  {Astrophys. J.}\ }\textbf {\bibinfo {volume} {932}},\ \bibinfo {pages} {131}
  (\bibinfo {year} {2022})},\ \Eprint {http://arxiv.org/abs/2203.01926}
  {arXiv:2203.01926 [astro-ph.CO]} \BibitemShut {NoStop}%
\bibitem [{\citenamefont {Nojiri}\ \emph {et~al.}(2022)\citenamefont {Nojiri},
  \citenamefont {Odintsov},\ and\ \citenamefont {Oikonomou}}]{Nojiri:2022ski}%
  \BibitemOpen
  \bibfield  {author} {\bibinfo {author} {\bibfnamefont {S.}~\bibnamefont
  {Nojiri}}, \bibinfo {author} {\bibfnamefont {S.~D.}\ \bibnamefont
  {Odintsov}}, \ and\ \bibinfo {author} {\bibfnamefont {V.~K.}\ \bibnamefont
  {Oikonomou}},\ }\href {\doibase 10.1016/j.nuclphysb.2022.115850} {\bibfield
  {journal} {\bibinfo  {journal} {Nucl. Phys. B}\ }\textbf {\bibinfo {volume}
  {980}},\ \bibinfo {pages} {115850} (\bibinfo {year} {2022})},\ \Eprint
  {http://arxiv.org/abs/2205.11681} {arXiv:2205.11681 [gr-qc]} \BibitemShut
  {NoStop}%
\bibitem [{\citenamefont {Sch{\"o}neberg}\ and\ \citenamefont
  {Franco~Abell{\'a}n}(2022)}]{Schoneberg:2022grr}%
  \BibitemOpen
  \bibfield  {author} {\bibinfo {author} {\bibfnamefont {N.}~\bibnamefont
  {Sch{\"o}neberg}}\ and\ \bibinfo {author} {\bibfnamefont {G.}~\bibnamefont
  {Franco~Abell{\'a}n}},\ }\href {\doibase 10.1088/1475-7516/2022/12/001}
  {\bibfield  {journal} {\bibinfo  {journal} {JCAP}\ }\textbf {\bibinfo
  {volume} {12}},\ \bibinfo {pages} {001} (\bibinfo {year} {2022})},\ \Eprint
  {http://arxiv.org/abs/2206.11276} {arXiv:2206.11276 [astro-ph.CO]}
  \BibitemShut {NoStop}%
\bibitem [{\citenamefont {Moshafi}\ \emph {et~al.}(2022)\citenamefont
  {Moshafi}, \citenamefont {Firouzjahi},\ and\ \citenamefont
  {Talebian}}]{Moshafi:2022mva}%
  \BibitemOpen
  \bibfield  {author} {\bibinfo {author} {\bibfnamefont {H.}~\bibnamefont
  {Moshafi}}, \bibinfo {author} {\bibfnamefont {H.}~\bibnamefont {Firouzjahi}},
  \ and\ \bibinfo {author} {\bibfnamefont {A.}~\bibnamefont {Talebian}},\
  }\href {\doibase 10.3847/1538-4357/ac9c58} {\bibfield  {journal} {\bibinfo
  {journal} {Astrophys. J.}\ }\textbf {\bibinfo {volume} {940}},\ \bibinfo
  {pages} {121} (\bibinfo {year} {2022})},\ \Eprint
  {http://arxiv.org/abs/2208.05583} {arXiv:2208.05583 [astro-ph.CO]}
  \BibitemShut {NoStop}%
\bibitem [{\citenamefont {Rezazadeh}\ \emph {et~al.}(2024)\citenamefont
  {Rezazadeh}, \citenamefont {Ashoorioon},\ and\ \citenamefont
  {Grin}}]{Rezazadeh:2022lsf}%
  \BibitemOpen
  \bibfield  {author} {\bibinfo {author} {\bibfnamefont {K.}~\bibnamefont
  {Rezazadeh}}, \bibinfo {author} {\bibfnamefont {A.}~\bibnamefont
  {Ashoorioon}}, \ and\ \bibinfo {author} {\bibfnamefont {D.}~\bibnamefont
  {Grin}},\ }\href {\doibase 10.3847/1538-4357/ad7b16} {\bibfield  {journal}
  {\bibinfo  {journal} {Astrophys. J.}\ }\textbf {\bibinfo {volume} {975}},\
  \bibinfo {pages} {137} (\bibinfo {year} {2024})},\ \Eprint
  {http://arxiv.org/abs/2208.07631} {arXiv:2208.07631 [astro-ph.CO]}
  \BibitemShut {NoStop}%
\bibitem [{\citenamefont {Banerjee}\ \emph {et~al.}(2023)\citenamefont
  {Banerjee}, \citenamefont {Petronikolou},\ and\ \citenamefont
  {Saridakis}}]{Banerjee:2022ynv}%
  \BibitemOpen
  \bibfield  {author} {\bibinfo {author} {\bibfnamefont {S.}~\bibnamefont
  {Banerjee}}, \bibinfo {author} {\bibfnamefont {M.}~\bibnamefont
  {Petronikolou}}, \ and\ \bibinfo {author} {\bibfnamefont {E.~N.}\
  \bibnamefont {Saridakis}},\ }\href {\doibase 10.1103/PhysRevD.108.024012}
  {\bibfield  {journal} {\bibinfo  {journal} {Phys. Rev. D}\ }\textbf {\bibinfo
  {volume} {108}},\ \bibinfo {pages} {024012} (\bibinfo {year} {2023})},\
  \Eprint {http://arxiv.org/abs/2209.02426} {arXiv:2209.02426 [gr-qc]}
  \BibitemShut {NoStop}%
\bibitem [{\citenamefont {de~S{\'a}}\ \emph {et~al.}(2022)\citenamefont
  {de~S{\'a}}, \citenamefont {Benetti},\ and\ \citenamefont
  {Graef}}]{deSa:2022hsh}%
  \BibitemOpen
  \bibfield  {author} {\bibinfo {author} {\bibfnamefont {R.}~\bibnamefont
  {de~S{\'a}}}, \bibinfo {author} {\bibfnamefont {M.}~\bibnamefont {Benetti}},
  \ and\ \bibinfo {author} {\bibfnamefont {L.~L.}\ \bibnamefont {Graef}},\
  }\href {\doibase 10.1140/epjp/s13360-022-03343-w} {\bibfield  {journal}
  {\bibinfo  {journal} {Eur. Phys. J. Plus}\ }\textbf {\bibinfo {volume}
  {137}},\ \bibinfo {pages} {1129} (\bibinfo {year} {2022})},\ \Eprint
  {http://arxiv.org/abs/2209.11476} {arXiv:2209.11476 [astro-ph.CO]}
  \BibitemShut {NoStop}%
\bibitem [{\citenamefont {Akarsu}\ \emph {et~al.}(2023)\citenamefont {Akarsu},
  \citenamefont {Kumar}, \citenamefont {{\"O}z{\"u}lker}, \citenamefont
  {Vazquez},\ and\ \citenamefont {Yadav}}]{Akarsu:2022typ}%
  \BibitemOpen
  \bibfield  {author} {\bibinfo {author} {\bibfnamefont {O.}~\bibnamefont
  {Akarsu}}, \bibinfo {author} {\bibfnamefont {S.}~\bibnamefont {Kumar}},
  \bibinfo {author} {\bibfnamefont {E.}~\bibnamefont {{\"O}z{\"u}lker}},
  \bibinfo {author} {\bibfnamefont {J.~A.}\ \bibnamefont {Vazquez}}, \ and\
  \bibinfo {author} {\bibfnamefont {A.}~\bibnamefont {Yadav}},\ }\href
  {\doibase 10.1103/PhysRevD.108.023513} {\bibfield  {journal} {\bibinfo
  {journal} {Phys. Rev. D}\ }\textbf {\bibinfo {volume} {108}},\ \bibinfo
  {pages} {023513} (\bibinfo {year} {2023})},\ \Eprint
  {http://arxiv.org/abs/2211.05742} {arXiv:2211.05742 [astro-ph.CO]}
  \BibitemShut {NoStop}%
\bibitem [{\citenamefont {Lee}\ \emph {et~al.}(2023)\citenamefont {Lee},
  \citenamefont {Ali-Ha{\"\i}moud}, \citenamefont {Sch{\"o}neberg},\ and\
  \citenamefont {Poulin}}]{Lee:2022gzh}%
  \BibitemOpen
  \bibfield  {author} {\bibinfo {author} {\bibfnamefont {N.}~\bibnamefont
  {Lee}}, \bibinfo {author} {\bibfnamefont {Y.}~\bibnamefont
  {Ali-Ha{\"\i}moud}}, \bibinfo {author} {\bibfnamefont {N.}~\bibnamefont
  {Sch{\"o}neberg}}, \ and\ \bibinfo {author} {\bibfnamefont {V.}~\bibnamefont
  {Poulin}},\ }\href {\doibase 10.1103/PhysRevLett.130.161003} {\bibfield
  {journal} {\bibinfo  {journal} {Phys. Rev. Lett.}\ }\textbf {\bibinfo
  {volume} {130}},\ \bibinfo {pages} {161003} (\bibinfo {year} {2023})},\
  \Eprint {http://arxiv.org/abs/2212.04494} {arXiv:2212.04494 [astro-ph.CO]}
  \BibitemShut {NoStop}%
\bibitem [{\citenamefont {Khodadi}\ and\ \citenamefont
  {Schreck}(2023)}]{Khodadi:2023ezj}%
  \BibitemOpen
  \bibfield  {author} {\bibinfo {author} {\bibfnamefont {M.}~\bibnamefont
  {Khodadi}}\ and\ \bibinfo {author} {\bibfnamefont {M.}~\bibnamefont
  {Schreck}},\ }\href {\doibase 10.1016/j.dark.2023.101170} {\bibfield
  {journal} {\bibinfo  {journal} {Phys. Dark Univ.}\ }\textbf {\bibinfo
  {volume} {39}},\ \bibinfo {pages} {101170} (\bibinfo {year} {2023})},\
  \Eprint {http://arxiv.org/abs/2301.03883} {arXiv:2301.03883 [gr-qc]}
  \BibitemShut {NoStop}%
\bibitem [{\citenamefont {Bernui}\ \emph {et~al.}(2023)\citenamefont {Bernui},
  \citenamefont {Di~Valentino}, \citenamefont {Giar{\`e}}, \citenamefont
  {Kumar},\ and\ \citenamefont {Nunes}}]{Bernui:2023byc}%
  \BibitemOpen
  \bibfield  {author} {\bibinfo {author} {\bibfnamefont {A.}~\bibnamefont
  {Bernui}}, \bibinfo {author} {\bibfnamefont {E.}~\bibnamefont
  {Di~Valentino}}, \bibinfo {author} {\bibfnamefont {W.}~\bibnamefont
  {Giar{\`e}}}, \bibinfo {author} {\bibfnamefont {S.}~\bibnamefont {Kumar}}, \
  and\ \bibinfo {author} {\bibfnamefont {R.~C.}\ \bibnamefont {Nunes}},\ }\href
  {\doibase 10.1103/PhysRevD.107.103531} {\bibfield  {journal} {\bibinfo
  {journal} {Phys. Rev. D}\ }\textbf {\bibinfo {volume} {107}},\ \bibinfo
  {pages} {103531} (\bibinfo {year} {2023})},\ \Eprint
  {http://arxiv.org/abs/2301.06097} {arXiv:2301.06097 [astro-ph.CO]}
  \BibitemShut {NoStop}%
\bibitem [{\citenamefont {Ben-Dayan}\ and\ \citenamefont
  {Kumar}(2023)}]{Ben-Dayan:2023rgt}%
  \BibitemOpen
  \bibfield  {author} {\bibinfo {author} {\bibfnamefont {I.}~\bibnamefont
  {Ben-Dayan}}\ and\ \bibinfo {author} {\bibfnamefont {U.}~\bibnamefont
  {Kumar}},\ }\href {\doibase 10.1088/1475-7516/2023/12/047} {\bibfield
  {journal} {\bibinfo  {journal} {JCAP}\ }\textbf {\bibinfo {volume} {12}},\
  \bibinfo {pages} {047} (\bibinfo {year} {2023})},\ \Eprint
  {http://arxiv.org/abs/2302.00067} {arXiv:2302.00067 [astro-ph.CO]}
  \BibitemShut {NoStop}%
\bibitem [{\citenamefont {G{\'o}mez-Valent}\ \emph
  {et~al.}(2024{\natexlab{a}})\citenamefont {G{\'o}mez-Valent}, \citenamefont
  {Mavromatos},\ and\ \citenamefont
  {Sol{\`a}~Peracaula}}]{Gomez-Valent:2023hov}%
  \BibitemOpen
  \bibfield  {author} {\bibinfo {author} {\bibfnamefont {A.}~\bibnamefont
  {G{\'o}mez-Valent}}, \bibinfo {author} {\bibfnamefont {N.~E.}\ \bibnamefont
  {Mavromatos}}, \ and\ \bibinfo {author} {\bibfnamefont {J.}~\bibnamefont
  {Sol{\`a}~Peracaula}},\ }\href {\doibase 10.1088/1361-6382/ad0fb8} {\bibfield
   {journal} {\bibinfo  {journal} {Class. Quant. Grav.}\ }\textbf {\bibinfo
  {volume} {41}},\ \bibinfo {pages} {015026} (\bibinfo {year}
  {2024}{\natexlab{a}})},\ \Eprint {http://arxiv.org/abs/2305.15774}
  {arXiv:2305.15774 [gr-qc]} \BibitemShut {NoStop}%
\bibitem [{\citenamefont {Ruchika}\ \emph {et~al.}(2024)\citenamefont
  {Ruchika}, \citenamefont {Rathore}, \citenamefont {Roy~Choudhury},\ and\
  \citenamefont {Rentala}}]{Ruchika:2023ugh}%
  \BibitemOpen
  \bibfield  {author} {\bibinfo {author} {\bibnamefont {Ruchika}}, \bibinfo
  {author} {\bibfnamefont {H.}~\bibnamefont {Rathore}}, \bibinfo {author}
  {\bibfnamefont {S.}~\bibnamefont {Roy~Choudhury}}, \ and\ \bibinfo {author}
  {\bibfnamefont {V.}~\bibnamefont {Rentala}},\ }\href {\doibase
  10.1088/1475-7516/2024/06/056} {\bibfield  {journal} {\bibinfo  {journal}
  {JCAP}\ }\textbf {\bibinfo {volume} {06}},\ \bibinfo {pages} {056} (\bibinfo
  {year} {2024})},\ \Eprint {http://arxiv.org/abs/2306.05450} {arXiv:2306.05450
  [astro-ph.CO]} \BibitemShut {NoStop}%
\bibitem [{\citenamefont {Adil}\ \emph {et~al.}(2024)\citenamefont {Adil},
  \citenamefont {Akarsu}, \citenamefont {Di~Valentino}, \citenamefont {Nunes},
  \citenamefont {{\"O}z{\"u}lker}, \citenamefont {Sen},\ and\ \citenamefont
  {Specogna}}]{Adil:2023exv}%
  \BibitemOpen
  \bibfield  {author} {\bibinfo {author} {\bibfnamefont {S.~A.}\ \bibnamefont
  {Adil}}, \bibinfo {author} {\bibfnamefont {{\"O}.}~\bibnamefont {Akarsu}},
  \bibinfo {author} {\bibfnamefont {E.}~\bibnamefont {Di~Valentino}}, \bibinfo
  {author} {\bibfnamefont {R.~C.}\ \bibnamefont {Nunes}}, \bibinfo {author}
  {\bibfnamefont {E.}~\bibnamefont {{\"O}z{\"u}lker}}, \bibinfo {author}
  {\bibfnamefont {A.~A.}\ \bibnamefont {Sen}}, \ and\ \bibinfo {author}
  {\bibfnamefont {E.}~\bibnamefont {Specogna}},\ }\href {\doibase
  10.1103/PhysRevD.109.023527} {\bibfield  {journal} {\bibinfo  {journal}
  {Phys. Rev. D}\ }\textbf {\bibinfo {volume} {109}},\ \bibinfo {pages}
  {023527} (\bibinfo {year} {2024})},\ \Eprint
  {http://arxiv.org/abs/2306.08046} {arXiv:2306.08046 [astro-ph.CO]}
  \BibitemShut {NoStop}%
\bibitem [{\citenamefont {Frion}\ \emph {et~al.}(2023)\citenamefont {Frion},
  \citenamefont {Camarena}, \citenamefont {Giani}, \citenamefont {Miranda},
  \citenamefont {Bertacca}, \citenamefont {Marra},\ and\ \citenamefont
  {Piattella}}]{Frion:2023xwq}%
  \BibitemOpen
  \bibfield  {author} {\bibinfo {author} {\bibfnamefont {E.}~\bibnamefont
  {Frion}}, \bibinfo {author} {\bibfnamefont {D.}~\bibnamefont {Camarena}},
  \bibinfo {author} {\bibfnamefont {L.}~\bibnamefont {Giani}}, \bibinfo
  {author} {\bibfnamefont {T.}~\bibnamefont {Miranda}}, \bibinfo {author}
  {\bibfnamefont {D.}~\bibnamefont {Bertacca}}, \bibinfo {author}
  {\bibfnamefont {V.}~\bibnamefont {Marra}}, \ and\ \bibinfo {author}
  {\bibfnamefont {O.~F.}\ \bibnamefont {Piattella}},\ }\href {\doibase
  10.21105/astro.2307.06320} {\  (\bibinfo {year} {2023}),\
  10.21105/astro.2307.06320},\ \Eprint {http://arxiv.org/abs/2307.06320}
  {arXiv:2307.06320 [astro-ph.CO]} \BibitemShut {NoStop}%
\bibitem [{\citenamefont {G{\'o}mez-Valent}\ \emph
  {et~al.}(2024{\natexlab{b}})\citenamefont {G{\'o}mez-Valent}, \citenamefont
  {Favale}, \citenamefont {Migliaccio},\ and\ \citenamefont
  {Sen}}]{Gomez-Valent:2023uof}%
  \BibitemOpen
  \bibfield  {author} {\bibinfo {author} {\bibfnamefont {A.}~\bibnamefont
  {G{\'o}mez-Valent}}, \bibinfo {author} {\bibfnamefont {A.}~\bibnamefont
  {Favale}}, \bibinfo {author} {\bibfnamefont {M.}~\bibnamefont {Migliaccio}},
  \ and\ \bibinfo {author} {\bibfnamefont {A.~A.}\ \bibnamefont {Sen}},\ }\href
  {\doibase 10.1103/PhysRevD.109.023525} {\bibfield  {journal} {\bibinfo
  {journal} {Phys. Rev. D}\ }\textbf {\bibinfo {volume} {109}},\ \bibinfo
  {pages} {023525} (\bibinfo {year} {2024}{\natexlab{b}})},\ \Eprint
  {http://arxiv.org/abs/2309.07795} {arXiv:2309.07795 [astro-ph.CO]}
  \BibitemShut {NoStop}%
\bibitem [{\citenamefont {Akarsu}\ \emph {et~al.}(2024)\citenamefont {Akarsu},
  \citenamefont {Colg{\'a}in}, \citenamefont {Sen},\ and\ \citenamefont
  {Sheikh-Jabbari}}]{Akarsu:2024qiq}%
  \BibitemOpen
  \bibfield  {author} {\bibinfo {author} {\bibfnamefont {{\"O}.}~\bibnamefont
  {Akarsu}}, \bibinfo {author} {\bibfnamefont {E.~{\'O}.}\ \bibnamefont
  {Colg{\'a}in}}, \bibinfo {author} {\bibfnamefont {A.~A.}\ \bibnamefont
  {Sen}}, \ and\ \bibinfo {author} {\bibfnamefont {M.~M.}\ \bibnamefont
  {Sheikh-Jabbari}},\ }\href {\doibase 10.3390/universe10080305} {\bibfield
  {journal} {\bibinfo  {journal} {Universe}\ }\textbf {\bibinfo {volume}
  {10}},\ \bibinfo {pages} {305} (\bibinfo {year} {2024})},\ \Eprint
  {http://arxiv.org/abs/2402.04767} {arXiv:2402.04767 [astro-ph.CO]}
  \BibitemShut {NoStop}%
\bibitem [{\citenamefont {Giar{\`e}}\ \emph
  {et~al.}(2024{\natexlab{b}})\citenamefont {Giar{\`e}}, \citenamefont {Zhai},
  \citenamefont {Pan}, \citenamefont {Di~Valentino}, \citenamefont {Nunes},\
  and\ \citenamefont {van~de Bruck}}]{Giare:2024ytc}%
  \BibitemOpen
  \bibfield  {author} {\bibinfo {author} {\bibfnamefont {W.}~\bibnamefont
  {Giar{\`e}}}, \bibinfo {author} {\bibfnamefont {Y.}~\bibnamefont {Zhai}},
  \bibinfo {author} {\bibfnamefont {S.}~\bibnamefont {Pan}}, \bibinfo {author}
  {\bibfnamefont {E.}~\bibnamefont {Di~Valentino}}, \bibinfo {author}
  {\bibfnamefont {R.~C.}\ \bibnamefont {Nunes}}, \ and\ \bibinfo {author}
  {\bibfnamefont {C.}~\bibnamefont {van~de Bruck}},\ }\href {\doibase
  10.1103/PhysRevD.110.063527} {\bibfield  {journal} {\bibinfo  {journal}
  {Phys. Rev. D}\ }\textbf {\bibinfo {volume} {110}},\ \bibinfo {pages}
  {063527} (\bibinfo {year} {2024}{\natexlab{b}})},\ \Eprint
  {http://arxiv.org/abs/2404.02110} {arXiv:2404.02110 [astro-ph.CO]}
  \BibitemShut {NoStop}%
\bibitem [{\citenamefont {Lynch}\ \emph
  {et~al.}(2024{\natexlab{a}})\citenamefont {Lynch}, \citenamefont {Knox},\
  and\ \citenamefont {Chluba}}]{Lynch:2024gmp}%
  \BibitemOpen
  \bibfield  {author} {\bibinfo {author} {\bibfnamefont {G.~P.}\ \bibnamefont
  {Lynch}}, \bibinfo {author} {\bibfnamefont {L.}~\bibnamefont {Knox}}, \ and\
  \bibinfo {author} {\bibfnamefont {J.}~\bibnamefont {Chluba}},\ }\href
  {\doibase 10.1103/PhysRevD.110.063518} {\bibfield  {journal} {\bibinfo
  {journal} {Phys. Rev. D}\ }\textbf {\bibinfo {volume} {110}},\ \bibinfo
  {pages} {063518} (\bibinfo {year} {2024}{\natexlab{a}})},\ \Eprint
  {http://arxiv.org/abs/2404.05715} {arXiv:2404.05715 [astro-ph.CO]}
  \BibitemShut {NoStop}%
\bibitem [{\citenamefont {Giar{\`e}}\ \emph
  {et~al.}(2024{\natexlab{c}})\citenamefont {Giar{\`e}}, \citenamefont
  {Sabogal}, \citenamefont {Nunes},\ and\ \citenamefont
  {Di~Valentino}}]{Giare:2024smz}%
  \BibitemOpen
  \bibfield  {author} {\bibinfo {author} {\bibfnamefont {W.}~\bibnamefont
  {Giar{\`e}}}, \bibinfo {author} {\bibfnamefont {M.~A.}\ \bibnamefont
  {Sabogal}}, \bibinfo {author} {\bibfnamefont {R.~C.}\ \bibnamefont {Nunes}},
  \ and\ \bibinfo {author} {\bibfnamefont {E.}~\bibnamefont {Di~Valentino}},\
  }\href {\doibase 10.1103/PhysRevLett.133.251003} {\bibfield  {journal}
  {\bibinfo  {journal} {Phys. Rev. Lett.}\ }\textbf {\bibinfo {volume} {133}},\
  \bibinfo {pages} {251003} (\bibinfo {year} {2024}{\natexlab{c}})},\ \Eprint
  {http://arxiv.org/abs/2404.15232} {arXiv:2404.15232 [astro-ph.CO]}
  \BibitemShut {NoStop}%
\bibitem [{\citenamefont {Lynch}\ \emph
  {et~al.}(2024{\natexlab{b}})\citenamefont {Lynch}, \citenamefont {Knox},\
  and\ \citenamefont {Chluba}}]{Lynch:2024hzh}%
  \BibitemOpen
  \bibfield  {author} {\bibinfo {author} {\bibfnamefont {G.~P.}\ \bibnamefont
  {Lynch}}, \bibinfo {author} {\bibfnamefont {L.}~\bibnamefont {Knox}}, \ and\
  \bibinfo {author} {\bibfnamefont {J.}~\bibnamefont {Chluba}},\ }\href
  {\doibase 10.1103/PhysRevD.110.083538} {\bibfield  {journal} {\bibinfo
  {journal} {Phys. Rev. D}\ }\textbf {\bibinfo {volume} {110}},\ \bibinfo
  {pages} {083538} (\bibinfo {year} {2024}{\natexlab{b}})},\ \Eprint
  {http://arxiv.org/abs/2406.10202} {arXiv:2406.10202 [astro-ph.CO]}
  \BibitemShut {NoStop}%
\bibitem [{\citenamefont {Nozari}\ \emph {et~al.}(2024)\citenamefont {Nozari},
  \citenamefont {Saghafi},\ and\ \citenamefont {Hajebrahimi}}]{Nozari:2024wir}%
  \BibitemOpen
  \bibfield  {author} {\bibinfo {author} {\bibfnamefont {K.}~\bibnamefont
  {Nozari}}, \bibinfo {author} {\bibfnamefont {S.}~\bibnamefont {Saghafi}}, \
  and\ \bibinfo {author} {\bibfnamefont {M.}~\bibnamefont {Hajebrahimi}},\
  }\href {\doibase 10.1016/j.dark.2024.101571} {\bibfield  {journal} {\bibinfo
  {journal} {Phys. Dark Univ.}\ }\textbf {\bibinfo {volume} {46}},\ \bibinfo
  {pages} {101571} (\bibinfo {year} {2024})},\ \Eprint
  {http://arxiv.org/abs/2407.01961} {arXiv:2407.01961 [gr-qc]} \BibitemShut
  {NoStop}%
\bibitem [{\citenamefont {Li}\ \emph {et~al.}(2024)\citenamefont {Li},
  \citenamefont {Wu}, \citenamefont {Du}, \citenamefont {Jin}, \citenamefont
  {Li}, \citenamefont {Zhang},\ and\ \citenamefont {Zhang}}]{Li:2024qso}%
  \BibitemOpen
  \bibfield  {author} {\bibinfo {author} {\bibfnamefont {T.-N.}\ \bibnamefont
  {Li}}, \bibinfo {author} {\bibfnamefont {P.-J.}\ \bibnamefont {Wu}}, \bibinfo
  {author} {\bibfnamefont {G.-H.}\ \bibnamefont {Du}}, \bibinfo {author}
  {\bibfnamefont {S.-J.}\ \bibnamefont {Jin}}, \bibinfo {author} {\bibfnamefont
  {H.-L.}\ \bibnamefont {Li}}, \bibinfo {author} {\bibfnamefont {J.-F.}\
  \bibnamefont {Zhang}}, \ and\ \bibinfo {author} {\bibfnamefont
  {X.}~\bibnamefont {Zhang}},\ }\href {\doibase 10.3847/1538-4357/ad87f0}
  {\bibfield  {journal} {\bibinfo  {journal} {Astrophys. J.}\ }\textbf
  {\bibinfo {volume} {976}},\ \bibinfo {pages} {1} (\bibinfo {year} {2024})},\
  \Eprint {http://arxiv.org/abs/2407.14934} {arXiv:2407.14934 [astro-ph.CO]}
  \BibitemShut {NoStop}%
\bibitem [{\citenamefont {Escamilla}\ \emph {et~al.}(2024)\citenamefont
  {Escamilla}, \citenamefont {Fiorucci}, \citenamefont {Montani},\ and\
  \citenamefont {Di~Valentino}}]{Escamilla:2024xmz}%
  \BibitemOpen
  \bibfield  {author} {\bibinfo {author} {\bibfnamefont {L.~A.}\ \bibnamefont
  {Escamilla}}, \bibinfo {author} {\bibfnamefont {D.}~\bibnamefont {Fiorucci}},
  \bibinfo {author} {\bibfnamefont {G.}~\bibnamefont {Montani}}, \ and\
  \bibinfo {author} {\bibfnamefont {E.}~\bibnamefont {Di~Valentino}},\ }\href
  {\doibase 10.1016/j.dark.2024.101652} {\bibfield  {journal} {\bibinfo
  {journal} {Phys. Dark Univ.}\ }\textbf {\bibinfo {volume} {46}},\ \bibinfo
  {pages} {101652} (\bibinfo {year} {2024})},\ \Eprint
  {http://arxiv.org/abs/2408.04354} {arXiv:2408.04354 [astro-ph.CO]}
  \BibitemShut {NoStop}%
\bibitem [{\citenamefont {Roy~Choudhury}\ and\ \citenamefont
  {Okumura}(2024)}]{RoyChoudhury:2024wri}%
  \BibitemOpen
  \bibfield  {author} {\bibinfo {author} {\bibfnamefont {S.}~\bibnamefont
  {Roy~Choudhury}}\ and\ \bibinfo {author} {\bibfnamefont {T.}~\bibnamefont
  {Okumura}},\ }\href {\doibase 10.3847/2041-8213/ad8c26} {\bibfield  {journal}
  {\bibinfo  {journal} {Astrophys. J. Lett.}\ }\textbf {\bibinfo {volume}
  {976}},\ \bibinfo {pages} {L11} (\bibinfo {year} {2024})},\ \Eprint
  {http://arxiv.org/abs/2409.13022} {arXiv:2409.13022 [astro-ph.CO]}
  \BibitemShut {NoStop}%
\bibitem [{\citenamefont {Mirpoorian}\ \emph {et~al.}(2025)\citenamefont
  {Mirpoorian}, \citenamefont {Jedamzik},\ and\ \citenamefont
  {Pogosian}}]{Mirpoorian:2024fka}%
  \BibitemOpen
  \bibfield  {author} {\bibinfo {author} {\bibfnamefont {S.~H.}\ \bibnamefont
  {Mirpoorian}}, \bibinfo {author} {\bibfnamefont {K.}~\bibnamefont
  {Jedamzik}}, \ and\ \bibinfo {author} {\bibfnamefont {L.}~\bibnamefont
  {Pogosian}},\ }\href {\doibase 10.1103/PhysRevD.111.083519} {\bibfield
  {journal} {\bibinfo  {journal} {Phys. Rev. D}\ }\textbf {\bibinfo {volume}
  {111}},\ \bibinfo {pages} {083519} (\bibinfo {year} {2025})},\ \Eprint
  {http://arxiv.org/abs/2411.16678} {arXiv:2411.16678 [astro-ph.CO]}
  \BibitemShut {NoStop}%
\bibitem [{\citenamefont {G{\'o}mez-Valent}\ and\ \citenamefont
  {Sol{\`a}~Peracaula}(2025)}]{Gomez-Valent:2024ejh}%
  \BibitemOpen
  \bibfield  {author} {\bibinfo {author} {\bibfnamefont {A.}~\bibnamefont
  {G{\'o}mez-Valent}}\ and\ \bibinfo {author} {\bibfnamefont {J.}~\bibnamefont
  {Sol{\`a}~Peracaula}},\ }\href {\doibase 10.1016/j.physletb.2025.139391}
  {\bibfield  {journal} {\bibinfo  {journal} {Phys. Lett. B}\ }\textbf
  {\bibinfo {volume} {864}},\ \bibinfo {pages} {139391} (\bibinfo {year}
  {2025})},\ \Eprint {http://arxiv.org/abs/2412.15124} {arXiv:2412.15124
  [astro-ph.CO]} \BibitemShut {NoStop}%
\bibitem [{\citenamefont {Li}\ \emph {et~al.}(2026{\natexlab{a}})\citenamefont
  {Li}, \citenamefont {Du}, \citenamefont {Li}, \citenamefont {Wu},
  \citenamefont {Jin}, \citenamefont {Zhang},\ and\ \citenamefont
  {Zhang}}]{Li:2025owk}%
  \BibitemOpen
  \bibfield  {author} {\bibinfo {author} {\bibfnamefont {T.-N.}\ \bibnamefont
  {Li}}, \bibinfo {author} {\bibfnamefont {G.-H.}\ \bibnamefont {Du}}, \bibinfo
  {author} {\bibfnamefont {Y.-H.}\ \bibnamefont {Li}}, \bibinfo {author}
  {\bibfnamefont {P.-J.}\ \bibnamefont {Wu}}, \bibinfo {author} {\bibfnamefont
  {S.-J.}\ \bibnamefont {Jin}}, \bibinfo {author} {\bibfnamefont {J.-F.}\
  \bibnamefont {Zhang}}, \ and\ \bibinfo {author} {\bibfnamefont
  {X.}~\bibnamefont {Zhang}},\ }\href {\doibase 10.1007/s11433-025-2771-5}
  {\bibfield  {journal} {\bibinfo  {journal} {Sci. China Phys. Mech. Astron.}\
  }\textbf {\bibinfo {volume} {69}},\ \bibinfo {pages} {210413} (\bibinfo
  {year} {2026}{\natexlab{a}})},\ \Eprint {http://arxiv.org/abs/2501.07361}
  {arXiv:2501.07361 [astro-ph.CO]} \BibitemShut {NoStop}%
\bibitem [{\citenamefont {Yang}\ \emph {et~al.}(2025)\citenamefont {Yang},
  \citenamefont {Wang},\ and\ \citenamefont {Dai}}]{Yang:2025vnm}%
  \BibitemOpen
  \bibfield  {author} {\bibinfo {author} {\bibfnamefont {Y.}~\bibnamefont
  {Yang}}, \bibinfo {author} {\bibfnamefont {Y.}~\bibnamefont {Wang}}, \ and\
  \bibinfo {author} {\bibfnamefont {X.}~\bibnamefont {Dai}},\ }\href {\doibase
  10.1140/epjc/s10052-025-13990-9} {\bibfield  {journal} {\bibinfo  {journal}
  {Eur. Phys. J. C}\ }\textbf {\bibinfo {volume} {85}},\ \bibinfo {pages} {224}
  (\bibinfo {year} {2025})},\ \Eprint {http://arxiv.org/abs/2502.17792}
  {arXiv:2502.17792 [astro-ph.CO]} \BibitemShut {NoStop}%
\bibitem [{\citenamefont {Lee}\ \emph {et~al.}(2025)\citenamefont {Lee},
  \citenamefont {Braglia},\ and\ \citenamefont
  {Ali-Ha{\"\i}moud}}]{Lee:2025yah}%
  \BibitemOpen
  \bibfield  {author} {\bibinfo {author} {\bibfnamefont {N.}~\bibnamefont
  {Lee}}, \bibinfo {author} {\bibfnamefont {M.}~\bibnamefont {Braglia}}, \ and\
  \bibinfo {author} {\bibfnamefont {Y.}~\bibnamefont {Ali-Ha{\"\i}moud}},\
  }\href {\doibase 10.1103/9q3f-5zrd} {\bibfield  {journal} {\bibinfo
  {journal} {Phys. Rev. D}\ }\textbf {\bibinfo {volume} {112}},\ \bibinfo
  {pages} {083506} (\bibinfo {year} {2025})},\ \Eprint
  {http://arxiv.org/abs/2504.07966} {arXiv:2504.07966 [astro-ph.CO]}
  \BibitemShut {NoStop}%
\bibitem [{\citenamefont {Wang}\ \emph
  {et~al.}(2026{\natexlab{a}})\citenamefont {Wang}, \citenamefont {Lei},
  \citenamefont {Tang},\ and\ \citenamefont {Fan}}]{Wang:2025dzn}%
  \BibitemOpen
  \bibfield  {author} {\bibinfo {author} {\bibfnamefont {Y.-Y.}\ \bibnamefont
  {Wang}}, \bibinfo {author} {\bibfnamefont {L.}~\bibnamefont {Lei}}, \bibinfo
  {author} {\bibfnamefont {S.-P.}\ \bibnamefont {Tang}}, \ and\ \bibinfo
  {author} {\bibfnamefont {Y.-Z.}\ \bibnamefont {Fan}},\ }\href {\doibase
  10.1088/1475-7516/2026/01/009} {\bibfield  {journal} {\bibinfo  {journal}
  {JCAP}\ }\textbf {\bibinfo {volume} {01}},\ \bibinfo {pages} {009} (\bibinfo
  {year} {2026}{\natexlab{a}})},\ \Eprint {http://arxiv.org/abs/2508.19081}
  {arXiv:2508.19081 [astro-ph.CO]} \BibitemShut {NoStop}%
\bibitem [{\citenamefont {Zhang}\ \emph {et~al.}(2025)\citenamefont {Zhang},
  \citenamefont {Li}, \citenamefont {Du}, \citenamefont {Zhou}, \citenamefont
  {Gao}, \citenamefont {Zhang},\ and\ \citenamefont {Zhang}}]{Zhang:2025dwu}%
  \BibitemOpen
  \bibfield  {author} {\bibinfo {author} {\bibfnamefont {Y.-M.}\ \bibnamefont
  {Zhang}}, \bibinfo {author} {\bibfnamefont {T.-N.}\ \bibnamefont {Li}},
  \bibinfo {author} {\bibfnamefont {G.-H.}\ \bibnamefont {Du}}, \bibinfo
  {author} {\bibfnamefont {S.-H.}\ \bibnamefont {Zhou}}, \bibinfo {author}
  {\bibfnamefont {L.-Y.}\ \bibnamefont {Gao}}, \bibinfo {author} {\bibfnamefont
  {J.-F.}\ \bibnamefont {Zhang}}, \ and\ \bibinfo {author} {\bibfnamefont
  {X.}~\bibnamefont {Zhang}},\ }\href@noop {} {\  (\bibinfo {year} {2025})},\
  \Eprint {http://arxiv.org/abs/2510.12627} {arXiv:2510.12627 [astro-ph.CO]}
  \BibitemShut {NoStop}%
\bibitem [{\citenamefont {Chen}\ \emph {et~al.}(2025)\citenamefont {Chen},
  \citenamefont {Xu}, \citenamefont {Li},\ and\ \citenamefont
  {Han}}]{Chen:2025nfy}%
  \BibitemOpen
  \bibfield  {author} {\bibinfo {author} {\bibfnamefont {L.}~\bibnamefont
  {Chen}}, \bibinfo {author} {\bibfnamefont {P.}~\bibnamefont {Xu}}, \bibinfo
  {author} {\bibfnamefont {G.}~\bibnamefont {Li}}, \ and\ \bibinfo {author}
  {\bibfnamefont {Y.}~\bibnamefont {Han}},\ }\href@noop {} {\  (\bibinfo {year}
  {2025})},\ \Eprint {http://arxiv.org/abs/2511.15087} {arXiv:2511.15087
  [astro-ph.CO]} \BibitemShut {NoStop}%
\bibitem [{\citenamefont {Kumar}(2026)}]{Kumar:2025obb}%
  \BibitemOpen
  \bibfield  {author} {\bibinfo {author} {\bibfnamefont {S.}~\bibnamefont
  {Kumar}},\ }\href {\doibase 10.1016/j.dark.2026.102248} {\bibfield  {journal}
  {\bibinfo  {journal} {Phys. Dark Univ.}\ }\textbf {\bibinfo {volume} {52}},\
  \bibinfo {pages} {102248} (\bibinfo {year} {2026})},\ \Eprint
  {http://arxiv.org/abs/2512.19000} {arXiv:2512.19000 [astro-ph.CO]}
  \BibitemShut {NoStop}%
\bibitem [{\citenamefont {H{\"o}g{\r{a}}s}\ \emph {et~al.}(2025)\citenamefont
  {H{\"o}g{\r{a}}s}, \citenamefont {M{\"o}rtsell}, \citenamefont {Desmond},\
  and\ \citenamefont {Riess}}]{Hogas:2025mii}%
  \BibitemOpen
  \bibfield  {author} {\bibinfo {author} {\bibfnamefont {M.}~\bibnamefont
  {H{\"o}g{\r{a}}s}}, \bibinfo {author} {\bibfnamefont {E.}~\bibnamefont
  {M{\"o}rtsell}}, \bibinfo {author} {\bibfnamefont {H.}~\bibnamefont
  {Desmond}}, \ and\ \bibinfo {author} {\bibfnamefont {A.}~\bibnamefont
  {Riess}},\ }\href@noop {} {\  (\bibinfo {year} {2025})},\ \Eprint
  {http://arxiv.org/abs/2512.14814} {arXiv:2512.14814 [astro-ph.CO]}
  \BibitemShut {NoStop}%
\bibitem [{\citenamefont {Li}\ \emph {et~al.}(2026{\natexlab{b}})\citenamefont
  {Li}, \citenamefont {Giar{\`e}}, \citenamefont {Du}, \citenamefont {Li},
  \citenamefont {Di~Valentino}, \citenamefont {Zhang},\ and\ \citenamefont
  {Zhang}}]{Li:2026xaz}%
  \BibitemOpen
  \bibfield  {author} {\bibinfo {author} {\bibfnamefont {T.-N.}\ \bibnamefont
  {Li}}, \bibinfo {author} {\bibfnamefont {W.}~\bibnamefont {Giar{\`e}}},
  \bibinfo {author} {\bibfnamefont {G.-H.}\ \bibnamefont {Du}}, \bibinfo
  {author} {\bibfnamefont {Y.-H.}\ \bibnamefont {Li}}, \bibinfo {author}
  {\bibfnamefont {E.}~\bibnamefont {Di~Valentino}}, \bibinfo {author}
  {\bibfnamefont {J.-F.}\ \bibnamefont {Zhang}}, \ and\ \bibinfo {author}
  {\bibfnamefont {X.}~\bibnamefont {Zhang}},\ }\href@noop {} {\  (\bibinfo
  {year} {2026}{\natexlab{b}})},\ \Eprint {http://arxiv.org/abs/2601.07361}
  {arXiv:2601.07361 [astro-ph.CO]} \BibitemShut {NoStop}%
\bibitem [{\citenamefont {Bouhmadi-L{\'o}pez}\ \emph
  {et~al.}(2026)\citenamefont {Bouhmadi-L{\'o}pez}, \citenamefont {Boiza},
  \citenamefont {Petronikolou},\ and\ \citenamefont
  {Saridakis}}]{Bouhmadi-Lopez:2026dte}%
  \BibitemOpen
  \bibfield  {author} {\bibinfo {author} {\bibfnamefont {M.}~\bibnamefont
  {Bouhmadi-L{\'o}pez}}, \bibinfo {author} {\bibfnamefont {C.~G.}\ \bibnamefont
  {Boiza}}, \bibinfo {author} {\bibfnamefont {M.}~\bibnamefont {Petronikolou}},
  \ and\ \bibinfo {author} {\bibfnamefont {E.~N.}\ \bibnamefont {Saridakis}},\
  }\href {\doibase 10.3390/universe12030081} {\bibfield  {journal} {\bibinfo
  {journal} {Universe}\ }\textbf {\bibinfo {volume} {12}},\ \bibinfo {pages}
  {81} (\bibinfo {year} {2026})},\ \Eprint {http://arxiv.org/abs/2601.22225}
  {arXiv:2601.22225 [gr-qc]} \BibitemShut {NoStop}%
\bibitem [{\citenamefont {Dai}\ \emph {et~al.}(2026)\citenamefont {Dai},
  \citenamefont {Yang}, \citenamefont {Wang}, \citenamefont {Qu}, \citenamefont
  {Yi},\ and\ \citenamefont {Wang}}]{Dai:2026pvx}%
  \BibitemOpen
  \bibfield  {author} {\bibinfo {author} {\bibfnamefont {X.}~\bibnamefont
  {Dai}}, \bibinfo {author} {\bibfnamefont {Y.}~\bibnamefont {Yang}}, \bibinfo
  {author} {\bibfnamefont {Y.}~\bibnamefont {Wang}}, \bibinfo {author}
  {\bibfnamefont {Y.}~\bibnamefont {Qu}}, \bibinfo {author} {\bibfnamefont
  {S.}~\bibnamefont {Yi}}, \ and\ \bibinfo {author} {\bibfnamefont
  {F.}~\bibnamefont {Wang}},\ }\href {\doibase 10.1103/zg3l-yt32} {\bibfield
  {journal} {\bibinfo  {journal} {Phys. Rev. D}\ }\textbf {\bibinfo {volume}
  {113}},\ \bibinfo {pages} {063514} (\bibinfo {year} {2026})},\ \Eprint
  {http://arxiv.org/abs/2602.22840} {arXiv:2602.22840 [astro-ph.CO]}
  \BibitemShut {NoStop}%
\bibitem [{\citenamefont {Wang}\ \emph
  {et~al.}(2026{\natexlab{b}})\citenamefont {Wang}, \citenamefont {Yang},
  \citenamefont {Dai}, \citenamefont {Yi}, \citenamefont {Qu},\ and\
  \citenamefont {Wang}}]{Wang:2026rkx}%
  \BibitemOpen
  \bibfield  {author} {\bibinfo {author} {\bibfnamefont {Y.}~\bibnamefont
  {Wang}}, \bibinfo {author} {\bibfnamefont {Y.}~\bibnamefont {Yang}}, \bibinfo
  {author} {\bibfnamefont {X.}~\bibnamefont {Dai}}, \bibinfo {author}
  {\bibfnamefont {S.}~\bibnamefont {Yi}}, \bibinfo {author} {\bibfnamefont
  {Y.}~\bibnamefont {Qu}}, \ and\ \bibinfo {author} {\bibfnamefont
  {F.}~\bibnamefont {Wang}},\ }\href {\doibase 10.1103/k1nl-rxsy} {\bibfield
  {journal} {\bibinfo  {journal} {Phys. Rev. D}\ }\textbf {\bibinfo {volume}
  {113}},\ \bibinfo {pages} {063551} (\bibinfo {year} {2026}{\natexlab{b}})},\
  \Eprint {http://arxiv.org/abs/2603.25999} {arXiv:2603.25999 [astro-ph.CO]}
  \BibitemShut {NoStop}%
\bibitem [{\citenamefont {Mukhopadhyay}\ \emph {et~al.}(2026)\citenamefont
  {Mukhopadhyay}, \citenamefont {Mukherjee},\ and\ \citenamefont
  {Tkatchenko}}]{Mukhopadhyay:2026fyk}%
  \BibitemOpen
  \bibfield  {author} {\bibinfo {author} {\bibfnamefont {U.}~\bibnamefont
  {Mukhopadhyay}}, \bibinfo {author} {\bibfnamefont {P.}~\bibnamefont
  {Mukherjee}}, \ and\ \bibinfo {author} {\bibfnamefont {A.}~\bibnamefont
  {Tkatchenko}},\ }\href@noop {} {\  (\bibinfo {year} {2026})},\ \Eprint
  {http://arxiv.org/abs/2603.28391} {arXiv:2603.28391 [astro-ph.CO]}
  \BibitemShut {NoStop}%
\bibitem [{\citenamefont {Pantos}\ and\ \citenamefont
  {Perivolaropoulos}(2026)}]{Pantos:2026rpe}%
  \BibitemOpen
  \bibfield  {author} {\bibinfo {author} {\bibfnamefont {I.}~\bibnamefont
  {Pantos}}\ and\ \bibinfo {author} {\bibfnamefont {L.}~\bibnamefont
  {Perivolaropoulos}},\ }\href {\doibase 10.1016/j.dark.2026.102347} {\bibfield
   {journal} {\bibinfo  {journal} {Phys. Dark Univ.}\ }\textbf {\bibinfo
  {volume} {52}},\ \bibinfo {pages} {102347} (\bibinfo {year} {2026})},\
  \Eprint {http://arxiv.org/abs/2604.11106} {arXiv:2604.11106 [astro-ph.CO]}
  \BibitemShut {NoStop}%
\bibitem [{\citenamefont {Solanki}(2026)}]{Solanki:2026tpi}%
  \BibitemOpen
  \bibfield  {author} {\bibinfo {author} {\bibfnamefont {R.}~\bibnamefont
  {Solanki}},\ }\href {\doibase 10.1016/j.physletb.2026.140413} {\bibfield
  {journal} {\bibinfo  {journal} {Phys. Lett. B}\ }\textbf {\bibinfo {volume}
  {876}},\ \bibinfo {pages} {140413} (\bibinfo {year} {2026})},\ \Eprint
  {http://arxiv.org/abs/2604.11821} {arXiv:2604.11821 [gr-qc]} \BibitemShut
  {NoStop}%
\bibitem [{\citenamefont {Wang}\ and\ \citenamefont
  {Wang}(2026)}]{Wang:2026kor}%
  \BibitemOpen
  \bibfield  {author} {\bibinfo {author} {\bibfnamefont {J.-Q.}\ \bibnamefont
  {Wang}}\ and\ \bibinfo {author} {\bibfnamefont {S.-J.}\ \bibnamefont
  {Wang}},\ }in\ \href@noop {} {\emph {\bibinfo {booktitle} {{25th Hellenic
  School and Workshops on Elementary Particle Physics and Gravity}}}}\
  (\bibinfo {year} {2026})\ \Eprint {http://arxiv.org/abs/2604.28013}
  {arXiv:2604.28013 [astro-ph.CO]} \BibitemShut {NoStop}%
\bibitem [{\citenamefont {Jusufi}\ \emph {et~al.}(2026)\citenamefont {Jusufi},
  \citenamefont {Khodahami}, \citenamefont {Sheykhi}, \citenamefont
  {Levi~Said},\ and\ \citenamefont {Saridakis}}]{Jusufi:2026rfd}%
  \BibitemOpen
  \bibfield  {author} {\bibinfo {author} {\bibfnamefont {K.}~\bibnamefont
  {Jusufi}}, \bibinfo {author} {\bibfnamefont {A.~A.}\ \bibnamefont
  {Khodahami}}, \bibinfo {author} {\bibfnamefont {A.}~\bibnamefont {Sheykhi}},
  \bibinfo {author} {\bibfnamefont {J.}~\bibnamefont {Levi~Said}}, \ and\
  \bibinfo {author} {\bibfnamefont {E.~N.}\ \bibnamefont {Saridakis}},\
  }\href@noop {} {\  (\bibinfo {year} {2026})},\ \Eprint
  {http://arxiv.org/abs/2605.14977} {arXiv:2605.14977 [gr-qc]} \BibitemShut
  {NoStop}%
\bibitem [{\citenamefont {Dhyani}\ \emph {et~al.}(2026)\citenamefont {Dhyani},
  \citenamefont {Mukherjee}, \citenamefont {Chatterjee},\ and\ \citenamefont
  {Sen}}]{Dhyani:2026trw}%
  \BibitemOpen
  \bibfield  {author} {\bibinfo {author} {\bibfnamefont {R.}~\bibnamefont
  {Dhyani}}, \bibinfo {author} {\bibfnamefont {P.}~\bibnamefont {Mukherjee}},
  \bibinfo {author} {\bibfnamefont {A.}~\bibnamefont {Chatterjee}}, \ and\
  \bibinfo {author} {\bibfnamefont {A.~A.}\ \bibnamefont {Sen}},\ }\href@noop
  {} {\  (\bibinfo {year} {2026})},\ \Eprint {http://arxiv.org/abs/2605.26050}
  {arXiv:2605.26050 [astro-ph.CO]} \BibitemShut {NoStop}%
\bibitem [{\citenamefont {Li}\ \emph {et~al.}(2026{\natexlab{c}})\citenamefont
  {Li}, \citenamefont {Liu}, \citenamefont {Li}, \citenamefont {Du},
  \citenamefont {Shafieloo},\ and\ \citenamefont {Biesiada}}]{Li:2026hwq}%
  \BibitemOpen
  \bibfield  {author} {\bibinfo {author} {\bibfnamefont {X.}~\bibnamefont
  {Li}}, \bibinfo {author} {\bibfnamefont {T.}~\bibnamefont {Liu}}, \bibinfo
  {author} {\bibfnamefont {T.-N.}\ \bibnamefont {Li}}, \bibinfo {author}
  {\bibfnamefont {G.-H.}\ \bibnamefont {Du}}, \bibinfo {author} {\bibfnamefont
  {A.}~\bibnamefont {Shafieloo}}, \ and\ \bibinfo {author} {\bibfnamefont
  {M.}~\bibnamefont {Biesiada}},\ }\href {\doibase 10.3847/2041-8213/ae5a36}
  {\bibfield  {journal} {\bibinfo  {journal} {Astrophys. J. Lett.}\ }\textbf
  {\bibinfo {volume} {1001}},\ \bibinfo {pages} {L21} (\bibinfo {year}
  {2026}{\natexlab{c}})},\ \Eprint {http://arxiv.org/abs/2606.05723}
  {arXiv:2606.05723 [astro-ph.CO]} \BibitemShut {NoStop}%
\bibitem [{\citenamefont {Akarsu}\ \emph {et~al.}(2026)\citenamefont {Akarsu},
  \citenamefont {Perivolaropoulos}, \citenamefont {Y{\"u}kselci},\ and\
  \citenamefont {Zhuk}}]{Akarsu:2026lva}%
  \BibitemOpen
  \bibfield  {author} {\bibinfo {author} {\bibfnamefont {{\"O}.}~\bibnamefont
  {Akarsu}}, \bibinfo {author} {\bibfnamefont {L.}~\bibnamefont
  {Perivolaropoulos}}, \bibinfo {author} {\bibfnamefont {A.~E.}\ \bibnamefont
  {Y{\"u}kselci}}, \ and\ \bibinfo {author} {\bibfnamefont {A.}~\bibnamefont
  {Zhuk}},\ }\href@noop {} {\  (\bibinfo {year} {2026})},\ \Eprint
  {http://arxiv.org/abs/2606.11062} {arXiv:2606.11062 [gr-qc]} \BibitemShut
  {NoStop}%
\bibitem [{\citenamefont {Jia}\ \emph {et~al.}(2026)\citenamefont {Jia},
  \citenamefont {Dai}, \citenamefont {Yang},\ and\ \citenamefont
  {Wang}}]{Jia:2026vdt}%
  \BibitemOpen
  \bibfield  {author} {\bibinfo {author} {\bibfnamefont {X.-D.}\ \bibnamefont
  {Jia}}, \bibinfo {author} {\bibfnamefont {X.-Y.}\ \bibnamefont {Dai}},
  \bibinfo {author} {\bibfnamefont {Y.-P.}\ \bibnamefont {Yang}}, \ and\
  \bibinfo {author} {\bibfnamefont {F.-Y.}\ \bibnamefont {Wang}},\ }\href
  {\doibase 10.3390/galaxies14030055} {\bibfield  {journal} {\bibinfo
  {journal} {Galaxies}\ }\textbf {\bibinfo {volume} {14}},\ \bibinfo {pages}
  {55} (\bibinfo {year} {2026})},\ \Eprint {http://arxiv.org/abs/2606.12980}
  {arXiv:2606.12980 [astro-ph.CO]} \BibitemShut {NoStop}%
\bibitem [{\citenamefont {Li}\ \emph {et~al.}(2026{\natexlab{d}})\citenamefont
  {Li}, \citenamefont {Du}, \citenamefont {Wang}, \citenamefont {Li},
  \citenamefont {Zhang},\ and\ \citenamefont {Zhang}}]{Li:2026asg}%
  \BibitemOpen
  \bibfield  {author} {\bibinfo {author} {\bibfnamefont {T.-N.}\ \bibnamefont
  {Li}}, \bibinfo {author} {\bibfnamefont {G.-H.}\ \bibnamefont {Du}}, \bibinfo
  {author} {\bibfnamefont {H.}~\bibnamefont {Wang}}, \bibinfo {author}
  {\bibfnamefont {Y.-H.}\ \bibnamefont {Li}}, \bibinfo {author} {\bibfnamefont
  {J.-F.}\ \bibnamefont {Zhang}}, \ and\ \bibinfo {author} {\bibfnamefont
  {X.}~\bibnamefont {Zhang}}\ }(\bibinfo {year} {2026})\ \Eprint
  {http://arxiv.org/abs/2606.21826} {arXiv:2606.21826 [astro-ph.CO]}
  \BibitemShut {NoStop}%
\bibitem [{\citenamefont {Efstathiou}(2020)}]{Efstathiou:2020wxn}%
  \BibitemOpen
  \bibfield  {author} {\bibinfo {author} {\bibfnamefont {G.}~\bibnamefont
  {Efstathiou}},\ }\href@noop {} {\  (\bibinfo {year} {2020})},\ \Eprint
  {http://arxiv.org/abs/2007.10716} {arXiv:2007.10716 [astro-ph.CO]}
  \BibitemShut {NoStop}%
\bibitem [{\citenamefont {Wojtak}\ and\ \citenamefont
  {Hjorth}(2022)}]{Wojtak:2022bct}%
  \BibitemOpen
  \bibfield  {author} {\bibinfo {author} {\bibfnamefont {R.}~\bibnamefont
  {Wojtak}}\ and\ \bibinfo {author} {\bibfnamefont {J.}~\bibnamefont
  {Hjorth}},\ }\href {\doibase 10.1093/mnras/stac1878} {\bibfield  {journal}
  {\bibinfo  {journal} {Mon. Not. Roy. Astron. Soc.}\ }\textbf {\bibinfo
  {volume} {515}},\ \bibinfo {pages} {2790} (\bibinfo {year} {2022})},\ \Eprint
  {http://arxiv.org/abs/2206.08160} {arXiv:2206.08160 [astro-ph.CO]}
  \BibitemShut {NoStop}%
\bibitem [{\citenamefont {Giani}\ \emph {et~al.}(2024)\citenamefont {Giani},
  \citenamefont {Howlett}, \citenamefont {Said}, \citenamefont {Davis},\ and\
  \citenamefont {Vagnozzi}}]{Giani:2023aor}%
  \BibitemOpen
  \bibfield  {author} {\bibinfo {author} {\bibfnamefont {L.}~\bibnamefont
  {Giani}}, \bibinfo {author} {\bibfnamefont {C.}~\bibnamefont {Howlett}},
  \bibinfo {author} {\bibfnamefont {K.}~\bibnamefont {Said}}, \bibinfo {author}
  {\bibfnamefont {T.}~\bibnamefont {Davis}}, \ and\ \bibinfo {author}
  {\bibfnamefont {S.}~\bibnamefont {Vagnozzi}},\ }\href {\doibase
  10.1088/1475-7516/2024/01/071} {\bibfield  {journal} {\bibinfo  {journal}
  {JCAP}\ }\textbf {\bibinfo {volume} {01}},\ \bibinfo {pages} {071} (\bibinfo
  {year} {2024})},\ \Eprint {http://arxiv.org/abs/2311.00215} {arXiv:2311.00215
  [astro-ph.CO]} \BibitemShut {NoStop}%
\bibitem [{\citenamefont {Perivolaropoulos}(2024)}]{Perivolaropoulos:2024yxv}%
  \BibitemOpen
  \bibfield  {author} {\bibinfo {author} {\bibfnamefont {L.}~\bibnamefont
  {Perivolaropoulos}},\ }\href {\doibase 10.1103/PhysRevD.110.123518}
  {\bibfield  {journal} {\bibinfo  {journal} {Phys. Rev. D}\ }\textbf {\bibinfo
  {volume} {110}},\ \bibinfo {pages} {123518} (\bibinfo {year} {2024})},\
  \Eprint {http://arxiv.org/abs/2408.11031} {arXiv:2408.11031 [astro-ph.CO]}
  \BibitemShut {NoStop}%
\bibitem [{\citenamefont {H{\"o}g{\r{a}}s}\ and\ \citenamefont
  {M{\"o}rtsell}(2026)}]{Hogas:2026urs}%
  \BibitemOpen
  \bibfield  {author} {\bibinfo {author} {\bibfnamefont {M.}~\bibnamefont
  {H{\"o}g{\r{a}}s}}\ and\ \bibinfo {author} {\bibfnamefont {E.}~\bibnamefont
  {M{\"o}rtsell}},\ }\href {\doibase 10.1093/mnras/stag724} {\  (\bibinfo
  {year} {2026}),\ 10.1093/mnras/stag724},\ \Eprint
  {http://arxiv.org/abs/2601.22215} {arXiv:2601.22215 [astro-ph.CO]}
  \BibitemShut {NoStop}%
\bibitem [{\citenamefont {Bernal}\ \emph {et~al.}(2016)\citenamefont {Bernal},
  \citenamefont {Verde},\ and\ \citenamefont {Riess}}]{Bernal:2016gxb}%
  \BibitemOpen
  \bibfield  {author} {\bibinfo {author} {\bibfnamefont {J.~L.}\ \bibnamefont
  {Bernal}}, \bibinfo {author} {\bibfnamefont {L.}~\bibnamefont {Verde}}, \
  and\ \bibinfo {author} {\bibfnamefont {A.~G.}\ \bibnamefont {Riess}},\ }\href
  {\doibase 10.1088/1475-7516/2016/10/019} {\bibfield  {journal} {\bibinfo
  {journal} {JCAP}\ }\textbf {\bibinfo {volume} {10}},\ \bibinfo {pages} {019}
  (\bibinfo {year} {2016})},\ \Eprint {http://arxiv.org/abs/1607.05617}
  {arXiv:1607.05617 [astro-ph.CO]} \BibitemShut {NoStop}%
\bibitem [{\citenamefont {Addison}\ \emph {et~al.}(2018)\citenamefont
  {Addison}, \citenamefont {Watts}, \citenamefont {Bennett}, \citenamefont
  {Halpern}, \citenamefont {Hinshaw},\ and\ \citenamefont
  {Weiland}}]{Addison:2017fdm}%
  \BibitemOpen
  \bibfield  {author} {\bibinfo {author} {\bibfnamefont {G.~E.}\ \bibnamefont
  {Addison}}, \bibinfo {author} {\bibfnamefont {D.~J.}\ \bibnamefont {Watts}},
  \bibinfo {author} {\bibfnamefont {C.~L.}\ \bibnamefont {Bennett}}, \bibinfo
  {author} {\bibfnamefont {M.}~\bibnamefont {Halpern}}, \bibinfo {author}
  {\bibfnamefont {G.}~\bibnamefont {Hinshaw}}, \ and\ \bibinfo {author}
  {\bibfnamefont {J.~L.}\ \bibnamefont {Weiland}},\ }\href {\doibase
  10.3847/1538-4357/aaa1ed} {\bibfield  {journal} {\bibinfo  {journal}
  {Astrophys. J.}\ }\textbf {\bibinfo {volume} {853}},\ \bibinfo {pages} {119}
  (\bibinfo {year} {2018})},\ \Eprint {http://arxiv.org/abs/1707.06547}
  {arXiv:1707.06547 [astro-ph.CO]} \BibitemShut {NoStop}%
\bibitem [{\citenamefont {Lemos}\ \emph {et~al.}(2019)\citenamefont {Lemos},
  \citenamefont {Lee}, \citenamefont {Efstathiou},\ and\ \citenamefont
  {Gratton}}]{Lemos:2018smw}%
  \BibitemOpen
  \bibfield  {author} {\bibinfo {author} {\bibfnamefont {P.}~\bibnamefont
  {Lemos}}, \bibinfo {author} {\bibfnamefont {E.}~\bibnamefont {Lee}}, \bibinfo
  {author} {\bibfnamefont {G.}~\bibnamefont {Efstathiou}}, \ and\ \bibinfo
  {author} {\bibfnamefont {S.}~\bibnamefont {Gratton}},\ }\href {\doibase
  10.1093/mnras/sty3082} {\bibfield  {journal} {\bibinfo  {journal} {Mon. Not.
  Roy. Astron. Soc.}\ }\textbf {\bibinfo {volume} {483}},\ \bibinfo {pages}
  {4803} (\bibinfo {year} {2019})},\ \Eprint {http://arxiv.org/abs/1806.06781}
  {arXiv:1806.06781 [astro-ph.CO]} \BibitemShut {NoStop}%
\bibitem [{\citenamefont {Aylor}\ \emph {et~al.}(2019)\citenamefont {Aylor},
  \citenamefont {Joy}, \citenamefont {Knox}, \citenamefont {Millea},
  \citenamefont {Raghunathan},\ and\ \citenamefont {Wu}}]{Aylor:2018drw}%
  \BibitemOpen
  \bibfield  {author} {\bibinfo {author} {\bibfnamefont {K.}~\bibnamefont
  {Aylor}}, \bibinfo {author} {\bibfnamefont {M.}~\bibnamefont {Joy}}, \bibinfo
  {author} {\bibfnamefont {L.}~\bibnamefont {Knox}}, \bibinfo {author}
  {\bibfnamefont {M.}~\bibnamefont {Millea}}, \bibinfo {author} {\bibfnamefont
  {S.}~\bibnamefont {Raghunathan}}, \ and\ \bibinfo {author} {\bibfnamefont
  {W.~L.~K.}\ \bibnamefont {Wu}},\ }\href {\doibase 10.3847/1538-4357/ab0898}
  {\bibfield  {journal} {\bibinfo  {journal} {Astrophys. J.}\ }\textbf
  {\bibinfo {volume} {874}},\ \bibinfo {pages} {4} (\bibinfo {year} {2019})},\
  \Eprint {http://arxiv.org/abs/1811.00537} {arXiv:1811.00537 [astro-ph.CO]}
  \BibitemShut {NoStop}%
\bibitem [{\citenamefont {Sch\"oneberg}\ \emph {et~al.}(2019)\citenamefont
  {Sch\"oneberg}, \citenamefont {Lesgourgues},\ and\ \citenamefont
  {Hooper}}]{Schoneberg:2019wmt}%
  \BibitemOpen
  \bibfield  {author} {\bibinfo {author} {\bibfnamefont {N.}~\bibnamefont
  {Sch\"oneberg}}, \bibinfo {author} {\bibfnamefont {J.}~\bibnamefont
  {Lesgourgues}}, \ and\ \bibinfo {author} {\bibfnamefont {D.~C.}\ \bibnamefont
  {Hooper}},\ }\href {\doibase 10.1088/1475-7516/2019/10/029} {\bibfield
  {journal} {\bibinfo  {journal} {JCAP}\ }\textbf {\bibinfo {volume} {10}},\
  \bibinfo {pages} {029} (\bibinfo {year} {2019})},\ \Eprint
  {http://arxiv.org/abs/1907.11594} {arXiv:1907.11594 [astro-ph.CO]}
  \BibitemShut {NoStop}%
\bibitem [{\citenamefont {Knox}\ and\ \citenamefont
  {Millea}(2020)}]{Knox:2019rjx}%
  \BibitemOpen
  \bibfield  {author} {\bibinfo {author} {\bibfnamefont {L.}~\bibnamefont
  {Knox}}\ and\ \bibinfo {author} {\bibfnamefont {M.}~\bibnamefont {Millea}},\
  }\href {\doibase 10.1103/PhysRevD.101.043533} {\bibfield  {journal} {\bibinfo
   {journal} {Phys. Rev. D}\ }\textbf {\bibinfo {volume} {101}},\ \bibinfo
  {pages} {043533} (\bibinfo {year} {2020})},\ \Eprint
  {http://arxiv.org/abs/1908.03663} {arXiv:1908.03663 [astro-ph.CO]}
  \BibitemShut {NoStop}%
\bibitem [{\citenamefont {Arendse}\ \emph {et~al.}(2020)\citenamefont {Arendse}
  \emph {et~al.}}]{Arendse:2019hev}%
  \BibitemOpen
  \bibfield  {author} {\bibinfo {author} {\bibfnamefont {N.}~\bibnamefont
  {Arendse}} \emph {et~al.},\ }\href {\doibase 10.1051/0004-6361/201936720}
  {\bibfield  {journal} {\bibinfo  {journal} {Astron. Astrophys.}\ }\textbf
  {\bibinfo {volume} {639}},\ \bibinfo {pages} {A57} (\bibinfo {year}
  {2020})},\ \Eprint {http://arxiv.org/abs/1909.07986} {arXiv:1909.07986
  [astro-ph.CO]} \BibitemShut {NoStop}%
\bibitem [{\citenamefont {Efstathiou}(2021)}]{Efstathiou:2021ocp}%
  \BibitemOpen
  \bibfield  {author} {\bibinfo {author} {\bibfnamefont {G.}~\bibnamefont
  {Efstathiou}},\ }\href {\doibase 10.1093/mnras/stab1588} {\bibfield
  {journal} {\bibinfo  {journal} {Mon. Not. Roy. Astron. Soc.}\ }\textbf
  {\bibinfo {volume} {505}},\ \bibinfo {pages} {3866} (\bibinfo {year}
  {2021})},\ \Eprint {http://arxiv.org/abs/2103.08723} {arXiv:2103.08723
  [astro-ph.CO]} \BibitemShut {NoStop}%
\bibitem [{\citenamefont {Cai}\ \emph {et~al.}(2022)\citenamefont {Cai},
  \citenamefont {Guo}, \citenamefont {Wang}, \citenamefont {Yu},\ and\
  \citenamefont {Zhou}}]{Cai:2021weh}%
  \BibitemOpen
  \bibfield  {author} {\bibinfo {author} {\bibfnamefont {R.-G.}\ \bibnamefont
  {Cai}}, \bibinfo {author} {\bibfnamefont {Z.-K.}\ \bibnamefont {Guo}},
  \bibinfo {author} {\bibfnamefont {S.-J.}\ \bibnamefont {Wang}}, \bibinfo
  {author} {\bibfnamefont {W.-W.}\ \bibnamefont {Yu}}, \ and\ \bibinfo {author}
  {\bibfnamefont {Y.}~\bibnamefont {Zhou}},\ }\href {\doibase
  10.1103/PhysRevD.105.L021301} {\bibfield  {journal} {\bibinfo  {journal}
  {Phys. Rev. D}\ }\textbf {\bibinfo {volume} {105}},\ \bibinfo {pages}
  {L021301} (\bibinfo {year} {2022})},\ \Eprint
  {http://arxiv.org/abs/2107.13286} {arXiv:2107.13286 [astro-ph.CO]}
  \BibitemShut {NoStop}%
\bibitem [{\citenamefont {Keeley}\ and\ \citenamefont
  {Shafieloo}(2023)}]{Keeley:2022ojz}%
  \BibitemOpen
  \bibfield  {author} {\bibinfo {author} {\bibfnamefont {R.~E.}\ \bibnamefont
  {Keeley}}\ and\ \bibinfo {author} {\bibfnamefont {A.}~\bibnamefont
  {Shafieloo}},\ }\href {\doibase 10.1103/PhysRevLett.131.111002} {\bibfield
  {journal} {\bibinfo  {journal} {Phys. Rev. Lett.}\ }\textbf {\bibinfo
  {volume} {131}},\ \bibinfo {pages} {111002} (\bibinfo {year} {2023})},\
  \Eprint {http://arxiv.org/abs/2206.08440} {arXiv:2206.08440 [astro-ph.CO]}
  \BibitemShut {NoStop}%
\bibitem [{\citenamefont {K.}\ and\ \citenamefont
  {Sreenath}(2026)}]{K:2025juc}%
  \BibitemOpen
  \bibfield  {author} {\bibinfo {author} {\bibfnamefont {R.}~\bibnamefont
  {K.}}\ and\ \bibinfo {author} {\bibfnamefont {V.}~\bibnamefont {Sreenath}},\
  }\href {\doibase 10.1103/ksdc-pb3h} {\bibfield  {journal} {\bibinfo
  {journal} {Phys. Rev. D}\ }\textbf {\bibinfo {volume} {113}},\ \bibinfo
  {pages} {023548} (\bibinfo {year} {2026})},\ \Eprint
  {http://arxiv.org/abs/2504.21386} {arXiv:2504.21386 [astro-ph.CO]}
  \BibitemShut {NoStop}%
\bibitem [{\citenamefont {Chluba}\ \emph {et~al.}(2012)\citenamefont {Chluba},
  \citenamefont {Erickcek},\ and\ \citenamefont {Ben-Dayan}}]{Chluba:2012we}%
  \BibitemOpen
  \bibfield  {author} {\bibinfo {author} {\bibfnamefont {J.}~\bibnamefont
  {Chluba}}, \bibinfo {author} {\bibfnamefont {A.~L.}\ \bibnamefont
  {Erickcek}}, \ and\ \bibinfo {author} {\bibfnamefont {I.}~\bibnamefont
  {Ben-Dayan}},\ }\href {\doibase 10.1088/0004-637X/758/2/76} {\bibfield
  {journal} {\bibinfo  {journal} {Astrophys. J.}\ }\textbf {\bibinfo {volume}
  {758}},\ \bibinfo {pages} {76} (\bibinfo {year} {2012})},\ \Eprint
  {http://arxiv.org/abs/1203.2681} {arXiv:1203.2681 [astro-ph.CO]} \BibitemShut
  {NoStop}%
\bibitem [{\citenamefont {Chluba}\ and\ \citenamefont
  {Grin}(2013)}]{Chluba:2013dna}%
  \BibitemOpen
  \bibfield  {author} {\bibinfo {author} {\bibfnamefont {J.}~\bibnamefont
  {Chluba}}\ and\ \bibinfo {author} {\bibfnamefont {D.}~\bibnamefont {Grin}},\
  }\href {\doibase 10.1093/mnras/stt1129} {\bibfield  {journal} {\bibinfo
  {journal} {Mon. Not. Roy. Astron. Soc.}\ }\textbf {\bibinfo {volume} {434}},\
  \bibinfo {pages} {1619} (\bibinfo {year} {2013})},\ \Eprint
  {http://arxiv.org/abs/1304.4596} {arXiv:1304.4596 [astro-ph.CO]} \BibitemShut
  {NoStop}%
\bibitem [{\citenamefont {Frolovsky}\ and\ \citenamefont
  {Ketov}(2025)}]{Frolovsky:2025iao}%
  \BibitemOpen
  \bibfield  {author} {\bibinfo {author} {\bibfnamefont {D.}~\bibnamefont
  {Frolovsky}}\ and\ \bibinfo {author} {\bibfnamefont {S.~V.}\ \bibnamefont
  {Ketov}},\ }\href {\doibase 10.1142/S0217732325501822} {\bibfield  {journal}
  {\bibinfo  {journal} {Mod. Phys. Lett. A}\ }\textbf {\bibinfo {volume}
  {40}},\ \bibinfo {pages} {2550182} (\bibinfo {year} {2025})},\ \Eprint
  {http://arxiv.org/abs/2505.17514} {arXiv:2505.17514 [astro-ph.CO]}
  \BibitemShut {NoStop}%
\bibitem [{\citenamefont {Louis}\ \emph {et~al.}(2025)\citenamefont {Louis}
  \emph {et~al.}}]{AtacamaCosmologyTelescope:2025blo}%
  \BibitemOpen
  \bibfield  {author} {\bibinfo {author} {\bibfnamefont {T.}~\bibnamefont
  {Louis}} \emph {et~al.} (\bibinfo {collaboration} {Atacama Cosmology
  Telescope}),\ }\href {\doibase 10.1088/1475-7516/2025/11/062} {\bibfield
  {journal} {\bibinfo  {journal} {JCAP}\ }\textbf {\bibinfo {volume} {11}},\
  \bibinfo {pages} {062} (\bibinfo {year} {2025})},\ \Eprint
  {http://arxiv.org/abs/2503.14452} {arXiv:2503.14452 [astro-ph.CO]}
  \BibitemShut {NoStop}%
\bibitem [{\citenamefont {Jiang}\ \emph
  {et~al.}(2024{\natexlab{b}})\citenamefont {Jiang}, \citenamefont {Pedrotti},
  \citenamefont {da~Costa},\ and\ \citenamefont {Vagnozzi}}]{Jiang:2024xnu}%
  \BibitemOpen
  \bibfield  {author} {\bibinfo {author} {\bibfnamefont {J.-Q.}\ \bibnamefont
  {Jiang}}, \bibinfo {author} {\bibfnamefont {D.}~\bibnamefont {Pedrotti}},
  \bibinfo {author} {\bibfnamefont {S.~S.}\ \bibnamefont {da~Costa}}, \ and\
  \bibinfo {author} {\bibfnamefont {S.}~\bibnamefont {Vagnozzi}},\ }\href
  {\doibase 10.1103/PhysRevD.110.123519} {\bibfield  {journal} {\bibinfo
  {journal} {Phys. Rev. D}\ }\textbf {\bibinfo {volume} {110}},\ \bibinfo
  {pages} {123519} (\bibinfo {year} {2024}{\natexlab{b}})},\ \Eprint
  {http://arxiv.org/abs/2408.02365} {arXiv:2408.02365 [astro-ph.CO]}
  \BibitemShut {NoStop}%
\bibitem [{\citenamefont {Benevento}\ \emph {et~al.}(2020)\citenamefont
  {Benevento}, \citenamefont {Hu},\ and\ \citenamefont
  {Raveri}}]{Benevento:2020fev}%
  \BibitemOpen
  \bibfield  {author} {\bibinfo {author} {\bibfnamefont {G.}~\bibnamefont
  {Benevento}}, \bibinfo {author} {\bibfnamefont {W.}~\bibnamefont {Hu}}, \
  and\ \bibinfo {author} {\bibfnamefont {M.}~\bibnamefont {Raveri}},\ }\href
  {\doibase 10.1103/PhysRevD.101.103517} {\bibfield  {journal} {\bibinfo
  {journal} {Phys. Rev. D}\ }\textbf {\bibinfo {volume} {101}},\ \bibinfo
  {pages} {103517} (\bibinfo {year} {2020})},\ \Eprint
  {http://arxiv.org/abs/2002.11707} {arXiv:2002.11707 [astro-ph.CO]}
  \BibitemShut {NoStop}%
\bibitem [{\citenamefont {Teixeira}\ \emph {et~al.}(2025)\citenamefont
  {Teixeira}, \citenamefont {Giar{\`e}}, \citenamefont {Hogg}, \citenamefont
  {Montandon}, \citenamefont {Poudou},\ and\ \citenamefont
  {Poulin}}]{Teixeira:2025czm}%
  \BibitemOpen
  \bibfield  {author} {\bibinfo {author} {\bibfnamefont {E.~M.}\ \bibnamefont
  {Teixeira}}, \bibinfo {author} {\bibfnamefont {W.}~\bibnamefont {Giar{\`e}}},
  \bibinfo {author} {\bibfnamefont {N.~B.}\ \bibnamefont {Hogg}}, \bibinfo
  {author} {\bibfnamefont {T.}~\bibnamefont {Montandon}}, \bibinfo {author}
  {\bibfnamefont {A.}~\bibnamefont {Poudou}}, \ and\ \bibinfo {author}
  {\bibfnamefont {V.}~\bibnamefont {Poulin}},\ }\href {\doibase
  10.1103/zzmp-rxrh} {\bibfield  {journal} {\bibinfo  {journal} {Phys. Rev. D}\
  }\textbf {\bibinfo {volume} {112}},\ \bibinfo {pages} {023515} (\bibinfo
  {year} {2025})},\ \Eprint {http://arxiv.org/abs/2504.10464} {arXiv:2504.10464
  [astro-ph.CO]} \BibitemShut {NoStop}%
\bibitem [{\citenamefont {Zhou}\ \emph
  {et~al.}(2025{\natexlab{a}})\citenamefont {Zhou}, \citenamefont {Miao},
  \citenamefont {Bi}, \citenamefont {Ai},\ and\ \citenamefont
  {Zhang}}]{Zhou:2025kws}%
  \BibitemOpen
  \bibfield  {author} {\bibinfo {author} {\bibfnamefont {Z.}~\bibnamefont
  {Zhou}}, \bibinfo {author} {\bibfnamefont {Z.}~\bibnamefont {Miao}}, \bibinfo
  {author} {\bibfnamefont {S.}~\bibnamefont {Bi}}, \bibinfo {author}
  {\bibfnamefont {C.}~\bibnamefont {Ai}}, \ and\ \bibinfo {author}
  {\bibfnamefont {H.}~\bibnamefont {Zhang}},\ }\href {\doibase
  10.1103/bc81-3fj5} {\bibfield  {journal} {\bibinfo  {journal} {Phys. Rev. D}\
  }\textbf {\bibinfo {volume} {112}},\ \bibinfo {pages} {103502} (\bibinfo
  {year} {2025}{\natexlab{a}})},\ \Eprint {http://arxiv.org/abs/2506.23556}
  {arXiv:2506.23556 [astro-ph.CO]} \BibitemShut {NoStop}%
\bibitem [{\citenamefont {Pedrotti}\ \emph {et~al.}(2026)\citenamefont
  {Pedrotti}, \citenamefont {Escamilla}, \citenamefont {Marra}, \citenamefont
  {Perivolaropoulos},\ and\ \citenamefont {Vagnozzi}}]{Pedrotti:2025ccw}%
  \BibitemOpen
  \bibfield  {author} {\bibinfo {author} {\bibfnamefont {D.}~\bibnamefont
  {Pedrotti}}, \bibinfo {author} {\bibfnamefont {L.~A.}\ \bibnamefont
  {Escamilla}}, \bibinfo {author} {\bibfnamefont {V.}~\bibnamefont {Marra}},
  \bibinfo {author} {\bibfnamefont {L.}~\bibnamefont {Perivolaropoulos}}, \
  and\ \bibinfo {author} {\bibfnamefont {S.}~\bibnamefont {Vagnozzi}},\ }\href
  {\doibase 10.1103/pn9j-8whx} {\bibfield  {journal} {\bibinfo  {journal}
  {Phys. Rev. D}\ }\textbf {\bibinfo {volume} {113}},\ \bibinfo {pages}
  {043507} (\bibinfo {year} {2026})},\ \Eprint
  {http://arxiv.org/abs/2510.01974} {arXiv:2510.01974 [astro-ph.CO]}
  \BibitemShut {NoStop}%
\bibitem [{\citenamefont {Zhou}\ \emph
  {et~al.}(2025{\natexlab{b}})\citenamefont {Zhou}, \citenamefont {Miao},
  \citenamefont {Zhang}, \citenamefont {Yang}, \citenamefont {Fu},\ and\
  \citenamefont {Ai}}]{Zhou:2025dxo}%
  \BibitemOpen
  \bibfield  {author} {\bibinfo {author} {\bibfnamefont {Z.}~\bibnamefont
  {Zhou}}, \bibinfo {author} {\bibfnamefont {Z.}~\bibnamefont {Miao}}, \bibinfo
  {author} {\bibfnamefont {R.}~\bibnamefont {Zhang}}, \bibinfo {author}
  {\bibfnamefont {H.}~\bibnamefont {Yang}}, \bibinfo {author} {\bibfnamefont
  {P.}~\bibnamefont {Fu}}, \ and\ \bibinfo {author} {\bibfnamefont
  {C.}~\bibnamefont {Ai}},\ }\href@noop {} {\  (\bibinfo {year}
  {2025}{\natexlab{b}})},\ \Eprint {http://arxiv.org/abs/2511.02357}
  {arXiv:2511.02357 [astro-ph.CO]} \BibitemShut {NoStop}%
\bibitem [{\citenamefont {Bansal}\ and\ \citenamefont
  {Huterer}(2026)}]{Bansal:2026axl}%
  \BibitemOpen
  \bibfield  {author} {\bibinfo {author} {\bibfnamefont {P.}~\bibnamefont
  {Bansal}}\ and\ \bibinfo {author} {\bibfnamefont {D.}~\bibnamefont
  {Huterer}},\ }\href {\doibase 10.1103/ydnj-myzb} {\bibfield  {journal}
  {\bibinfo  {journal} {Phys. Rev. D}\ }\textbf {\bibinfo {volume} {113}},\
  \bibinfo {pages} {103539} (\bibinfo {year} {2026})},\ \Eprint
  {http://arxiv.org/abs/2602.06293} {arXiv:2602.06293 [astro-ph.CO]}
  \BibitemShut {NoStop}%
\bibitem [{\citenamefont {Zhou}\ \emph {et~al.}(2026)\citenamefont {Zhou},
  \citenamefont {Miao}, \citenamefont {Bi}, \citenamefont {Ai},\ and\
  \citenamefont {Zhang}}]{Zhou:2026iar}%
  \BibitemOpen
  \bibfield  {author} {\bibinfo {author} {\bibfnamefont {Z.}~\bibnamefont
  {Zhou}}, \bibinfo {author} {\bibfnamefont {Z.}~\bibnamefont {Miao}}, \bibinfo
  {author} {\bibfnamefont {S.}~\bibnamefont {Bi}}, \bibinfo {author}
  {\bibfnamefont {C.}~\bibnamefont {Ai}}, \ and\ \bibinfo {author}
  {\bibfnamefont {H.}~\bibnamefont {Zhang}},\ }\href@noop {} {\  (\bibinfo
  {year} {2026})},\ \Eprint {http://arxiv.org/abs/2606.12822} {arXiv:2606.12822
  [astro-ph.CO]} \BibitemShut {NoStop}%
\bibitem [{\citenamefont {Toda}\ \emph {et~al.}(2024)\citenamefont {Toda},
  \citenamefont {Giar{\`e}}, \citenamefont {{\"O}z{\"u}lker}, \citenamefont
  {Di~Valentino},\ and\ \citenamefont {Vagnozzi}}]{Toda:2024ncp}%
  \BibitemOpen
  \bibfield  {author} {\bibinfo {author} {\bibfnamefont {Y.}~\bibnamefont
  {Toda}}, \bibinfo {author} {\bibfnamefont {W.}~\bibnamefont {Giar{\`e}}},
  \bibinfo {author} {\bibfnamefont {E.}~\bibnamefont {{\"O}z{\"u}lker}},
  \bibinfo {author} {\bibfnamefont {E.}~\bibnamefont {Di~Valentino}}, \ and\
  \bibinfo {author} {\bibfnamefont {S.}~\bibnamefont {Vagnozzi}},\ }\href
  {\doibase 10.1016/j.dark.2024.101676} {\bibfield  {journal} {\bibinfo
  {journal} {Phys. Dark Univ.}\ }\textbf {\bibinfo {volume} {46}},\ \bibinfo
  {pages} {101676} (\bibinfo {year} {2024})},\ \Eprint
  {http://arxiv.org/abs/2407.01173} {arXiv:2407.01173 [astro-ph.CO]}
  \BibitemShut {NoStop}%
\bibitem [{\citenamefont {Pedrotti}(2026)}]{Pedrotti:2026dwj}%
  \BibitemOpen
  \bibfield  {author} {\bibinfo {author} {\bibfnamefont {D.}~\bibnamefont
  {Pedrotti}},\ }\href@noop {} {\  (\bibinfo {year} {2026})},\ \Eprint
  {http://arxiv.org/abs/2604.25813} {arXiv:2604.25813 [astro-ph.CO]}
  \BibitemShut {NoStop}%
\bibitem [{\citenamefont {Vagnozzi}(2021{\natexlab{b}})}]{Vagnozzi:2021gjh}%
  \BibitemOpen
  \bibfield  {author} {\bibinfo {author} {\bibfnamefont {S.}~\bibnamefont
  {Vagnozzi}},\ }\href {\doibase 10.1103/PhysRevD.104.063524} {\bibfield
  {journal} {\bibinfo  {journal} {Phys. Rev. D}\ }\textbf {\bibinfo {volume}
  {104}},\ \bibinfo {pages} {063524} (\bibinfo {year} {2021}{\natexlab{b}})},\
  \Eprint {http://arxiv.org/abs/2105.10425} {arXiv:2105.10425 [astro-ph.CO]}
  \BibitemShut {NoStop}%
\bibitem [{\citenamefont {Vagnozzi}\ \emph {et~al.}(2022)\citenamefont
  {Vagnozzi}, \citenamefont {Pacucci},\ and\ \citenamefont
  {Loeb}}]{Vagnozzi:2021tjv}%
  \BibitemOpen
  \bibfield  {author} {\bibinfo {author} {\bibfnamefont {S.}~\bibnamefont
  {Vagnozzi}}, \bibinfo {author} {\bibfnamefont {F.}~\bibnamefont {Pacucci}}, \
  and\ \bibinfo {author} {\bibfnamefont {A.}~\bibnamefont {Loeb}},\ }\href
  {\doibase 10.1016/j.jheap.2022.07.004} {\bibfield  {journal} {\bibinfo
  {journal} {JHEAp}\ }\textbf {\bibinfo {volume} {36}},\ \bibinfo {pages} {27}
  (\bibinfo {year} {2022})},\ \Eprint {http://arxiv.org/abs/2105.10421}
  {arXiv:2105.10421 [astro-ph.CO]} \BibitemShut {NoStop}%
\bibitem [{\citenamefont {Poulin}\ \emph {et~al.}(2025)\citenamefont {Poulin},
  \citenamefont {Smith}, \citenamefont {Calder{\'o}n},\ and\ \citenamefont
  {Simon}}]{Poulin:2024ken}%
  \BibitemOpen
  \bibfield  {author} {\bibinfo {author} {\bibfnamefont {V.}~\bibnamefont
  {Poulin}}, \bibinfo {author} {\bibfnamefont {T.~L.}\ \bibnamefont {Smith}},
  \bibinfo {author} {\bibfnamefont {R.}~\bibnamefont {Calder{\'o}n}}, \ and\
  \bibinfo {author} {\bibfnamefont {T.}~\bibnamefont {Simon}},\ }\href
  {\doibase 10.1103/PhysRevD.111.083552} {\bibfield  {journal} {\bibinfo
  {journal} {Phys. Rev. D}\ }\textbf {\bibinfo {volume} {111}},\ \bibinfo
  {pages} {083552} (\bibinfo {year} {2025})},\ \Eprint
  {http://arxiv.org/abs/2407.18292} {arXiv:2407.18292 [astro-ph.CO]}
  \BibitemShut {NoStop}%
\bibitem [{\citenamefont {Pedrotti}\ \emph {et~al.}(2025)\citenamefont
  {Pedrotti}, \citenamefont {Jiang}, \citenamefont {Escamilla}, \citenamefont
  {da~Costa},\ and\ \citenamefont {Vagnozzi}}]{Pedrotti:2024kpn}%
  \BibitemOpen
  \bibfield  {author} {\bibinfo {author} {\bibfnamefont {D.}~\bibnamefont
  {Pedrotti}}, \bibinfo {author} {\bibfnamefont {J.-Q.}\ \bibnamefont {Jiang}},
  \bibinfo {author} {\bibfnamefont {L.~A.}\ \bibnamefont {Escamilla}}, \bibinfo
  {author} {\bibfnamefont {S.~S.}\ \bibnamefont {da~Costa}}, \ and\ \bibinfo
  {author} {\bibfnamefont {S.}~\bibnamefont {Vagnozzi}},\ }\href {\doibase
  10.1103/PhysRevD.111.023506} {\bibfield  {journal} {\bibinfo  {journal}
  {Phys. Rev. D}\ }\textbf {\bibinfo {volume} {111}},\ \bibinfo {pages}
  {023506} (\bibinfo {year} {2025})},\ \Eprint
  {http://arxiv.org/abs/2408.04530} {arXiv:2408.04530 [astro-ph.CO]}
  \BibitemShut {NoStop}%
\bibitem [{\citenamefont {Giovanetti}(2026)}]{Giovanetti:2026aku}%
  \BibitemOpen
  \bibfield  {author} {\bibinfo {author} {\bibfnamefont {C.}~\bibnamefont
  {Giovanetti}},\ }\href@noop {} {\  (\bibinfo {year} {2026})},\ \Eprint
  {http://arxiv.org/abs/2604.05095} {arXiv:2604.05095 [astro-ph.CO]}
  \BibitemShut {NoStop}%
\bibitem [{\citenamefont {Kamionkowski}\ and\ \citenamefont
  {Riess}(2023)}]{Kamionkowski:2022pkx}%
  \BibitemOpen
  \bibfield  {author} {\bibinfo {author} {\bibfnamefont {M.}~\bibnamefont
  {Kamionkowski}}\ and\ \bibinfo {author} {\bibfnamefont {A.~G.}\ \bibnamefont
  {Riess}},\ }\href {\doibase 10.1146/annurev-nucl-111422-024107} {\bibfield
  {journal} {\bibinfo  {journal} {Ann. Rev. Nucl. Part. Sci.}\ }\textbf
  {\bibinfo {volume} {73}},\ \bibinfo {pages} {153} (\bibinfo {year} {2023})},\
  \Eprint {http://arxiv.org/abs/2211.04492} {arXiv:2211.04492 [astro-ph.CO]}
  \BibitemShut {NoStop}%
\bibitem [{\citenamefont {Poulin}\ \emph {et~al.}(2023)\citenamefont {Poulin},
  \citenamefont {Smith},\ and\ \citenamefont {Karwal}}]{Poulin:2023lkg}%
  \BibitemOpen
  \bibfield  {author} {\bibinfo {author} {\bibfnamefont {V.}~\bibnamefont
  {Poulin}}, \bibinfo {author} {\bibfnamefont {T.~L.}\ \bibnamefont {Smith}}, \
  and\ \bibinfo {author} {\bibfnamefont {T.}~\bibnamefont {Karwal}},\ }\href
  {\doibase 10.1016/j.dark.2023.101348} {\bibfield  {journal} {\bibinfo
  {journal} {Phys. Dark Univ.}\ }\textbf {\bibinfo {volume} {42}},\ \bibinfo
  {pages} {101348} (\bibinfo {year} {2023})},\ \Eprint
  {http://arxiv.org/abs/2302.09032} {arXiv:2302.09032 [astro-ph.CO]}
  \BibitemShut {NoStop}%
\bibitem [{\citenamefont {McDonough}\ \emph {et~al.}(2024)\citenamefont
  {McDonough}, \citenamefont {Hill}, \citenamefont {Ivanov}, \citenamefont
  {La~Posta},\ and\ \citenamefont {Toomey}}]{McDonough:2023qcu}%
  \BibitemOpen
  \bibfield  {author} {\bibinfo {author} {\bibfnamefont {E.}~\bibnamefont
  {McDonough}}, \bibinfo {author} {\bibfnamefont {J.~C.}\ \bibnamefont {Hill}},
  \bibinfo {author} {\bibfnamefont {M.~M.}\ \bibnamefont {Ivanov}}, \bibinfo
  {author} {\bibfnamefont {A.}~\bibnamefont {La~Posta}}, \ and\ \bibinfo
  {author} {\bibfnamefont {M.~W.}\ \bibnamefont {Toomey}},\ }\href {\doibase
  10.1142/S0218271824300039} {\bibfield  {journal} {\bibinfo  {journal} {Int.
  J. Mod. Phys. D}\ }\textbf {\bibinfo {volume} {33}},\ \bibinfo {pages}
  {2430003} (\bibinfo {year} {2024})},\ \Eprint
  {http://arxiv.org/abs/2310.19899} {arXiv:2310.19899 [astro-ph.CO]}
  \BibitemShut {NoStop}%
\bibitem [{\citenamefont {Karwal}\ and\ \citenamefont
  {Kamionkowski}(2016)}]{Karwal:2016vyq}%
  \BibitemOpen
  \bibfield  {author} {\bibinfo {author} {\bibfnamefont {T.}~\bibnamefont
  {Karwal}}\ and\ \bibinfo {author} {\bibfnamefont {M.}~\bibnamefont
  {Kamionkowski}},\ }\href {\doibase 10.1103/PhysRevD.94.103523} {\bibfield
  {journal} {\bibinfo  {journal} {Phys. Rev. D}\ }\textbf {\bibinfo {volume}
  {94}},\ \bibinfo {pages} {103523} (\bibinfo {year} {2016})},\ \Eprint
  {http://arxiv.org/abs/1608.01309} {arXiv:1608.01309 [astro-ph.CO]}
  \BibitemShut {NoStop}%
\bibitem [{\citenamefont {Poulin}\ \emph {et~al.}(2018)\citenamefont {Poulin},
  \citenamefont {Smith}, \citenamefont {Grin}, \citenamefont {Karwal},\ and\
  \citenamefont {Kamionkowski}}]{Poulin:2018dzj}%
  \BibitemOpen
  \bibfield  {author} {\bibinfo {author} {\bibfnamefont {V.}~\bibnamefont
  {Poulin}}, \bibinfo {author} {\bibfnamefont {T.~L.}\ \bibnamefont {Smith}},
  \bibinfo {author} {\bibfnamefont {D.}~\bibnamefont {Grin}}, \bibinfo {author}
  {\bibfnamefont {T.}~\bibnamefont {Karwal}}, \ and\ \bibinfo {author}
  {\bibfnamefont {M.}~\bibnamefont {Kamionkowski}},\ }\href {\doibase
  10.1103/PhysRevD.98.083525} {\bibfield  {journal} {\bibinfo  {journal} {Phys.
  Rev. D}\ }\textbf {\bibinfo {volume} {98}},\ \bibinfo {pages} {083525}
  (\bibinfo {year} {2018})},\ \Eprint {http://arxiv.org/abs/1806.10608}
  {arXiv:1806.10608 [astro-ph.CO]} \BibitemShut {NoStop}%
\bibitem [{\citenamefont {Poulin}\ \emph {et~al.}(2019)\citenamefont {Poulin},
  \citenamefont {Smith}, \citenamefont {Karwal},\ and\ \citenamefont
  {Kamionkowski}}]{Poulin:2018cxd}%
  \BibitemOpen
  \bibfield  {author} {\bibinfo {author} {\bibfnamefont {V.}~\bibnamefont
  {Poulin}}, \bibinfo {author} {\bibfnamefont {T.~L.}\ \bibnamefont {Smith}},
  \bibinfo {author} {\bibfnamefont {T.}~\bibnamefont {Karwal}}, \ and\ \bibinfo
  {author} {\bibfnamefont {M.}~\bibnamefont {Kamionkowski}},\ }\href {\doibase
  10.1103/PhysRevLett.122.221301} {\bibfield  {journal} {\bibinfo  {journal}
  {Phys. Rev. Lett.}\ }\textbf {\bibinfo {volume} {122}},\ \bibinfo {pages}
  {221301} (\bibinfo {year} {2019})},\ \Eprint
  {http://arxiv.org/abs/1811.04083} {arXiv:1811.04083 [astro-ph.CO]}
  \BibitemShut {NoStop}%
\bibitem [{\citenamefont {Agrawal}\ \emph {et~al.}(2023)\citenamefont
  {Agrawal}, \citenamefont {Cyr-Racine}, \citenamefont {Pinner},\ and\
  \citenamefont {Randall}}]{Agrawal:2019lmo}%
  \BibitemOpen
  \bibfield  {author} {\bibinfo {author} {\bibfnamefont {P.}~\bibnamefont
  {Agrawal}}, \bibinfo {author} {\bibfnamefont {F.-Y.}\ \bibnamefont
  {Cyr-Racine}}, \bibinfo {author} {\bibfnamefont {D.}~\bibnamefont {Pinner}},
  \ and\ \bibinfo {author} {\bibfnamefont {L.}~\bibnamefont {Randall}},\ }\href
  {\doibase 10.1016/j.dark.2023.101347} {\bibfield  {journal} {\bibinfo
  {journal} {Phys. Dark Univ.}\ }\textbf {\bibinfo {volume} {42}},\ \bibinfo
  {pages} {101347} (\bibinfo {year} {2023})},\ \Eprint
  {http://arxiv.org/abs/1904.01016} {arXiv:1904.01016 [astro-ph.CO]}
  \BibitemShut {NoStop}%
\bibitem [{\citenamefont {Lin}\ \emph {et~al.}(2019)\citenamefont {Lin},
  \citenamefont {Benevento}, \citenamefont {Hu},\ and\ \citenamefont
  {Raveri}}]{Lin:2019qug}%
  \BibitemOpen
  \bibfield  {author} {\bibinfo {author} {\bibfnamefont {M.-X.}\ \bibnamefont
  {Lin}}, \bibinfo {author} {\bibfnamefont {G.}~\bibnamefont {Benevento}},
  \bibinfo {author} {\bibfnamefont {W.}~\bibnamefont {Hu}}, \ and\ \bibinfo
  {author} {\bibfnamefont {M.}~\bibnamefont {Raveri}},\ }\href {\doibase
  10.1103/PhysRevD.100.063542} {\bibfield  {journal} {\bibinfo  {journal}
  {Phys. Rev. D}\ }\textbf {\bibinfo {volume} {100}},\ \bibinfo {pages}
  {063542} (\bibinfo {year} {2019})},\ \Eprint
  {http://arxiv.org/abs/1905.12618} {arXiv:1905.12618 [astro-ph.CO]}
  \BibitemShut {NoStop}%
\bibitem [{\citenamefont {Smith}\ \emph {et~al.}(2020)\citenamefont {Smith},
  \citenamefont {Poulin},\ and\ \citenamefont {Amin}}]{Smith:2019ihp}%
  \BibitemOpen
  \bibfield  {author} {\bibinfo {author} {\bibfnamefont {T.~L.}\ \bibnamefont
  {Smith}}, \bibinfo {author} {\bibfnamefont {V.}~\bibnamefont {Poulin}}, \
  and\ \bibinfo {author} {\bibfnamefont {M.~A.}\ \bibnamefont {Amin}},\ }\href
  {\doibase 10.1103/PhysRevD.101.063523} {\bibfield  {journal} {\bibinfo
  {journal} {Phys. Rev. D}\ }\textbf {\bibinfo {volume} {101}},\ \bibinfo
  {pages} {063523} (\bibinfo {year} {2020})},\ \Eprint
  {http://arxiv.org/abs/1908.06995} {arXiv:1908.06995 [astro-ph.CO]}
  \BibitemShut {NoStop}%
\bibitem [{\citenamefont {Niedermann}\ and\ \citenamefont
  {Sloth}(2021{\natexlab{a}})}]{Niedermann:2019olb}%
  \BibitemOpen
  \bibfield  {author} {\bibinfo {author} {\bibfnamefont {F.}~\bibnamefont
  {Niedermann}}\ and\ \bibinfo {author} {\bibfnamefont {M.~S.}\ \bibnamefont
  {Sloth}},\ }\href {\doibase 10.1103/PhysRevD.103.L041303} {\bibfield
  {journal} {\bibinfo  {journal} {Phys. Rev. D}\ }\textbf {\bibinfo {volume}
  {103}},\ \bibinfo {pages} {L041303} (\bibinfo {year} {2021}{\natexlab{a}})},\
  \Eprint {http://arxiv.org/abs/1910.10739} {arXiv:1910.10739 [astro-ph.CO]}
  \BibitemShut {NoStop}%
\bibitem [{\citenamefont {Berghaus}\ and\ \citenamefont
  {Karwal}(2020)}]{Berghaus:2019cls}%
  \BibitemOpen
  \bibfield  {author} {\bibinfo {author} {\bibfnamefont {K.~V.}\ \bibnamefont
  {Berghaus}}\ and\ \bibinfo {author} {\bibfnamefont {T.}~\bibnamefont
  {Karwal}},\ }\href {\doibase 10.1103/PhysRevD.101.083537} {\bibfield
  {journal} {\bibinfo  {journal} {Phys. Rev. D}\ }\textbf {\bibinfo {volume}
  {101}},\ \bibinfo {pages} {083537} (\bibinfo {year} {2020})},\ \Eprint
  {http://arxiv.org/abs/1911.06281} {arXiv:1911.06281 [astro-ph.CO]}
  \BibitemShut {NoStop}%
\bibitem [{\citenamefont {Sakstein}\ and\ \citenamefont
  {Trodden}(2020)}]{Sakstein:2019fmf}%
  \BibitemOpen
  \bibfield  {author} {\bibinfo {author} {\bibfnamefont {J.}~\bibnamefont
  {Sakstein}}\ and\ \bibinfo {author} {\bibfnamefont {M.}~\bibnamefont
  {Trodden}},\ }\href {\doibase 10.1103/PhysRevLett.124.161301} {\bibfield
  {journal} {\bibinfo  {journal} {Phys. Rev. Lett.}\ }\textbf {\bibinfo
  {volume} {124}},\ \bibinfo {pages} {161301} (\bibinfo {year} {2020})},\
  \Eprint {http://arxiv.org/abs/1911.11760} {arXiv:1911.11760 [astro-ph.CO]}
  \BibitemShut {NoStop}%
\bibitem [{\citenamefont {Nojiri}\ \emph {et~al.}(2020)\citenamefont {Nojiri},
  \citenamefont {Odintsov},\ and\ \citenamefont {Oikonomou}}]{Nojiri:2019fft}%
  \BibitemOpen
  \bibfield  {author} {\bibinfo {author} {\bibfnamefont {S.}~\bibnamefont
  {Nojiri}}, \bibinfo {author} {\bibfnamefont {S.~D.}\ \bibnamefont
  {Odintsov}}, \ and\ \bibinfo {author} {\bibfnamefont {V.~K.}\ \bibnamefont
  {Oikonomou}},\ }\href {\doibase 10.1016/j.dark.2020.100602} {\bibfield
  {journal} {\bibinfo  {journal} {Phys. Dark Univ.}\ }\textbf {\bibinfo
  {volume} {29}},\ \bibinfo {pages} {100602} (\bibinfo {year} {2020})},\
  \Eprint {http://arxiv.org/abs/1912.13128} {arXiv:1912.13128 [gr-qc]}
  \BibitemShut {NoStop}%
\bibitem [{\citenamefont {Ye}\ and\ \citenamefont
  {Piao}(2020{\natexlab{a}})}]{Ye:2020btb}%
  \BibitemOpen
  \bibfield  {author} {\bibinfo {author} {\bibfnamefont {G.}~\bibnamefont
  {Ye}}\ and\ \bibinfo {author} {\bibfnamefont {Y.-S.}\ \bibnamefont {Piao}},\
  }\href {\doibase 10.1103/PhysRevD.101.083507} {\bibfield  {journal} {\bibinfo
   {journal} {Phys. Rev. D}\ }\textbf {\bibinfo {volume} {101}},\ \bibinfo
  {pages} {083507} (\bibinfo {year} {2020}{\natexlab{a}})},\ \Eprint
  {http://arxiv.org/abs/2001.02451} {arXiv:2001.02451 [astro-ph.CO]}
  \BibitemShut {NoStop}%
\bibitem [{\citenamefont {Zumalacarregui}(2020)}]{Zumalacarregui:2020cjh}%
  \BibitemOpen
  \bibfield  {author} {\bibinfo {author} {\bibfnamefont {M.}~\bibnamefont
  {Zumalacarregui}},\ }\href {\doibase 10.1103/PhysRevD.102.023523} {\bibfield
  {journal} {\bibinfo  {journal} {Phys. Rev. D}\ }\textbf {\bibinfo {volume}
  {102}},\ \bibinfo {pages} {023523} (\bibinfo {year} {2020})},\ \Eprint
  {http://arxiv.org/abs/2003.06396} {arXiv:2003.06396 [astro-ph.CO]}
  \BibitemShut {NoStop}%
\bibitem [{\citenamefont {Ballesteros}\ \emph
  {et~al.}(2020{\natexlab{b}})\citenamefont {Ballesteros}, \citenamefont
  {Notari},\ and\ \citenamefont {Rompineve}}]{Ballesteros:2020sik}%
  \BibitemOpen
  \bibfield  {author} {\bibinfo {author} {\bibfnamefont {G.}~\bibnamefont
  {Ballesteros}}, \bibinfo {author} {\bibfnamefont {A.}~\bibnamefont {Notari}},
  \ and\ \bibinfo {author} {\bibfnamefont {F.}~\bibnamefont {Rompineve}},\
  }\href {\doibase 10.1088/1475-7516/2020/11/024} {\bibfield  {journal}
  {\bibinfo  {journal} {JCAP}\ }\textbf {\bibinfo {volume} {11}},\ \bibinfo
  {pages} {024} (\bibinfo {year} {2020}{\natexlab{b}})},\ \Eprint
  {http://arxiv.org/abs/2004.05049} {arXiv:2004.05049 [astro-ph.CO]}
  \BibitemShut {NoStop}%
\bibitem [{\citenamefont {Braglia}\ \emph
  {et~al.}(2020{\natexlab{a}})\citenamefont {Braglia}, \citenamefont
  {Ballardini}, \citenamefont {Emond}, \citenamefont {Finelli}, \citenamefont
  {Gumrukcuoglu}, \citenamefont {Koyama},\ and\ \citenamefont
  {Paoletti}}]{Braglia:2020iik}%
  \BibitemOpen
  \bibfield  {author} {\bibinfo {author} {\bibfnamefont {M.}~\bibnamefont
  {Braglia}}, \bibinfo {author} {\bibfnamefont {M.}~\bibnamefont {Ballardini}},
  \bibinfo {author} {\bibfnamefont {W.~T.}\ \bibnamefont {Emond}}, \bibinfo
  {author} {\bibfnamefont {F.}~\bibnamefont {Finelli}}, \bibinfo {author}
  {\bibfnamefont {A.~E.}\ \bibnamefont {Gumrukcuoglu}}, \bibinfo {author}
  {\bibfnamefont {K.}~\bibnamefont {Koyama}}, \ and\ \bibinfo {author}
  {\bibfnamefont {D.}~\bibnamefont {Paoletti}},\ }\href {\doibase
  10.1103/PhysRevD.102.023529} {\bibfield  {journal} {\bibinfo  {journal}
  {Phys. Rev. D}\ }\textbf {\bibinfo {volume} {102}},\ \bibinfo {pages}
  {023529} (\bibinfo {year} {2020}{\natexlab{a}})},\ \Eprint
  {http://arxiv.org/abs/2004.11161} {arXiv:2004.11161 [astro-ph.CO]}
  \BibitemShut {NoStop}%
\bibitem [{\citenamefont {Ballardini}\ \emph {et~al.}(2020)\citenamefont
  {Ballardini}, \citenamefont {Braglia}, \citenamefont {Finelli}, \citenamefont
  {Paoletti}, \citenamefont {Starobinsky},\ and\ \citenamefont
  {Umilt{\`a}}}]{Ballardini:2020iws}%
  \BibitemOpen
  \bibfield  {author} {\bibinfo {author} {\bibfnamefont {M.}~\bibnamefont
  {Ballardini}}, \bibinfo {author} {\bibfnamefont {M.}~\bibnamefont {Braglia}},
  \bibinfo {author} {\bibfnamefont {F.}~\bibnamefont {Finelli}}, \bibinfo
  {author} {\bibfnamefont {D.}~\bibnamefont {Paoletti}}, \bibinfo {author}
  {\bibfnamefont {A.~A.}\ \bibnamefont {Starobinsky}}, \ and\ \bibinfo {author}
  {\bibfnamefont {C.}~\bibnamefont {Umilt{\`a}}},\ }\href {\doibase
  10.1088/1475-7516/2020/10/044} {\bibfield  {journal} {\bibinfo  {journal}
  {JCAP}\ }\textbf {\bibinfo {volume} {10}},\ \bibinfo {pages} {044} (\bibinfo
  {year} {2020})},\ \Eprint {http://arxiv.org/abs/2004.14349} {arXiv:2004.14349
  [astro-ph.CO]} \BibitemShut {NoStop}%
\bibitem [{\citenamefont {Gogoi}\ \emph {et~al.}(2021)\citenamefont {Gogoi},
  \citenamefont {Sharma}, \citenamefont {Chanda},\ and\ \citenamefont
  {Das}}]{Gogoi:2020qif}%
  \BibitemOpen
  \bibfield  {author} {\bibinfo {author} {\bibfnamefont {A.}~\bibnamefont
  {Gogoi}}, \bibinfo {author} {\bibfnamefont {R.~K.}\ \bibnamefont {Sharma}},
  \bibinfo {author} {\bibfnamefont {P.}~\bibnamefont {Chanda}}, \ and\ \bibinfo
  {author} {\bibfnamefont {S.}~\bibnamefont {Das}},\ }\href {\doibase
  10.3847/1538-4357/abfe5b} {\bibfield  {journal} {\bibinfo  {journal}
  {Astrophys. J.}\ }\textbf {\bibinfo {volume} {915}},\ \bibinfo {pages} {132}
  (\bibinfo {year} {2021})},\ \Eprint {http://arxiv.org/abs/2005.11889}
  {arXiv:2005.11889 [astro-ph.CO]} \BibitemShut {NoStop}%
\bibitem [{\citenamefont {Braglia}\ \emph
  {et~al.}(2020{\natexlab{b}})\citenamefont {Braglia}, \citenamefont {Emond},
  \citenamefont {Finelli}, \citenamefont {Gumrukcuoglu},\ and\ \citenamefont
  {Koyama}}]{Braglia:2020bym}%
  \BibitemOpen
  \bibfield  {author} {\bibinfo {author} {\bibfnamefont {M.}~\bibnamefont
  {Braglia}}, \bibinfo {author} {\bibfnamefont {W.~T.}\ \bibnamefont {Emond}},
  \bibinfo {author} {\bibfnamefont {F.}~\bibnamefont {Finelli}}, \bibinfo
  {author} {\bibfnamefont {A.~E.}\ \bibnamefont {Gumrukcuoglu}}, \ and\
  \bibinfo {author} {\bibfnamefont {K.}~\bibnamefont {Koyama}},\ }\href
  {\doibase 10.1103/PhysRevD.102.083513} {\bibfield  {journal} {\bibinfo
  {journal} {Phys. Rev. D}\ }\textbf {\bibinfo {volume} {102}},\ \bibinfo
  {pages} {083513} (\bibinfo {year} {2020}{\natexlab{b}})},\ \Eprint
  {http://arxiv.org/abs/2005.14053} {arXiv:2005.14053 [astro-ph.CO]}
  \BibitemShut {NoStop}%
\bibitem [{\citenamefont {Gonzalez}\ \emph {et~al.}(2020)\citenamefont
  {Gonzalez}, \citenamefont {Hertzberg},\ and\ \citenamefont
  {Rompineve}}]{Gonzalez:2020fdy}%
  \BibitemOpen
  \bibfield  {author} {\bibinfo {author} {\bibfnamefont {M.}~\bibnamefont
  {Gonzalez}}, \bibinfo {author} {\bibfnamefont {M.~P.}\ \bibnamefont
  {Hertzberg}}, \ and\ \bibinfo {author} {\bibfnamefont {F.}~\bibnamefont
  {Rompineve}},\ }\href {\doibase 10.1088/1475-7516/2020/10/028} {\bibfield
  {journal} {\bibinfo  {journal} {JCAP}\ }\textbf {\bibinfo {volume} {10}},\
  \bibinfo {pages} {028} (\bibinfo {year} {2020})},\ \Eprint
  {http://arxiv.org/abs/2006.13959} {arXiv:2006.13959 [astro-ph.CO]}
  \BibitemShut {NoStop}%
\bibitem [{\citenamefont {Ye}\ and\ \citenamefont
  {Piao}(2020{\natexlab{b}})}]{Ye:2020oix}%
  \BibitemOpen
  \bibfield  {author} {\bibinfo {author} {\bibfnamefont {G.}~\bibnamefont
  {Ye}}\ and\ \bibinfo {author} {\bibfnamefont {Y.-S.}\ \bibnamefont {Piao}},\
  }\href {\doibase 10.1103/PhysRevD.102.083523} {\bibfield  {journal} {\bibinfo
   {journal} {Phys. Rev. D}\ }\textbf {\bibinfo {volume} {102}},\ \bibinfo
  {pages} {083523} (\bibinfo {year} {2020}{\natexlab{b}})},\ \Eprint
  {http://arxiv.org/abs/2008.10832} {arXiv:2008.10832 [astro-ph.CO]}
  \BibitemShut {NoStop}%
\bibitem [{\citenamefont {Niedermann}\ and\ \citenamefont
  {Sloth}(2021{\natexlab{b}})}]{Niedermann:2020qbw}%
  \BibitemOpen
  \bibfield  {author} {\bibinfo {author} {\bibfnamefont {F.}~\bibnamefont
  {Niedermann}}\ and\ \bibinfo {author} {\bibfnamefont {M.~S.}\ \bibnamefont
  {Sloth}},\ }\href {\doibase 10.1103/PhysRevD.103.103537} {\bibfield
  {journal} {\bibinfo  {journal} {Phys. Rev. D}\ }\textbf {\bibinfo {volume}
  {103}},\ \bibinfo {pages} {103537} (\bibinfo {year} {2021}{\natexlab{b}})},\
  \Eprint {http://arxiv.org/abs/2009.00006} {arXiv:2009.00006 [astro-ph.CO]}
  \BibitemShut {NoStop}%
\bibitem [{\citenamefont {Murgia}\ \emph {et~al.}(2021)\citenamefont {Murgia},
  \citenamefont {Abell{\'a}n},\ and\ \citenamefont {Poulin}}]{Murgia:2020ryi}%
  \BibitemOpen
  \bibfield  {author} {\bibinfo {author} {\bibfnamefont {R.}~\bibnamefont
  {Murgia}}, \bibinfo {author} {\bibfnamefont {G.~F.}\ \bibnamefont
  {Abell{\'a}n}}, \ and\ \bibinfo {author} {\bibfnamefont {V.}~\bibnamefont
  {Poulin}},\ }\href {\doibase 10.1103/PhysRevD.103.063502} {\bibfield
  {journal} {\bibinfo  {journal} {Phys. Rev. D}\ }\textbf {\bibinfo {volume}
  {103}},\ \bibinfo {pages} {063502} (\bibinfo {year} {2021})},\ \Eprint
  {http://arxiv.org/abs/2009.10733} {arXiv:2009.10733 [astro-ph.CO]}
  \BibitemShut {NoStop}%
\bibitem [{\citenamefont {Smith}\ \emph {et~al.}(2021)\citenamefont {Smith},
  \citenamefont {Poulin}, \citenamefont {Bernal}, \citenamefont {Boddy},
  \citenamefont {Kamionkowski},\ and\ \citenamefont {Murgia}}]{Smith:2020rxx}%
  \BibitemOpen
  \bibfield  {author} {\bibinfo {author} {\bibfnamefont {T.~L.}\ \bibnamefont
  {Smith}}, \bibinfo {author} {\bibfnamefont {V.}~\bibnamefont {Poulin}},
  \bibinfo {author} {\bibfnamefont {J.~L.}\ \bibnamefont {Bernal}}, \bibinfo
  {author} {\bibfnamefont {K.~K.}\ \bibnamefont {Boddy}}, \bibinfo {author}
  {\bibfnamefont {M.}~\bibnamefont {Kamionkowski}}, \ and\ \bibinfo {author}
  {\bibfnamefont {R.}~\bibnamefont {Murgia}},\ }\href {\doibase
  10.1103/PhysRevD.103.123542} {\bibfield  {journal} {\bibinfo  {journal}
  {Phys. Rev. D}\ }\textbf {\bibinfo {volume} {103}},\ \bibinfo {pages}
  {123542} (\bibinfo {year} {2021})},\ \Eprint
  {http://arxiv.org/abs/2009.10740} {arXiv:2009.10740 [astro-ph.CO]}
  \BibitemShut {NoStop}%
\bibitem [{\citenamefont {Carrillo~Gonz{\'a}lez}\ \emph
  {et~al.}(2021)\citenamefont {Carrillo~Gonz{\'a}lez}, \citenamefont {Liang},
  \citenamefont {Sakstein},\ and\ \citenamefont
  {Trodden}}]{CarrilloGonzalez:2020oac}%
  \BibitemOpen
  \bibfield  {author} {\bibinfo {author} {\bibfnamefont {M.}~\bibnamefont
  {Carrillo~Gonz{\'a}lez}}, \bibinfo {author} {\bibfnamefont {Q.}~\bibnamefont
  {Liang}}, \bibinfo {author} {\bibfnamefont {J.}~\bibnamefont {Sakstein}}, \
  and\ \bibinfo {author} {\bibfnamefont {M.}~\bibnamefont {Trodden}},\ }\href
  {\doibase 10.1088/1475-7516/2021/04/063} {\bibfield  {journal} {\bibinfo
  {journal} {JCAP}\ }\textbf {\bibinfo {volume} {04}},\ \bibinfo {pages} {063}
  (\bibinfo {year} {2021})},\ \Eprint {http://arxiv.org/abs/2011.09895}
  {arXiv:2011.09895 [astro-ph.CO]} \BibitemShut {NoStop}%
\bibitem [{\citenamefont {Braglia}\ \emph {et~al.}(2021)\citenamefont
  {Braglia}, \citenamefont {Ballardini}, \citenamefont {Finelli},\ and\
  \citenamefont {Koyama}}]{Braglia:2020auw}%
  \BibitemOpen
  \bibfield  {author} {\bibinfo {author} {\bibfnamefont {M.}~\bibnamefont
  {Braglia}}, \bibinfo {author} {\bibfnamefont {M.}~\bibnamefont {Ballardini}},
  \bibinfo {author} {\bibfnamefont {F.}~\bibnamefont {Finelli}}, \ and\
  \bibinfo {author} {\bibfnamefont {K.}~\bibnamefont {Koyama}},\ }\href
  {\doibase 10.1103/PhysRevD.103.043528} {\bibfield  {journal} {\bibinfo
  {journal} {Phys. Rev. D}\ }\textbf {\bibinfo {volume} {103}},\ \bibinfo
  {pages} {043528} (\bibinfo {year} {2021})},\ \Eprint
  {http://arxiv.org/abs/2011.12934} {arXiv:2011.12934 [astro-ph.CO]}
  \BibitemShut {NoStop}%
\bibitem [{\citenamefont {Adi}\ and\ \citenamefont
  {Kovetz}(2021)}]{Adi:2020qqf}%
  \BibitemOpen
  \bibfield  {author} {\bibinfo {author} {\bibfnamefont {T.}~\bibnamefont
  {Adi}}\ and\ \bibinfo {author} {\bibfnamefont {E.~D.}\ \bibnamefont
  {Kovetz}},\ }\href {\doibase 10.1103/PhysRevD.103.023530} {\bibfield
  {journal} {\bibinfo  {journal} {Phys. Rev. D}\ }\textbf {\bibinfo {volume}
  {103}},\ \bibinfo {pages} {023530} (\bibinfo {year} {2021})},\ \Eprint
  {http://arxiv.org/abs/2011.13853} {arXiv:2011.13853 [astro-ph.CO]}
  \BibitemShut {NoStop}%
\bibitem [{\citenamefont {Oikonomou}(2021{\natexlab{a}})}]{Oikonomou:2020qah}%
  \BibitemOpen
  \bibfield  {author} {\bibinfo {author} {\bibfnamefont {V.~K.}\ \bibnamefont
  {Oikonomou}},\ }\href {\doibase 10.1103/PhysRevD.103.044036} {\bibfield
  {journal} {\bibinfo  {journal} {Phys. Rev. D}\ }\textbf {\bibinfo {volume}
  {103}},\ \bibinfo {pages} {044036} (\bibinfo {year} {2021}{\natexlab{a}})},\
  \Eprint {http://arxiv.org/abs/2012.00586} {arXiv:2012.00586 [astro-ph.CO]}
  \BibitemShut {NoStop}%
\bibitem [{\citenamefont {Oikonomou}(2021{\natexlab{b}})}]{Oikonomou:2020oex}%
  \BibitemOpen
  \bibfield  {author} {\bibinfo {author} {\bibfnamefont {V.~K.}\ \bibnamefont
  {Oikonomou}},\ }\href {\doibase 10.1103/PhysRevD.103.124028} {\bibfield
  {journal} {\bibinfo  {journal} {Phys. Rev. D}\ }\textbf {\bibinfo {volume}
  {103}},\ \bibinfo {pages} {124028} (\bibinfo {year} {2021}{\natexlab{b}})},\
  \Eprint {http://arxiv.org/abs/2012.01312} {arXiv:2012.01312 [gr-qc]}
  \BibitemShut {NoStop}%
\bibitem [{\citenamefont {Karwal}\ \emph {et~al.}(2022)\citenamefont {Karwal},
  \citenamefont {Raveri}, \citenamefont {Jain}, \citenamefont {Khoury},\ and\
  \citenamefont {Trodden}}]{Karwal:2021vpk}%
  \BibitemOpen
  \bibfield  {author} {\bibinfo {author} {\bibfnamefont {T.}~\bibnamefont
  {Karwal}}, \bibinfo {author} {\bibfnamefont {M.}~\bibnamefont {Raveri}},
  \bibinfo {author} {\bibfnamefont {B.}~\bibnamefont {Jain}}, \bibinfo {author}
  {\bibfnamefont {J.}~\bibnamefont {Khoury}}, \ and\ \bibinfo {author}
  {\bibfnamefont {M.}~\bibnamefont {Trodden}},\ }\href {\doibase
  10.1103/PhysRevD.105.063535} {\bibfield  {journal} {\bibinfo  {journal}
  {Phys. Rev. D}\ }\textbf {\bibinfo {volume} {105}},\ \bibinfo {pages}
  {063535} (\bibinfo {year} {2022})},\ \Eprint
  {http://arxiv.org/abs/2106.13290} {arXiv:2106.13290 [astro-ph.CO]}
  \BibitemShut {NoStop}%
\bibitem [{\citenamefont {Jiang}\ and\ \citenamefont
  {Piao}(2021)}]{Jiang:2021bab}%
  \BibitemOpen
  \bibfield  {author} {\bibinfo {author} {\bibfnamefont {J.-Q.}\ \bibnamefont
  {Jiang}}\ and\ \bibinfo {author} {\bibfnamefont {Y.-S.}\ \bibnamefont
  {Piao}},\ }\href {\doibase 10.1103/PhysRevD.104.103524} {\bibfield  {journal}
  {\bibinfo  {journal} {Phys. Rev. D}\ }\textbf {\bibinfo {volume} {104}},\
  \bibinfo {pages} {103524} (\bibinfo {year} {2021})},\ \Eprint
  {http://arxiv.org/abs/2107.07128} {arXiv:2107.07128 [astro-ph.CO]}
  \BibitemShut {NoStop}%
\bibitem [{\citenamefont {G{\'o}mez-Valent}\ \emph {et~al.}(2021)\citenamefont
  {G{\'o}mez-Valent}, \citenamefont {Zheng}, \citenamefont {Amendola},
  \citenamefont {Pettorino},\ and\ \citenamefont
  {Wetterich}}]{Gomez-Valent:2021cbe}%
  \BibitemOpen
  \bibfield  {author} {\bibinfo {author} {\bibfnamefont {A.}~\bibnamefont
  {G{\'o}mez-Valent}}, \bibinfo {author} {\bibfnamefont {Z.}~\bibnamefont
  {Zheng}}, \bibinfo {author} {\bibfnamefont {L.}~\bibnamefont {Amendola}},
  \bibinfo {author} {\bibfnamefont {V.}~\bibnamefont {Pettorino}}, \ and\
  \bibinfo {author} {\bibfnamefont {C.}~\bibnamefont {Wetterich}},\ }\href
  {\doibase 10.1103/PhysRevD.104.083536} {\bibfield  {journal} {\bibinfo
  {journal} {Phys. Rev. D}\ }\textbf {\bibinfo {volume} {104}},\ \bibinfo
  {pages} {083536} (\bibinfo {year} {2021})},\ \Eprint
  {http://arxiv.org/abs/2107.11065} {arXiv:2107.11065 [astro-ph.CO]}
  \BibitemShut {NoStop}%
\bibitem [{\citenamefont {Ye}\ \emph {et~al.}(2023{\natexlab{a}})\citenamefont
  {Ye}, \citenamefont {Zhang},\ and\ \citenamefont {Piao}}]{Ye:2021iwa}%
  \BibitemOpen
  \bibfield  {author} {\bibinfo {author} {\bibfnamefont {G.}~\bibnamefont
  {Ye}}, \bibinfo {author} {\bibfnamefont {J.}~\bibnamefont {Zhang}}, \ and\
  \bibinfo {author} {\bibfnamefont {Y.-S.}\ \bibnamefont {Piao}},\ }\href
  {\doibase 10.1016/j.physletb.2023.137770} {\bibfield  {journal} {\bibinfo
  {journal} {Phys. Lett. B}\ }\textbf {\bibinfo {volume} {839}},\ \bibinfo
  {pages} {137770} (\bibinfo {year} {2023}{\natexlab{a}})},\ \Eprint
  {http://arxiv.org/abs/2107.13391} {arXiv:2107.13391 [astro-ph.CO]}
  \BibitemShut {NoStop}%
\bibitem [{\citenamefont {Poulin}\ \emph {et~al.}(2021)\citenamefont {Poulin},
  \citenamefont {Smith},\ and\ \citenamefont {Bartlett}}]{Poulin:2021bjr}%
  \BibitemOpen
  \bibfield  {author} {\bibinfo {author} {\bibfnamefont {V.}~\bibnamefont
  {Poulin}}, \bibinfo {author} {\bibfnamefont {T.~L.}\ \bibnamefont {Smith}}, \
  and\ \bibinfo {author} {\bibfnamefont {A.}~\bibnamefont {Bartlett}},\ }\href
  {\doibase 10.1103/PhysRevD.104.123550} {\bibfield  {journal} {\bibinfo
  {journal} {Phys. Rev. D}\ }\textbf {\bibinfo {volume} {104}},\ \bibinfo
  {pages} {123550} (\bibinfo {year} {2021})},\ \Eprint
  {http://arxiv.org/abs/2109.06229} {arXiv:2109.06229 [astro-ph.CO]}
  \BibitemShut {NoStop}%
\bibitem [{\citenamefont {Niedermann}\ and\ \citenamefont
  {Sloth}(2022{\natexlab{a}})}]{Niedermann:2021ijp}%
  \BibitemOpen
  \bibfield  {author} {\bibinfo {author} {\bibfnamefont {F.}~\bibnamefont
  {Niedermann}}\ and\ \bibinfo {author} {\bibfnamefont {M.~S.}\ \bibnamefont
  {Sloth}},\ }\href {\doibase 10.1016/j.physletb.2022.137555} {\bibfield
  {journal} {\bibinfo  {journal} {Phys. Lett. B}\ }\textbf {\bibinfo {volume}
  {835}},\ \bibinfo {pages} {137555} (\bibinfo {year} {2022}{\natexlab{a}})},\
  \Eprint {http://arxiv.org/abs/2112.00759} {arXiv:2112.00759 [hep-ph]}
  \BibitemShut {NoStop}%
\bibitem [{\citenamefont {Niedermann}\ and\ \citenamefont
  {Sloth}(2022{\natexlab{b}})}]{Niedermann:2021vgd}%
  \BibitemOpen
  \bibfield  {author} {\bibinfo {author} {\bibfnamefont {F.}~\bibnamefont
  {Niedermann}}\ and\ \bibinfo {author} {\bibfnamefont {M.~S.}\ \bibnamefont
  {Sloth}},\ }\href {\doibase 10.1103/PhysRevD.105.063509} {\bibfield
  {journal} {\bibinfo  {journal} {Phys. Rev. D}\ }\textbf {\bibinfo {volume}
  {105}},\ \bibinfo {pages} {063509} (\bibinfo {year} {2022}{\natexlab{b}})},\
  \Eprint {http://arxiv.org/abs/2112.00770} {arXiv:2112.00770 [hep-ph]}
  \BibitemShut {NoStop}%
\bibitem [{\citenamefont {Herold}\ \emph {et~al.}(2022)\citenamefont {Herold},
  \citenamefont {Ferreira},\ and\ \citenamefont {Komatsu}}]{Herold:2021ksg}%
  \BibitemOpen
  \bibfield  {author} {\bibinfo {author} {\bibfnamefont {L.}~\bibnamefont
  {Herold}}, \bibinfo {author} {\bibfnamefont {E.~G.~M.}\ \bibnamefont
  {Ferreira}}, \ and\ \bibinfo {author} {\bibfnamefont {E.}~\bibnamefont
  {Komatsu}},\ }\href {\doibase 10.3847/2041-8213/ac63a3} {\bibfield  {journal}
  {\bibinfo  {journal} {Astrophys. J. Lett.}\ }\textbf {\bibinfo {volume}
  {929}},\ \bibinfo {pages} {L16} (\bibinfo {year} {2022})},\ \Eprint
  {http://arxiv.org/abs/2112.12140} {arXiv:2112.12140 [astro-ph.CO]}
  \BibitemShut {NoStop}%
\bibitem [{\citenamefont {Wang}\ and\ \citenamefont
  {Piao}(2022)}]{Wang:2022jpo}%
  \BibitemOpen
  \bibfield  {author} {\bibinfo {author} {\bibfnamefont {H.}~\bibnamefont
  {Wang}}\ and\ \bibinfo {author} {\bibfnamefont {Y.-S.}\ \bibnamefont
  {Piao}},\ }\href {\doibase 10.1016/j.physletb.2022.137244} {\bibfield
  {journal} {\bibinfo  {journal} {Phys. Lett. B}\ }\textbf {\bibinfo {volume}
  {832}},\ \bibinfo {pages} {137244} (\bibinfo {year} {2022})},\ \Eprint
  {http://arxiv.org/abs/2201.07079} {arXiv:2201.07079 [astro-ph.CO]}
  \BibitemShut {NoStop}%
\bibitem [{\citenamefont {Smith}\ \emph {et~al.}(2022)\citenamefont {Smith},
  \citenamefont {Lucca}, \citenamefont {Poulin}, \citenamefont {Abellan},
  \citenamefont {Balkenhol}, \citenamefont {Benabed}, \citenamefont {Galli},\
  and\ \citenamefont {Murgia}}]{Smith:2022hwi}%
  \BibitemOpen
  \bibfield  {author} {\bibinfo {author} {\bibfnamefont {T.~L.}\ \bibnamefont
  {Smith}}, \bibinfo {author} {\bibfnamefont {M.}~\bibnamefont {Lucca}},
  \bibinfo {author} {\bibfnamefont {V.}~\bibnamefont {Poulin}}, \bibinfo
  {author} {\bibfnamefont {G.~F.}\ \bibnamefont {Abellan}}, \bibinfo {author}
  {\bibfnamefont {L.}~\bibnamefont {Balkenhol}}, \bibinfo {author}
  {\bibfnamefont {K.}~\bibnamefont {Benabed}}, \bibinfo {author} {\bibfnamefont
  {S.}~\bibnamefont {Galli}}, \ and\ \bibinfo {author} {\bibfnamefont
  {R.}~\bibnamefont {Murgia}},\ }\href {\doibase 10.1103/PhysRevD.106.043526}
  {\bibfield  {journal} {\bibinfo  {journal} {Phys. Rev. D}\ }\textbf {\bibinfo
  {volume} {106}},\ \bibinfo {pages} {043526} (\bibinfo {year} {2022})},\
  \Eprint {http://arxiv.org/abs/2202.09379} {arXiv:2202.09379 [astro-ph.CO]}
  \BibitemShut {NoStop}%
\bibitem [{\citenamefont {Oikonomou}\ and\ \citenamefont
  {Lymperiadou}(2022)}]{Oikonomou:2022yle}%
  \BibitemOpen
  \bibfield  {author} {\bibinfo {author} {\bibfnamefont {V.~K.}\ \bibnamefont
  {Oikonomou}}\ and\ \bibinfo {author} {\bibfnamefont {E.~C.}\ \bibnamefont
  {Lymperiadou}},\ }\href {\doibase 10.3390/sym14061143} {\bibfield  {journal}
  {\bibinfo  {journal} {Symmetry}\ }\textbf {\bibinfo {volume} {14}},\ \bibinfo
  {pages} {1143} (\bibinfo {year} {2022})},\ \Eprint
  {http://arxiv.org/abs/2206.00721} {arXiv:2206.00721 [gr-qc]} \BibitemShut
  {NoStop}%
\bibitem [{\citenamefont {Reeves}\ \emph {et~al.}(2023)\citenamefont {Reeves},
  \citenamefont {Herold}, \citenamefont {Vagnozzi}, \citenamefont {Sherwin},\
  and\ \citenamefont {Ferreira}}]{Reeves:2022aoi}%
  \BibitemOpen
  \bibfield  {author} {\bibinfo {author} {\bibfnamefont {A.}~\bibnamefont
  {Reeves}}, \bibinfo {author} {\bibfnamefont {L.}~\bibnamefont {Herold}},
  \bibinfo {author} {\bibfnamefont {S.}~\bibnamefont {Vagnozzi}}, \bibinfo
  {author} {\bibfnamefont {B.~D.}\ \bibnamefont {Sherwin}}, \ and\ \bibinfo
  {author} {\bibfnamefont {E.~G.~M.}\ \bibnamefont {Ferreira}},\ }\href
  {\doibase 10.1093/mnras/stad317} {\bibfield  {journal} {\bibinfo  {journal}
  {Mon. Not. Roy. Astron. Soc.}\ }\textbf {\bibinfo {volume} {520}},\ \bibinfo
  {pages} {3688} (\bibinfo {year} {2023})},\ \Eprint
  {http://arxiv.org/abs/2207.01501} {arXiv:2207.01501 [astro-ph.CO]}
  \BibitemShut {NoStop}%
\bibitem [{\citenamefont {G{\'o}mez-Valent}\ \emph {et~al.}(2022)\citenamefont
  {G{\'o}mez-Valent}, \citenamefont {Zheng}, \citenamefont {Amendola},
  \citenamefont {Wetterich},\ and\ \citenamefont
  {Pettorino}}]{Gomez-Valent:2022bku}%
  \BibitemOpen
  \bibfield  {author} {\bibinfo {author} {\bibfnamefont {A.}~\bibnamefont
  {G{\'o}mez-Valent}}, \bibinfo {author} {\bibfnamefont {Z.}~\bibnamefont
  {Zheng}}, \bibinfo {author} {\bibfnamefont {L.}~\bibnamefont {Amendola}},
  \bibinfo {author} {\bibfnamefont {C.}~\bibnamefont {Wetterich}}, \ and\
  \bibinfo {author} {\bibfnamefont {V.}~\bibnamefont {Pettorino}},\ }\href
  {\doibase 10.1103/PhysRevD.106.103522} {\bibfield  {journal} {\bibinfo
  {journal} {Phys. Rev. D}\ }\textbf {\bibinfo {volume} {106}},\ \bibinfo
  {pages} {103522} (\bibinfo {year} {2022})},\ \Eprint
  {http://arxiv.org/abs/2207.14487} {arXiv:2207.14487 [astro-ph.CO]}
  \BibitemShut {NoStop}%
\bibitem [{\citenamefont {Escudero}\ \emph {et~al.}(2022)\citenamefont
  {Escudero}, \citenamefont {Kuo}, \citenamefont {Keeley},\ and\ \citenamefont
  {Abazajian}}]{Escudero:2022rbq}%
  \BibitemOpen
  \bibfield  {author} {\bibinfo {author} {\bibfnamefont {H.~G.}\ \bibnamefont
  {Escudero}}, \bibinfo {author} {\bibfnamefont {J.-L.}\ \bibnamefont {Kuo}},
  \bibinfo {author} {\bibfnamefont {R.~E.}\ \bibnamefont {Keeley}}, \ and\
  \bibinfo {author} {\bibfnamefont {K.~N.}\ \bibnamefont {Abazajian}},\ }\href
  {\doibase 10.1103/PhysRevD.106.103517} {\bibfield  {journal} {\bibinfo
  {journal} {Phys. Rev. D}\ }\textbf {\bibinfo {volume} {106}},\ \bibinfo
  {pages} {103517} (\bibinfo {year} {2022})},\ \Eprint
  {http://arxiv.org/abs/2208.14435} {arXiv:2208.14435 [astro-ph.CO]}
  \BibitemShut {NoStop}%
\bibitem [{\citenamefont {Cruz}\ \emph
  {et~al.}(2023{\natexlab{a}})\citenamefont {Cruz}, \citenamefont
  {Niedermann},\ and\ \citenamefont {Sloth}}]{Cruz:2022oqk}%
  \BibitemOpen
  \bibfield  {author} {\bibinfo {author} {\bibfnamefont {J.~S.}\ \bibnamefont
  {Cruz}}, \bibinfo {author} {\bibfnamefont {F.}~\bibnamefont {Niedermann}}, \
  and\ \bibinfo {author} {\bibfnamefont {M.~S.}\ \bibnamefont {Sloth}},\ }\href
  {\doibase 10.1088/1475-7516/2023/02/041} {\bibfield  {journal} {\bibinfo
  {journal} {JCAP}\ }\textbf {\bibinfo {volume} {02}},\ \bibinfo {pages} {041}
  (\bibinfo {year} {2023}{\natexlab{a}})},\ \Eprint
  {http://arxiv.org/abs/2209.02708} {arXiv:2209.02708 [astro-ph.CO]}
  \BibitemShut {NoStop}%
\bibitem [{\citenamefont {Wang}\ and\ \citenamefont
  {Piao}(2023)}]{Wang:2022bmk}%
  \BibitemOpen
  \bibfield  {author} {\bibinfo {author} {\bibfnamefont {H.}~\bibnamefont
  {Wang}}\ and\ \bibinfo {author} {\bibfnamefont {Y.-S.}\ \bibnamefont
  {Piao}},\ }\href {\doibase 10.1103/PhysRevD.108.083516} {\bibfield  {journal}
  {\bibinfo  {journal} {Phys. Rev. D}\ }\textbf {\bibinfo {volume} {108}},\
  \bibinfo {pages} {083516} (\bibinfo {year} {2023})},\ \Eprint
  {http://arxiv.org/abs/2209.09685} {arXiv:2209.09685 [astro-ph.CO]}
  \BibitemShut {NoStop}%
\bibitem [{\citenamefont {Herold}\ and\ \citenamefont
  {Ferreira}(2023)}]{Herold:2022iib}%
  \BibitemOpen
  \bibfield  {author} {\bibinfo {author} {\bibfnamefont {L.}~\bibnamefont
  {Herold}}\ and\ \bibinfo {author} {\bibfnamefont {E.~G.~M.}\ \bibnamefont
  {Ferreira}},\ }\href {\doibase 10.1103/PhysRevD.108.043513} {\bibfield
  {journal} {\bibinfo  {journal} {Phys. Rev. D}\ }\textbf {\bibinfo {volume}
  {108}},\ \bibinfo {pages} {043513} (\bibinfo {year} {2023})},\ \Eprint
  {http://arxiv.org/abs/2210.16296} {arXiv:2210.16296 [astro-ph.CO]}
  \BibitemShut {NoStop}%
\bibitem [{\citenamefont {Cruz}\ \emph
  {et~al.}(2023{\natexlab{b}})\citenamefont {Cruz}, \citenamefont {Hannestad},
  \citenamefont {Holm}, \citenamefont {Niedermann}, \citenamefont {Sloth},\
  and\ \citenamefont {Tram}}]{Cruz:2023cxy}%
  \BibitemOpen
  \bibfield  {author} {\bibinfo {author} {\bibfnamefont {J.~S.}\ \bibnamefont
  {Cruz}}, \bibinfo {author} {\bibfnamefont {S.}~\bibnamefont {Hannestad}},
  \bibinfo {author} {\bibfnamefont {E.~B.}\ \bibnamefont {Holm}}, \bibinfo
  {author} {\bibfnamefont {F.}~\bibnamefont {Niedermann}}, \bibinfo {author}
  {\bibfnamefont {M.~S.}\ \bibnamefont {Sloth}}, \ and\ \bibinfo {author}
  {\bibfnamefont {T.}~\bibnamefont {Tram}},\ }\href {\doibase
  10.1103/PhysRevD.108.023518} {\bibfield  {journal} {\bibinfo  {journal}
  {Phys. Rev. D}\ }\textbf {\bibinfo {volume} {108}},\ \bibinfo {pages}
  {023518} (\bibinfo {year} {2023}{\natexlab{b}})},\ \Eprint
  {http://arxiv.org/abs/2302.07934} {arXiv:2302.07934 [astro-ph.CO]}
  \BibitemShut {NoStop}%
\bibitem [{\citenamefont {Goldstein}\ \emph {et~al.}(2023)\citenamefont
  {Goldstein}, \citenamefont {Hill}, \citenamefont {Ir{\v{s}}i{\v{c}}},\ and\
  \citenamefont {Sherwin}}]{Goldstein:2023gnw}%
  \BibitemOpen
  \bibfield  {author} {\bibinfo {author} {\bibfnamefont {S.}~\bibnamefont
  {Goldstein}}, \bibinfo {author} {\bibfnamefont {J.~C.}\ \bibnamefont {Hill}},
  \bibinfo {author} {\bibfnamefont {V.}~\bibnamefont {Ir{\v{s}}i{\v{c}}}}, \
  and\ \bibinfo {author} {\bibfnamefont {B.~D.}\ \bibnamefont {Sherwin}},\
  }\href {\doibase 10.1103/PhysRevLett.131.201001} {\bibfield  {journal}
  {\bibinfo  {journal} {Phys. Rev. Lett.}\ }\textbf {\bibinfo {volume} {131}},\
  \bibinfo {pages} {201001} (\bibinfo {year} {2023})},\ \Eprint
  {http://arxiv.org/abs/2303.00746} {arXiv:2303.00746 [astro-ph.CO]}
  \BibitemShut {NoStop}%
\bibitem [{\citenamefont {Eskilt}\ \emph {et~al.}(2023)\citenamefont {Eskilt},
  \citenamefont {Herold}, \citenamefont {Komatsu}, \citenamefont {Murai},
  \citenamefont {Namikawa},\ and\ \citenamefont {Naokawa}}]{Eskilt:2023nxm}%
  \BibitemOpen
  \bibfield  {author} {\bibinfo {author} {\bibfnamefont {J.~R.}\ \bibnamefont
  {Eskilt}}, \bibinfo {author} {\bibfnamefont {L.}~\bibnamefont {Herold}},
  \bibinfo {author} {\bibfnamefont {E.}~\bibnamefont {Komatsu}}, \bibinfo
  {author} {\bibfnamefont {K.}~\bibnamefont {Murai}}, \bibinfo {author}
  {\bibfnamefont {T.}~\bibnamefont {Namikawa}}, \ and\ \bibinfo {author}
  {\bibfnamefont {F.}~\bibnamefont {Naokawa}},\ }\href {\doibase
  10.1103/PhysRevLett.131.121001} {\bibfield  {journal} {\bibinfo  {journal}
  {Phys. Rev. Lett.}\ }\textbf {\bibinfo {volume} {131}},\ \bibinfo {pages}
  {121001} (\bibinfo {year} {2023})},\ \Eprint
  {http://arxiv.org/abs/2303.15369} {arXiv:2303.15369 [astro-ph.CO]}
  \BibitemShut {NoStop}%
\bibitem [{\citenamefont {Cruz}\ \emph
  {et~al.}(2023{\natexlab{c}})\citenamefont {Cruz}, \citenamefont
  {Niedermann},\ and\ \citenamefont {Sloth}}]{Cruz:2023lmn}%
  \BibitemOpen
  \bibfield  {author} {\bibinfo {author} {\bibfnamefont {J.~S.}\ \bibnamefont
  {Cruz}}, \bibinfo {author} {\bibfnamefont {F.}~\bibnamefont {Niedermann}}, \
  and\ \bibinfo {author} {\bibfnamefont {M.~S.}\ \bibnamefont {Sloth}},\ }\href
  {\doibase 10.1088/1475-7516/2023/11/033} {\bibfield  {journal} {\bibinfo
  {journal} {JCAP}\ }\textbf {\bibinfo {volume} {11}},\ \bibinfo {pages} {033}
  (\bibinfo {year} {2023}{\natexlab{c}})},\ \Eprint
  {http://arxiv.org/abs/2305.08895} {arXiv:2305.08895 [astro-ph.CO]}
  \BibitemShut {NoStop}%
\bibitem [{\citenamefont {Odintsov}\ \emph {et~al.}(2023)\citenamefont
  {Odintsov}, \citenamefont {Oikonomou},\ and\ \citenamefont
  {Sharov}}]{Odintsov:2023cli}%
  \BibitemOpen
  \bibfield  {author} {\bibinfo {author} {\bibfnamefont {S.~D.}\ \bibnamefont
  {Odintsov}}, \bibinfo {author} {\bibfnamefont {V.~K.}\ \bibnamefont
  {Oikonomou}}, \ and\ \bibinfo {author} {\bibfnamefont {G.~S.}\ \bibnamefont
  {Sharov}},\ }\href {\doibase 10.1016/j.physletb.2023.137988} {\bibfield
  {journal} {\bibinfo  {journal} {Phys. Lett. B}\ }\textbf {\bibinfo {volume}
  {843}},\ \bibinfo {pages} {137988} (\bibinfo {year} {2023})},\ \Eprint
  {http://arxiv.org/abs/2305.17513} {arXiv:2305.17513 [gr-qc]} \BibitemShut
  {NoStop}%
\bibitem [{\citenamefont {Ye}\ \emph {et~al.}(2023{\natexlab{b}})\citenamefont
  {Ye}, \citenamefont {Jiang},\ and\ \citenamefont {Piao}}]{Ye:2023zel}%
  \BibitemOpen
  \bibfield  {author} {\bibinfo {author} {\bibfnamefont {G.}~\bibnamefont
  {Ye}}, \bibinfo {author} {\bibfnamefont {J.-Q.}\ \bibnamefont {Jiang}}, \
  and\ \bibinfo {author} {\bibfnamefont {Y.-S.}\ \bibnamefont {Piao}},\ }\href
  {\doibase 10.1103/PhysRevD.108.063512} {\bibfield  {journal} {\bibinfo
  {journal} {Phys. Rev. D}\ }\textbf {\bibinfo {volume} {108}},\ \bibinfo
  {pages} {063512} (\bibinfo {year} {2023}{\natexlab{b}})},\ \Eprint
  {http://arxiv.org/abs/2305.18873} {arXiv:2305.18873 [astro-ph.CO]}
  \BibitemShut {NoStop}%
\bibitem [{\citenamefont {Sharma}\ \emph {et~al.}(2024)\citenamefont {Sharma},
  \citenamefont {Das},\ and\ \citenamefont {Poulin}}]{Sharma:2023kzr}%
  \BibitemOpen
  \bibfield  {author} {\bibinfo {author} {\bibfnamefont {R.~K.}\ \bibnamefont
  {Sharma}}, \bibinfo {author} {\bibfnamefont {S.}~\bibnamefont {Das}}, \ and\
  \bibinfo {author} {\bibfnamefont {V.}~\bibnamefont {Poulin}},\ }\href
  {\doibase 10.1103/PhysRevD.109.043530} {\bibfield  {journal} {\bibinfo
  {journal} {Phys. Rev. D}\ }\textbf {\bibinfo {volume} {109}},\ \bibinfo
  {pages} {043530} (\bibinfo {year} {2024})},\ \Eprint
  {http://arxiv.org/abs/2309.00401} {arXiv:2309.00401 [astro-ph.CO]}
  \BibitemShut {NoStop}%
\bibitem [{\citenamefont {Efstathiou}\ \emph {et~al.}(2024)\citenamefont
  {Efstathiou}, \citenamefont {Rosenberg},\ and\ \citenamefont
  {Poulin}}]{Efstathiou:2023fbn}%
  \BibitemOpen
  \bibfield  {author} {\bibinfo {author} {\bibfnamefont {G.}~\bibnamefont
  {Efstathiou}}, \bibinfo {author} {\bibfnamefont {E.}~\bibnamefont
  {Rosenberg}}, \ and\ \bibinfo {author} {\bibfnamefont {V.}~\bibnamefont
  {Poulin}},\ }\href {\doibase 10.1103/PhysRevLett.132.221002} {\bibfield
  {journal} {\bibinfo  {journal} {Phys. Rev. Lett.}\ }\textbf {\bibinfo
  {volume} {132}},\ \bibinfo {pages} {221002} (\bibinfo {year} {2024})},\
  \Eprint {http://arxiv.org/abs/2311.00524} {arXiv:2311.00524 [astro-ph.CO]}
  \BibitemShut {NoStop}%
\bibitem [{\citenamefont {Pedreira}\ \emph {et~al.}(2024)\citenamefont
  {Pedreira}, \citenamefont {Benetti}, \citenamefont {Ferreira}, \citenamefont
  {Graef},\ and\ \citenamefont {Herold}}]{Pedreira:2023qqt}%
  \BibitemOpen
  \bibfield  {author} {\bibinfo {author} {\bibfnamefont {I.~d. O.~C.}\
  \bibnamefont {Pedreira}}, \bibinfo {author} {\bibfnamefont {M.}~\bibnamefont
  {Benetti}}, \bibinfo {author} {\bibfnamefont {E.~G.~M.}\ \bibnamefont
  {Ferreira}}, \bibinfo {author} {\bibfnamefont {L.~L.}\ \bibnamefont {Graef}},
  \ and\ \bibinfo {author} {\bibfnamefont {L.}~\bibnamefont {Herold}},\ }\href
  {\doibase 10.1103/PhysRevD.109.103525} {\bibfield  {journal} {\bibinfo
  {journal} {Phys. Rev. D}\ }\textbf {\bibinfo {volume} {109}},\ \bibinfo
  {pages} {103525} (\bibinfo {year} {2024})},\ \Eprint
  {http://arxiv.org/abs/2311.04977} {arXiv:2311.04977 [astro-ph.CO]}
  \BibitemShut {NoStop}%
\bibitem [{\citenamefont {Khalife}\ \emph {et~al.}(2024)\citenamefont
  {Khalife}, \citenamefont {Zanjani}, \citenamefont {Galli}, \citenamefont
  {G{\"u}nther}, \citenamefont {Lesgourgues},\ and\ \citenamefont
  {Benabed}}]{Khalife:2023qbu}%
  \BibitemOpen
  \bibfield  {author} {\bibinfo {author} {\bibfnamefont {A.~R.}\ \bibnamefont
  {Khalife}}, \bibinfo {author} {\bibfnamefont {M.~B.}\ \bibnamefont
  {Zanjani}}, \bibinfo {author} {\bibfnamefont {S.}~\bibnamefont {Galli}},
  \bibinfo {author} {\bibfnamefont {S.}~\bibnamefont {G{\"u}nther}}, \bibinfo
  {author} {\bibfnamefont {J.}~\bibnamefont {Lesgourgues}}, \ and\ \bibinfo
  {author} {\bibfnamefont {K.}~\bibnamefont {Benabed}},\ }\href {\doibase
  10.1088/1475-7516/2024/04/059} {\bibfield  {journal} {\bibinfo  {journal}
  {JCAP}\ }\textbf {\bibinfo {volume} {04}},\ \bibinfo {pages} {059} (\bibinfo
  {year} {2024})},\ \Eprint {http://arxiv.org/abs/2312.09814} {arXiv:2312.09814
  [astro-ph.CO]} \BibitemShut {NoStop}%
\bibitem [{\citenamefont {Wang}\ and\ \citenamefont
  {Piao}(2024)}]{Wang:2024jug}%
  \BibitemOpen
  \bibfield  {author} {\bibinfo {author} {\bibfnamefont {H.}~\bibnamefont
  {Wang}}\ and\ \bibinfo {author} {\bibfnamefont {Y.-S.}\ \bibnamefont
  {Piao}},\ }\href {\doibase 10.1016/j.physletb.2024.138914} {\bibfield
  {journal} {\bibinfo  {journal} {Phys. Lett. B}\ }\textbf {\bibinfo {volume}
  {856}},\ \bibinfo {pages} {138914} (\bibinfo {year} {2024})},\ \Eprint
  {http://arxiv.org/abs/2401.08812} {arXiv:2401.08812 [gr-qc]} \BibitemShut
  {NoStop}%
\bibitem [{\citenamefont {Garny}\ \emph {et~al.}(2024)\citenamefont {Garny},
  \citenamefont {Niedermann}, \citenamefont {Rubira},\ and\ \citenamefont
  {Sloth}}]{Garny:2024ums}%
  \BibitemOpen
  \bibfield  {author} {\bibinfo {author} {\bibfnamefont {M.}~\bibnamefont
  {Garny}}, \bibinfo {author} {\bibfnamefont {F.}~\bibnamefont {Niedermann}},
  \bibinfo {author} {\bibfnamefont {H.}~\bibnamefont {Rubira}}, \ and\ \bibinfo
  {author} {\bibfnamefont {M.~S.}\ \bibnamefont {Sloth}},\ }\href {\doibase
  10.1103/PhysRevD.110.023531} {\bibfield  {journal} {\bibinfo  {journal}
  {Phys. Rev. D}\ }\textbf {\bibinfo {volume} {110}},\ \bibinfo {pages}
  {023531} (\bibinfo {year} {2024})},\ \Eprint
  {http://arxiv.org/abs/2404.07256} {arXiv:2404.07256 [astro-ph.CO]}
  \BibitemShut {NoStop}%
\bibitem [{\citenamefont {Chatrchyan}\ \emph {et~al.}(2025)\citenamefont
  {Chatrchyan}, \citenamefont {Niedermann}, \citenamefont {Poulin},\ and\
  \citenamefont {Sloth}}]{Chatrchyan:2024xjj}%
  \BibitemOpen
  \bibfield  {author} {\bibinfo {author} {\bibfnamefont {A.}~\bibnamefont
  {Chatrchyan}}, \bibinfo {author} {\bibfnamefont {F.}~\bibnamefont
  {Niedermann}}, \bibinfo {author} {\bibfnamefont {V.}~\bibnamefont {Poulin}},
  \ and\ \bibinfo {author} {\bibfnamefont {M.~S.}\ \bibnamefont {Sloth}},\
  }\href {\doibase 10.1103/PhysRevD.111.043536} {\bibfield  {journal} {\bibinfo
   {journal} {Phys. Rev. D}\ }\textbf {\bibinfo {volume} {111}},\ \bibinfo
  {pages} {043536} (\bibinfo {year} {2025})},\ \Eprint
  {http://arxiv.org/abs/2408.14537} {arXiv:2408.14537 [astro-ph.CO]}
  \BibitemShut {NoStop}%
\bibitem [{\citenamefont {Jiang}\ \emph
  {et~al.}(2025{\natexlab{a}})\citenamefont {Jiang}, \citenamefont {Liu},
  \citenamefont {Zhan},\ and\ \citenamefont {Hu}}]{Jiang:2024tll}%
  \BibitemOpen
  \bibfield  {author} {\bibinfo {author} {\bibfnamefont {J.-Q.}\ \bibnamefont
  {Jiang}}, \bibinfo {author} {\bibfnamefont {W.}~\bibnamefont {Liu}}, \bibinfo
  {author} {\bibfnamefont {H.}~\bibnamefont {Zhan}}, \ and\ \bibinfo {author}
  {\bibfnamefont {B.}~\bibnamefont {Hu}},\ }\href {\doibase
  10.1103/PhysRevD.111.023519} {\bibfield  {journal} {\bibinfo  {journal}
  {Phys. Rev. D}\ }\textbf {\bibinfo {volume} {111}},\ \bibinfo {pages}
  {023519} (\bibinfo {year} {2025}{\natexlab{a}})},\ \Eprint
  {http://arxiv.org/abs/2409.19941} {arXiv:2409.19941 [astro-ph.CO]}
  \BibitemShut {NoStop}%
\bibitem [{\citenamefont {Jiang}(2025{\natexlab{a}})}]{Jiang:2024nha}%
  \BibitemOpen
  \bibfield  {author} {\bibinfo {author} {\bibfnamefont {J.-Q.}\ \bibnamefont
  {Jiang}},\ }\href {\doibase 10.1103/PhysRevD.111.043528} {\bibfield
  {journal} {\bibinfo  {journal} {Phys. Rev. D}\ }\textbf {\bibinfo {volume}
  {111}},\ \bibinfo {pages} {043528} (\bibinfo {year} {2025}{\natexlab{a}})},\
  \Eprint {http://arxiv.org/abs/2410.10559} {arXiv:2410.10559 [astro-ph.CO]}
  \BibitemShut {NoStop}%
\bibitem [{\citenamefont {Simon}\ \emph {et~al.}(2025)\citenamefont {Simon},
  \citenamefont {Adi}, \citenamefont {Bernal}, \citenamefont {Kovetz},
  \citenamefont {Poulin},\ and\ \citenamefont {Smith}}]{Simon:2024jmu}%
  \BibitemOpen
  \bibfield  {author} {\bibinfo {author} {\bibfnamefont {T.}~\bibnamefont
  {Simon}}, \bibinfo {author} {\bibfnamefont {T.}~\bibnamefont {Adi}}, \bibinfo
  {author} {\bibfnamefont {J.~L.}\ \bibnamefont {Bernal}}, \bibinfo {author}
  {\bibfnamefont {E.~D.}\ \bibnamefont {Kovetz}}, \bibinfo {author}
  {\bibfnamefont {V.}~\bibnamefont {Poulin}}, \ and\ \bibinfo {author}
  {\bibfnamefont {T.~L.}\ \bibnamefont {Smith}},\ }\href {\doibase
  10.1103/PhysRevD.111.023523} {\bibfield  {journal} {\bibinfo  {journal}
  {Phys. Rev. D}\ }\textbf {\bibinfo {volume} {111}},\ \bibinfo {pages}
  {023523} (\bibinfo {year} {2025})},\ \Eprint
  {http://arxiv.org/abs/2410.21459} {arXiv:2410.21459 [astro-ph.CO]}
  \BibitemShut {NoStop}%
\bibitem [{\citenamefont {Forconi}\ \emph {et~al.}(2025)\citenamefont
  {Forconi}, \citenamefont {Favale},\ and\ \citenamefont
  {G{\'o}mez-Valent}}]{Forconi:2025cwp}%
  \BibitemOpen
  \bibfield  {author} {\bibinfo {author} {\bibfnamefont {M.}~\bibnamefont
  {Forconi}}, \bibinfo {author} {\bibfnamefont {A.}~\bibnamefont {Favale}}, \
  and\ \bibinfo {author} {\bibfnamefont {A.}~\bibnamefont {G{\'o}mez-Valent}},\
  }\href {\doibase 10.1103/rpf5-ldks} {\bibfield  {journal} {\bibinfo
  {journal} {Phys. Rev. D}\ }\textbf {\bibinfo {volume} {112}},\ \bibinfo
  {pages} {023517} (\bibinfo {year} {2025})},\ \Eprint
  {http://arxiv.org/abs/2501.11571} {arXiv:2501.11571 [astro-ph.CO]}
  \BibitemShut {NoStop}%
\bibitem [{\citenamefont {Jiang}\ and\ \citenamefont
  {Piao}(2025)}]{Jiang:2025ylr}%
  \BibitemOpen
  \bibfield  {author} {\bibinfo {author} {\bibfnamefont {J.-Q.}\ \bibnamefont
  {Jiang}}\ and\ \bibinfo {author} {\bibfnamefont {Y.-S.}\ \bibnamefont
  {Piao}},\ }\href {\doibase 10.1103/PhysRevD.111.103505} {\bibfield  {journal}
  {\bibinfo  {journal} {Phys. Rev. D}\ }\textbf {\bibinfo {volume} {111}},\
  \bibinfo {pages} {103505} (\bibinfo {year} {2025})},\ \Eprint
  {http://arxiv.org/abs/2501.16883} {arXiv:2501.16883 [astro-ph.CO]}
  \BibitemShut {NoStop}%
\bibitem [{\citenamefont {Stahl}\ \emph {et~al.}(2025)\citenamefont {Stahl},
  \citenamefont {Poulin}, \citenamefont {Famaey},\ and\ \citenamefont
  {Ibata}}]{Stahl:2025czl}%
  \BibitemOpen
  \bibfield  {author} {\bibinfo {author} {\bibfnamefont {C.}~\bibnamefont
  {Stahl}}, \bibinfo {author} {\bibfnamefont {V.}~\bibnamefont {Poulin}},
  \bibinfo {author} {\bibfnamefont {B.}~\bibnamefont {Famaey}}, \ and\ \bibinfo
  {author} {\bibfnamefont {R.}~\bibnamefont {Ibata}},\ }\href {\doibase
  10.1088/1475-7516/2025/06/021} {\bibfield  {journal} {\bibinfo  {journal}
  {JCAP}\ }\textbf {\bibinfo {volume} {06}},\ \bibinfo {pages} {021} (\bibinfo
  {year} {2025})},\ \Eprint {http://arxiv.org/abs/2502.14608} {arXiv:2502.14608
  [astro-ph.CO]} \BibitemShut {NoStop}%
\bibitem [{\citenamefont {Jiang}(2025{\natexlab{b}})}]{Jiang:2025hco}%
  \BibitemOpen
  \bibfield  {author} {\bibinfo {author} {\bibfnamefont {J.-Q.}\ \bibnamefont
  {Jiang}},\ }\href {\doibase 10.1016/j.dark.2025.101902} {\bibfield  {journal}
  {\bibinfo  {journal} {Phys. Dark Univ.}\ }\textbf {\bibinfo {volume} {48}},\
  \bibinfo {pages} {101902} (\bibinfo {year} {2025}{\natexlab{b}})},\ \Eprint
  {http://arxiv.org/abs/2502.15541} {arXiv:2502.15541 [astro-ph.CO]}
  \BibitemShut {NoStop}%
\bibitem [{\citenamefont {Smith}\ \emph {et~al.}(2025)\citenamefont {Smith},
  \citenamefont {Brax}, \citenamefont {van~de Bruck}, \citenamefont {Burgess},\
  and\ \citenamefont {Davis}}]{Smith:2025grk}%
  \BibitemOpen
  \bibfield  {author} {\bibinfo {author} {\bibfnamefont {A.}~\bibnamefont
  {Smith}}, \bibinfo {author} {\bibfnamefont {P.}~\bibnamefont {Brax}},
  \bibinfo {author} {\bibfnamefont {C.}~\bibnamefont {van~de Bruck}}, \bibinfo
  {author} {\bibfnamefont {C.~P.}\ \bibnamefont {Burgess}}, \ and\ \bibinfo
  {author} {\bibfnamefont {A.-C.}\ \bibnamefont {Davis}},\ }\href {\doibase
  10.1140/epjc/s10052-025-14735-4} {\bibfield  {journal} {\bibinfo  {journal}
  {Eur. Phys. J. C}\ }\textbf {\bibinfo {volume} {85}},\ \bibinfo {pages}
  {1062} (\bibinfo {year} {2025})},\ \Eprint {http://arxiv.org/abs/2505.05450}
  {arXiv:2505.05450 [hep-th]} \BibitemShut {NoStop}%
\bibitem [{\citenamefont {Poulin}\ \emph {et~al.}(2026)\citenamefont {Poulin},
  \citenamefont {Smith}, \citenamefont {Calder{\'o}n},\ and\ \citenamefont
  {Simon}}]{Poulin:2025nfb}%
  \BibitemOpen
  \bibfield  {author} {\bibinfo {author} {\bibfnamefont {V.}~\bibnamefont
  {Poulin}}, \bibinfo {author} {\bibfnamefont {T.~L.}\ \bibnamefont {Smith}},
  \bibinfo {author} {\bibfnamefont {R.}~\bibnamefont {Calder{\'o}n}}, \ and\
  \bibinfo {author} {\bibfnamefont {T.}~\bibnamefont {Simon}},\ }\href
  {\doibase 10.1103/bx25-1g5d} {\bibfield  {journal} {\bibinfo  {journal}
  {Phys. Rev. D}\ }\textbf {\bibinfo {volume} {113}},\ \bibinfo {pages}
  {063519} (\bibinfo {year} {2026})},\ \Eprint
  {http://arxiv.org/abs/2505.08051} {arXiv:2505.08051 [astro-ph.CO]}
  \BibitemShut {NoStop}%
\bibitem [{\citenamefont {Yashiki}(2025)}]{Yashiki:2025loj}%
  \BibitemOpen
  \bibfield  {author} {\bibinfo {author} {\bibfnamefont {M.}~\bibnamefont
  {Yashiki}},\ }\href {\doibase 10.1103/qw1d-mdrz} {\bibfield  {journal}
  {\bibinfo  {journal} {Phys. Rev. D}\ }\textbf {\bibinfo {volume} {112}},\
  \bibinfo {pages} {063517} (\bibinfo {year} {2025})},\ \Eprint
  {http://arxiv.org/abs/2505.23382} {arXiv:2505.23382 [astro-ph.CO]}
  \BibitemShut {NoStop}%
\bibitem [{\citenamefont {Garny}\ \emph {et~al.}(2025)\citenamefont {Garny},
  \citenamefont {Niedermann}, \citenamefont {Rubira},\ and\ \citenamefont
  {Sloth}}]{Garny:2025kqj}%
  \BibitemOpen
  \bibfield  {author} {\bibinfo {author} {\bibfnamefont {M.}~\bibnamefont
  {Garny}}, \bibinfo {author} {\bibfnamefont {F.}~\bibnamefont {Niedermann}},
  \bibinfo {author} {\bibfnamefont {H.}~\bibnamefont {Rubira}}, \ and\ \bibinfo
  {author} {\bibfnamefont {M.~S.}\ \bibnamefont {Sloth}},\ }\href@noop {} {\
  (\bibinfo {year} {2025})},\ \Eprint {http://arxiv.org/abs/2508.03795}
  {arXiv:2508.03795 [astro-ph.CO]} \BibitemShut {NoStop}%
\bibitem [{\citenamefont {Toda}\ and\ \citenamefont
  {Seto}(2026)}]{Toda:2025kcq}%
  \BibitemOpen
  \bibfield  {author} {\bibinfo {author} {\bibfnamefont {Y.}~\bibnamefont
  {Toda}}\ and\ \bibinfo {author} {\bibfnamefont {O.}~\bibnamefont {Seto}},\
  }\href {\doibase 10.1088/1475-7516/2026/02/019} {\bibfield  {journal}
  {\bibinfo  {journal} {JCAP}\ }\textbf {\bibinfo {volume} {02}},\ \bibinfo
  {pages} {019} (\bibinfo {year} {2026})},\ \Eprint
  {http://arxiv.org/abs/2508.09025} {arXiv:2508.09025 [astro-ph.CO]}
  \BibitemShut {NoStop}%
\bibitem [{\citenamefont {Wang}\ and\ \citenamefont
  {Piao}(2025)}]{Wang:2025djw}%
  \BibitemOpen
  \bibfield  {author} {\bibinfo {author} {\bibfnamefont {H.}~\bibnamefont
  {Wang}}\ and\ \bibinfo {author} {\bibfnamefont {Y.-S.}\ \bibnamefont
  {Piao}},\ }\href@noop {} {\  (\bibinfo {year} {2025})},\ \Eprint
  {http://arxiv.org/abs/2511.16606} {arXiv:2511.16606 [astro-ph.CO]}
  \BibitemShut {NoStop}%
\bibitem [{\citenamefont {Yin}\ \emph {et~al.}(2026)\citenamefont {Yin},
  \citenamefont {Du}, \citenamefont {Li},\ and\ \citenamefont
  {Zhang}}]{Yin:2026gss}%
  \BibitemOpen
  \bibfield  {author} {\bibinfo {author} {\bibfnamefont {L.}~\bibnamefont
  {Yin}}, \bibinfo {author} {\bibfnamefont {G.-H.}\ \bibnamefont {Du}},
  \bibinfo {author} {\bibfnamefont {T.-N.}\ \bibnamefont {Li}}, \ and\ \bibinfo
  {author} {\bibfnamefont {X.}~\bibnamefont {Zhang}},\ }\href@noop {} {\
  (\bibinfo {year} {2026})},\ \Eprint {http://arxiv.org/abs/2601.13624}
  {arXiv:2601.13624 [astro-ph.CO]} \BibitemShut {NoStop}%
\bibitem [{\citenamefont {Gonz{\'a}lez-Fuentes}\ and\ \citenamefont
  {G{\'o}mez-Valent}(2026)}]{Gonzalez-Fuentes:2026rgu}%
  \BibitemOpen
  \bibfield  {author} {\bibinfo {author} {\bibfnamefont {A.}~\bibnamefont
  {Gonz{\'a}lez-Fuentes}}\ and\ \bibinfo {author} {\bibfnamefont
  {A.}~\bibnamefont {G{\'o}mez-Valent}},\ }\href@noop {} {\  (\bibinfo {year}
  {2026})},\ \Eprint {http://arxiv.org/abs/2603.26560} {arXiv:2603.26560
  [astro-ph.CO]} \BibitemShut {NoStop}%
\bibitem [{\citenamefont {Jhaveri}\ \emph {et~al.}(2026)\citenamefont
  {Jhaveri}, \citenamefont {Karwal}, \citenamefont {Crawford}, \citenamefont
  {Hu}, \citenamefont {Khalife}, \citenamefont {Balkenhol},\ and\ \citenamefont
  {Ge}}]{Jhaveri:2026bla}%
  \BibitemOpen
  \bibfield  {author} {\bibinfo {author} {\bibfnamefont {T.}~\bibnamefont
  {Jhaveri}}, \bibinfo {author} {\bibfnamefont {T.}~\bibnamefont {Karwal}},
  \bibinfo {author} {\bibfnamefont {T.}~\bibnamefont {Crawford}}, \bibinfo
  {author} {\bibfnamefont {W.}~\bibnamefont {Hu}}, \bibinfo {author}
  {\bibfnamefont {A.~R.}\ \bibnamefont {Khalife}}, \bibinfo {author}
  {\bibfnamefont {L.}~\bibnamefont {Balkenhol}}, \ and\ \bibinfo {author}
  {\bibfnamefont {F.}~\bibnamefont {Ge}},\ }\href@noop {} {\  (\bibinfo {year}
  {2026})},\ \Eprint {http://arxiv.org/abs/2604.08530} {arXiv:2604.08530
  [astro-ph.CO]} \BibitemShut {NoStop}%
\bibitem [{\citenamefont {Bella}\ \emph {et~al.}(2026)\citenamefont {Bella},
  \citenamefont {Poulin}, \citenamefont {Vagnozzi},\ and\ \citenamefont
  {Knox}}]{Bella:2026zuk}%
  \BibitemOpen
  \bibfield  {author} {\bibinfo {author} {\bibfnamefont {M.}~\bibnamefont
  {Bella}}, \bibinfo {author} {\bibfnamefont {V.}~\bibnamefont {Poulin}},
  \bibinfo {author} {\bibfnamefont {S.}~\bibnamefont {Vagnozzi}}, \ and\
  \bibinfo {author} {\bibfnamefont {L.}~\bibnamefont {Knox}},\ }\href@noop {}
  {\  (\bibinfo {year} {2026})},\ \Eprint {http://arxiv.org/abs/2604.13535}
  {arXiv:2604.13535 [astro-ph.CO]} \BibitemShut {NoStop}%
\bibitem [{\citenamefont {Carloni}\ and\ \citenamefont
  {Luongo}(2026)}]{Carloni:2026yut}%
  \BibitemOpen
  \bibfield  {author} {\bibinfo {author} {\bibfnamefont {Y.}~\bibnamefont
  {Carloni}}\ and\ \bibinfo {author} {\bibfnamefont {O.}~\bibnamefont
  {Luongo}},\ }\href@noop {} {\  (\bibinfo {year} {2026})},\ \Eprint
  {http://arxiv.org/abs/2604.18053} {arXiv:2604.18053 [astro-ph.CO]}
  \BibitemShut {NoStop}%
\bibitem [{\citenamefont {Zhang}\ and\ \citenamefont
  {Yin}(2026)}]{Zhang:2026fzj}%
  \BibitemOpen
  \bibfield  {author} {\bibinfo {author} {\bibfnamefont {K.}~\bibnamefont
  {Zhang}}\ and\ \bibinfo {author} {\bibfnamefont {L.}~\bibnamefont {Yin}},\
  }\href@noop {} {\  (\bibinfo {year} {2026})},\ \Eprint
  {http://arxiv.org/abs/2605.24341} {arXiv:2605.24341 [astro-ph.CO]}
  \BibitemShut {NoStop}%
\bibitem [{\citenamefont {Giar{\`e}}\ and\ \citenamefont
  {Sakstein}(2026)}]{Giare:2026tyk}%
  \BibitemOpen
  \bibfield  {author} {\bibinfo {author} {\bibfnamefont {W.}~\bibnamefont
  {Giar{\`e}}}\ and\ \bibinfo {author} {\bibfnamefont {J.}~\bibnamefont
  {Sakstein}},\ }\href@noop {} {\  (\bibinfo {year} {2026})},\ \Eprint
  {http://arxiv.org/abs/2605.26116} {arXiv:2605.26116 [astro-ph.CO]}
  \BibitemShut {NoStop}%
\bibitem [{\citenamefont {Du}\ \emph {et~al.}(2026)\citenamefont {Du},
  \citenamefont {Li}, \citenamefont {Yin}, \citenamefont {Zhou}, \citenamefont
  {Wang}, \citenamefont {Zhang},\ and\ \citenamefont {Zhang}}]{Du:2026qtq}%
  \BibitemOpen
  \bibfield  {author} {\bibinfo {author} {\bibfnamefont {G.-H.}\ \bibnamefont
  {Du}}, \bibinfo {author} {\bibfnamefont {T.-N.}\ \bibnamefont {Li}}, \bibinfo
  {author} {\bibfnamefont {L.}~\bibnamefont {Yin}}, \bibinfo {author}
  {\bibfnamefont {S.-H.}\ \bibnamefont {Zhou}}, \bibinfo {author}
  {\bibfnamefont {H.}~\bibnamefont {Wang}}, \bibinfo {author} {\bibfnamefont
  {J.-F.}\ \bibnamefont {Zhang}}, \ and\ \bibinfo {author} {\bibfnamefont
  {X.}~\bibnamefont {Zhang}},\ }\href@noop {} {\  (\bibinfo {year} {2026})},\
  \Eprint {http://arxiv.org/abs/2606.19090} {arXiv:2606.19090 [astro-ph.CO]}
  \BibitemShut {NoStop}%
\bibitem [{\citenamefont {Simon}(2024)}]{Simon:2023hlp}%
  \BibitemOpen
  \bibfield  {author} {\bibinfo {author} {\bibfnamefont {T.}~\bibnamefont
  {Simon}},\ }\href {\doibase 10.1103/PhysRevD.110.023528} {\bibfield
  {journal} {\bibinfo  {journal} {Phys. Rev. D}\ }\textbf {\bibinfo {volume}
  {110}},\ \bibinfo {pages} {023528} (\bibinfo {year} {2024})},\ \Eprint
  {http://arxiv.org/abs/2310.16800} {arXiv:2310.16800 [astro-ph.CO]}
  \BibitemShut {NoStop}%
\bibitem [{\citenamefont {McDonough}\ and\ \citenamefont
  {Scalisi}(2023)}]{McDonough:2022pku}%
  \BibitemOpen
  \bibfield  {author} {\bibinfo {author} {\bibfnamefont {E.}~\bibnamefont
  {McDonough}}\ and\ \bibinfo {author} {\bibfnamefont {M.}~\bibnamefont
  {Scalisi}},\ }\href {\doibase 10.1007/JHEP10(2023)118} {\bibfield  {journal}
  {\bibinfo  {journal} {JHEP}\ }\textbf {\bibinfo {volume} {10}},\ \bibinfo
  {pages} {118} (\bibinfo {year} {2023})},\ \Eprint
  {http://arxiv.org/abs/2209.00011} {arXiv:2209.00011 [hep-th]} \BibitemShut
  {NoStop}%
\bibitem [{\citenamefont {Cicoli}\ \emph {et~al.}(2023)\citenamefont {Cicoli},
  \citenamefont {Licheri}, \citenamefont {Mahanta}, \citenamefont {McDonough},
  \citenamefont {Pedro},\ and\ \citenamefont {Scalisi}}]{Cicoli:2023qri}%
  \BibitemOpen
  \bibfield  {author} {\bibinfo {author} {\bibfnamefont {M.}~\bibnamefont
  {Cicoli}}, \bibinfo {author} {\bibfnamefont {M.}~\bibnamefont {Licheri}},
  \bibinfo {author} {\bibfnamefont {R.}~\bibnamefont {Mahanta}}, \bibinfo
  {author} {\bibfnamefont {E.}~\bibnamefont {McDonough}}, \bibinfo {author}
  {\bibfnamefont {F.~G.}\ \bibnamefont {Pedro}}, \ and\ \bibinfo {author}
  {\bibfnamefont {M.}~\bibnamefont {Scalisi}},\ }\href {\doibase
  10.1007/JHEP06(2023)052} {\bibfield  {journal} {\bibinfo  {journal} {JHEP}\
  }\textbf {\bibinfo {volume} {06}},\ \bibinfo {pages} {052} (\bibinfo {year}
  {2023})},\ \Eprint {http://arxiv.org/abs/2303.03414} {arXiv:2303.03414
  [hep-th]} \BibitemShut {NoStop}%
\bibitem [{\citenamefont {Akrami}\ \emph {et~al.}(2020)\citenamefont {Akrami}
  \emph {et~al.}}]{Planck:2020olo}%
  \BibitemOpen
  \bibfield  {author} {\bibinfo {author} {\bibfnamefont {Y.}~\bibnamefont
  {Akrami}} \emph {et~al.} (\bibinfo {collaboration} {Planck}),\ }\href
  {\doibase 10.1051/0004-6361/202038073} {\bibfield  {journal} {\bibinfo
  {journal} {Astron. Astrophys.}\ }\textbf {\bibinfo {volume} {643}},\ \bibinfo
  {pages} {A42} (\bibinfo {year} {2020})},\ \Eprint
  {http://arxiv.org/abs/2007.04997} {arXiv:2007.04997 [astro-ph.CO]}
  \BibitemShut {NoStop}%
\bibitem [{\citenamefont {Rosenberg}\ \emph {et~al.}(2022)\citenamefont
  {Rosenberg}, \citenamefont {Gratton},\ and\ \citenamefont
  {Efstathiou}}]{Rosenberg:2022sdy}%
  \BibitemOpen
  \bibfield  {author} {\bibinfo {author} {\bibfnamefont {E.}~\bibnamefont
  {Rosenberg}}, \bibinfo {author} {\bibfnamefont {S.}~\bibnamefont {Gratton}},
  \ and\ \bibinfo {author} {\bibfnamefont {G.}~\bibnamefont {Efstathiou}},\
  }\href {\doibase 10.1093/mnras/stac2744} {\bibfield  {journal} {\bibinfo
  {journal} {Mon. Not. Roy. Astron. Soc.}\ }\textbf {\bibinfo {volume} {517}},\
  \bibinfo {pages} {4620} (\bibinfo {year} {2022})},\ \Eprint
  {http://arxiv.org/abs/2205.10869} {arXiv:2205.10869 [astro-ph.CO]}
  \BibitemShut {NoStop}%
\bibitem [{\citenamefont {Aghanim}\ \emph {et~al.}(2020)\citenamefont {Aghanim}
  \emph {et~al.}}]{Planck:2019nip}%
  \BibitemOpen
  \bibfield  {author} {\bibinfo {author} {\bibfnamefont {N.}~\bibnamefont
  {Aghanim}} \emph {et~al.} (\bibinfo {collaboration} {Planck}),\ }\href
  {\doibase 10.1051/0004-6361/201936386} {\bibfield  {journal} {\bibinfo
  {journal} {Astron. Astrophys.}\ }\textbf {\bibinfo {volume} {641}},\ \bibinfo
  {pages} {A5} (\bibinfo {year} {2020})},\ \Eprint
  {http://arxiv.org/abs/1907.12875} {arXiv:1907.12875 [astro-ph.CO]}
  \BibitemShut {NoStop}%
\bibitem [{\citenamefont {Delouis}\ \emph {et~al.}(2019)\citenamefont
  {Delouis}, \citenamefont {Pagano}, \citenamefont {Mottet}, \citenamefont
  {Puget},\ and\ \citenamefont {Vibert}}]{Delouis:2019bub}%
  \BibitemOpen
  \bibfield  {author} {\bibinfo {author} {\bibfnamefont {J.~M.}\ \bibnamefont
  {Delouis}}, \bibinfo {author} {\bibfnamefont {L.}~\bibnamefont {Pagano}},
  \bibinfo {author} {\bibfnamefont {S.}~\bibnamefont {Mottet}}, \bibinfo
  {author} {\bibfnamefont {J.~L.}\ \bibnamefont {Puget}}, \ and\ \bibinfo
  {author} {\bibfnamefont {L.}~\bibnamefont {Vibert}},\ }\href {\doibase
  10.1051/0004-6361/201834882} {\bibfield  {journal} {\bibinfo  {journal}
  {Astron. Astrophys.}\ }\textbf {\bibinfo {volume} {629}},\ \bibinfo {pages}
  {A38} (\bibinfo {year} {2019})},\ \Eprint {http://arxiv.org/abs/1901.11386}
  {arXiv:1901.11386 [astro-ph.CO]} \BibitemShut {NoStop}%
\bibitem [{\citenamefont {Carron}\ \emph {et~al.}(2022)\citenamefont {Carron},
  \citenamefont {Mirmelstein},\ and\ \citenamefont {Lewis}}]{Carron:2022eyg}%
  \BibitemOpen
  \bibfield  {author} {\bibinfo {author} {\bibfnamefont {J.}~\bibnamefont
  {Carron}}, \bibinfo {author} {\bibfnamefont {M.}~\bibnamefont {Mirmelstein}},
  \ and\ \bibinfo {author} {\bibfnamefont {A.}~\bibnamefont {Lewis}},\ }\href
  {\doibase 10.1088/1475-7516/2022/09/039} {\bibfield  {journal} {\bibinfo
  {journal} {JCAP}\ }\textbf {\bibinfo {volume} {09}},\ \bibinfo {pages} {039}
  (\bibinfo {year} {2022})},\ \Eprint {http://arxiv.org/abs/2206.07773}
  {arXiv:2206.07773 [astro-ph.CO]} \BibitemShut {NoStop}%
\bibitem [{\citenamefont {Camphuis}\ \emph {et~al.}(2026)\citenamefont
  {Camphuis} \emph {et~al.}}]{SPT-3G:2025bzu}%
  \BibitemOpen
  \bibfield  {author} {\bibinfo {author} {\bibfnamefont {E.}~\bibnamefont
  {Camphuis}} \emph {et~al.} (\bibinfo {collaboration} {SPT-3G}),\ }\href
  {\doibase 10.1103/7wt3-9v2y} {\bibfield  {journal} {\bibinfo  {journal}
  {Phys. Rev. D}\ }\textbf {\bibinfo {volume} {113}},\ \bibinfo {pages}
  {083504} (\bibinfo {year} {2026})},\ \Eprint
  {http://arxiv.org/abs/2506.20707} {arXiv:2506.20707 [astro-ph.CO]}
  \BibitemShut {NoStop}%
\bibitem [{\citenamefont {Ge}\ \emph {et~al.}(2025)\citenamefont {Ge} \emph
  {et~al.}}]{SPT-3G:2024atg}%
  \BibitemOpen
  \bibfield  {author} {\bibinfo {author} {\bibfnamefont {F.}~\bibnamefont {Ge}}
  \emph {et~al.} (\bibinfo {collaboration} {SPT-3G}),\ }\href {\doibase
  10.1103/PhysRevD.111.083534} {\bibfield  {journal} {\bibinfo  {journal}
  {Phys. Rev. D}\ }\textbf {\bibinfo {volume} {111}},\ \bibinfo {pages}
  {083534} (\bibinfo {year} {2025})},\ \Eprint
  {http://arxiv.org/abs/2411.06000} {arXiv:2411.06000 [astro-ph.CO]}
  \BibitemShut {NoStop}%
\bibitem [{\citenamefont {Abdul~Karim}\ \emph {et~al.}(2025)\citenamefont
  {Abdul~Karim} \emph {et~al.}}]{DESI:2025zgx}%
  \BibitemOpen
  \bibfield  {author} {\bibinfo {author} {\bibfnamefont {M.}~\bibnamefont
  {Abdul~Karim}} \emph {et~al.} (\bibinfo {collaboration} {DESI}),\ }\href
  {\doibase 10.1103/tr6y-kpc6} {\bibfield  {journal} {\bibinfo  {journal}
  {Phys. Rev. D}\ }\textbf {\bibinfo {volume} {112}},\ \bibinfo {pages}
  {083515} (\bibinfo {year} {2025})},\ \Eprint
  {http://arxiv.org/abs/2503.14738} {arXiv:2503.14738 [astro-ph.CO]}
  \BibitemShut {NoStop}%
\bibitem [{\citenamefont {McDonough}\ and\ \citenamefont
  {Ferreira}(2025)}]{McDonough:2025lzo}%
  \BibitemOpen
  \bibfield  {author} {\bibinfo {author} {\bibfnamefont {E.}~\bibnamefont
  {McDonough}}\ and\ \bibinfo {author} {\bibfnamefont {E.~G.~M.}\ \bibnamefont
  {Ferreira}},\ }\href@noop {} {\  (\bibinfo {year} {2025})},\ \Eprint
  {http://arxiv.org/abs/2512.05108} {arXiv:2512.05108 [astro-ph.CO]}
  \BibitemShut {NoStop}%
\bibitem [{\citenamefont {Vagnozzi}\ \emph {et~al.}(2017)\citenamefont
  {Vagnozzi}, \citenamefont {Giusarma}, \citenamefont {Mena}, \citenamefont
  {Freese}, \citenamefont {Gerbino}, \citenamefont {Ho},\ and\ \citenamefont
  {Lattanzi}}]{Vagnozzi:2017ovm}%
  \BibitemOpen
  \bibfield  {author} {\bibinfo {author} {\bibfnamefont {S.}~\bibnamefont
  {Vagnozzi}}, \bibinfo {author} {\bibfnamefont {E.}~\bibnamefont {Giusarma}},
  \bibinfo {author} {\bibfnamefont {O.}~\bibnamefont {Mena}}, \bibinfo {author}
  {\bibfnamefont {K.}~\bibnamefont {Freese}}, \bibinfo {author} {\bibfnamefont
  {M.}~\bibnamefont {Gerbino}}, \bibinfo {author} {\bibfnamefont
  {S.}~\bibnamefont {Ho}}, \ and\ \bibinfo {author} {\bibfnamefont
  {M.}~\bibnamefont {Lattanzi}},\ }\href {\doibase 10.1103/PhysRevD.96.123503}
  {\bibfield  {journal} {\bibinfo  {journal} {Phys. Rev. D}\ }\textbf {\bibinfo
  {volume} {96}},\ \bibinfo {pages} {123503} (\bibinfo {year} {2017})},\
  \Eprint {http://arxiv.org/abs/1701.08172} {arXiv:1701.08172 [astro-ph.CO]}
  \BibitemShut {NoStop}%
\bibitem [{\citenamefont {Tanseri}\ \emph {et~al.}(2022)\citenamefont
  {Tanseri}, \citenamefont {Hagstotz}, \citenamefont {Vagnozzi}, \citenamefont
  {Giusarma},\ and\ \citenamefont {Freese}}]{Tanseri:2022zfe}%
  \BibitemOpen
  \bibfield  {author} {\bibinfo {author} {\bibfnamefont {I.}~\bibnamefont
  {Tanseri}}, \bibinfo {author} {\bibfnamefont {S.}~\bibnamefont {Hagstotz}},
  \bibinfo {author} {\bibfnamefont {S.}~\bibnamefont {Vagnozzi}}, \bibinfo
  {author} {\bibfnamefont {E.}~\bibnamefont {Giusarma}}, \ and\ \bibinfo
  {author} {\bibfnamefont {K.}~\bibnamefont {Freese}},\ }\href {\doibase
  10.1016/j.jheap.2022.07.002} {\bibfield  {journal} {\bibinfo  {journal}
  {JHEAp}\ }\textbf {\bibinfo {volume} {36}},\ \bibinfo {pages} {1} (\bibinfo
  {year} {2022})},\ \Eprint {http://arxiv.org/abs/2207.01913} {arXiv:2207.01913
  [astro-ph.CO]} \BibitemShut {NoStop}%
\bibitem [{\citenamefont {Jiang}\ \emph
  {et~al.}(2025{\natexlab{b}})\citenamefont {Jiang}, \citenamefont {Giar{\`e}},
  \citenamefont {Gariazzo}, \citenamefont {Dainotti}, \citenamefont
  {Di~Valentino}, \citenamefont {Mena}, \citenamefont {Pedrotti}, \citenamefont
  {da~Costa},\ and\ \citenamefont {Vagnozzi}}]{Jiang:2024viw}%
  \BibitemOpen
  \bibfield  {author} {\bibinfo {author} {\bibfnamefont {J.-Q.}\ \bibnamefont
  {Jiang}}, \bibinfo {author} {\bibfnamefont {W.}~\bibnamefont {Giar{\`e}}},
  \bibinfo {author} {\bibfnamefont {S.}~\bibnamefont {Gariazzo}}, \bibinfo
  {author} {\bibfnamefont {M.~G.}\ \bibnamefont {Dainotti}}, \bibinfo {author}
  {\bibfnamefont {E.}~\bibnamefont {Di~Valentino}}, \bibinfo {author}
  {\bibfnamefont {O.}~\bibnamefont {Mena}}, \bibinfo {author} {\bibfnamefont
  {D.}~\bibnamefont {Pedrotti}}, \bibinfo {author} {\bibfnamefont {S.~S.}\
  \bibnamefont {da~Costa}}, \ and\ \bibinfo {author} {\bibfnamefont
  {S.}~\bibnamefont {Vagnozzi}},\ }\href {\doibase
  10.1088/1475-7516/2025/01/153} {\bibfield  {journal} {\bibinfo  {journal}
  {JCAP}\ }\textbf {\bibinfo {volume} {01}},\ \bibinfo {pages} {153} (\bibinfo
  {year} {2025}{\natexlab{b}})},\ \Eprint {http://arxiv.org/abs/2407.18047}
  {arXiv:2407.18047 [astro-ph.CO]} \BibitemShut {NoStop}%
\bibitem [{\citenamefont {Roy~Choudhury}(2025)}]{RoyChoudhury:2025dhe}%
  \BibitemOpen
  \bibfield  {author} {\bibinfo {author} {\bibfnamefont {S.}~\bibnamefont
  {Roy~Choudhury}},\ }\href {\doibase 10.3847/2041-8213/ade1cc} {\bibfield
  {journal} {\bibinfo  {journal} {Astrophys. J. Lett.}\ }\textbf {\bibinfo
  {volume} {986}},\ \bibinfo {pages} {L31} (\bibinfo {year} {2025})},\ \bibinfo
  {note} {[Erratum: Astrophys.J.Lett. 1001, L25 (2026), Erratum: Astrophys.J.
  1001, L25 (2026)]},\ \Eprint {http://arxiv.org/abs/2504.15340}
  {arXiv:2504.15340 [astro-ph.CO]} \BibitemShut {NoStop}%
\bibitem [{\citenamefont {Roy~Choudhury}\ \emph {et~al.}(2025)\citenamefont
  {Roy~Choudhury}, \citenamefont {Okumura},\ and\ \citenamefont
  {Umetsu}}]{RoyChoudhury:2025iis}%
  \BibitemOpen
  \bibfield  {author} {\bibinfo {author} {\bibfnamefont {S.}~\bibnamefont
  {Roy~Choudhury}}, \bibinfo {author} {\bibfnamefont {T.}~\bibnamefont
  {Okumura}}, \ and\ \bibinfo {author} {\bibfnamefont {K.}~\bibnamefont
  {Umetsu}},\ }\href {\doibase 10.3847/2041-8213/ae1a64} {\bibfield  {journal}
  {\bibinfo  {journal} {Astrophys. J. Lett.}\ }\textbf {\bibinfo {volume}
  {994}},\ \bibinfo {pages} {L26} (\bibinfo {year} {2025})},\ \Eprint
  {http://arxiv.org/abs/2509.26144} {arXiv:2509.26144 [astro-ph.CO]}
  \BibitemShut {NoStop}%
\bibitem [{\citenamefont {Hill}\ \emph {et~al.}(2020)\citenamefont {Hill},
  \citenamefont {McDonough}, \citenamefont {Toomey},\ and\ \citenamefont
  {Alexander}}]{Hill:2020osr}%
  \BibitemOpen
  \bibfield  {author} {\bibinfo {author} {\bibfnamefont {J.~C.}\ \bibnamefont
  {Hill}}, \bibinfo {author} {\bibfnamefont {E.}~\bibnamefont {McDonough}},
  \bibinfo {author} {\bibfnamefont {M.~W.}\ \bibnamefont {Toomey}}, \ and\
  \bibinfo {author} {\bibfnamefont {S.}~\bibnamefont {Alexander}},\ }\href
  {\doibase 10.1103/PhysRevD.102.043507} {\bibfield  {journal} {\bibinfo
  {journal} {Phys. Rev. D}\ }\textbf {\bibinfo {volume} {102}},\ \bibinfo
  {pages} {043507} (\bibinfo {year} {2020})},\ \Eprint
  {http://arxiv.org/abs/2003.07355} {arXiv:2003.07355 [astro-ph.CO]}
  \BibitemShut {NoStop}%
\bibitem [{\citenamefont {Blas}\ \emph {et~al.}(2011)\citenamefont {Blas},
  \citenamefont {Lesgourgues},\ and\ \citenamefont {Tram}}]{Blas:2011rf}%
  \BibitemOpen
  \bibfield  {author} {\bibinfo {author} {\bibfnamefont {D.}~\bibnamefont
  {Blas}}, \bibinfo {author} {\bibfnamefont {J.}~\bibnamefont {Lesgourgues}}, \
  and\ \bibinfo {author} {\bibfnamefont {T.}~\bibnamefont {Tram}},\ }\href
  {\doibase 10.1088/1475-7516/2011/07/034} {\bibfield  {journal} {\bibinfo
  {journal} {JCAP}\ }\textbf {\bibinfo {volume} {07}},\ \bibinfo {pages} {034}
  (\bibinfo {year} {2011})},\ \Eprint {http://arxiv.org/abs/1104.2933}
  {arXiv:1104.2933 [astro-ph.CO]} \BibitemShut {NoStop}%
\bibitem [{\citenamefont {Torrado}\ and\ \citenamefont
  {Lewis}(2021)}]{Torrado:2020dgo}%
  \BibitemOpen
  \bibfield  {author} {\bibinfo {author} {\bibfnamefont {J.}~\bibnamefont
  {Torrado}}\ and\ \bibinfo {author} {\bibfnamefont {A.}~\bibnamefont
  {Lewis}},\ }\href {\doibase 10.1088/1475-7516/2021/05/057} {\bibfield
  {journal} {\bibinfo  {journal} {JCAP}\ }\textbf {\bibinfo {volume} {05}},\
  \bibinfo {pages} {057} (\bibinfo {year} {2021})},\ \Eprint
  {http://arxiv.org/abs/2005.05290} {arXiv:2005.05290 [astro-ph.IM]}
  \BibitemShut {NoStop}%
\bibitem [{\citenamefont {Gelman}\ and\ \citenamefont
  {Rubin}(1992)}]{Gelman:1992zz}%
  \BibitemOpen
  \bibfield  {author} {\bibinfo {author} {\bibfnamefont {A.}~\bibnamefont
  {Gelman}}\ and\ \bibinfo {author} {\bibfnamefont {D.~B.}\ \bibnamefont
  {Rubin}},\ }\href {\doibase 10.1214/ss/1177011136} {\bibfield  {journal}
  {\bibinfo  {journal} {Statist. Sci.}\ }\textbf {\bibinfo {volume} {7}},\
  \bibinfo {pages} {457} (\bibinfo {year} {1992})}\BibitemShut {NoStop}%
\bibitem [{\citenamefont {Lewis}(2025)}]{Lewis:2019xzd}%
  \BibitemOpen
  \bibfield  {author} {\bibinfo {author} {\bibfnamefont {A.}~\bibnamefont
  {Lewis}},\ }\href {\doibase 10.1088/1475-7516/2025/08/025} {\bibfield
  {journal} {\bibinfo  {journal} {JCAP}\ }\textbf {\bibinfo {volume} {08}},\
  \bibinfo {pages} {025} (\bibinfo {year} {2025})},\ \Eprint
  {http://arxiv.org/abs/1910.13970} {arXiv:1910.13970 [astro-ph.IM]}
  \BibitemShut {NoStop}%
\bibitem [{\citenamefont {Casertano}\ \emph {et~al.}(2026)\citenamefont
  {Casertano} \emph {et~al.}}]{H0DN:2025lyy}%
  \BibitemOpen
  \bibfield  {author} {\bibinfo {author} {\bibfnamefont {S.}~\bibnamefont
  {Casertano}} \emph {et~al.} (\bibinfo {collaboration} {H0DN}),\ }\href
  {\doibase 10.1051/0004-6361/202557993} {\bibfield  {journal} {\bibinfo
  {journal} {Astron. Astrophys.}\ }\textbf {\bibinfo {volume} {708}},\ \bibinfo
  {pages} {A166} (\bibinfo {year} {2026})},\ \Eprint
  {http://arxiv.org/abs/2510.23823} {arXiv:2510.23823 [astro-ph.CO]}
  \BibitemShut {NoStop}%
\bibitem [{\citenamefont {Frosina}\ and\ \citenamefont
  {Urbano}(2023)}]{Frosina:2023nxu}%
  \BibitemOpen
  \bibfield  {author} {\bibinfo {author} {\bibfnamefont {L.}~\bibnamefont
  {Frosina}}\ and\ \bibinfo {author} {\bibfnamefont {A.}~\bibnamefont
  {Urbano}},\ }\href {\doibase 10.1103/PhysRevD.108.103544} {\bibfield
  {journal} {\bibinfo  {journal} {Phys. Rev. D}\ }\textbf {\bibinfo {volume}
  {108}},\ \bibinfo {pages} {103544} (\bibinfo {year} {2023})},\ \Eprint
  {http://arxiv.org/abs/2308.06915} {arXiv:2308.06915 [astro-ph.CO]}
  \BibitemShut {NoStop}%
\bibitem [{\citenamefont {Garcia-Bellido}\ \emph {et~al.}(2016)\citenamefont
  {Garcia-Bellido}, \citenamefont {Peloso},\ and\ \citenamefont
  {Unal}}]{Garcia-Bellido:2016dkw}%
  \BibitemOpen
  \bibfield  {author} {\bibinfo {author} {\bibfnamefont {J.}~\bibnamefont
  {Garcia-Bellido}}, \bibinfo {author} {\bibfnamefont {M.}~\bibnamefont
  {Peloso}}, \ and\ \bibinfo {author} {\bibfnamefont {C.}~\bibnamefont
  {Unal}},\ }\href {\doibase 10.1088/1475-7516/2016/12/031} {\bibfield
  {journal} {\bibinfo  {journal} {JCAP}\ }\textbf {\bibinfo {volume} {12}},\
  \bibinfo {pages} {031} (\bibinfo {year} {2016})},\ \Eprint
  {http://arxiv.org/abs/1610.03763} {arXiv:1610.03763 [astro-ph.CO]}
  \BibitemShut {NoStop}%
\bibitem [{\citenamefont {Garcia-Bellido}\ \emph {et~al.}(2017)\citenamefont
  {Garcia-Bellido}, \citenamefont {Peloso},\ and\ \citenamefont
  {Unal}}]{Garcia-Bellido:2017aan}%
  \BibitemOpen
  \bibfield  {author} {\bibinfo {author} {\bibfnamefont {J.}~\bibnamefont
  {Garcia-Bellido}}, \bibinfo {author} {\bibfnamefont {M.}~\bibnamefont
  {Peloso}}, \ and\ \bibinfo {author} {\bibfnamefont {C.}~\bibnamefont
  {Unal}},\ }\href {\doibase 10.1088/1475-7516/2017/09/013} {\bibfield
  {journal} {\bibinfo  {journal} {JCAP}\ }\textbf {\bibinfo {volume} {09}},\
  \bibinfo {pages} {013} (\bibinfo {year} {2017})},\ \Eprint
  {http://arxiv.org/abs/1707.02441} {arXiv:1707.02441 [astro-ph.CO]}
  \BibitemShut {NoStop}%
\bibitem [{\citenamefont {Bartolo}\ \emph
  {et~al.}(2019{\natexlab{a}})\citenamefont {Bartolo}, \citenamefont {De~Luca},
  \citenamefont {Franciolini}, \citenamefont {Lewis}, \citenamefont {Peloso},\
  and\ \citenamefont {Riotto}}]{Bartolo:2018evs}%
  \BibitemOpen
  \bibfield  {author} {\bibinfo {author} {\bibfnamefont {N.}~\bibnamefont
  {Bartolo}}, \bibinfo {author} {\bibfnamefont {V.}~\bibnamefont {De~Luca}},
  \bibinfo {author} {\bibfnamefont {G.}~\bibnamefont {Franciolini}}, \bibinfo
  {author} {\bibfnamefont {A.}~\bibnamefont {Lewis}}, \bibinfo {author}
  {\bibfnamefont {M.}~\bibnamefont {Peloso}}, \ and\ \bibinfo {author}
  {\bibfnamefont {A.}~\bibnamefont {Riotto}},\ }\href {\doibase
  10.1103/PhysRevLett.122.211301} {\bibfield  {journal} {\bibinfo  {journal}
  {Phys. Rev. Lett.}\ }\textbf {\bibinfo {volume} {122}},\ \bibinfo {pages}
  {211301} (\bibinfo {year} {2019}{\natexlab{a}})},\ \Eprint
  {http://arxiv.org/abs/1810.12218} {arXiv:1810.12218 [astro-ph.CO]}
  \BibitemShut {NoStop}%
\bibitem [{\citenamefont {Bartolo}\ \emph
  {et~al.}(2019{\natexlab{b}})\citenamefont {Bartolo}, \citenamefont {De~Luca},
  \citenamefont {Franciolini}, \citenamefont {Peloso}, \citenamefont {Racco},\
  and\ \citenamefont {Riotto}}]{Bartolo:2018rku}%
  \BibitemOpen
  \bibfield  {author} {\bibinfo {author} {\bibfnamefont {N.}~\bibnamefont
  {Bartolo}}, \bibinfo {author} {\bibfnamefont {V.}~\bibnamefont {De~Luca}},
  \bibinfo {author} {\bibfnamefont {G.}~\bibnamefont {Franciolini}}, \bibinfo
  {author} {\bibfnamefont {M.}~\bibnamefont {Peloso}}, \bibinfo {author}
  {\bibfnamefont {D.}~\bibnamefont {Racco}}, \ and\ \bibinfo {author}
  {\bibfnamefont {A.}~\bibnamefont {Riotto}},\ }\href {\doibase
  10.1103/PhysRevD.99.103521} {\bibfield  {journal} {\bibinfo  {journal} {Phys.
  Rev. D}\ }\textbf {\bibinfo {volume} {99}},\ \bibinfo {pages} {103521}
  (\bibinfo {year} {2019}{\natexlab{b}})},\ \Eprint
  {http://arxiv.org/abs/1810.12224} {arXiv:1810.12224 [astro-ph.CO]}
  \BibitemShut {NoStop}%
\bibitem [{\citenamefont {Iovino}\ \emph {et~al.}(2025)\citenamefont {Iovino},
  \citenamefont {Perna},\ and\ \citenamefont {Veerm{\"a}e}}]{Iovino:2025cdy}%
  \BibitemOpen
  \bibfield  {author} {\bibinfo {author} {\bibfnamefont {A.}~\bibnamefont
  {Iovino}, \bibfnamefont {Junior.}}, \bibinfo {author} {\bibfnamefont
  {G.}~\bibnamefont {Perna}}, \ and\ \bibinfo {author} {\bibfnamefont
  {H.}~\bibnamefont {Veerm{\"a}e}},\ }\href@noop {} {\  (\bibinfo {year}
  {2025})},\ \Eprint {http://arxiv.org/abs/2512.13648} {arXiv:2512.13648
  [astro-ph.CO]} \BibitemShut {NoStop}%
\bibitem [{\citenamefont {Hong}\ \emph {et~al.}(2026)\citenamefont {Hong},
  \citenamefont {Pi}, \citenamefont {Wang},\ and\ \citenamefont
  {Zhang}}]{Hong:2026rcl}%
  \BibitemOpen
  \bibfield  {author} {\bibinfo {author} {\bibfnamefont {W.}~\bibnamefont
  {Hong}}, \bibinfo {author} {\bibfnamefont {S.}~\bibnamefont {Pi}}, \bibinfo
  {author} {\bibfnamefont {A.}~\bibnamefont {Wang}}, \ and\ \bibinfo {author}
  {\bibfnamefont {Z.}~\bibnamefont {Zhang}},\ }\href@noop {} {\  (\bibinfo
  {year} {2026})},\ \Eprint {http://arxiv.org/abs/2601.05069} {arXiv:2601.05069
  [astro-ph.CO]} \BibitemShut {NoStop}%
\bibitem [{\citenamefont {Blas}\ \emph {et~al.}(2026)\citenamefont {Blas},
  \citenamefont {Foster}, \citenamefont {Gouttenoire}, \citenamefont {Iovino},
  \citenamefont {Musco}, \citenamefont {Trifinopoulos},\ and\ \citenamefont
  {Vanvlasselaer}}]{Blas:2026xws}%
  \BibitemOpen
  \bibfield  {author} {\bibinfo {author} {\bibfnamefont {D.}~\bibnamefont
  {Blas}}, \bibinfo {author} {\bibfnamefont {J.~W.}\ \bibnamefont {Foster}},
  \bibinfo {author} {\bibfnamefont {Y.}~\bibnamefont {Gouttenoire}}, \bibinfo
  {author} {\bibfnamefont {A.~J.}\ \bibnamefont {Iovino}}, \bibinfo {author}
  {\bibfnamefont {I.}~\bibnamefont {Musco}}, \bibinfo {author} {\bibfnamefont
  {S.}~\bibnamefont {Trifinopoulos}}, \ and\ \bibinfo {author} {\bibfnamefont
  {M.}~\bibnamefont {Vanvlasselaer}},\ }\href@noop {} {\  (\bibinfo {year}
  {2026})},\ \Eprint {http://arxiv.org/abs/2602.12252} {arXiv:2602.12252
  [astro-ph.CO]} \BibitemShut {NoStop}%
\bibitem [{\citenamefont {Foster}\ \emph {et~al.}(2025)\citenamefont {Foster},
  \citenamefont {Blas}, \citenamefont {Bourgoin}, \citenamefont {Hees},
  \citenamefont {Herrero-Valea}, \citenamefont {Jenkins},\ and\ \citenamefont
  {Xue}}]{Foster:2025nzf}%
  \BibitemOpen
  \bibfield  {author} {\bibinfo {author} {\bibfnamefont {J.~W.}\ \bibnamefont
  {Foster}}, \bibinfo {author} {\bibfnamefont {D.}~\bibnamefont {Blas}},
  \bibinfo {author} {\bibfnamefont {A.}~\bibnamefont {Bourgoin}}, \bibinfo
  {author} {\bibfnamefont {A.}~\bibnamefont {Hees}}, \bibinfo {author}
  {\bibfnamefont {M.}~\bibnamefont {Herrero-Valea}}, \bibinfo {author}
  {\bibfnamefont {A.~C.}\ \bibnamefont {Jenkins}}, \ and\ \bibinfo {author}
  {\bibfnamefont {X.}~\bibnamefont {Xue}},\ }\href@noop {} {\  (\bibinfo {year}
  {2025})},\ \Eprint {http://arxiv.org/abs/2504.15334} {arXiv:2504.15334
  [astro-ph.CO]} \BibitemShut {NoStop}%
\bibitem [{\citenamefont {Alcock}\ \emph {et~al.}(2000)\citenamefont {Alcock}
  \emph {et~al.}}]{MACHO:2000qbb}%
  \BibitemOpen
  \bibfield  {author} {\bibinfo {author} {\bibfnamefont {C.}~\bibnamefont
  {Alcock}} \emph {et~al.} (\bibinfo {collaboration} {MACHO}),\ }\href
  {\doibase 10.1086/309512} {\bibfield  {journal} {\bibinfo  {journal}
  {Astrophys. J.}\ }\textbf {\bibinfo {volume} {542}},\ \bibinfo {pages} {281}
  (\bibinfo {year} {2000})},\ \Eprint {http://arxiv.org/abs/astro-ph/0001272}
  {arXiv:astro-ph/0001272} \BibitemShut {NoStop}%
\bibitem [{\citenamefont {Jedamzik}(1998)}]{Jedamzik:1998hc}%
  \BibitemOpen
  \bibfield  {author} {\bibinfo {author} {\bibfnamefont {K.}~\bibnamefont
  {Jedamzik}},\ }\href {\doibase 10.1016/S0370-1573(98)00067-2} {\bibfield
  {journal} {\bibinfo  {journal} {Phys. Rept.}\ }\textbf {\bibinfo {volume}
  {307}},\ \bibinfo {pages} {155} (\bibinfo {year} {1998})},\ \Eprint
  {http://arxiv.org/abs/astro-ph/9805147} {arXiv:astro-ph/9805147} \BibitemShut
  {NoStop}%
\bibitem [{\citenamefont {Niikura}\ \emph
  {et~al.}(2019{\natexlab{a}})\citenamefont {Niikura}, \citenamefont {Takada},
  \citenamefont {Yokoyama}, \citenamefont {Sumi},\ and\ \citenamefont
  {Masaki}}]{Niikura:2019kqi}%
  \BibitemOpen
  \bibfield  {author} {\bibinfo {author} {\bibfnamefont {H.}~\bibnamefont
  {Niikura}}, \bibinfo {author} {\bibfnamefont {M.}~\bibnamefont {Takada}},
  \bibinfo {author} {\bibfnamefont {S.}~\bibnamefont {Yokoyama}}, \bibinfo
  {author} {\bibfnamefont {T.}~\bibnamefont {Sumi}}, \ and\ \bibinfo {author}
  {\bibfnamefont {S.}~\bibnamefont {Masaki}},\ }\href {\doibase
  10.1103/PhysRevD.99.083503} {\bibfield  {journal} {\bibinfo  {journal} {Phys.
  Rev. D}\ }\textbf {\bibinfo {volume} {99}},\ \bibinfo {pages} {083503}
  (\bibinfo {year} {2019}{\natexlab{a}})},\ \Eprint
  {http://arxiv.org/abs/1901.07120} {arXiv:1901.07120 [astro-ph.CO]}
  \BibitemShut {NoStop}%
\bibitem [{\citenamefont {Niikura}\ \emph
  {et~al.}(2019{\natexlab{b}})\citenamefont {Niikura} \emph
  {et~al.}}]{Niikura:2017zjd}%
  \BibitemOpen
  \bibfield  {author} {\bibinfo {author} {\bibfnamefont {H.}~\bibnamefont
  {Niikura}} \emph {et~al.},\ }\href {\doibase 10.1038/s41550-019-0723-1}
  {\bibfield  {journal} {\bibinfo  {journal} {Nature Astron.}\ }\textbf
  {\bibinfo {volume} {3}},\ \bibinfo {pages} {524} (\bibinfo {year}
  {2019}{\natexlab{b}})},\ \Eprint {http://arxiv.org/abs/1701.02151}
  {arXiv:1701.02151 [astro-ph.CO]} \BibitemShut {NoStop}%
\bibitem [{\citenamefont {Hawkins}(2020)}]{Hawkins:2020zie}%
  \BibitemOpen
  \bibfield  {author} {\bibinfo {author} {\bibfnamefont {M.~R.~S.}\
  \bibnamefont {Hawkins}},\ }\href {\doibase 10.1051/0004-6361/201936462}
  {\bibfield  {journal} {\bibinfo  {journal} {Astron. Astrophys.}\ }\textbf
  {\bibinfo {volume} {633}},\ \bibinfo {pages} {A107} (\bibinfo {year}
  {2020})},\ \Eprint {http://arxiv.org/abs/2001.07633} {arXiv:2001.07633
  [astro-ph.GA]} \BibitemShut {NoStop}%
\bibitem [{\citenamefont {Gorton}\ and\ \citenamefont
  {Green}(2022)}]{Gorton:2022fyb}%
  \BibitemOpen
  \bibfield  {author} {\bibinfo {author} {\bibfnamefont {M.}~\bibnamefont
  {Gorton}}\ and\ \bibinfo {author} {\bibfnamefont {A.~M.}\ \bibnamefont
  {Green}},\ }\href {\doibase 10.1088/1475-7516/2022/08/035} {\bibfield
  {journal} {\bibinfo  {journal} {JCAP}\ }\textbf {\bibinfo {volume} {08}},\
  \bibinfo {pages} {035} (\bibinfo {year} {2022})},\ \Eprint
  {http://arxiv.org/abs/2203.04209} {arXiv:2203.04209 [astro-ph.CO]}
  \BibitemShut {NoStop}%
\bibitem [{\citenamefont {Hawkins}(2022)}]{Hawkins:2022vqo}%
  \BibitemOpen
  \bibfield  {author} {\bibinfo {author} {\bibfnamefont {M.~R.~S.}\
  \bibnamefont {Hawkins}},\ }\href {\doibase 10.1093/mnras/stac863} {\bibfield
  {journal} {\bibinfo  {journal} {Mon. Not. Roy. Astron. Soc.}\ }\textbf
  {\bibinfo {volume} {512}},\ \bibinfo {pages} {5706} (\bibinfo {year}
  {2022})},\ \Eprint {http://arxiv.org/abs/2204.09143} {arXiv:2204.09143
  [astro-ph.CO]} \BibitemShut {NoStop}%
\bibitem [{\citenamefont {Hawkins}\ and\ \citenamefont
  {Garc{\'\i}a-Bellido}(2025)}]{Hawkins:2025mlo}%
  \BibitemOpen
  \bibfield  {author} {\bibinfo {author} {\bibfnamefont {M.~R.~S.}\
  \bibnamefont {Hawkins}}\ and\ \bibinfo {author} {\bibfnamefont
  {J.}~\bibnamefont {Garc{\'\i}a-Bellido}},\ }\href@noop {} {\  (\bibinfo
  {year} {2025})},\ \Eprint {http://arxiv.org/abs/2509.05400} {arXiv:2509.05400
  [astro-ph.GA]} \BibitemShut {NoStop}%
\bibitem [{\citenamefont {Fairbairn}\ \emph {et~al.}(2025)\citenamefont
  {Fairbairn}, \citenamefont {Heurtier},\ and\ \citenamefont
  {Olea-Romacho}}]{Fairbairn:2025fko}%
  \BibitemOpen
  \bibfield  {author} {\bibinfo {author} {\bibfnamefont {M.}~\bibnamefont
  {Fairbairn}}, \bibinfo {author} {\bibfnamefont {L.}~\bibnamefont {Heurtier}},
  \ and\ \bibinfo {author} {\bibfnamefont {M.~O.}\ \bibnamefont
  {Olea-Romacho}},\ }\href@noop {} {\  (\bibinfo {year} {2025})},\ \Eprint
  {http://arxiv.org/abs/2511.01612} {arXiv:2511.01612 [astro-ph.CO]}
  \BibitemShut {NoStop}%
\bibitem [{\citenamefont {Linde}\ and\ \citenamefont
  {Mukhanov}(1997)}]{Linde:1996gt}%
  \BibitemOpen
  \bibfield  {author} {\bibinfo {author} {\bibfnamefont {A.~D.}\ \bibnamefont
  {Linde}}\ and\ \bibinfo {author} {\bibfnamefont {V.~F.}\ \bibnamefont
  {Mukhanov}},\ }\href {\doibase 10.1103/PhysRevD.56.R535} {\bibfield
  {journal} {\bibinfo  {journal} {Phys. Rev. D}\ }\textbf {\bibinfo {volume}
  {56}},\ \bibinfo {pages} {R535} (\bibinfo {year} {1997})},\ \Eprint
  {http://arxiv.org/abs/astro-ph/9610219} {arXiv:astro-ph/9610219} \BibitemShut
  {NoStop}%
\bibitem [{\citenamefont {Enqvist}\ and\ \citenamefont
  {Sloth}(2002)}]{Enqvist:2001zp}%
  \BibitemOpen
  \bibfield  {author} {\bibinfo {author} {\bibfnamefont {K.}~\bibnamefont
  {Enqvist}}\ and\ \bibinfo {author} {\bibfnamefont {M.~S.}\ \bibnamefont
  {Sloth}},\ }\href {\doibase 10.1016/S0550-3213(02)00043-3} {\bibfield
  {journal} {\bibinfo  {journal} {Nucl. Phys. B}\ }\textbf {\bibinfo {volume}
  {626}},\ \bibinfo {pages} {395} (\bibinfo {year} {2002})},\ \Eprint
  {http://arxiv.org/abs/hep-ph/0109214} {arXiv:hep-ph/0109214} \BibitemShut
  {NoStop}%
\bibitem [{\citenamefont {Lyth}\ and\ \citenamefont
  {Wands}(2002)}]{Lyth:2001nq}%
  \BibitemOpen
  \bibfield  {author} {\bibinfo {author} {\bibfnamefont {D.~H.}\ \bibnamefont
  {Lyth}}\ and\ \bibinfo {author} {\bibfnamefont {D.}~\bibnamefont {Wands}},\
  }\href {\doibase 10.1016/S0370-2693(01)01366-1} {\bibfield  {journal}
  {\bibinfo  {journal} {Phys. Lett. B}\ }\textbf {\bibinfo {volume} {524}},\
  \bibinfo {pages} {5} (\bibinfo {year} {2002})},\ \Eprint
  {http://arxiv.org/abs/hep-ph/0110002} {arXiv:hep-ph/0110002} \BibitemShut
  {NoStop}%
\bibitem [{\citenamefont {Moroi}\ and\ \citenamefont
  {Takahashi}(2001)}]{Moroi:2001ct}%
  \BibitemOpen
  \bibfield  {author} {\bibinfo {author} {\bibfnamefont {T.}~\bibnamefont
  {Moroi}}\ and\ \bibinfo {author} {\bibfnamefont {T.}~\bibnamefont
  {Takahashi}},\ }\href {\doibase 10.1016/S0370-2693(01)01295-3} {\bibfield
  {journal} {\bibinfo  {journal} {Phys. Lett. B}\ }\textbf {\bibinfo {volume}
  {522}},\ \bibinfo {pages} {215} (\bibinfo {year} {2001})},\ \bibinfo {note}
  {[Erratum: Phys.Lett.B 539, 303--303 (2002)]},\ \Eprint
  {http://arxiv.org/abs/hep-ph/0110096} {arXiv:hep-ph/0110096} \BibitemShut
  {NoStop}%
\bibitem [{\citenamefont {Allegrini}\ \emph
  {et~al.}(2026{\natexlab{b}})\citenamefont {Allegrini}, \citenamefont
  {Iovino}, \citenamefont {Perna},\ and\ \citenamefont
  {Veerm{\"a}e}}]{Allegrini:2026jqt}%
  \BibitemOpen
  \bibfield  {author} {\bibinfo {author} {\bibfnamefont {S.}~\bibnamefont
  {Allegrini}}, \bibinfo {author} {\bibfnamefont {A.~J.}\ \bibnamefont
  {Iovino}}, \bibinfo {author} {\bibfnamefont {G.}~\bibnamefont {Perna}}, \
  and\ \bibinfo {author} {\bibfnamefont {H.}~\bibnamefont {Veerm{\"a}e}},\
  }\href@noop {} {\  (\bibinfo {year} {2026}{\natexlab{b}})},\ \Eprint
  {http://arxiv.org/abs/2606.28296} {arXiv:2606.28296 [astro-ph.CO]}
  \BibitemShut {NoStop}%
\bibitem [{\citenamefont {Colg{\'a}in}\ \emph {et~al.}(2025)\citenamefont
  {Colg{\'a}in}, \citenamefont {Pourojaghi},\ and\ \citenamefont
  {Sheikh-Jabbari}}]{Colgain:2024ksa}%
  \BibitemOpen
  \bibfield  {author} {\bibinfo {author} {\bibfnamefont {E.~{\'O}.}\
  \bibnamefont {Colg{\'a}in}}, \bibinfo {author} {\bibfnamefont
  {S.}~\bibnamefont {Pourojaghi}}, \ and\ \bibinfo {author} {\bibfnamefont
  {M.~M.}\ \bibnamefont {Sheikh-Jabbari}},\ }\href {\doibase
  10.1140/epjc/s10052-025-13995-4} {\bibfield  {journal} {\bibinfo  {journal}
  {Eur. Phys. J. C}\ }\textbf {\bibinfo {volume} {85}},\ \bibinfo {pages} {286}
  (\bibinfo {year} {2025})},\ \Eprint {http://arxiv.org/abs/2406.06389}
  {arXiv:2406.06389 [astro-ph.CO]} \BibitemShut {NoStop}%
\bibitem [{\citenamefont {Weiner}(2026)}]{Weiner:2026sfm}%
  \BibitemOpen
  \bibfield  {author} {\bibinfo {author} {\bibfnamefont {Z.~J.}\ \bibnamefont
  {Weiner}},\ }\href@noop {} {\  (\bibinfo {year} {2026})},\ \Eprint
  {http://arxiv.org/abs/2603.18131} {arXiv:2603.18131 [astro-ph.CO]}
  \BibitemShut {NoStop}%
\bibitem [{\citenamefont {Shlivko}\ and\ \citenamefont
  {Poulin}(2026)}]{Shlivko:2026jxa}%
  \BibitemOpen
  \bibfield  {author} {\bibinfo {author} {\bibfnamefont {D.}~\bibnamefont
  {Shlivko}}\ and\ \bibinfo {author} {\bibfnamefont {V.}~\bibnamefont
  {Poulin}},\ }\href@noop {} {\  (\bibinfo {year} {2026})},\ \Eprint
  {http://arxiv.org/abs/2603.22406} {arXiv:2603.22406 [astro-ph.CO]}
  \BibitemShut {NoStop}%
\bibitem [{\citenamefont {Bostan}\ and\ \citenamefont
  {Roy~Choudhury}(2024)}]{Bostan:2023ped}%
  \BibitemOpen
  \bibfield  {author} {\bibinfo {author} {\bibfnamefont {N.}~\bibnamefont
  {Bostan}}\ and\ \bibinfo {author} {\bibfnamefont {S.}~\bibnamefont
  {Roy~Choudhury}},\ }\href {\doibase 10.1088/1475-7516/2024/07/032} {\bibfield
   {journal} {\bibinfo  {journal} {JCAP}\ }\textbf {\bibinfo {volume} {07}},\
  \bibinfo {pages} {032} (\bibinfo {year} {2024})},\ \Eprint
  {http://arxiv.org/abs/2310.01491} {arXiv:2310.01491 [astro-ph.CO]}
  \BibitemShut {NoStop}%
\bibitem [{\citenamefont {Ade}\ \emph {et~al.}(2019)\citenamefont {Ade} \emph
  {et~al.}}]{SimonsObservatory:2018koc}%
  \BibitemOpen
  \bibfield  {author} {\bibinfo {author} {\bibfnamefont {P.}~\bibnamefont
  {Ade}} \emph {et~al.} (\bibinfo {collaboration} {Simons Observatory}),\
  }\href {\doibase 10.1088/1475-7516/2019/02/056} {\bibfield  {journal}
  {\bibinfo  {journal} {JCAP}\ }\textbf {\bibinfo {volume} {02}},\ \bibinfo
  {pages} {056} (\bibinfo {year} {2019})},\ \Eprint
  {http://arxiv.org/abs/1808.07445} {arXiv:1808.07445 [astro-ph.CO]}
  \BibitemShut {NoStop}%
\bibitem [{\citenamefont {Abitbol}\ \emph {et~al.}(2019)\citenamefont {Abitbol}
  \emph {et~al.}}]{SimonsObservatory:2019qwx}%
  \BibitemOpen
  \bibfield  {author} {\bibinfo {author} {\bibfnamefont {M.~H.}\ \bibnamefont
  {Abitbol}} \emph {et~al.} (\bibinfo {collaboration} {Simons Observatory}),\
  }\href@noop {} {\bibfield  {journal} {\bibinfo  {journal} {Bull. Am. Astron.
  Soc.}\ }\textbf {\bibinfo {volume} {51}},\ \bibinfo {pages} {147} (\bibinfo
  {year} {2019})},\ \Eprint {http://arxiv.org/abs/1907.08284} {arXiv:1907.08284
  [astro-ph.IM]} \BibitemShut {NoStop}%
\end{thebibliography}%

\end{document}